\documentclass[twocolumn,appendixfloats,trackchanges]{aastex631}

\usepackage{natbib}
\usepackage{multirow}
\usepackage{amssymb}
\usepackage{amsmath}
\usepackage{verbatim}
\usepackage{xspace}
\usepackage{listings}

\newcommand{\fiveyr}{NG5}
\newcommand{\nineyr}{NG9}
\newcommand{\elevenyr}{NG11}
\newcommand{\twelveyr}{NG12}
\newcommand{\twelvewb}{NG12WB}
\newcommand{\fifteenyr}{NG15}

\newcommand{\nbin}{n_{\rm bin}}

\newcommand{\equad}{\rm EQUAD\xspace}
\newcommand{\efac}{\rm EFAC\xspace}
\newcommand{\ecorr}{\rm ECORR\xspace}
\newcommand{\eq}{\mathcal{Q}}
\newcommand{\ef}{\mathcal{F}}
\newcommand{\ec}{\mathcal{J}}

\newcommand{\tempo}{{\tt Tempo}}
\newcommand{\tempotwo}{{\tt Tempo2}}
\newcommand{\pint}{{\tt PINT}}
\newcommand{\enterprise}{{\tt ENTERPRISE}}
\newcommand{\pintpal}{{\tt PINT\_Pal}\xspace}

\newcommand{\noisepaper}{NG15detchar}
\newcommand{\timingpaper}{NG15}
\newcommand{\gwbpaper}{NG15gwb}

\NewPageAfterKeywords

\begin{document}

\title{The NANOGrav 15-year Data Set: Observations and Timing of 68 Millisecond Pulsars}
\shorttitle{The NANOGrav 15-year Data Set}
\shortauthors{The NANOGrav Collaboration}

\author[0000-0001-5134-3925]{Gabriella Agazie}
\affiliation{Center for Gravitation, Cosmology and Astrophysics, Department of Physics, University of Wisconsin-Milwaukee,\\ P.O. Box 413, Milwaukee, WI 53201, USA}
\author[0000-0003-2449-5426]{Md Faisal Alam}
\affiliation{Department of Physics and Astronomy, Franklin \& Marshall College, P.O. Box 3003, Lancaster, PA 17604, USA}
\affiliation{Department of Physics, 1110 West Green Street, Urbana, IL 61801-3003, USA}
\author[0000-0002-8935-9882]{Akash Anumarlapudi}
\affiliation{Center for Gravitation, Cosmology and Astrophysics, Department of Physics, University of Wisconsin-Milwaukee,\\ P.O. Box 413, Milwaukee, WI 53201, USA}
\author[0000-0003-0638-3340]{Anne M. Archibald}
\affiliation{Newcastle University, NE1 7RU, UK}
\author{Zaven Arzoumanian}
\affiliation{X-Ray Astrophysics Laboratory, NASA Goddard Space Flight Center, Code 662, Greenbelt, MD 20771, USA}
\author[0000-0003-2745-753X]{Paul T. Baker}
\affiliation{Department of Physics and Astronomy, Widener University, One University Place, Chester, PA 19013, USA}
\author[0000-0002-2183-1087]{Laura Blecha}
\affiliation{Physics Department, University of Florida, Gainesville, FL 32611, USA}
\author{Victoria Bonidie}
\affiliation{Department of Physics and Astronomy, Franklin \& Marshall College, P.O. Box 3003, Lancaster, PA 17604, USA}
\affiliation{Department of Physics and Astronomy, University of Pittsburgh, 3941 O'Hara Street, Pittsburgh, PA 15260, USA}
\affiliation{Pittsburgh Particle Physics, Astrophysics, and Cosmology Center (PITT PACC), University of Pittsburgh, Pittsburgh, PA 15260, USA}
\author[0000-0001-6341-7178]{Adam Brazier}
\affiliation{Cornell Center for Astrophysics and Planetary Science and Department of Astronomy, Cornell University, Ithaca, NY 14853, USA}
\affiliation{Cornell Center for Advanced Computing, Cornell University, Ithaca, NY 14853, USA}
\author[0000-0003-3053-6538]{Paul R. Brook}
\affiliation{Institute for Gravitational Wave Astronomy and School of Physics and Astronomy, University of Birmingham, Edgbaston, Birmingham B15 2TT, UK}
% ...

\begin{abstract}
We present observations and timing analyses of 68 millisecond pulsars (MSPs) comprising the 15-year data set of the North American Nanohertz Observatory for Gravitational Waves (NANOGrav).
NANOGrav is a pulsar timing array (PTA) experiment that is sensitive to low-frequency gravitational waves.
This is NANOGrav's fifth public data release, including both ``narrowband'' and ``wideband'' time-of-arrival (TOA) measurements and corresponding pulsar timing models.
We have added 21 MSPs and extended our timing baselines by three years, now spanning nearly 16~years for some of our sources.
The data were collected using the Arecibo Observatory, the Green Bank Telescope, and the Very Large Array between frequencies of 327~MHz and 3~GHz, with most sources observed approximately monthly.
A number of notable methodological and procedural changes were made compared to our previous data sets. These improve the overall quality of the TOA data set and are part of the transition to new pulsar timing and PTA analysis software packages.
For the first time, our data products are accompanied by a full suite of software to reproduce data reduction, analysis, and results.
Our timing models include a variety of newly detected astrometric and binary pulsar parameters, including several significant improvements to pulsar mass constraints. We find that the time series of 23 pulsars contain detectable levels of red noise, 10 of which are new measurements. In this data set, we find evidence for a stochastic gravitational-wave background.
\end{abstract}


\section{Introduction}
\label{sec:intro}
Pulsar timing arrays (PTAs) use high-precision timing of pulsars to attempt to directly detect and study nHz-frequency gravitational waves (GWs). The idea was first proposed in the 1970s \citep{sazhin78,detweiler79}, but it was not until the discovery of millisecond pulsars \citep[MSPs;][]{bkh+82,acr+82}, and the possibility of long-term microsecond-level timing with many MSPs that the idea became practical \citep[e.g.,][]{fb90}. For recent reviews on the technique and the science that can result, see \citet{hd17,Tiburzi2018,taylor2021,vob21}.

PTA data are meticulously acquired for dozens of MSPs on cadences of days to months using the world's largest radio telescopes. Sophisticated digital hardware synchronously averages the faint pulsar signals at multiple radio frequencies from a few hundred MHz to several GHz. After calculation of times of arrival (TOAs), PTAs regularly analyze these data and release them to the public.

The North American Nanohertz Observatory for Gravitational Waves \citep[NANOGrav;][]{Ransom2019} was formed in 2007 and has previously published four data releases (see below). Other PTA collaborations include the European PTA \citep[EPTA;][]{ccg+21}, the Parkes PTA \citep[PPTA;][]{rsc+21}, and the Indian PTA \citep[InPTA;][]{inpta22}. All four collaborations comprise the International PTA (IPTA) and earlier versions of NANOGrav,  EPTA, and  PPTA data sets have been combined into IPTA data releases  known as IPTA DR1 \citep{IPTA2010} and IPTA DR2 \citep{IPTADR2}. The IPTA continues to grow, and soon the InPTA and MeerKAT PTA project will  begin to contribute data to IPTA data sets \citep{inpta22,meerkat2023}.  

This paper describes the latest release of NANOGrav data, the ``15-year Data Set,'' which we have collected over more than 15\,years using the Arecibo Observatory, the Green Bank Telescope (GBT), and the Very Large Array (VLA). The data and analysis procedures we describe are based on and extend those from our earlier releases, which are known as our 5-year \citep[][herein \fiveyr]{Demorest2013}, 9-year \citep[][herein \nineyr]{Arzoumanian2015b}, 11-year \citep[][herein \elevenyr]{Arzoumanian2018a}, and 12.5-year ``narrowband'' \citep[][herein \twelveyr]{Alam21} and 12.5-year ``wideband'' data sets \citep[][herein \twelvewb]{alam21wb}. We will similarly refer to the present 15-year data set as \fifteenyr.

The data release reported on in this paper is an improvement upon previous \twelveyr\ and \twelvewb\ data sets in several ways. It includes 21 additional MSPs (for a total of 68~MSPs)  and an additional $\sim$2.9\,years of timing baseline for the 47~continuing MSPs. We have used a new Python-based timing pipeline, utilizing Jupyter notebooks, that relies on \pint\footnote{\url{https://github.com/nanograv/PINT}} as the primary timing software \citep{2021PINT}. We are concurrently releasing both the traditional ``narrowband'' TOAs derived from many subbands of our radio observing bands, and ``wideband'' TOAs derived using the methodology of \citet{PDR14} and \citet{Pennucci19}. We have added VLA data on six pulsars, one of which is a new addition to the PTA, and have included all of the Arecibo data available up until we were no longer able to use the telescope in 2020 August, four months before its  tragic collapse in 2020 December. The data included in this release will also be included in the third data release (DR3) of the IPTA.
 
This work is presented alongside a GW analysis searching for a GWB in the 15-yr data set \citep[][hereafter \gwbpaper]{NG15gwb}, detailed noise analysis of our PTA detector \citep[][\noisepaper]{NG15detchar}, and astrophysical interpretation of the GWB results \citep{NG15interp, NG15cosmo}. NANOGrav's most recent GW results for the \twelveyr\ data set can be found in \citet{arzoumanian2020} for the stochastic gravitational wave background (GWB), \citet{arzoumanian2021} for non-Einsteinian polarization modes in the GWB, \citet{arzoumanian2021b} for constraints on cosmological phase transitions, and \citet{arzoumanian2023} for GWs from continuous wave sources.

The plan for this paper is as follows.
In Sections~\ref{sec:obs} and \ref{sec:datared}, we describe the observations and data reduction, respectively.
In Section~\ref{sec:timing}, we describe timing models fit to the TOAs
for each pulsar, including both deterministic astrophysical
phenomena and stochastic noise terms.
In Section~\ref{sec:compare}, we compare timing models from this
data set with those from \twelveyr\ and report on any
newly-significant astrometric, binary, and noise parameters.
In Section~\ref{sec:newpsr}, we acknowledge pulsar surveys that have
discovered new MSPs included in this data set and we present flux density 
measurements for these new pulsars at two or more radio frequencies.
In Section~\ref{sec:conclusion}, we summarize the work.

The NANOGrav 15-year data set files include narrowband
and wideband TOAs developed in the present paper, parameterized timing
models for all pulsars for each of the TOA sets, configuration files
used to run our timing analysis notebooks, and support files
such as telescope clock offset measurements.  The data set presented here is
preserved on Zenodo at doi:\url{10.5281/zenodo.7967585}.\footnote{All of NANOGrav's data sets
are available at \url{http://data.nanograv.org}, including the 15-yr data set presented here.
Raw telescope data products are also available from the same website. The 15-yr data
set has been preserved on Zenodo at doi:\url{10.5281/zenodo.7967585}.}
Raw telescope data
products are also available from the same website, as is code used to
do all of the analysis described here. A living repository of \pint-based
timing analysis software born out of this work can be found at
\url{https://github.com/nanograv/pint_pal}.

\section{Observations}
\label{sec:obs}
The NANOGrav 15-year data set contains data from the 100-m GBT, the 305-m telescope at the Arecibo Observatory (Arecibo) prior to its cable failure and eventual collapse, and the 27 25-m antennae of the VLA. The procedures we used are nearly identical to those in \nineyr, \elevenyr, and \twelveyr. As in \twelveyr, we have generated and analyzed both narrowband and wideband TOAs; we present both narrowband (NB) and wideband (WB) versions of the data set in this work. While the fundamental data reduction and timing procedures have not changed significantly from our previous releases, for this data set substantial effort was put into software development, with a major goal of improving versioning and reproducibility of results.  Our new timing pipeline streamlines the iterative nature of our analyses and allows documentation of changes made to the software version, TOAs, timing solution, or noise model. The pipeline will be discussed further in Section~\ref{sec:timing}. Here we describe the observations and data reduction used to generate the TOAs in this data set.

\subsection{Data Collection \label{sec:datacollection}}

\begin{figure*}
    \centering
    \includegraphics[scale=0.84]{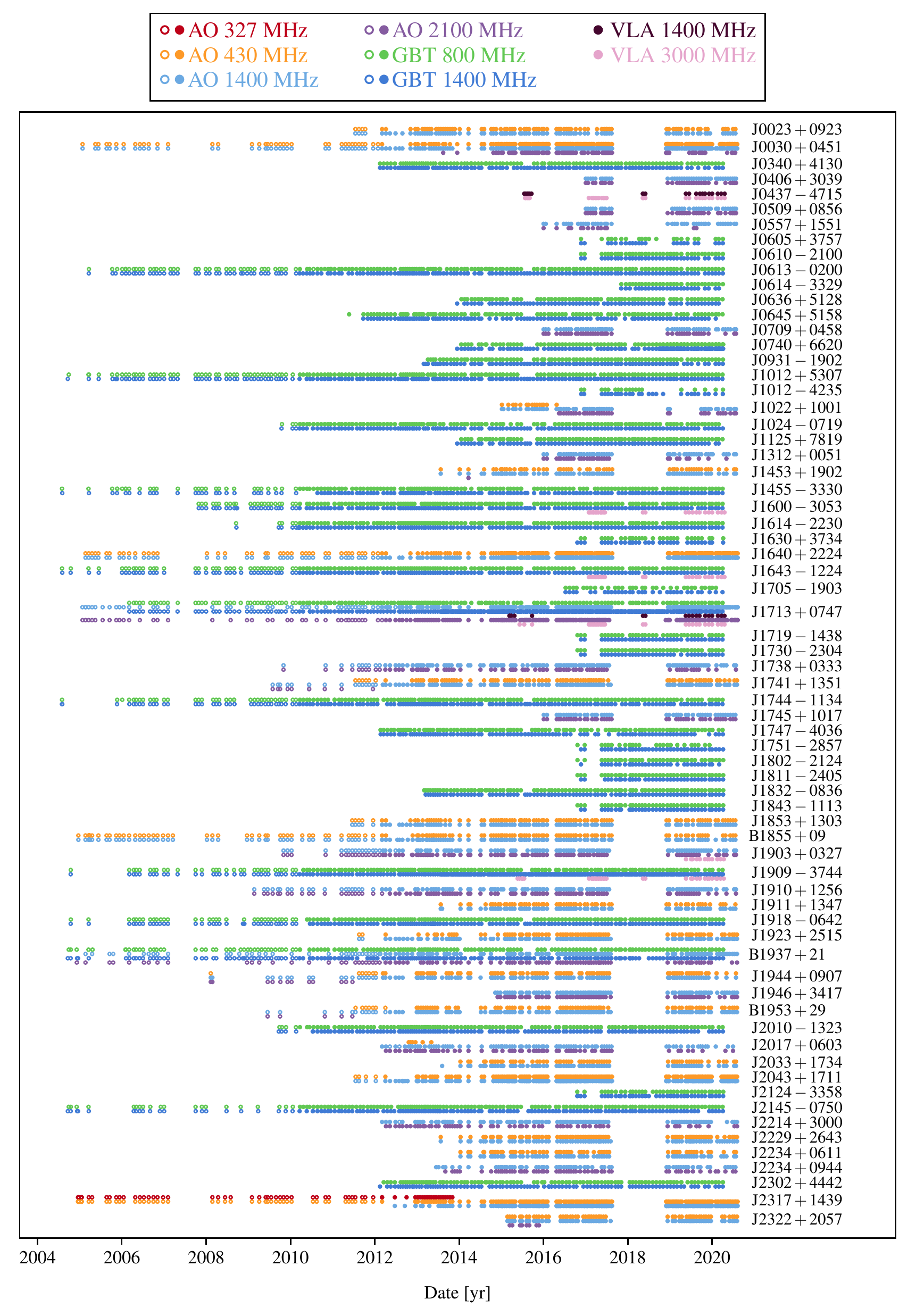}
    \caption{Epochs of all TOAs in the data set. The observatory and observing frequency are indicated by color: Arecibo observations are red (327\,MHz), orange (430\,MHz), light blue (1.4\,GHz), and purple (2.1\,GHz); GBT observations are green (800\,MHz) and dark blue (1.4\,GHz); and VLA observations are brown (1.4\,GHz) and pink (3\,GHz). The Arecibo and GBT data acquisition systems are indicated by symbols: open circles are ASP or GASP, and closed circles are PUPPI or GUPPI. Only a single backend (YUPPI) was used at the VLA.}
    \label{fig:epochs}
\end{figure*}

The present data set contains timing data and solutions for 68 MSPs, 21 more  than were included in \twelveyr. A summary of the pulsars and their timing baselines for this data set is given in Figure~\ref{fig:epochs}. This large increase in the number of MSPs in this data set results from two factors: first, NANOGrav tested the suitability of many newly discovered MSPs and included high-impact PTA sources; second, NANOGrav revisited testing previously unsuitable (known) MSPs with new instrumentation for inclusion in the PTA. The MSPs with suitable timing precision ($\lesssim1$\,$\mu$s scatter in the daily-averaged residuals over 2--3 test observations) were incorporated into the regular NANOGrav observing schedule prior to 2018 August, such that they have $>$\,2-yr timing baselines in this data set. Pulsars with shorter data spans are not included here as it is difficult to reliably model red noise over a shorter timing baseline, and they do not contribute significantly to the  goal of low-frequency GW detection.

Data were collected using the GBT, Arecibo, and the VLA. A summary of the observing systems can be found in Table~\ref{tab:observing_systems}. The maximum timing baseline for an individual pulsar in this data release is 16 years.
GBT and VLA data through 2020 April 4 are included in this data set, corresponding to the last date when the GUPPI backend on the GBT was used.  Arecibo data are included through 2020 August 10, the date of the first cable breaking at Arecibo and the end of regular observations using the telescope. Of the 68 MSPs in \fifteenyr, 30 MSPs with declinations $0^{\circ} < \delta < +39^{\circ}$ were solely observed with Arecibo, and 31 MSPs with $\delta > -45^{\circ}$ (and outside the declination range accessible to Arecibo) with the GBT. Two pulsars, PSRs~J1713+0747 and B1937+21, were observed with the GBT, Arecibo, and the VLA; PSRs~J1600$-$3053, J1643$-$1224, and J1909$-$3744 were observed at both the GBT and the VLA; PSR J1903+0327 was observed with Arecibo and the VLA; and PSR~J0437$-$4715 was observed only at the VLA.

We adhered to a roughly monthly cadence in our observations: every $\sim$\,three weeks at Arecibo and four weeks at the GBT and VLA. As described in \twelveyr, with the GBT and Arecibo we additionally performed high-cadence observations (every $\sim$\,five days) of six of the highest timing precision pulsars for increased sensitivity to single sources emitting continuous GWs. There were some interruptions in data-taking, in particular: telescope painting at Arecibo and azimuth track refurbishment at the GBT in 2007, earthquake damage at Arecibo in 2014, and a brief pause in observing at Arecibo following Hurricane Maria in 2017. The transition from test observations to incorporating new MSPs into our regular observing campaign at the GBT looks like a gap in data collection for those pulsars in early 2017. The lack of Arecibo data in 2018 is due to an instrumental issue described further in Section~\ref{sec:badrange}.

Each MSP was observed with at least two widely-separated frequencies at each epoch (with a few exceptions), with the observing frequencies for a given pulsar chosen to maximize its timing precision. At Arecibo, MSPs were observed at 430\,MHz and 1.4\,GHz or at 1.4\,GHz and 2.1\,GHz. Several Arecibo pulsars were observed at all three frequencies (Table~\ref{tab:psrtoastats}) and
PSR~J2317+1439 was originally observed at 327 and 430\,MHz, but was migrated to 430\,MHz and 1.4\,GHz at roughly the midpoint in its timing baseline. At the GBT, all MSPs were observed at 820\,MHz and 1.4\,GHz. The VLA is expected to provide improved sensitivity and frequency coverage in the 2--4\,GHz range, so it was used for 3-GHz observations of five MSPs predicted to give optimal timing precision at these frequencies \citep{Lam18b}: PSRs~J1713+0747, J1600$-$3053, J1643+1224, J1903+0327, and J1909$-$3744. The VLA was additionally used for dual-frequency (1.4 and 3\,GHz) observations of PSR~J0437$-$4715, a very bright MSP that lies just below the declination range accessible to the GBT, and for PSR~J1713$+$0747. The latter is observed at 1.4\,GHz with all three telescopes in order to provide a cross-check for possible instrumental systematics. The exceptions to dual-frequency observations were the high-cadence observations at the GBT, which were carried out at only 1.4\,GHz but with a large bandwidth.

At Arecibo and the GBT, over the course of our experiment we have used two generations of data acquisition instrumentation (i.e., pulsar ``backends''). For the first approximately six years, the ASP and GASP instruments with 64\,MHz bandwidth \citep{Demorest2007} were used to acquire data from Arecibo and the GBT respectively, after which we transitioned to PUPPI (Puerto Rican Ultimate Pulsar Processing Instrument; at Arecibo, in 2012) and GUPPI (Green Bank Ultimate Pulsar Processing Instrument; at the GBT, in 2010). The PUPPI and GUPPI backends processed bandwidths of up to 800\,MHz \citep{DuPlain2008} and provided significantly improved timing compared to ASP and GASP. Simultaneous observations using both generations of instruments were used to measure offsets between the old and new backends (Appendix A of \nineyr), and we continue to include these offsets in our timing models\footnote{These offsets are included in the ASP/GASP lines of the TOA files, after the ``-to'' flag, for pulsars with ASP or GASP data.} as we have in previous data releases. The YUPPI backend\footnote{\url{https://science.nrao.edu/facilities/vla/docs/manuals/oss/performance/pulsar}} was used to acquire all VLA data in this data release. The resulting raw data consist of folded, full-Stokes pulse profiles, with 2048 pulse phase bins, radio frequency resolution of 4\,MHz (ASP/GASP), $\sim$\,1.5\,MHz (GUPPI/PUPPI) or 1\,MHz (YUPPI), and sub-integrations of 1\,s (with PUPPI at 1.4 and 2.1\,GHz) or 10\,s (with all other receiver/back-end pairings). Hereafter, if a description applies to all three of the GUPPI, PUPPI, and YUPPI backends, we will refer to them in shorthand as the ``UPPI backends.''

VLA observations employed the ``phased array'' mode of telescope operation.  In this mode, the voltage data streams of the individual antennae are coherently added, resulting in a single data stream with properties roughly equivalent to those of a single dish of equal total collecting area.  This summed signal is then subdivided into frequency channels, coherently dedispersed, and folded in the same manner as typical single-dish pulsar data.  The final results are data products (folded profiles) that in subsequent processing can be treated identically as single-dish data.  In order to maximize sensitivity of the array sum, radio-wave phase fluctuations due to atmospheric variations across the array must be corrected in real time.  During observations, phase corrections were measured using a bright continuum calibrator, typically within $\sim$5\,deg of the target pulsar, and applied to the data.  This re-phasing process was repeated every 10 minutes.

For all telescopes/backends, at the start of each pulsar observation, a pulsed noise-diode signal was injected in order to later calibrate the pulsar signal amplitude during the data reduction steps. For GUPPI and PUPPI data, the noise signal amplitude was calibrated into equivalent flux density units (i.e., Janskys) approximately monthly using on/off observations of an unpolarized continuum radio source of known flux density. Details about the continuum sources can be found in Section~8 of \twelveyr. For the VLA with YUPPI, this second flux-calibration step was not done, and the resulting data are scaled in units relative to the noise diode power.

\begin{deluxetable*}{@{\extracolsep{0pt}}cccccccccc}[ht!]
\centerwidetable
\tabletypesize{\footnotesize}
\tablewidth{0pt}
\tablecaption{Observing Frequencies and Bandwidths\tablenotemark{\footnotesize{a}}\label{tab:observing_systems}}
\tablehead{\\[-5pt] & \multicolumn{9}{c}{Backends} \\[0pt] \cline{2-10}\\[-13pt] \colhead{} & \multicolumn{4}{c}{ASP/GASP} & & \multicolumn{4}{c}{PUPPI/GUPPI/YUPPI} \\[-2pt]
           \cline{2-5}\cline{7-10}
           \rule{0pt}{1pt}
           Telescope      &                             & Frequency                         &  Usable                    &  $\Delta$DM &&                            & Frequency                         &   Usable                 &  $\Delta$DM \\
           Receiver       &  Data Span\tablenotemark{\tiny{b}} &        Range\tablenotemark{\tiny{c}} & Bandwidth\tablenotemark{\tiny{d}} & Delay\tablenotemark{\tiny{e}} && Data Span\tablenotemark{\tiny{b}} &        Range\tablenotemark{\tiny{c}} & Bandwidth\tablenotemark{\tiny{d}} & Delay\tablenotemark{\tiny{e}} \\
                          &            &     [MHz]        &  [MHz]    & [$\mu$s] &&           &    [MHz]         &   [MHz] & [$\mu$s]
           }
\startdata
\multicolumn{10}{l}{\phn Arecibo} \\[0pt]
\hline
\rule[0pt]{0pt}{11pt}%
  327    &  $2005.0-2012.0$ & $315-339$   &  34  & 2.86 && $2012.2-2020.6$ &  $302-352$  &  50  & 6.00 \\
  430    &  $2005.0-2012.3$ & $422-442$   &  20  & 1.03 && $2012.2-2020.6$ &  $421-445$  &  24  & 1.23 \\
  L-wide &  $2004.9-2012.3$ & $1380-1444$ &  64  & 0.09 && $2012.2-2020.6$ & $1147-1765$ & 600  & 0.91\\
  S-wide &  $2004.9-2012.6$ & $2316-2380$ &  64  & 0.02 && $2012.2-2020.6$ & \tablenotemark{\tiny{\phantom{f}}}$1700-2404$\tablenotemark{\tiny{f}}  &   460  & 0.36 \\[0pt]
\hline
\multicolumn{1}{c}{\rule[-3pt]{0pt}{11pt}GBT} & \multicolumn{9}{c}{} \\[0pt]
\hline
\rule[0pt]{0pt}{11pt}%
  Rcvr\_800 &  $2004.6-2011.0$ & $822-866$   &  64  & 0.30 && $2010.2-2020.3$ &  $722-919$    &   180  & 1.52 \\
  Rcvr1\_2  &  $2004.6-2010.8$ & $1386-1434$ &  64  & 0.07 && $2010.2-2020.3$ &  $1151-1885$  &   640  & 0.98 \\
\hline
\multicolumn{1}{c}{\rule[-3pt]{0pt}{11pt}VLA} & \multicolumn{9}{c}{} \\[0pt]
\hline
\rule[0pt]{0pt}{11pt}%
  1.5GHz &  $\cdots$ & $\cdots$ & $\cdots$ & $\cdots$ && $2015.2-2020.3$ &  $1026-2014$  &  800 & 1.46 \\
  3GHz   &  $\cdots$ & $\cdots$ & $\cdots$ & $\cdots$ && $2015.3-2020.0$ &  $2012-3988$  & 1700 & 0.38
\enddata
\tablenotetext{a}{Table reproduced and modified from \twelvewb, including adding VLA/YUPPI information.\vspace{-0.1in}}
\tablenotetext{b}{Dates of instrument use.  Observation dates of individual pulsars vary; see Figure~\ref{fig:epochs}.\vspace{-0.1in}}
\tablenotetext{c}{Typical values; some observations differed. Some frequencies unusable due to radio frequency interference.\vspace{-0.1in}}
\tablenotetext{d}{Approximate and representative values after excluding narrow subbands with radio frequency interference.\vspace{-0.1in}}
\tablenotetext{e}{Representative dispersive delay between profiles at the extrema frequencies listed in the Frequency Range column induced by a $\Delta$DM$ = 5 \times 10^{-4}$~cm$^{-3}$~pc, which is approximately the median uncertainty across all wideband DM measurements in the data set; for scale, 1~$\mu$s~$\sim$~1~phase~bin for a 2~ms pulsar with our configuration of $\nbin = 2048$.\vspace{-0.1in}}
\tablenotetext{f}{Non-contiguous usable bands at $1700-1880$ and $2050-2404$~MHz.\vspace{-0.1in}}
\end{deluxetable*}

\section{Data Reduction and Times-of-Arrival Generation}
\label{sec:datared}
\subsection{Data Reduction and Procedure \label{sec:datareduction}}

We followed a standardized process to use the raw data files to calculate calibrated profiles and TOAs. Some steps can be iterative in nature; for example, initial timing must be done before an analysis identifying outlying TOAs. With this caveat in mind, we list the steps that we took to produce science-ready pulse profiles and to generate TOAs from those profiles:

\begin{enumerate}
    \item \textit{Exclusion of known bad/corrupted files or time ranges.} From previous data sets, we were already aware of a number of data files that were not suitable for inclusion in the data set due to various instrumental issues. In addition, in newer data contributing only to \fifteenyr, there is a $\sim10$-month time range with unusable Arecibo data (see Section~\ref{sec:badrange}). We excluded all of these files from \fifteenyr.
    \item \textit{Artifact removal.} This step removes artifacts from the interleaved analog-to-digital converter (ADC) scheme used by the UPPI receivers to achieve their wide bandwidths. The artifacts appear as negatively dispersed pulses and impact TOA generation if not removed. Details about these artifacts and their removal are given in Section 2.3.1 and Figure 2 of \twelveyr.
    \item \textit{Radio frequency interference excision and calibration.} We performed the standard radio frequency interference (RFI) excision and calibration procedures detailed in \elevenyr, with the additional step of excising RFI from the calibration files along with the data files as described in Section 2.3.2 of \twelveyr.
    \item \textit{Time- and frequency-averaging.} Details for Arecibo and GBT data are given in Section~3.1 of \nineyr. 
    The UPPI data were frequency-averaged (``scrunched'') into 1.6 to 32\,MHz-bandwidth channels, depending on the receiver. ASP/GASP data were not averaged, and remained at the native 4\,MHz resolution. The cleaned, calibrated profiles were time-averaged into sub-integrations of up to 30\,minutes. For the few pulsars with very short binary periods, we time-averaged into sub-integrations no longer than 2.5\% of the orbital period.
    \item \textit{TOA generation.} This process is described in Section~\ref{sec:toagen}, with further details in \nineyr\ and \elevenyr; details for wideband-specific TOA generation can be found in \twelvewb\ and references therein.
    \item \textit{Outlier analysis of TOA residuals.} The automated process by which we identified and removed outlier TOA residuals is described in Section~\ref{sec:outlier}.
    \item \textit{Further, often iterative, data cleaning.} Further data quality checks and cleaning are described in Section~\ref{sec:dataquality}.
\end{enumerate}

Steps 2--5 were performed using the \texttt{nanopipe}\footnote{\href{https://github.com/demorest/nanopipe}{https://github.com/demorest/nanopipe}, v1.2.4} data processing pipeline \citep{nanopipe}.  This in turn uses the \texttt{PSRCHIVE}\footnote{\href{http://psrchive.sf.net}{http://psrchive.sf.net}, v2021-06-03} pulsar data analysis software package \citep{Hotan04} to perform most of the processing steps outlined here, plus \texttt{PulsePortraiture} for wideband template generation and TOA measurement \citep{pulseportraiture}.  All software package versions and specific data processing settings used in this analysis, along with the resulting TOA data (step 5), were tracked in a git-based version control repository (``\texttt{toagen}'').  A version tag and corresponding git hash linking the TOA to a specific git commit are included in the final TOA lines, allowing a complete description of the TOA provenance to be reconstructed as needed.

\subsection{Time-of-Arrival Generation \label{sec:toagen}}

TOAs are derived from folded pulse-profile data by measuring the time shift of the observed profile relative to a model pulse shape for each pulsar, known as a template; this basic approach has been standard in pulsar timing analyses for decades \citep[e.g.,][]{Taylor92}.  This data release includes both narrowband and wideband TOAs, as did \twelveyr.  In the narrowband approach, a separate TOA is measured for each frequency channel, at the final time and frequency resolution described in Section~\ref{sec:datareduction}, step 4; dispersion measure (DM) is fit to these TOAs along with other timing parameters.  In the wideband approach, a single TOA and DM value are measured for the full receiver band, using a frequency-dependent template (also called a portrait) that accounts for pulse shape evolution \citep[for details, see][]{PDR14}.

All narrowband template profiles used in this data set were regenerated from the UPPI data, using the procedures described in \fiveyr.  All profiles for a given pulsar and receiver band are aligned, weighted by signal-to-noise ratio, and summed to create a final average.  This average profile is then ``denoised'' using wavelet decomposition and thresholding.  The process is iterated several times to converge on the final template.  All narrowband TOAs, including those from ASP and GASP data, were then generated using this set of templates using procedures described in more detail in \nineyr\ and \elevenyr.  As in \twelveyr, we used an improved algorithm to calculate TOA uncertainties, in which we numerically integrate the TOA probability distribution (Equation 12 in Appendix B of \nineyr) to mitigate underestimation of uncertainties for low S/N TOAs.

Wideband template portaits were created following \twelvewb.  Again, all data for a given pulsar and receiver are aligned and summed, while in this case preserving (rather than averaging over) frequency channels.  The final average portrait is then modeled and denoised as described in \citet{Pennucci19} to create a template that preserves profile evolution with frequency.  For the 47 MSPs in \twelveyr, we re-used the previous GUPPI and PUPPI wideband templates from that analysis.  New templates were generated for the YUPPI data and the 21 pulsars new to the present data set.  New wideband TOAs for all data were then generated as described in \twelvewb.

\subsection{Cleaning the Data Set for Improved Data Quality \label{sec:dataquality}} 
In Section 2.5 of \twelveyr, we detailed a number of steps taken to improve the quality of our data set. Many of the same steps were taken in this data set as well, in particular the following data cuts (and corresponding \twelveyr\ sections): S/N cut (2.5.1), bad DM range cut (2.5.2), outlier residual TOA cut (2.5.3, except a different method was used in the present work, as described in Appendix~\ref{sec:outlier} below), epoch $F$-test cut (2.5.7), and orphan data cut (2.5.9). The corrupt calibration cut (\twelveyr\ Section 2.5.5) was not explicitly repeated in this data set, but the files that were cut from the \twelveyr\ were also cut from \fifteenyr. Manual removal of individual TOAs or data files was done similarly as in \twelveyr\ Sections 2.5.4 and 2.5.8, with some differences, so these cuts are described below in Appendix~\ref{sec:badfiletoa}.

Table \ref{tab:toa_flags} summarizes each of the cuts used in \fifteenyr, including the order in which they were applied, and the number of TOAs affected by each cut in the narrowband and wideband data sets respectively. After all TOA cutting was complete, the final narrowband data set had 676,465 TOAs (34.6\% cut) and the wideband data set had 20,290 TOAs (21.6\% cut).

\begin{deluxetable*}{lp{10cm}rr}
\centerwidetable
\tabletypesize{\scriptsize}
\tablewidth{\textwidth}
\tablecolumns{4}
\tablecaption{TOA Removal Flags\label{tab:toa_flags}}
\tablehead{\colhead{Flag} & \colhead{Reason for TOA Removal} & \colhead{$N_{\rm nb}$} & \colhead{$N_{\rm wb}$}}
\startdata
\cutinhead{Initial Cuts (Loading TOAs)}
\texttt{-cut simul} & ASP/GASP TOA taken at the same time as a PUPPI/GUPPI TOA & 6,883 & 648 \\
\texttt{-cut orphaned} & Insignificant data volume ($\leq3$ epochs for a given receiver band; usually test observations) & 2,837 & 13 \\
\texttt{-cut badrange} & Arecibo data affected by malfunctioning local oscilator (MJDs 57984--58447) & 56,658 & 2,036 \\
\texttt{-cut snr} & Profile data used to generate TOA does not meet signal-to-noise ratio threshold ($\mathrm{S/N}_{\rm nb}<8$ and $\mathrm{S/N}_{\rm wb}<25$) & 255,118 & 1,324 \\
\texttt{-cut poorfebe} & Poor quality data from given frontend/backend combination & 512 & 5 \\
\texttt{-cut eclipsing} & For pulsars showing signs of eclipses, TOAs near superior conjunction (within $10-15\%$ of an orbital phase) were automatically excised & 4,551 & 109 \\
\texttt{-cut dmx}\tablenotemark{a} & Ratio of maximum to minimum frequency in an observing epoch (in a single DMX bin) $f_{\rm{max}}/f_{\rm{min}} < 1.1$ & 13,006 & 1,086 \\
Initial Totals & & 339,565 & 5,221 \\
\cutinhead{Automated Outlier Analysis}
\texttt{-cut outlier10} & TOA has outlier probability $p_{i,{\rm out}} > 0.1$ & 5,374 & -- \\
\texttt{-cut maxout} & Entire file removed if a significant percentage ($>8\%$) of TOAs flagged as outliers & 5,072 & -- \\
\texttt{-cut epochdrop} & Entire epoch removed with \texttt{epochalyptica()}, based on an $F$-test $p$-value $< 10^{-6}$ & 6,027 & 188 \\
Auto Totals & & 16,473 & 188 \\
\cutinhead{Manual Cuts (Human Inspection)}
\texttt{-cut badtoa} & Remaining TOAs identified by human inspection and removed & 249 & 1 \\
\texttt{-cut badfile} & Remaining files identified by human inspection and removed & 1,163 & 170 \\
Manual Totals & & 1,412 & 171 \\
Removed TOAs & & 357,450 & 5,580 \\
Remaining TOAs & & 676,465 & 20,290 \\
\enddata
\tablecomments{The flags are listed here in the order in which they were applied in our final pipeline run. A single cut flag is assigned to each TOA that is removed. Narrowband and wideband data sets were analyzed independently.}
\tablenotetext{a}{$f_{\rm{max}}$ and $f_{\rm{min}}$ are TOA reference frequencies in the narrowband data set and are separately calculated for each TOA in the wideband data set.}
\end{deluxetable*}

\section{Timing Analysis}
\label{sec:timing}
From the outset, our goal was to produce a self-consistent data set consisting of human-readable configuration ({\tt yaml}) files, which produced standard timing model parameter ({\tt .par}) files for each pulsar. Aside from the parameter values themselves, configuration files contain all necessary information to reproduce timing results and the files are organized to facilitate using version control to track any changes made along the way. All timing fits used the JPL DE440 solar system ephemeris \citep{Park_2021} and the TT(BIPM2019) timescale.

As in \twelveyr, we used standardized Jupyter notebooks \citep{Kluyver2016jupyter} to automate our timing procedures and we built new software tools to improve transparency, code readability, and reproducibility of our results. Our analysis pipeline is now entirely \pint-based \citep{2021PINT} --- previous releases used {\tt TEMPO}\footnote{\url{https://github.com/nanograv/tempo}} \citep{tempo} --- and flexible enough to accommodate working with both narrowband and wideband TOAs. A frozen version of code used to produce \fifteenyr\ results as well as intermediate data products can be found at \url{data.nanograv.org} (see Section \ref{sec:dataproducts} for more details). An open source version of our timing analysis package, \pintpal, is also available.\footnote{\url{https://github.com/nanograv/pint\_pal}}

\subsection{Timing Models and Parameters}
\label{sec:timing_models}

Pulsar timing requires determination of the TOAs and comparison to a model composed of many different physical effects. Our timing parameter sets fall into six categories, with the numbers of parameters in each category fit per pulsar shown in Table~\ref{tab:models}. 
\begin{enumerate}
\item {\it Spin:} We fit three parameters describing the rotational phase, frequency, and frequency derivative of each pulsar. In one case (long-period binary PSR~J1024$-$0719), we also fit the second frequency derivative. 
\item {\it Astrometry:} For each pulsar we fit five parameters describing the two-dimensional position on the sky in ecliptic coordinates, the two-dimensional proper motion, and parallax. As in our previous works, we fit all five of these regardless of the measurement significance.
\item {\it Binary:} For 50 of our observed systems we fit binary parameters describing the orbit with a companion star. With the exception of the long-period binary PSR~J1024$-$0719, we fit, at a minimum, the five Keplerian parameters fully characterizing the orbit. Several different models were used depending on the orbital characteristics, see Section~\ref{sec:binary}.
\item {\it Dispersion Measure:} Time-variations in the DM require that we fit the TOAs with a dispersive delay proportional to $\nu^{-2}$ and the unknown DM at that epoch. This is discussed in greater depth in Section~\ref{sec:DMt}.
\item {\it Frequency dependence:} Additional time-independent but frequency-dependent delays are fit for on a per-pulsar basis (``FD'' parameters; see \nineyr), with the number included determined via our $F$-test procedure discussed below in Section~\ref{sec:Ftest}. In the narrowband data, we take these to describe time offsets due to differences in the observed pulse shape at a given frequency compared to the template shape used in timing. In the wideband data, we expect the pulse portrait to encapsulate the changing frequency-dependence of the profiles. However, we still find several significant FD parameters. We fit for these values in the same manner regardless, and the significance of these will be explored more in future work. Though the physical mechanism has not been definitively determined, other time-varying chromatic processes could be picked up as time-independent FD parameters. 
\item {\it Jump:} We fit for ``jumps'' to account for unknown phase offsets between data observed by different receivers and/or telescopes.
\end{enumerate}

\subsubsection{Binary Models}
\label{sec:binary}

We used one of five binary models, depending on the orbital characteristics of the pulsar in question. Low-eccentricity orbits (see Section~\ref{subsec:binaries})
were fit using the ELL1 \citep{Lange2001} or ELL1H \citep{Freire2010} model, depending on whether or not Shapiro delay was marginally detected; orbits with higher eccentricity were modeled with the BT \citep{BlandfordTeukolsky1976} or DD \citep{Damour85, Damour86, Damour1992} model; and for PSR~J1713+0747, in which we measure the physical orientation of the orbit, we used the DDK model \citep{Kopeikin1996}. Pedagogical descriptions of these binary models can be found in \citet{LorimerKramer2012}. In all cases, at a minimum we included and fit five Keplerian parameters in the binary model: the orbital period $P_b$ or orbital frequency $f_b$; the projected semimajor axis $x$; and either the eccentricity $e$, longitude of periastron $\omega$, and epoch of periastron passage $T_0$ in the DD or DDK models, \textit{or} two Laplace-Lagrange parameters ($\epsilon_1, \epsilon_2$) and the epoch of the ascending node $T_{\mathrm{asc}}$ in the ELL1 or ELL1H models.

The ELL1 model describes the orbit with the Laplace-Lagrange eccentricity parameterization, with $\epsilon_1 = e\sin{\omega}$ and $\epsilon_2 = e\cos{\omega}$ \citep[for a more complete description of this parameterization, see][]{Lange2001}. This parameterization is needed because for a nearly circular orbit, one cannot reliably define the time and location of the periastron. If the Shapiro delay was marginally detected via the $F$-statistic as described below, we instead modeled low-eccentricity orbits with the ELL1H model, which is simply the ELL1 model with the addition of the $h_3$ and $h_4$ parameters from the orthometric Shapiro delay parameterization of \citet{Freire2010}. For a significant detection of Shapiro delay, the companion mass $m_c$ and orbital inclination $\sin{i}$ parameters were instead included in the ELL1 model.

The BT and DD models directly measure $e$, $x$, and $T_0$, and are more accurate than the ELL1 model for orbits with a large enough eccentricity to measure the longitude and epoch of periastron. The BT model is Newtonian, while the DD model is a theory-independent relativistic model that allows for the inclusion of Shapiro delay parameters, companion mass $m_c$ and inclination angle via $\sin{i}$. In our data set, the use of ELL1 and DD is split nearly evenly between pulsars. For one of our most precisely-timed pulsars, PSR~J1713+0747, we found in \twelveyr\ that we were sensitive to the annual-orbital parallax \citep{Kopeikin1995}, which allows the true orientation of the orbit to be measured. Thus we transitioned from the DD to DDK model for this pulsar. The DDK model incorporates the orbital inclination and the longitude of the ascending node of the orbit, $\Omega_{\mathrm{asc}}$ \citep{Kopeikin1996}. Appendix~\ref{sec:kopeikin} describes our use of the DDK model in detail.

For the majority of orbital models, we used the orbital period $P_b$, but for four short orbital period ($P_b < 0.5$\,d) pulsars we instead included the orbital frequency $f_b$ in the model. Using the orbital frequency instead of period allows us to include multiple orbital frequency derivatives to better describe the orbit, rather than being restricted to only the first period derivative if using $P_b$; it also allows us to easily test for additional derivatives using the $F$-statistic. All four of these pulsars are in ``black widow'' systems \citep{Swihart22}. Three of them are non-eclipsing (PSRs~J0023+0923, J0636+5128, and J2214+3000), while PSR~J1705$-$1903 eclipses for $\approx$\,30\% of each 4.4-hr orbit.

All four of these pulsars' orbits are low-eccentricity and were modeled with the ELL1 model. For PSR~J2214+3000, only $f_b$ (``FB0'' in the timing model) was needed, with no derivatives; J0636+5128 required three derivatives (FB0 through FB3); and J0023+0923 and J1705$-$1903 required five derivatives (FB0 through FB5).

In all cases, we tested for the presence of the Shapiro delay using the $F$-statistic, as described in Section~\ref{sec:Ftest} below. If $m_{\mathrm c}$ and $\sin{i}$ were significantly detected, we added these parameters to the timing model and re-fit.

\subsubsection{Parameter Inclusion and Significance Testing Criteria}
\label{sec:Ftest}
As in \twelveyr\ and our prior data releases, an $F$-statistic was used to determine what additional parameters should be included in the timing model beyond those describing pulsar spin, astrometry, red and white noise, and Keplerian binary motion. This test is valid in the case that one model is a subset of another; therefore, these tests were performed by fitting a nominal model, recording the resulting $\chi^2$, then adding or subtracting the parameter of interest and re-fitting. The $F$-statistic is:
\begin{equation}
    F = \frac{(\chi^2_0 - \chi^2)/(n_0 - n)}{\chi^2/n}
\end{equation}
where $\chi^2_0$ and $n_0$ are the best-fit chi-squared value and number of degrees of freedom, respectively, for the nominal (superset) model. Parameters were included if they induced a $p < 0.0027$ ($\sim$3$\sigma$) change in the fit. As described in \twelveyr, this test was applied to secularly-evolving binary parameters ($\dot{P}_{\text b}$, $\dot{\omega}$, and $\dot{x}$), Shapiro delay parameters, frequency-dependent (FD; see \nineyr) parameters, and higher-order orbital frequency derivatives (FB$n$, where $n=0$ corresponds to the orbital frequency and $n=\{1, 2, ...\}$ correspond to frequency derivatives) where applicable. For parameters with components of increasing complexity (like FD1-5, where a model cannot include non-contiguous combinations of parameters), the test was applied iteratively. All FB and FD terms below the highest-order term that induced a significant change in the fit were included, even if a subset of the lower-order parameters did not produce a significant change; for example, if FD3 is significant but FD2 is not, then FD1, FD2, and FD3 are included in the model. The summary plots generated with \pintpal\ provide the user with suggestions for parameters to add or exclude in the model. 

\subsection{Dispersion Measure Variations}
\label{sec:DMt}
Changes in the Earth-pulsar line of sight require that we account for variations in the DM on short (i.e., per-epoch) timescales \citep{Jones2017}. We measure the pulsar signals over a wide range of radio frequencies to estimate the time-varying DM for both narrowband and wideband data sets (see Figure~\ref{fig:epochs}). We used the DMX model, a piecewise constant function, to describe these variations, with each DMX parameter describing the offset from a nominal fixed value. Modeling of these variations requires up to hundreds of additional parameters (see Table~\ref{tab:models}), and for most of our pulsars the expected variations from interstellar turbulence alone are significant enough between epochs that we cannot reduce the number of parameters \citep{Jones2017}.

At Arecibo and the VLA, observations with different receivers were consecutive whereas with the GBT one or more days passed between observations for scheduling efficiencies\footnote{Switches between prime-focus receivers (i.e., ``Rcvr\_800'') and Gregorian-focus receivers (i.e., ``Rcvr1\_2''), or vice-versa, at the GBT take 10-15\,min.}. We assumed a constant value for the DM over a window of time 0.5 days long for any pulsars with Arecibo observations and longer windows for pulsars solely observed with the GBT, with a 15-day range for GASP data and 6.5-day range for GUPPI/YUPPI data. If the maximum-to-minimum frequency ratio of the TOAs did not satisfy $\nu_{\rm max}/\nu_{\rm min} > 1.1$, we excised the data as described above and did not fit a DMX parameter.

Besides DM variations due to the turbulent interstellar medium, another visually-apparent contribution is from the Earth-pulsar line of sight intersecting different sections of the solar wind. For a few specific pulsars close to the ecliptic plane, the change in DM can be significant even within the time range of a single window. We use the following criterion to find such windows.  As in \elevenyr\ and \twelveyr, for this purpose, we used a toy model of a spherically-symmetric and static solar wind electron density given by $n_e(r) = n_0(r/r_0)^{-2}$ where $r$ is the distance from the Sun and $n_0 = 5~\mathrm{cm}^{-3}$ is the fiducial electron density at a distance $r_0 = 1$~au \citep[e.g.,][]{Splaver2005}. If the projected DM variation due to the change in the line of sight through the solar wind would have induced a timing variation of greater than 100 ns, the DMX time ranges were divided into 0.5-day windows, with data still subject to the frequency cut described above.  Note that this toy solar wind model is used only for this test while setting DMX window sizes; it is not included in the pulsar timing models themselves.  Thus reported DMX values in the timing models incorporate dispersion by both the solar wind and the interstellar medium.

In addition to being our most-observed pulsar, PSR J1713+0747 experienced rapid chromatic variations over a period of several months in 2016 \citep[e.g.][]{Lam2018,ccg+21,grs+21} such that it was necessary to split the DMX windows to avoid additional excess timing noise from the event as in \twelveyr. Following the conclusion of the current data set, PSR J1713+0747 experienced a profile change, first reported in \cite{Xu2021}, confirmed by \cite{Meyers2021}, and further discussed by \cite{Singha2021} and \cite{Jennings2022}. As this event occurred outside of the range of data reported here, we do not need to correct for it in this release.

\subsection{Noise Modeling}
\label{sec:noise}
Pulsar timing uses a phenomenological noise model for the data that is separated into two broad components distinguished by the time scale of the correlations between the TOAs. The white noise (WN) model, i.e., noise that is uncorrelated in time and hence spectrally flat, inflates the values of the TOA uncertainties, $\sigma_{\rm S/N}$, derived from the pulse template matching process \citep{Taylor92}. The template matching process assumes that the data are a scaled and shifted version of the pulse profile, however other sources of noise can be best modeled as WN, but do not come from the template matching process \citep{Lommen2013}. 

Our noise modeling paradigm is discussed in detail in \noisepaper, and briefly summarized here. Three WN parameters -- \efac ($\ef$), \equad ($\eq$) and \ecorr ($\ec$) --  are used to adjust TOA uncertainties in order to accurately represent the WN present in the data.  \efac scales TOAs linearly, accounting for uncertainty induced by template matching errors or template mismatches. For well-characterized systems, \efac tends to 1.0. \equad adds an additional WN in quadrature, ensuring a minimum error size. \ecorr also adds in quadrature and accounts for uncertainty that is correlated among frequencies within an observation, most importantly pulse jitter, though other mechanisms can add as well \citep{lcc+17}. 

Various differences between pulsar timing backends and radio observatory receivers (e.g., the frequency ranges covered) make it necessary to give different values of these WN parameters to each receiver/backend combination.
These WN terms come together with receiver/backend combination, $re/be$, dependence as
\begin{equation}\label{eq:sigmaTotal}
C_{ij} = \ef^2(re/be)[\sigma_{{\rm S/N},i}^2+\eq^2(re/be)]\; \delta_{ij}+ \ec^2(re/be)\;\mathcal{U}_{ij} 
\end{equation}
where the $i,j$ denote TOA indices across all observing epochs,  $\delta_{ij}$ is the Kronecker delta and we omit the dependence on receiver and backend, $re/be$, from here on for simplicity. While \efac and \equad only add to the diagonal of $C$, where $C_{ij}$ are the elements of the covariance matrix, the \ecorr terms are block diagonal for single observing epochs. \ecorr is modeled using a block diagonal matrix, $\mathcal{U}$, with values of $1$ for TOAs from the same epoch and zeros for all other entries. 

Pulsar timing data often show evidence of correlations across longer timescales than can easily be modeled with ECORR terms. These correlations are instead modeled as a single stationary Gaussian process with a power-law spectrum
\begin{equation}
P(f) = A_{\rm red}^2 \left(\frac{f}{1~\mathrm{yr}^{-1}}\right)^{\gamma_{\rm red}},
\label{eqn:rn_spec}
\end{equation}
where $f$ is the Fourier frequency, $\gamma_{\rm red}$ is the spectral index\footnote{Note that in this paper, $\gamma_{\rm red}$ is the true spectral index of the red noise spectrum and could be positive or negative. In other papers this equation is often specified to have a negative spectral index such that the value of $\gamma_{\rm red}$ is positive.}, and $A_{\rm red}$ is the noise amplitude at a reference frequency of 1~yr$^{-1}$. Long-timescale correlated data are characterized as red noise (RN) in which $\gamma_{\rm red} < 0$. ECORR is not separately modeled in the wideband data set, as it is completely covariant with EQUAD. 

The timing model and noise analysis are performed iteratively over each pulsar data set, first fitting a preliminary pulsar ephemeris and then doing a Bayesian noise analysis with \enterprise\footnote{\url{https://github.com/nanograv/enterprise}} using a linearized timing model (see Section~\ref{s:linearity} below for details about the linearized version of the timing model). We then refit the timing model with the noise parameters obtained from \enterprise. This process is repeated until the timing and noise parameters stabilize and noise posteriors look nearly identical from one run to the next. Slight changes from \twelveyr\ in the implementation of the Bayesian noise analysis include increasing the number of samples obtained and matching the prior for the spectral index to those used in the GW detection pipeline.

The WN parameters EFAC, EQUAD, and ECORR are measured for each pulsar-backend-receiver combination. A small fraction of these WN parameters changed by $>$\,3\,$\sigma$ between \twelveyr\ and \fifteenyr\ (28/159, 17/159, and 5/159 for EFAC, EQUAD, and ECORR, respectively). Even so, these changes are not unexpected as the length of the data set increases, especially for pulsars that were newly added to \twelveyr. 

Twenty-three pulsars in \timingpaper\ were found to have significant levels of RN. Ten of these measurements are newly significant in \fifteenyr, while 13 of the 14 sources with significant RN in \twelveyr\ continued to favor its inclusion in their timing models. Only PSR~J2317+1439, which favored an $A_{\rm red}$ an order of magnitude lower than the next-lowest $A_{\rm red}$ in \twelveyr, as well as a steep $\gamma_{\rm red}$ comparable to that of PSR~J0030+0451, had RN parameters removed from its timing model in \fifteenyr. No $\gamma_{\rm red}$ or $A_{\rm red}$ values differed between the data sets by more than a factor of $\sim$\,3\,$\sigma$ for pulsars with significant RN in both \twelveyr\ and \timingpaper. 

\subsection{Linearity Testing for Pulsar Timing Models}\label{s:linearity}

\newcommand{\bxi}{\mathbf{p}}

A pulsar timing model is a deterministic function $t^{\rm det}$ that takes parameter values $\bxi$ and predicts observed pulse phases at the time of each TOA $t^{\rm obs}$ (in time units) such that the observed TOAs are the deterministic values plus noise $n$, i.e., $t^{\rm obs} = t^{\rm det}(\bxi_{\rm true}) + n$. We are able to simplify our GW detection calculations by using a linear approximation to this function \citep{esvh13}:
\begin{equation}
r(\bxi_{\rm best}) \equiv t^{\rm obs} -t^{\rm det}(\bxi_{\rm best}) \approx \left.\frac{\partial t^{\rm det}(\bxi_{\rm true})}{\partial \delta \bxi}\right|_{\delta\bxi=0} \delta \bxi + n,
\end{equation}
where $\bxi_{\rm best}$ is our vector of best-fit parameters from our generalized least-squares fit which form our residuals $r$, and $\delta\bxi \approx \bxi_{\rm true} - \bxi_{\rm best}$ is therefore small. The term with the partial derivative on the right-hand side is the design matrix. One generally expects this approximation to be good when the data constrain the parameters tightly, relative to the level of non-linearity intrinsic to each parameter; for example, some are entirely linear.  Since this level varies significantly between different parameters in the model, we developed tests to verify that the approximation of linearity is adequate.

In our reported models, the parameter uncertainties and the parameter correlations define an ellipsoid. If the linear model is adequate, then the plausible pulsar parameter values are constrained by the data to lie within or near this ellipsoid. Thus this ellipsoid is the region where we require the timing model to be well approximated by the linearized model. While we generally use analytical marginalization with flat priors to remove the (linearized) timing model parameters from the GW detection process, we expect that most samples from preliminary detection Markov chains will fall within or near this error ellipsoid (Kaiser et al. in prep). Thus we try to test the degree to which the full timing model differs from its linearized approximation over this error ellipsoid.

One direct approach is simply to shift the value of each parameter, one at a time, by one sigma from the best-fit $\bxi_{\rm best}$.  For each parameter, this produces a trial parameter set $\bxi_{\rm trial}$, from which we then compute the difference in residuals $r(\bxi_{\rm trial}) - t^{\rm det}(\bxi_{\rm trial})$ (in time units). This provides a measurement of the deviation from linearity for this parameter at the time of each TOA.  In the case of highly covariant parameters this approach may result in $\bxi_{\rm trial}$ falling outside the full (multi-dimensional) error ellipse.  Therefore, this test can be considered a conservative upper limit on the level of non-linearity; the true level may be lower.  For each pulsar and parameter we checked whether the root-mean-squared amplitude of this deviation was less than 100\,ns. We found only a few parameters on a few pulsars that exceeded this threshold. The most common example was the time ($T_0$) and longitude ($\omega$) of periastron. These two parameters form part of an essentially polar representation of the orbital shape, and when the orbit is nearly circular this coordinate system introduces nonlinearities; for the pulsars where this was a problem the nonlinearity was typically on the scale of a few microseconds. A second cause for nonlinearity is the parameter $\sin i$, which is constrained by the Shapiro delay. The shape of the Shapiro delay as a function of orbital phase is quite nonlinear in the parameter $\sin i$, and so for pulsars where the Shapiro delay is poorly constrained this nonlinearity introduces a few-microsecond discrepancy. 

For the pulsars and parameters where nonlinearities appear in the above test, we carried out a more stringent test. Since the detection tools analytically marginalize the timing parameters, any discrepancy that can be modeled as a linear combination of the timing model derivatives disappears. We therefore evaluated each of the problem cases, removing all such components. The discrepancy that remains is not absorbed during Bayesian fitting. This process resolved nearly all of the above nonlinearities, with the following exceptions: the parameter $\sin i$ for the pulsars J1630+3734, J1811$-$2405, J1946+3417, J2017+0603, and J2302+4442, and the parameters $\sin i$, $P_b$, $\dot \omega$, $\omega$, and $T_0$ for the pulsar J1630+3734. For the parameter $\sin i$, in each case the three-sigma confidence interval extends above the physical limit of 1, so nonlinearities must be expected. They take the form of a sharp peak around superior conjunction, and are mostly about 1 microsecond. PSR~J1630+3734 is a pulsar with only three years of data, and because of the orbital coordinate system, its $P_b$ is highly correlated with $\dot\omega$, along with the nonlinearities we observe in other pulsars with nearly-circular orbits.  The discrepancies in this case are of the order 10 microseconds.

The conclusion of this analysis is that the vast majority of pulsar parameters in our data set can be very well approximated by a linearized model, therefore applying this approximation when marginalizing over the timing model as part of GW analyses is not a significant source of error.  For the small number of parameters that may show significant nonlinearity, the effect occurs at the pulsar's orbital period.  These are generally days to weeks so are unlikely to interfere with the GW signals of interest on timescales of years.  The main result of nonlinearity is that parameter uncertainties estimated from the linearized model are potentially unreliable.  For these parameters additional analysis is required to determine posterior probability distributions; refer to Sec.~\ref{sec:SD} for discussion of Shapiro delay measurements and pulsar masses in our results.

\subsection{Data Products}\label{sec:dataproducts}
Our complete catalog of observations for 68 millisecond pulsars is available alongside this publication at \url{https://data.nanograv.org}. We also provide metadata and other useful intermediate files generated from pipeline runs there, including:
\begin{itemize}
    \item Configuration (\texttt{*.yaml}) files,
    \item Standardized pulsar parameter (\texttt{*.par}) files,
    \item Initial TOA (\texttt{*.tim}) files,
    \item TOA files with cut flags applied (\texttt{*excise.tim}),
    \item Noise modeling chains and parameter files,
    \item DMX timeseries (\texttt{*dmxparse*.out}).
\end{itemize}
For ease of use, timing models will be made available in both \tempotwo\footnote{\url{https://bitbucket.org/psrsoft/tempo2}} \citep{tempo2} and \pint\ formats. In addition, the correlation matrices for fit parameters of all pulsars are provided in the following formats:
\begin{itemize}
    \item Human-readable list of pairwise correlations (\texttt{*.txt}),
    \item Machine-readable NumPy compressed correlation matrices (\texttt{*.npz}),
    \item Standardized HDF5 compressed correlation matrices (\texttt{*.hdf5}).
\end{itemize}
Finally, we have containerized the production environments utilized in the production of \timingpaper. The following options will be released alongside the \timingpaper\ data set: 
\begin{itemize}
    \item An interactive Docker \citep{merkel2014docker} container, which is built off of the Official Jupyter images to ensure compatibility with the JupyterHub software stack \citep{Jupyter2021}. 
    \item A Singularity \citep{Singularity} container, which is optimized for HPC workloads.
    \item An Anaconda Environment, consisting of Conda-Forge \citep{conda_forge_community_2015_4774216} packages tested against Python 3.10. 
\end{itemize}

\section{Timing Model Comparisons \& Newly Measured Parameters}
\label{sec:compare}
\subsection{Comparison of \twelveyr\ and \fifteenyr\ Timing Models}
The careful analysis of changes to timing model parameters can provide a wealth of secondary science results and provide a measure of confidence in additions to the data set. Here, we present a basic summary of changes between the 12.5 and 15-year data releases, focusing specifically on improvements in astrometric and binary parameter measurements, changes in WN and RN parameters, and discrepancies between timing solutions obtained with \tempotwo\ and \pint.

\subsubsection{Astrometric Parameter Comparison \label{sec:astrometry}}

\begin{figure*}
    \centering
    \includegraphics[width=0.7\linewidth]{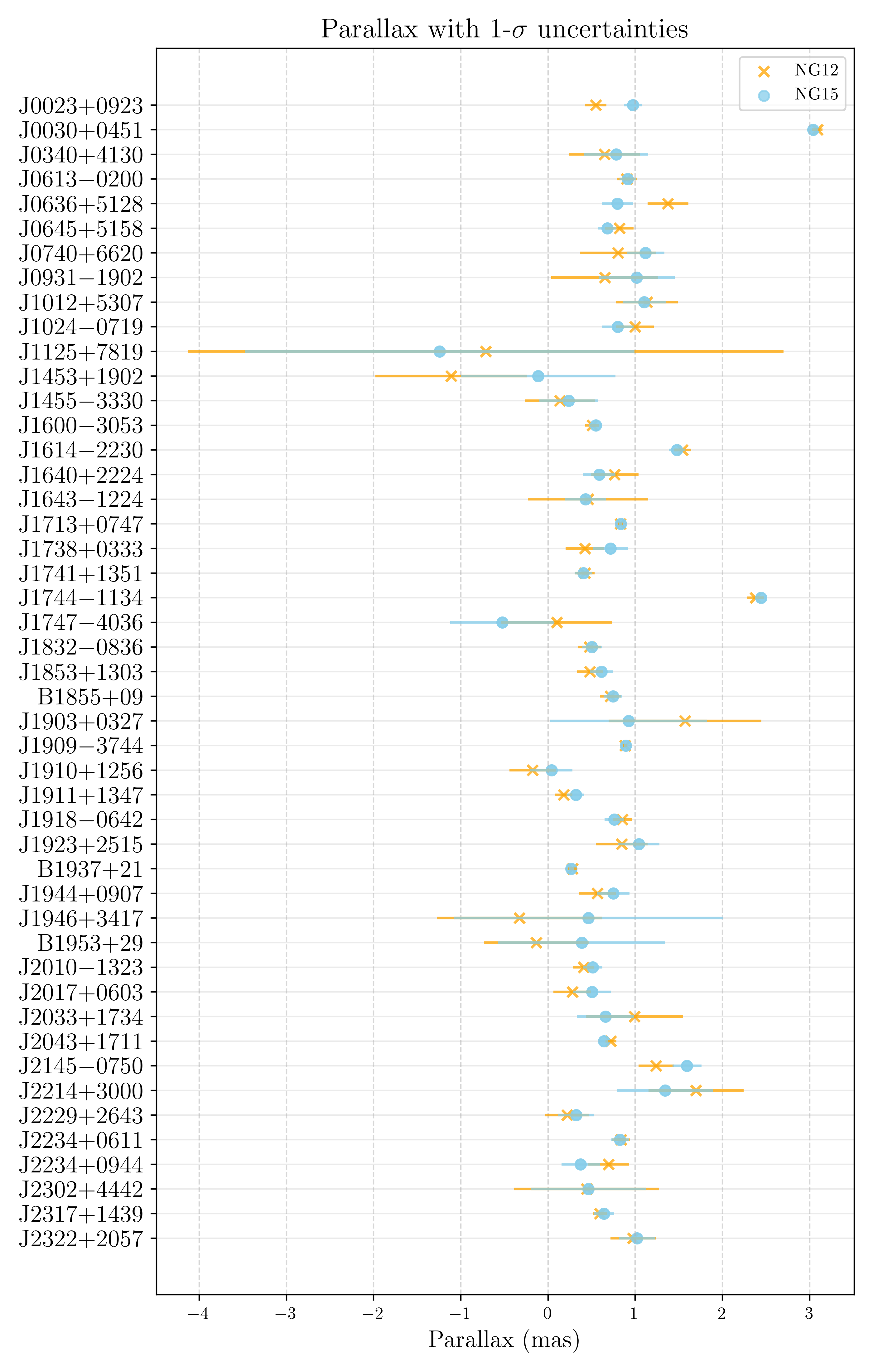}
    \caption{Measurements and uncertainties of parallax from the 12.5-yr and the current 15-yr data sets, showing the parallax values to be consistent across data sets. Orange Xs denote the \twelveyr\ measurement, while blue circles denote the \fifteenyr\ measurement. Horizontal lines of the corresponding color mark the extent of the 1-$\sigma$ measurement uncertainty. The timing algorithm allows both positive and negative parallaxes, even though only positive values are physically meaningful. As expected, none of the negative parallax values are significant within their uncertainties.}
    \label{fig:astrometry_parallax}
\end{figure*}

We begin by comparing the astrometric parameters, specifically the parallax and proper motion values. 

We are particularly interested in \fifteenyr\ parameters that have changed by $>3\sigma_{12.5}$ relative to \twelveyr, where $\sigma_{12.5}$ is the $1-\sigma$ uncertainty on the \twelveyr\ parameter
(i.e.~a conservative estimate of a 3$\sigma_{12.5}$ discrepancy).
We find three pulsars with $>3\sigma_{12.5}$ changes in the ecliptic longitude component of proper motion $\mu_{\lambda}$ (PSRs~J0636+5128, $3.5\sigma_{12.5}$; J1909$-$3744, $3.7\sigma_{12.5}$; and J1946+3417, $7.0\sigma_{12.5}$), and one pulsar with a $>3\sigma_{12.5}$ change in parallax (PSR~J0023+0923, $3.4\sigma_{12.5}$). It is known that variations in the red noise parameters, orbital parameters, and other parameters that require long timescales for measurement have covariances with astrometric parameters, such that a few pulsars will have $\gtrsim 3\sigma$ changes in some of these parameters. Such covariances can therefore explain the changes in astrometric parameters between \twelveyr\ and \fifteenyr. For example, PSR~J1946+3417 has red noise in \fifteenyr, but not in \twelveyr, explaining the change in proper motion; and the measured parallax of PSR~J0023+0923 changes significantly with the addition or removal of orbital frequency derivatives in its timing model.

In all cases, the uncertainty on the 15-yr parameter is smaller than that of the 12.5-yr, as is expected for a parameter measured with a longer data set. Figure~\ref{fig:astrometry_parallax} shows the 12.5- and 15-yr parallax values for all pulsars that were included in both data sets.

\subsubsection{Comparison of Narrowband \& Wideband Timing Models\label{subsec:NBWB}} 

As in \twelveyr\ and \twelvewb, we have generated both a narrowband and a wideband data set as part of the \fifteenyr\ process. However, it is critical to note that the wideband data set is not being used to derive any GW background or astrophysical interpretation-related results. Its primary purpose is to aid in refining the techniques used in wideband TOA generation and in the curation of wideband timing solutions in order to prepare for the eventual inclusion of data from wide-bandwidth receivers and cope with prohibitively large narrowband data volumes in future releases. In this work, the wideband data set is only used to derive preliminary pulsar masses for sources with significantly measured Shapiro delay parameters (see \ref{sec:binarycomparison}). 

A thorough comparison of all measured parameters common between the narrowband and wideband timing solutions for each source was conducted, as in \twelvewb. In that work, few discrepancies larger than 2\,$\sigma_{12.5}$ were found. However, the longer \fifteenyr\ wideband data set has shown a handful of discrepancies that are currently under investigation. In several cases, either the length of the wideband or the narrowband data set is sufficiently long (due to automatic outlier excision in the other data set) to significantly measure RN. In others, either the data set length or presence of RN caused a parameter (usually orbital) to be included in one of the data sets, but not both. Covariances between (usually orbital) parameters as discussed in Section~\ref{s:linearity} may also worsen these discrepancies. For example, a barely-significant measurement of the relativistic Shapiro delay would require the addition of the two post-Keplerian parameters that describe the effect. However, if the constraint is not strong, nonlinearities in binary orbital parameters may exacerbate discrepancies in these parameter measurements. In general, these covariances might lead to larger possible discrepancies in narrowband and wideband parameter constraints. The overall impact of these discrepancies is minimal, as the only analysis referencing the wideband data set is the preliminary measurement of pulsar masses. Further refinement of this procedure will proceed the publication of the next NANOGrav data release.

\subsubsection{Binary Model Comparison \label{sec:binarycomparison}}
\label{subsec:binaries}
The high cadence and long timing baseline of NANOGrav's data set enables the detailed study of a large number of MSP binary systems. \nineyr\ and its accompanying publication \cite{Fonseca2016}, as well as \elevenyr\ and \twelveyr, have included analyses of binary sources timed as part of the NANOGrav PTA. In this work, we present a basic overview of newly measured binary parameters, changes to timing models, and new constraints on pulsar masses from the measurement of post-Keplerian Shapiro delay orbital parameters. A manuscript describing our more in-depth analysis of the 15-year binary sources is in preparation. 

Of the 68 MSPs included in the current data set, 50 are in binary systems. Because these sources are selected for their high timing precision and stability, the vast majority are in near-circular orbits with white-dwarf companions. Of the 21 new MSPs added since \twelveyr, 18 are binaries. As more observations are added to the data set, sensitivity to secular binary parameters (i.e., $\dot{x}$, $\dot{\omega}$, and $\dot{P_{\text b}}$) improves. Additionally, parameters such as the relativistic Shapiro delay can be more precisely measured with better orbital phase coverage. We see a number of improvements and changes to binary parameter measurements resulting from the addition of $\sim$3 years of NANOGrav observations.

Consistency between the narrowband and wideband data sets is discussed in \ref{sec:astrometry}. Any discussion regarding the addition and removal of parameters, as well as their values, are based on the narrowband \twelveyr\ and 15-year data sets. However, because the wideband data set consists of significantly fewer TOAs (the \twelvewb\ data volume is $\sim$33 times smaller than that of \twelveyr), mass determinations based on post-Keplerian parameter measurements are based on the wideband data set to reduce computational burden.\\

\noindent
\emph{Discrepant binary parameters:} Several NANOGrav MSPs have shown significant changes ($>3\sigma$) in measured Keplerian orbital parameters since \twelveyr, as determined by the absolute value of the difference in parameter value between \twelveyr\ and now, divided by the \twelveyr\ uncertainty. However, these $>3\sigma$ discrepancies can generally be explained by changes to the pulsar's orbital or RN model. 

For example, the significance of $P_b$ for PSR~J1600$-$3053 decreased since \twelveyr\, primarily due to the addition of RN, a new measurement of $\dot{\omega}$, and an added frequency-dependent parameter (FD3). Relative to the recent, larger uncertainty, the value of $P_b$ increased by 4.3\,$\sigma$. Newly measured RN in PSR~J1614$-$2230 (one of the highest-mass neutron stars known; see \citealp[]{dem10}) was accompanied by a 3.8\,$\sigma$ increase in $P_b$.

Such changes are plausible as the PTA timing baseline increases and sensitivity to long-term orbital variations improves. Because secular evolution of parameters can happen on timescales similar to those of intrinsic RN, these two measurements can be covariant (see \twelveyr). 

Ten of the 50 binary sources show newly measured values for either $\dot{P_{\text b}}$, $\dot{x}$, or $\dot{e}$ (see Table~\ref{tab:newbinparams}). In the case of PSR~J0613$-$0200, $\dot{x}$ was removed and $\dot{P_{\text b}}$ was added.\\

\noindent
\emph{Changes to binary models:} Because the ELL1 binary model is only valid for sufficiently circular orbits, a test ($a\,\text{sin}(i)\,e^2/c \ll \sigma_{\mathrm {S/N}} / N_{\mathrm{TOA}}^{1/2}$; see Appendix A1 of \citealt{Lange2001}) was imposed on all ELL1 binaries to ensure its validity. ELL1 is no longer valid for PSR~J1918$-$0642, so this source is now parameterized by the DD binary model. 

The sources PSR~J1853+1303 and PSR~J1918$-$0642 required changes to their binary models as a result of the increased length of the data set; specifically, we have chosen to use the DD binary model instead of ELL1H and ELL1 for PSRs~J1853+1303 and J1918$-$0642, respectively (see Table~\ref{tab:newbinparams}). PSR~J1853+1303 was discussed in detail in \twelveyr\ because the orthometric Shapiro delay parameters $h_{\text 3}$ and $h_{\text 4}$ were newly constrained. We now significantly measure the traditional Shapiro delay parameters $m_{\text c}$ and $\sin i$ as determined by an $F$-test comparison; therefore, the DD parameterization is merited. Although these parameters are significantly measured, we do not yet meaningfully constrain the pulsar mass ($m_{\text p} = 3.3^{+12.3}_{-2.6}$\,$M_\odot$; see Section~\ref{sec:SD}). Continued observations will result in improved orbital coverage, potentially enabling a more precise $m_{\text p}$ measurement in future data releases.

\subsubsection{Pulsar Masses from the Relativistic Shapiro Delay}\label{sec:SD}
Because an extensive analysis of each of the 50 binary MSPs included here is beyond the scope of this data release, a more in-depth manuscript detailing our binary analysis similar to \nineyr\ and \cite{Fonseca2016} is in preparation. Here, we present pulsar mass ($m_{\text p}$) measurements obtained from measurement of the relativistic Shapiro delay in cases where the companion mass ($m_{\text c}$) and orbital inclination angle ($i$) are significantly constrained. When these two parameters are combined with the Keplerian mass function and the extremely well-determined $x$ and $P_b$ orbital parameters, the pulsar's mass and companion mass can be measured independently. In the rare case that other post-Keplerian parameters can also be constrained, one's ability to precisely measure $m_{\text p}$ is improved; however, our basic analysis does not take this additional information into account. 

\begin{figure}
    \centering
    \includegraphics[width=\linewidth]{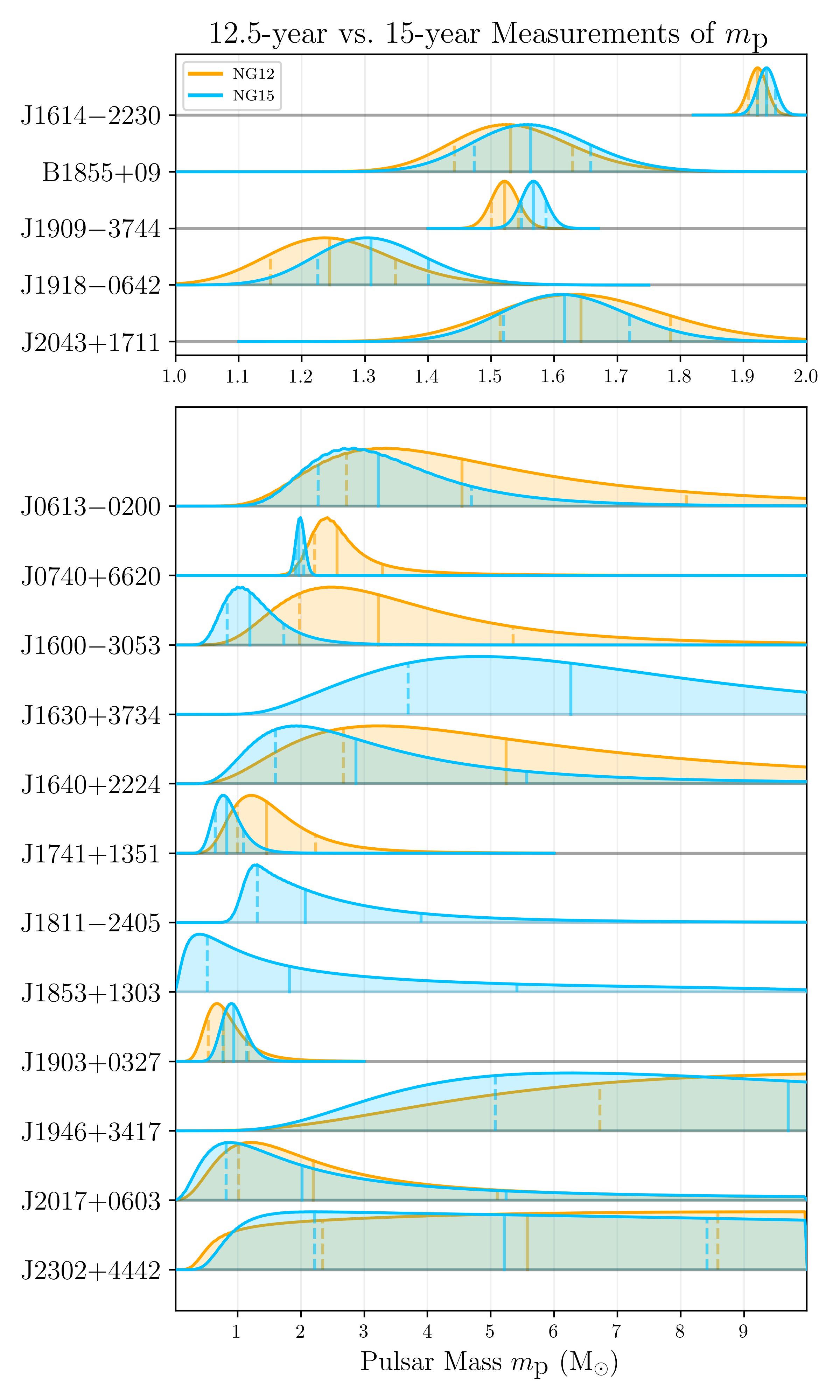}
    \caption{Posterior probability distributions for pulsar mass $m_{\text p}$ for each \fifteenyr\ binary with well-constrained traditional Shapiro delay parameters. Solid lines within the PDF curves indicate the median mass value, while the two dotted lines represent the 68.3\% confidence interval. Orange curves were derived using \twelveyr\ measurements; blue curves are from the updated \fifteenyr\ data set. Of the 14 Shapiro delay measurements common to \twelveyr\ and \fifteenyr, seven of the mean masses decreased, while seven either increased or did not appreciably change.}
    \label{fig:pulsarmass_comp}
\end{figure}

Posterior probability distribution functions (PDFs) for pulsar masses (see Figure~\ref{fig:pulsarmass_comp}) were derived from grid-based iteration over $m_{\text c}$ and sin($i$) using \pint. For each combination of $m_{\text c}$ and $\sin i$, a $\chi^2$ fit was performed without holding other model parameters (except the measured WN and RN values) fixed. Equations 4 and 5 and associated text in \cite{Fonseca2016} explain the Bayesian translation (with uniform priors) from this $\chi^2$ grid to marginalized posterior PDFs in $m_{\text c}$ and $\sin i$ in greater detail. While \cite{Fonseca2016} and \cite{fon21} (hereafter, FCP21) perform the $\chi^2$ model fits using \texttt{Tempo2}, the \pint-based results presented here were cross-checked with \texttt{Tempo2}-derived results. For assumed white-dwarf companions with poorly-constrained masses, grids spanned the 0--2\,$M_\odot$ range. All inclination angles (sampled uniformly in $\cos i$ rather than $\sin i$ to represent a random distribution of orbital orientations) were searched if that parameter was not already well constrained. Unless otherwise noted, reported uncertainties correspond to 68.3\% confidence intervals, and grids were 200 by 200 samples in size.

Two of the 18 binary MSPs added to the NANOGrav data set since \twelveyr, PSRs~J1630+3734 and J1811$-$2405, show (according to the $F$-test) significantly --- if not precisely --- measured Shapiro delay (see Table~\ref{tab:masses}). We measure the mass of PSR~J1630+3734 to be $6.7^{+5.0}_{-2.9}$\,$M_\odot$. While this mass is intriguingly high, this source has only three years of timing data and suffers from nonlinearities in multiple of its measured binary parameters (see Section~\ref{s:linearity}). PSR~J1811$-$2405 suffers from the same limitations and nonlinear tendencies as PSR~J1630+3734. A lengthened timing baseline will help resolve some of these covariances and improve our constraint of $m_{\text p}$.

A number of factors determine the precision of Shapiro delay measurements. Among these are the density of observational coverage around superior conjunction, a pulsar's timing precision and noise characteristics, and its orbital geometry. A number of $m_{\text p}$ measurements have improved between \twelveyr\ and the current data release, due in part to a longer timing baseline and improved orbital coverage (see Figure~\ref{fig:pulsarmass_comp} and Table~\ref{tab:masses}). The 68.3\% $m_{\text p}$ confidence intervals for PSRs~J0740+6620, J1600$-$3053, and J1741+1351, improved by a factor of $>2.5$ between \twelveyr\ and the present work. 

PSR~J0740+6620 is of particular interest to those wishing to constrain the poorly understood dense matter equation of state (EoS), as the measurement of unprecedentedly high-mass neutron stars provides support for a subset of ``stiff'' EoS. Timed by NANOGrav since 2014, this source is one of the most massive neutron stars known. \cite{cfr+20} reported its mass to be $2.14^{+0.10}_{-0.09}$\,$M_\odot$ after a series of follow-up campaigns around superior conjunction with the GBT. FCP21 provided an updated measurement of its mass, $2.08^{+0.07}_{-0.07}$\,$M_\odot$, after combining the aforementioned follow-up observations with a preliminary \fifteenyr\ and daily-cadence Canadian Hydrogen Intensity Mapping Experiment pulsar observing system \citep[CHIME/Pulsar; ][]{chimepulsaroverview} observations. Radio-derived neutron star mass constraints can be employed as priors in mass-to-radius measurements from X-ray lightcurve modeling (e.g., \citealp[]{ril21} and \citealp[]{mil21}).

This work demonstrates that based on NANOGrav observations alone, the mass of PSR~J0740+6620 is constrained $>$7 times better by \fifteenyr\ compared to \twelveyr, an improvement that can be attributed to the orbital-phase-targeted campaigns first presented in \cite{cfr+20} and the additional $\sim$3 years of regular NANOGrav observations. However, this analysis does not incorporate the high-cadence CHIME data, nor does it take into account the newly measured $\dot{P}_{\text b}$ (which is dependent on $m_{\text c}$, $m_{\text p}$, and the distance $d$ to the source), or optimized dispersion measure modeling as FCP21 does. \emph{We therefore do not regard the present measurement, which is consistent with the FCP21 to within 1$\sigma$, as superseding that result.}

PSR J1614$-$2230, the first directly observed 2-$M_\odot$ neutron star \citep{dem10}, is timed as part of the NANOGrav PTA. In this data release, the addition of newly measured RN to its timing model coincides with a $\sim$1-$\sigma$ increase in $m_{\text p}$. 

Conducting orbital-phase-specific observing campaigns around superior conjunction results in demonstrated improvements in $m_{\text p}$ constraints for sufficiently highly-inclined systems. For this reason, five MSPs currently timed by NANOGrav that show borderline $m_{\text p}$ constraints (including PSRs~J1630+3734, J1811$-$2405, and J1853+1303) were subject to such a targeted GBT campaign in spring 2022, the results of which will be included in a future manuscript. 

\begin{deluxetable*}{ccccc}
\centerwidetable
\tabletypesize{\scriptsize}
\tablecolumns{4}
\tablecaption{Binary Models \& Binary Model Changes for MSPs common to \twelveyr\ and \fifteenyr}\label{tab:newbinparams}
\tablehead{\colhead{Source} & \colhead{Binary Model} & \colhead{New Parameters} & \colhead{Removed Parameters}}
  \startdata
J0023+0923 & ELL1 & FB5 & - \\
J0613-0200 & ELL1 & $\dot{P_{\text b}}$ & $\dot{x}$ \\
J0636+5128 & ELL1 & FB2, FB3 & - \\
J0740+6620 & ELL1 & $\dot{P_{\text b}}$ & - \\
J1012+5307 & ELL1 & $\dot{x}$ & - \\
J1125+7819 & ELL1 & $\dot{x}$ & - \\
J1455-3330 & DD & - & - \\
J1600-3053 & DD & $\dot{\omega}$ & - \\
J1614-2230 & ELL1 & - & - \\
J1640+2224 & DD & $\dot{e}$ & - \\
J1643-1224 & DD & $\dot{e}$ & - \\
J1713+0747 & DDK & - & - \\
J1738+0333 & ELL1 & - & - \\
J1741+1351 & ELL1 & $\dot{P_{\text b}}$ & - \\
J1853+1303 & ELL1H to DD & - & - \\
B1855+09 & ELL1 & - & - \\
J1903+0327 & DD & - & - \\
J1909-3744 & ELL1 & - & - \\
J1910+1256 & DD & - & - \\
J1918-0642 & ELL1 to DD & $\dot{x}$ & - \\
J1946+3417 & DD & - & - \\
B1953+29 & DD & - & - \\
J2017+0603 & ELL1 & - & - \\
J2033+1734 & DD & - & - \\
J2043+1711 & ELL1 & $\dot{P_{\text b}}$ & - \\
J2145-0750 & ELL1H & - & - \\
J2214+3000 & ELL1 & - & - \\
J2229+2643 & DD & - & - \\
J2234+0611 & DD & $\dot{P_{\text b}}$ & - \\
J2234+0944 & ELL1 & - & - \\
J2302+4442 & DD & - & - \\
J2317+1439 & ELL1H & - & - \\
\enddata
\end{deluxetable*}

\begin{deluxetable*}{ccccc}
\centerwidetable
\tabletypesize{\scriptsize}
\tablecolumns{4}
\tablecaption{Shapiro delay-derived Mass Measurements}\label{tab:masses}
\tablehead{\colhead{Source} & \colhead{\twelveyr\ $m_{\text p}$ ($M_\odot$)} & \colhead{\fifteenyr\ $m_{\text p}$ ($M_\odot$)} & \colhead{$\sigma$ change\tablenotemark{a}}}

\startdata
J0613$-$0200 & $4.5^{+3.5}_{-1.8}$ & $3.2^{+1.4}_{-0.9}$ & $-0.5\,\sigma$\\
J0740+6620\tablenotemark{b} & $2.57^{+0.73}_{-0.35}$ & $1.99^{+0.07}_{-0.07}$ & $-1.1\,\sigma$\\
J1600$-$3053 & $3.2^{+2.1}_{-1.2}$ & $1.20^{+0.54}_{-0.36}$ & $-1.2\sigma$\\
J1614$-$2230 & $1.922^{+0.015}_{-0.015}$ & $1.937^{+0.014}_{-0.014}$ & $+1.0\sigma$\\
J1630+3734 & - & $6.3^{+4.2}_{-2.6}$ & - \\
J1640+2224 & $5.2^{+5.0}_{-2.6}$ & $2.8^{+2.3}_{-1.2}$ & $-0.6\sigma$\\
J1741+1351 & $1.5^{+0.77}_{-0.46}$ & $0.83^{+0.26}_{-0.19}$ & $-1.0\sigma$\\
J1811$-$2405 & - & $2.1^{+1.8}_{-0.76}$ & - \\
J1853+1303 & - & $1.8^{+3.6}_{-1.3}$ & - \\
B1855+09 & $1.531^{+0.098}_{-0.089}$ & $1.563^{+0.095}_{-0.089}$ & $+0.3\sigma$\\
J1903+0327 & $0.78^{+0.39}_{-0.24}$ & $0.94^{+0.21}_{-0.17}$ & $+0.5\sigma$\\
J1909$-$3744 & $1.52^{+0.022}_{-0.021}$ & $1.57^{+0.020}_{-0.019}$ & $+2.2\sigma$\\
J1918$-$0642 & $1.24^{+0.10}_{-0.09}$ & $1.31^{+0.09}_{-0.08}$ & $+0.7\sigma$\\
J1946+3417 & $3.9^{+0.8}_{-1.1}$ & $3.9^{+0.8}_{-1.1}$ & $0\sigma$\\
J2017+0603 & $2.2^{+2.9}_{-1.2}$ & $2.0^{+3.2}_{-1.2}$ & $0\sigma$\\
J2043+1711 & $1.64^{+0.14}_{-0.13}$ & $1.62^{+0.10}_{-0.10}$ & $-0.2\sigma$\\
J2302+4442 & $5.6^{+3.0}_{-3.2}$ & $5.2^{+3.2}_{-3.0}$ & $-0.1\sigma$
\enddata
\tablenotetext{a}{\fifteenyr\ $m_{\text p}$ $-$ \twelveyr\ $m_{\text p}$ divided by the average of the \twelveyr\ upper and lower 1-$\sigma$ uncertainties.}
\tablenotetext{b}{See Section~\ref{sec:SD}}
\end{deluxetable*}

\subsubsection{Changes in Noise Parameters Since the 12.5-Year Data Set} 

Section~\ref{sec:noise} and the detector characterization paper (\noisepaper) describe the WN and RN modeling conducted for each pulsar in the data set.
A small fraction of WN parameters (28/159, 17/159, and 5/159 for EFAC, EQUAD, and ECORR, respectively) changed by $>3\sigma$ between \twelveyr\ and \fifteenyr. Even so, these changes are not unexpected as the length of the data set increases, especially for pulsars that were newly added to \twelveyr. 

\begin{figure*}
    \centering
    \includegraphics[width=\linewidth]{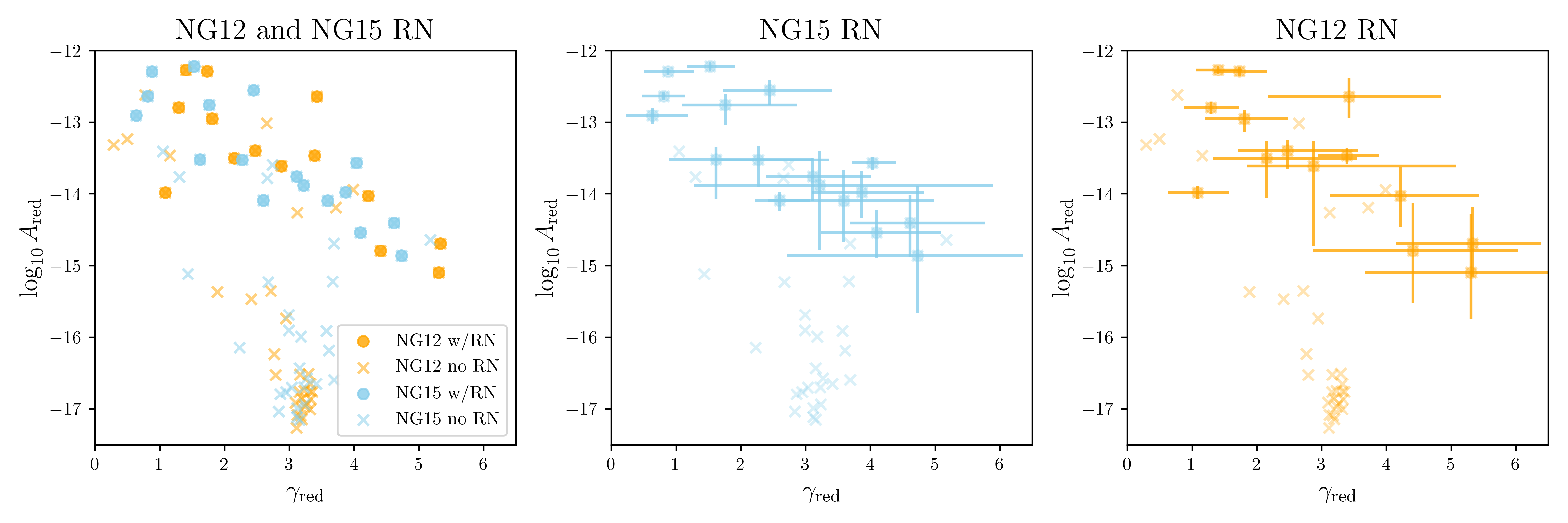}
    \caption{Comparisons of the two RN parameters (the spectral index $\gamma_{\rm red}$ and $\log_{10}$ amplitude, $\log_{10}A_{\rm red}$) between \fifteenyr\ and \twelveyr. \emph{Left:} $\log_{10}A_{\rm red}$ vs.~$\gamma_{\rm red}$ for \fifteenyr\ (blue) and \twelveyr\ (orange) without error bars. Solid square points are values for pulsars whose RN was found to be significant and is included in the timing model. Crosses are the measured RN parameters for pulsars without significant RN. \emph{Middle, right:} The same plot, split into the \fifteenyr\ values (middle) and \twelveyr\ values (right), including the 1-$\sigma$ confidence interval error bars. For crosses (insignificant RN) appearing in the ``high-$A_{\rm red}$'' group of points, the RN priors are the only information constraining those measurements; that is, the data provide no information about the RN content of those MSPs.}
    \label{fig:RN}
\end{figure*}

Twenty-three pulsars in \fifteenyr\ were found to have significant levels of RN (see solid points in Figure~\ref{fig:RN}). Ten of these measurements are newly significant in \fifteenyr, while 13 of the 14 sources with significant RN in \twelveyr\ continued to favor inclusion of RN in their timing models. Only PSR~J2317+1439, which favored RN amplitude an order of magnitude lower than the next-flattest spectral index in \twelveyr, as well as a steep spectral index comparable to that of PSR~J0030+0451, had RN parameters removed from its timing model in \fifteenyr. 
RN parameter values for pulsars with RN in both \twelveyr\ and \fifteenyr\ were checked for consistency. No spectral index or amplitude values differed between the data sets by more than a factor of just over $\sim$3$\sigma$. Not only are these inconsistencies small, but they are also expected to accompany a growing data set. 

\subsubsection{Kopeikin Parameter Analysis for PSR~J1713+0747}\label{sec:kopeikin1713}
\newcommand{\psr}{PSR~J1713+0747}
Unlike the majority of the binary systems analyzed here, for \psr\ we need to use the DDK model that incorporates the effects of proper motion and parallax on the binary orbit \citep[][and see Appendix~\ref{sec:kopeikin}]{Kopeikin1995,Kopeikin1996}.  However, in addition to the convention ambiguity discussed in Appendix~\ref{sec:kopeikin}, the nature of the analysis in \citet{Kopeikin1996} also leads to further ambiguity with nearly similar solutions possible for different choices of inclination $i$ and longitude of the ascending node $\Omega$.  These are locations where the change in projected semi-major axis $\delta x$ is equal to the best-fit value, but where $\delta \dot x$ is different.  The locations of these points can be found by first determining where $\delta \dot x$ is 0, which is where:
\begin{equation}
\Omega = \Omega_0 \equiv \tan^{-1} \left(\frac{\mu_\beta}{\mu_\lambda}\right)
\end{equation}
(defined in ecliptic coordinates).  With an initial  best-fit pair $(i,\Omega)$ the other points can then be identified as:
\begin{equation}
i,\Omega = \left\{ \begin{array}{cc}
i & \Omega \\
i & 2\Omega_0 - \Omega - 180{\degr} \\
180{\degr} - i & \Omega + 180{\degr} \\
180{\degr} - i & 2\Omega_0 - \Omega 
\end{array}
\right.
\label{eqn:ddk_alts}
\end{equation}
We wished to verify that the solution we had settled on was in fact a global and not just a local minimum.  To do this we computed a grid of $\chi^2$ over values of $(i,\Omega)$ with \pint, with all other parameters freely fit at each grid point.  We used a preliminary version of the narrow-band TOAs for this purpose, but subsequent changes and comparison with wide-band TOAs did not indicate any difference.

We stepped through all values of $\Omega$ from $-180\degr$ to $+180\degr$ in $1\degr$ increments, and similarly stepped through values of $i$ from $0\degr$ to $180\degr$ in $1\degr$ increments.  As mentioned in Appendix~\ref{sec:kopeikin} our final fit with \pint\ used ecliptic coordinates, but we wished to compare to results from the literature that were done in equatorial coordinates.  The value of $i$ will not change, but the value of $\Omega$ will change.  However, at this location in the sky the equatorial (ICRS) and ecliptic coordinate systems are almost aligned, with $\Omega_{\rm ICRS} = \Omega_{\rm Ecliptic} - 5.3\degr$.

\begin{figure}
\plotone{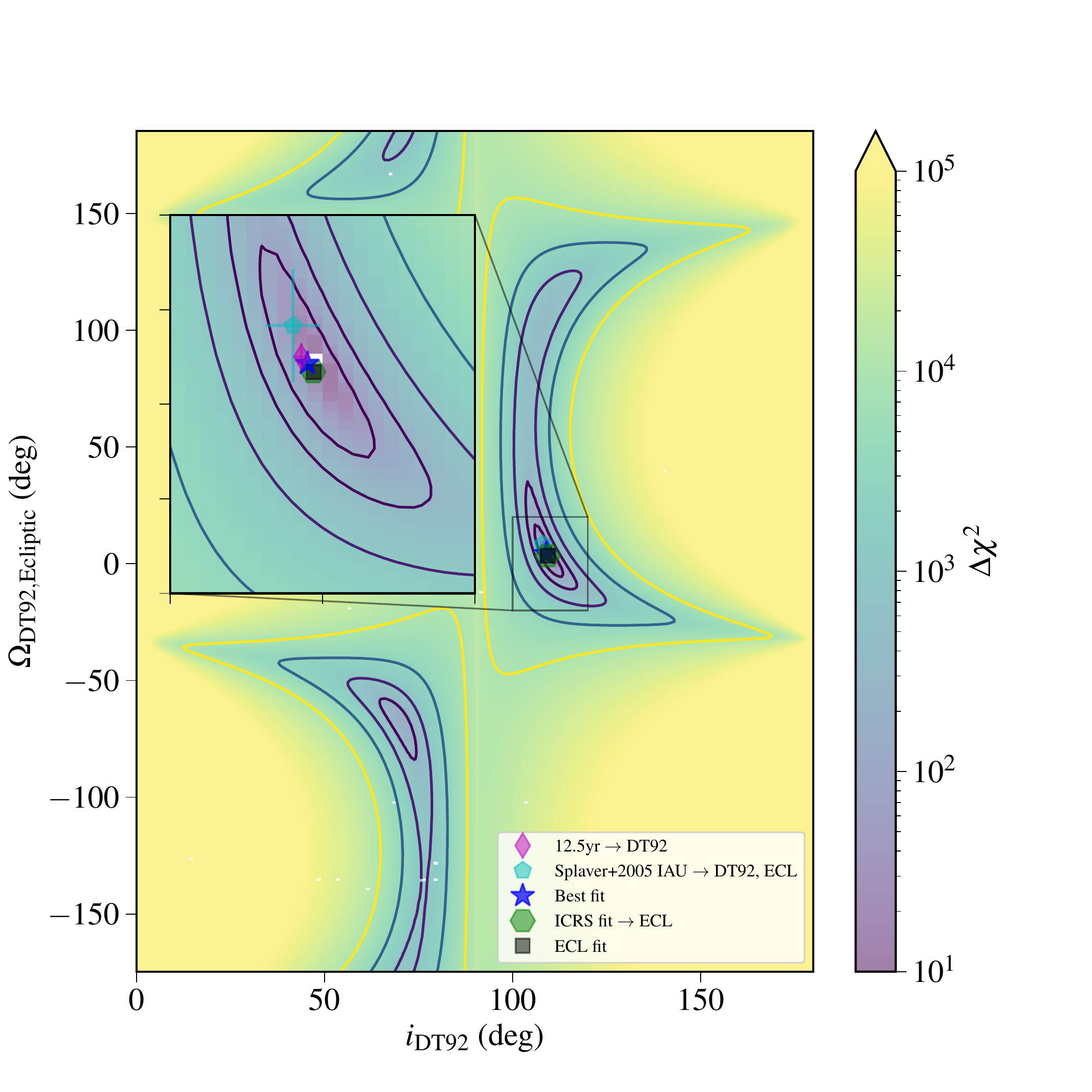}
    \caption{Grid of $\chi^2$ as a function of inclination $i$ and longitude of the ascending node $\Omega$ for \psr, computed with \pint.  All other parameters were fit at each grid point. The values are in the \citet[DT92; ][]{Damour1992} convention (see Appendix~\ref{sec:kopeikin}).  The color scale and contours are the logarithm of difference of $\chi^2$ compared to the best-fit value, with an inset zooming in to the best-fit region.  The magenta diamond is the value from the 12.5\,yr data release \citep{Alam21} converted from the IAU convention to DT92; these parameters will be slightly shifted due to the choice of timing model.  The cyan pentagon shows the value from \citet{Splaver2005} converted from the IAU convention to DT92, and additionally converted from equatorial to ecliptic coordinates.    The green hexagon is the best-fit when using equatorial coordinates, converted to ecliptic.  The black square is the best-fit when using ecliptic coordinates directly.  Finally the blue star is the minimum point in the $\chi^2$ grid.}
    \label{fig:ddk_grid}
\end{figure}

The results of this fit are shown in Figure~\ref{fig:ddk_grid}.  The four local minima given by Equation~\ref{eqn:ddk_alts} are clearly visible, and we actually found that our initial fit had ended up in the wrong local minimum.  However we have now verified that the minimum identified by our non-linear fitter is the global minimum, with $\chi^2$ differences of 261 to 904 for the alternatives (with approximately 59000 degrees of freedom).

This also agrees with the solutions from \citet{Splaver2005} and \citet{Alam21}, once they have been corrected to use the \citet[][hereafter DT92]{Damour1992} convention for defining $\Omega$ (Appendix~\ref{sec:kopeikin}) and ecliptic coordinates, again verifying that we have found the global minimum. We note that the \citet{Alam21} result was derived using the T2 timing model, and that the resulting values of $\Omega$ and $i$ may be shifted from the values found by the other models by roughly $0.1\sigma$, due to the use of unadjusted Keplerian parameters in computing binary delays (see Appendix~\ref{sec:kopeikin}).

Finally, we also get the same result if we fit using \tempo\ and convert from the IAU to DT92 convention for defining $\Omega$ (Appendix~\ref{sec:kopeikin}).

\subsection{Comparison of Split-Telescope Data Sets}
Most pulsars in the data set are observed by only one observatory, either GBT or Arecibo. However, two pulsars are observed by both GBT and Arecibo (PSR B1937+21 and PSR J1713+0747) and several are observed with the VLA in addition to GBT or Arecibo. 

For pulsars observed with GBT and Arecibo, we create timing solutions using TOAs from each telescope separately, in addition to the combined data set, and compare respective parameters. These data sets enable separate GBT and Arecibo ``split-telescope" gravitational wave background searches, as in \gwbpaper. Equally important, we use these separate data sets to check consistency between observatories. As both telescopes have extensive data sets for PSR B1937+21 and PSR J1713+0747, we expect the solutions to be statistically consistent with each other and with the combined solution. Comparisons of these solutions have confirmed that this is the case. 

Additionally, for pulsars observed with the VLA and Arecibo (J1903+0327) or the VLA and GBT (J1600-3053, J1643-1224, and J1909-3744), we create separate Arecibo or GBT-only data sets. Due to the small number of TOAs, we do not create VLA-only data sets for these sources. We again compare the Arecibo or GBT only solution to the complete solution. In all cases, we find parameters are consistent within $3-\sigma$, except where differences in the systems used, data spans, frequency ranges, or RN parameters would make changes expected.

\section{New Pulsars}
\label{sec:newpsr}
Since the 12.5-year data set, many new MSPs are now being observed and results for 21 new pulsars are included in \fifteenyr. Historically, NANOGrav has had a close relationship with pulsar survey groups using the Arecibo and Green Bank Observatories and in \fifteenyr, new sources came from three un-targeted surveys at these telescopes, the Arecibo 327\,MHz Drift-Scan Pulsar Survey \citep[PSRs~J0509+0856 and J0709+0458;][]{Martinez2019}, the Pulsar Arecibo L-band Feed Array survey \citep[PSR~J0557+1551;][]{skl+15}, and the Green Bank North Celestial Cap survey (PSR~J0406+3039).
Four more came from targeted surveys at Arecibo/Green Bank, guided by {\it Fermi} Large Area Telescope (LAT) unassociated $\gamma$-ray sources \citep[PSRs~J0605+3757, J1312+0051, J1630+3734, and J0614$-$3329;][]{SanpaArsa2016,rrc+11}.
We also draw on discoveries from similar targeted surveys carried out with the Parkes Telescope \citep[PSR~J1012$-$4235;][]{ckr+15} and the Effelsberg Telescope \cite[PSR~J1745+1017;][]{bgc+13} and additional blind surveys like the High Time Resolution Universe Survey \citep[PSRs~J1705$-$1903, J1719$-$1438, and J1811$-$2405;][]{mbc+19, nbb+14} and the Parkes Multibeam Pulsar Survey \citep[PSR~J1802$-$2124;][]{fsk+04}, which both use the Parkes Telescope.
Some sources added here are also monitored by the EPTA \citep[PSRs~J1751$-$2857 and J1843$-$1113;][]{Desvignes2016} and the PPTA \citep[PSR~J0437$-$4715;][]{rsc+21}. PSRs~J0610$-$2100, J1022+1001, J1730$-$2304, and J2124$-$3358 are also monitored by both EPTA and PPTA. While seemingly redundant, overlap in observing programs among regional PTAs like this is useful for International Pulsar Timing Array (IPTA) data combination and for diagnosing related issues \citep[e.g., see][]{IPTA2010,IPTADR2}. Overlap in observing programs can enhance our frequency coverage and cadence for these pulsars as well.

\section{Summary and Conclusions}
\label{sec:conclusion}
In this paper we have presented the 15-year data set from the NANOGrav collaboration, using data from the Green Bank, Arecibo, and VLA telescopes. Our longest timing baselines have increased by $\sim$3~yr over our previous data set. Even more impactful is the increase in the number of pulsars, from 47 MSPs in the 12.5-year data set to 68 MSPs in the present data set; this is the largest fractional increase ($\sim$50\%) in the number of pulsars since the size of the array doubled between \fiveyr\ and \nineyr. This large increase was made possible by NANOGrav's push to evaluate many new MSPs discovered in large-area sky surveys and targeted gamma-ray-guided surveys; it is  notable because a PTA's sensitivity to a GWB increases linearly with the number of pulsars in the array  and as the square root of the overall timing baseline \citep{Siemens2013}.

Another high-impact addition in this data set is the novel pipeline, \pintpal\, that automates the timing process. The user interfaces with a configuration file to set up the run by specifying which TOA files, noise chains, etc. to use, and the pipeline runs the timing analysis and outputs a series of diagnostic plots and tables with which the user determines whether or not the full timing model is sufficient. The timing analysis can easily be re-run by editing the configuration file, e.g. to conduct iterative noise analyses when parameters are added or removed from a given pulsar's timing model. This pipeline lowers the bar of entry for students and others who are relatively new to pulsar timing, and it will be of great use in future data sets, especially as the number of pulsars continues to grow. The intent is for this pipeline to also be adapted and used by other PTA collaborations and for more general pulsar timing analysis and IPTA data combination. The publicly available \pintpal\ can be found at \url{https://github.com/nanograv/pint_pal}.

In \twelveyr, we included a comparison between the \tempo\ and \pint\ timing results. A thorough comparison between timing software was also included in \citet{pint}. In the 15-yr data set, we have shifted to using \pint\ as the primary timing software. \pint\ is a modular, python-based software, and this transition decreases barriers to entry in understanding our pipeline and results. In \twelveyr, we conducted systematic comparisons between \tempo\ and \pint, demonstrating the consistency of our prior and current timing software.

In Section~\ref{sec:compare}, we performed a number of comparisons: we compared the astrometric, binary, and RN parameters measured in the 12.5-yr and 15-yr data sets, and the timing and RN parameter results between individual-telescope data sets for those pulsars observed with both Arecibo and GBT. Overall, the parameters are consistent between the various data sets. A small number of parallax, proper motion, and mass measurements have changed significantly between the 12.5-yr and 15-yr data releases, as detailed in Sections~\ref{sec:astrometry} and \ref{sec:binarycomparison}. In \twelveyr, we reported detectable levels of RN for 14 pulsars; in the 15-yr data set we measure RN for 23 pulsars, 13 of which were also found to show RN in \twelveyr. Aside from the one source for which we detected RN in \twelveyr\ but not in the 15-yr, RN parameters were consistent  between the data sets.

NANOGrav continues to be committed to public data releases for GW detection, as well as for individual pulsar studies.\footnote{Data from this paper are available at \url{http://data.nanograv.org} and preserved on Zenodo at doi:10.5281/zenodo.7967585.} In addition, we are committed to data-sharing and collaborative analysis within the IPTA, of which the EPTA, InPTA, and PPTA are also members. \fifteenyr, along with the most recent EPTA, InPTA, and PPTA data sets, have already been shared with the IPTA in order to create a combined  data set, which will be more sensitive to GW signals than any individual PTA data set. IPTA data combination analyses have begun.

We are also committed to continuing to increase the size of the NANOGrav PTA. The loss of Arecibo was significant, as new faint pulsars that fall in its declination range can no longer be added to the array, and we have been unable to continue observing a small number of faint pulsars previously observed at Arecibo. Despite this, NANOGrav will continue to evaluate new MSPs for inclusion in our GBT, VLA, and CHIME timing campaigns. With the proposed DSA-2000 telescope \citep{Hallinan2019}, NANOGrav will further expand its timing campaign, increasing the number of MSPs observed and thus our sensitivity to GWs.

\clearpage

{\it Author contributions.}

The alphabetical-order author list reflects the broad variety of
contributions of authors to the NANOGrav project.  Some specific
contributions to this paper, particularly incremental work
between \twelveyr\ and the present work, are as follows.

%
%
JKS and MED coordinated Arecibo observations.
PBD, JKS, and KSt coordinated GBT observations.
PBD and KSt coordinated VLA observations.
ZA,
PRB,
PTB,
HTC,
MED,
PBD,
TD,
ECF,
EF,
WF,
DCG,
PAG,
MLJ,
JL,
MTL,
DRL,
RSL,
AMc,
MAM,
NM,
CN,
NSP,
TTP,
SMR,
ASc,
BJS,
IHS,
JKS,
and
KSt
each ran at least 10 sessions and/or 20 hours of observations for this project.
MFA,
VB,
CC,
CJ,
AK,
YL,
KM,
EP,
LS,
and
SK
led observing programs run by undergraduate teams.
%
%
AMA,
BSA,
AB,
HTC,
PBD,
DCG,
JG,
NG,
RJJ,
MTL,
DJN,
NSP,
TTP,
and
JKS
developed and refined procedures and computational tools for the timing pipeline.
%
%
AMA,
HTC,
RJJ,
DLK,
MK,
JL,
AMc,
PSR,
SMR,
ASu,
BJS,
and
JKS
coordinated and implemented PINT development to facilitate \timingpaper\ analysis.
%
%
JKS coordinated development of the data set. 
AA,
GA,
AMA,
PRB,
HTC,
KC,
MED,
PBD,
EF,
WF,
DCG,
JG,
NG,
MLJ,
RJJ,
MK,
JL,
MTL,
MAM,
AMc,
BWM,
DJN,
BBPP,
NSP,
TTP,
HAR,
PSR,
SMR,
BJS,
CS,
IHS,
JKS,
and
HMW
generated and checked timing solutions for individual pulsars.
%
%
JKS coordinated the writing of this paper.
HTC and TTP also aided in coordination of data analyses and paper writing.
PBD wrote observing proposals and performed calibration and TOA
generation.
PBD and TTP developed wideband portraits to generate corresponding TOAs.
HTC, JG, NSP, and JKS
implemented and performed outlier analyses and data quality checks described in Section~\ref{sec:dataquality}.
GEF, AJ, SRT, and QW
assisted in implementing the outlier analysis code into the timing pipeline. 
NSP and JSH
assisted in interpreting the noise modeling results.
NSP undertook comparisons between \pint, \tempo, and \tempotwo\ results. DLK and IHS aided in comparisons specifically related to Kopeikin parameters, as described in Section~\ref{sec:kopeikin1713}.
MED performed the comparison of astrometric parameters between data sets.
HTC, KC, and JKS performed comparisons between the narrowband and wideband data sets.
HTC and EF undertook analysis of binary systems.
DCG compared split telescope results for the subset of pulsars observed with multiple telescopes.
Significant contributors to the text, tables, and figures in this paper include
HTC, MED, PBD, DCG, JG, JSH, DLK, MTL, MAM, DJN, SMR, IHS, and JKS.
HTC produced the timing residual plots.
JG and NG oversaw and maintained much of the computational infrastructure related and essential to this work and contributed to the design and public release of the \pintpal\ code repository.
We dedicate this paper to our wonderful colleague Jing Luo, whom we lost this past year.

The NANOGrav collaboration receives support from National Science Foundation (NSF) Physics Frontiers Center award numbers 1430284 and 2020265, the Gordon and Betty Moore Foundation, NSF AccelNet award number 2114721, an NSERC Discovery Grant, and CIFAR. The Arecibo Observatory is a facility of the NSF operated under cooperative agreement (AST-1744119) by the University of Central Florida (UCF) in alliance with Universidad Ana G. M{\'e}ndez (UAGM) and Yang Enterprises (YEI), Inc. The Green Bank Observatory is a facility of the NSF operated under cooperative agreement by Associated Universities, Inc. The National Radio Astronomy Observatory is a facility of the NSF operated under cooperative agreement by Associated Universities, Inc. This work was conducted using the Thorny Flat HPC Cluster at West Virginia University (WVU), which is funded in part by National Science Foundation (NSF) Major Research Instrumentation Program (MRI) Award number 1726534, and West Virginia University. This work was also conducted in part using the resources of the Advanced Computing Center for Research and Education (ACCRE) at Vanderbilt University, Nashville, TN.
NANOGrav is part of the International Pulsar Timing Array (IPTA); we would like to thank our IPTA colleagues for their help with this paper.

L.B. acknowledges support from the National Science Foundation under award AST-1909933 and from the Research Corporation for Science Advancement under Cottrell Scholar Award No. 27553.
P.R.B. is supported by the Science and Technology Facilities Council, grant number ST/W000946/1.
S.B. gratefully acknowledges the support of a Sloan Fellowship, and the support of NSF under award \#1815664.
M.C. and S.R.T. acknowledge support from NSF AST-2007993.
M.C. and N.S.P. were supported by the Vanderbilt Initiative in Data Intensive Astrophysics (VIDA) Fellowship.
Support for this work was provided by the NSF through the Grote Reber Fellowship Program administered by Associated Universities, Inc./National Radio Astronomy Observatory.
Support for H.T.C. is provided by NASA through the NASA Hubble Fellowship Program grant \#HST-HF2-51453.001 awarded by the Space Telescope Science Institute, which is operated by the Association of Universities for Research in Astronomy, Inc., for NASA, under contract NAS5-26555.
K.C. is supported by a UBC Four Year Fellowship (6456).
M.E.D. acknowledges support from the Naval Research Laboratory by NASA under contract S-15633Y.
T.D. and M.T.L. are supported by an NSF Astronomy and Astrophysics Grant (AAG) award number 2009468.
E.C.F. is supported by NASA under award number 80GSFC21M0002.
G.E.F., S.C.S., and S.J.V. are supported by NSF award PHY-2011772.
The Flatiron Institute is supported by the Simons Foundation.
A.D.J. and M.V. acknowledge support from the Caltech and Jet Propulsion Laboratory President's and Director's Research and Development Fund.
A.D.J. acknowledges support from the Sloan Foundation.
The work of N.La. and X.S. is partly supported by the George and Hannah Bolinger Memorial Fund in the College of Science at Oregon State University.
N.La. acknowledges the support from Larry W. Martin and Joyce B. O'Neill Endowed Fellowship in the College of Science at Oregon State University.
Part of this research was carried out at the Jet Propulsion Laboratory, California Institute of Technology, under a contract with the National Aeronautics and Space Administration (80NM0018D0004).
D.R.L. and M.A.M. are supported by NSF \#1458952.
M.A.M. is supported by NSF \#2009425.
C.M.F.M. was supported in part by the National Science Foundation under Grants No. NSF PHY-1748958 and AST-2106552.
A.Mi. is supported by the Deutsche Forschungsgemeinschaft under Germany's Excellence Strategy - EXC 2121 Quantum Universe - 390833306.
The Dunlap Institute is funded by an endowment established by the David Dunlap family and the University of Toronto.
K.D.O. was supported in part by NSF Grant No. 2207267.
T.T.P. acknowledges support from the Extragalactic Astrophysics Research Group at E\"{o}tv\"{o}s Lor\'{a}nd University, funded by the E\"{o}tv\"{o}s Lor\'{a}nd Research Network (ELKH), which was used during the development of this research.
S.M.R. and I.H.S. are CIFAR Fellows.
Portions of this work performed at NRL were supported by ONR 6.1 basic research funding.
J.D.R. also acknowledges support from start-up funds from Texas Tech University.
J.S. is supported by an NSF Astronomy and Astrophysics Postdoctoral Fellowship under award AST-2202388, and acknowledges previous support by the NSF under award 1847938.
S.R.T. acknowledges support from an NSF CAREER award \#2146016.
C.U. acknowledges support from BGU (Kreitman fellowship), and the Council for Higher Education and Israel Academy of Sciences and Humanities (Excellence fellowship).
C.A.W. acknowledges support from CIERA, the Adler Planetarium, and the Brinson Foundation through a CIERA-Adler postdoctoral fellowship.
O.Y. is supported by the National Science Foundation Graduate Research Fellowship under Grant No. DGE-2139292.

{\it Commonly used reference tags} for previous data set papers and others in the NANOGrav 15-year Data Set (\timingpaper) series:
\begin{itemize}
\item \fiveyr\ \citep{Demorest2013}: The NANOGrav 5-year Data Set
\item \nineyr\ \citep{Arzoumanian2015b}: The NANOGrav 9-year Data Set
\item \elevenyr\ \citep{Arzoumanian2018a}: The NANOGrav 11-year Data Set
\item \twelveyr\ \citep{Alam21}: The NANOGrav 12.5-year Narrowband Data Set
\item \twelvewb\ \citep{alam21wb}: The NANOGrav 12.5-year Wideband Data Set
\item \noisepaper\ \citep{NG15detchar}: The NANOGrav 15-year Detector Characterization
\item \gwbpaper\ \citep{NG15gwb}: The NANOGrav 15-year Gravitational Wave Background Analysis
\end{itemize}

\facilities{Arecibo, GBO, VLA}

\software{
    \texttt{astropy} \citep{astropy:2022}, 
    \texttt{Docker} \citep{merkel2014docker}, 
    \texttt{ENTERPRISE} \citep{enterprise}, 
    \texttt{enterprise\_extensions} \citep{enterprise_ext}, 
    \texttt{enterprise\_outliers} \citep{enterprise_outliers},
    \texttt{Jupyter} \citep{Kluyver2016jupyter, Jupyter2021},
    \texttt{libstempo} \citep{libstempo}, 
    \texttt{matplotlib} \citep{matplotlib}, 
    \texttt{nanopipe} \citep{nanopipe}, 
    \texttt{NumPy} \citep{NumPy}, 
    \texttt{PINT} \citep{pint}, 
    \texttt{PSRCHIVE} \citep{Hotan04}, 
    \texttt{PTMCMC} \citep{ptmcmc}, 
    \texttt{PyPulse} \citep{pypulse}, 
    \texttt{SciPy} \citep{SciPy},
    \texttt{Singularity} \citep{Singularity, singularity:2021}
    }

\clearpage

\startlongtable
\begin{deluxetable*}{crrrrr@{}lr@{}lr@{}lr@{}lr@{}lr@{}lr}
\tabletypesize{\footnotesize}
\tablecaption{Basic Pulsar Parameters and Narrowband TOA Statistics\label{tab:psrtoastats}}
\tablehead{
    \colhead{Source} &
    \colhead{$P$} &
    \colhead{$dP/dt$} &
    \colhead{DM} &
    \colhead{$P_b$} &
    \multicolumn{12}{c}{Median scaled TOA uncertainty\tablenotemark{a} ($\mu$s) / Number of epochs} &
    \colhead{Span} \\
    \cline{6-17} &
    \colhead{(ms)} &
    \colhead{($10^{-20}$)} &
    \colhead{(pc\,cm$^{-3}$)} &
    \colhead{(d)} &
    \multicolumn{2}{c}{327\,MHz} &
    \multicolumn{2}{c}{430\,MHz} &
    \multicolumn{2}{c}{820\,MHz} &
    \multicolumn{2}{c}{1.4\,GHz} &
    \multicolumn{2}{c}{2.1\,GHz} &
    \multicolumn{2}{c}{3.0\,GHz} &
    \colhead{(yr)}}
\startdata
J0023$+$0923 & 3.05 & 1.14 & 14.4 & 0.1 & \multicolumn{2}{c}{$\cdots$} & 0.067 &/81 & \multicolumn{2}{c}{$\cdots$} & 0.585 &/91 & \multicolumn{2}{c}{$\cdots$} & \multicolumn{2}{c}{$\cdots$} & 9.0 \\
J0030$+$0451 & 4.87 & 1.02 & 4.4 & $\cdots$ & \multicolumn{2}{c}{$\cdots$} & 0.188 &/257 & \multicolumn{2}{c}{$\cdots$} & 0.491 &/285 & 1.171 &/94 & \multicolumn{2}{c}{$\cdots$} & 15.5 \\
J0340$+$4130 & 3.30 & 0.70 & 49.6 & $\cdots$ & \multicolumn{2}{c}{$\cdots$} & \multicolumn{2}{c}{$\cdots$} & 0.888 &/102 & 2.240 &/101 & \multicolumn{2}{c}{$\cdots$} & \multicolumn{2}{c}{$\cdots$} & 8.1 \\
J0406$+$3039 & 2.61 & 0.83 & 49.4 & 7.0 & \multicolumn{2}{c}{$\cdots$} & \multicolumn{2}{c}{$\cdots$} & \multicolumn{2}{c}{$\cdots$} & 0.416 &/38 & 0.918 &/30 & \multicolumn{2}{c}{$\cdots$} & 3.6 \\
J0437$-$4715 & 5.76 & 5.73 & 2.6 & 5.7 & \multicolumn{2}{c}{$\cdots$} & \multicolumn{2}{c}{$\cdots$} & \multicolumn{2}{c}{$\cdots$} & 0.081 &/20 & \multicolumn{2}{c}{$\cdots$} & 0.080 &/28 & 4.8 \\
J0509$+$0856 & 4.06 & 0.44 & 38.3 & 4.9 & \multicolumn{2}{c}{$\cdots$} & \multicolumn{2}{c}{$\cdots$} & \multicolumn{2}{c}{$\cdots$} & 1.319 &/33 & 4.766 &/31 & \multicolumn{2}{c}{$\cdots$} & 3.6 \\
J0557$+$1551 & 2.56 & 0.72 & 102.6 & 4.8 & \multicolumn{2}{c}{$\cdots$} & \multicolumn{2}{c}{$\cdots$} & \multicolumn{2}{c}{$\cdots$} & 1.402 &/40 & 1.464 &/14 & \multicolumn{2}{c}{$\cdots$} & 4.6 \\
J0605$+$3757 & 2.73 & 0.47 & 20.9 & 55.7 & \multicolumn{2}{c}{$\cdots$} & \multicolumn{2}{c}{$\cdots$} & 1.205 &/20 & 1.778 &/23 & \multicolumn{2}{c}{$\cdots$} & \multicolumn{2}{c}{$\cdots$} & 3.4 \\
J0610$-$2100 & 3.86 & 1.23 & 61.3 & 0.3 & \multicolumn{2}{c}{$\cdots$} & \multicolumn{2}{c}{$\cdots$} & 0.894 &/37 & 1.319 &/36 & \multicolumn{2}{c}{$\cdots$} & \multicolumn{2}{c}{$\cdots$} & 3.4 \\
J0613$-$0200 & 3.06 & 0.96 & 38.8 & 1.2 & \multicolumn{2}{c}{$\cdots$} & \multicolumn{2}{c}{$\cdots$} & 0.111 &/165 & 0.654 &/165 & \multicolumn{2}{c}{$\cdots$} & \multicolumn{2}{c}{$\cdots$} & 15.0 \\
J0614$-$3329 & 3.15 & 1.74 & 37.1 & 53.6 & \multicolumn{2}{c}{$\cdots$} & \multicolumn{2}{c}{$\cdots$} & 0.554 &/29 & 1.044 &/26 & \multicolumn{2}{c}{$\cdots$} & \multicolumn{2}{c}{$\cdots$} & 2.4 \\
J0636$+$5128 & 2.87 & 0.34 & 11.1 & 0.1 & \multicolumn{2}{c}{$\cdots$} & \multicolumn{2}{c}{$\cdots$} & 0.269 &/72 & 0.596 &/73 & \multicolumn{2}{c}{$\cdots$} & \multicolumn{2}{c}{$\cdots$} & 6.3 \\
J0645$+$5158 & 8.85 & 0.49 & 18.2 & $\cdots$ & \multicolumn{2}{c}{$\cdots$} & \multicolumn{2}{c}{$\cdots$} & 0.299 &/98 & 0.823 &/100 & \multicolumn{2}{c}{$\cdots$} & \multicolumn{2}{c}{$\cdots$} & 8.9 \\
J0709$+$0458 & 34.43 & 38.02 & 44.3 & 4.4 & \multicolumn{2}{c}{$\cdots$} & \multicolumn{2}{c}{$\cdots$} & \multicolumn{2}{c}{$\cdots$} & 3.105 &/50 & 7.486 &/40 & \multicolumn{2}{c}{$\cdots$} & 4.6 \\
J0740$+$6620 & 2.89 & 1.22 & 15.0 & 4.8 & \multicolumn{2}{c}{$\cdots$} & \multicolumn{2}{c}{$\cdots$} & 0.489 &/127 & 0.792 &/165 & \multicolumn{2}{c}{$\cdots$} & \multicolumn{2}{c}{$\cdots$} & 6.3 \\
J0931$-$1902 & 4.64 & 0.36 & 41.5 & $\cdots$ & \multicolumn{2}{c}{$\cdots$} & \multicolumn{2}{c}{$\cdots$} & 1.020 &/84 & 1.840 &/83 & \multicolumn{2}{c}{$\cdots$} & \multicolumn{2}{c}{$\cdots$} & 7.1 \\
J1012$+$5307 & 5.26 & 1.71 & 8.9 & 0.6 & \multicolumn{2}{c}{$\cdots$} & \multicolumn{2}{c}{$\cdots$} & 0.406 &/169 & 0.781 &/175 & \multicolumn{2}{c}{$\cdots$} & \multicolumn{2}{c}{$\cdots$} & 15.5 \\
J1012$-$4235 & 3.10 & 0.66 & 71.7 & 38.0 & \multicolumn{2}{c}{$\cdots$} & \multicolumn{2}{c}{$\cdots$} & 1.129 &/18 & 2.081 &/27 & \multicolumn{2}{c}{$\cdots$} & \multicolumn{2}{c}{$\cdots$} & 3.4 \\
J1022$+$1001 & 16.45 & 4.33 & 9.4 & 7.8 & \multicolumn{2}{c}{$\cdots$} & 0.140 &/15 & \multicolumn{2}{c}{$\cdots$} & 0.428 &/50 & 0.605 &/35 & \multicolumn{2}{c}{$\cdots$} & 5.6 \\
J1024$-$0719 & 5.16 & 1.86 & 8.4 & $\cdots$ & \multicolumn{2}{c}{$\cdots$} & \multicolumn{2}{c}{$\cdots$} & 0.644 &/120 & 1.132 &/124 & \multicolumn{2}{c}{$\cdots$} & \multicolumn{2}{c}{$\cdots$} & 10.5 \\
J1125$+$7819 & 4.20 & 0.69 & 11.2 & 15.4 & \multicolumn{2}{c}{$\cdots$} & \multicolumn{2}{c}{$\cdots$} & 1.118 &/75 & 2.175 &/73 & \multicolumn{2}{c}{$\cdots$} & \multicolumn{2}{c}{$\cdots$} & 6.3 \\
J1312$+$0051 & 4.23 & 1.75 & 15.3 & 38.5 & \multicolumn{2}{c}{$\cdots$} & \multicolumn{2}{c}{$\cdots$} & \multicolumn{2}{c}{$\cdots$} & 2.012 &/45 & 2.319 &/32 & \multicolumn{2}{c}{$\cdots$} & 4.6 \\
J1453$+$1902 & 5.79 & 1.17 & 14.1 & $\cdots$ & \multicolumn{2}{c}{$\cdots$} & 1.211 &/54 & \multicolumn{2}{c}{$\cdots$} & 2.426 &/68 & 7.343 &/1 & \multicolumn{2}{c}{$\cdots$} & 7.0 \\
J1455$-$3330 & 7.99 & 2.43 & 13.6 & 76.2 & \multicolumn{2}{c}{$\cdots$} & \multicolumn{2}{c}{$\cdots$} & 1.224 &/149 & 2.100 &/145 & \multicolumn{2}{c}{$\cdots$} & \multicolumn{2}{c}{$\cdots$} & 15.7 \\
J1600$-$3053 & 3.60 & 0.95 & 52.3 & 14.3 & \multicolumn{2}{c}{$\cdots$} & \multicolumn{2}{c}{$\cdots$} & 0.285 &/144 & 0.253 &/148 & \multicolumn{2}{c}{$\cdots$} & 0.966 &/23 & 12.5 \\
J1614$-$2230 & 3.15 & 0.96 & 34.5 & 8.7 & \multicolumn{2}{c}{$\cdots$} & \multicolumn{2}{c}{$\cdots$} & 0.397 &/127 & 0.677 &/142 & \multicolumn{2}{c}{$\cdots$} & \multicolumn{2}{c}{$\cdots$} & 11.5 \\
J1630$+$3734 & 3.32 & 1.07 & 14.1 & 12.5 & \multicolumn{2}{c}{$\cdots$} & \multicolumn{2}{c}{$\cdots$} & 0.586 &/29 & 1.080 &/33 & \multicolumn{2}{c}{$\cdots$} & \multicolumn{2}{c}{$\cdots$} & 3.5 \\
J1640$+$2224 & 3.16 & 0.28 & 18.5 & 175.5 & \multicolumn{2}{c}{$\cdots$} & 0.057 &/264 & \multicolumn{2}{c}{$\cdots$} & 0.428 &/283 & \multicolumn{2}{c}{$\cdots$} & \multicolumn{2}{c}{$\cdots$} & 15.5 \\
J1643$-$1224 & 4.62 & 1.85 & 62.3 & 147.0 & \multicolumn{2}{c}{$\cdots$} & \multicolumn{2}{c}{$\cdots$} & 0.306 &/161 & 0.567 &/164 & \multicolumn{2}{c}{$\cdots$} & 2.012 &/23 & 15.7 \\
J1705$-$1903 & 2.48 & 2.15 & 57.5 & 0.2 & \multicolumn{2}{c}{$\cdots$} & \multicolumn{2}{c}{$\cdots$} & 0.288 &/32 & 0.192 &/31 & \multicolumn{2}{c}{$\cdots$} & \multicolumn{2}{c}{$\cdots$} & 3.7 \\
J1713$+$0747 & 4.57 & 0.85 & 16.0 & 67.8 & \multicolumn{2}{c}{$\cdots$} & \multicolumn{2}{c}{$\cdots$} & 0.195 &/161 & 0.089 &/676 & 0.080 &/263 & 0.364 &/27 & 15.5 \\
J1719$-$1438 & 5.79 & 0.80 & 36.8 & 0.1 & \multicolumn{2}{c}{$\cdots$} & \multicolumn{2}{c}{$\cdots$} & 0.958 &/35 & 1.474 &/37 & \multicolumn{2}{c}{$\cdots$} & \multicolumn{2}{c}{$\cdots$} & 3.4 \\
J1730$-$2304 & 8.12 & 2.02 & 9.6 & $\cdots$ & \multicolumn{2}{c}{$\cdots$} & \multicolumn{2}{c}{$\cdots$} & 0.561 &/37 & 1.187 &/36 & \multicolumn{2}{c}{$\cdots$} & \multicolumn{2}{c}{$\cdots$} & 3.4 \\
J1738$+$0333 & 5.85 & 2.41 & 33.8 & 0.4 & \multicolumn{2}{c}{$\cdots$} & \multicolumn{2}{c}{$\cdots$} & \multicolumn{2}{c}{$\cdots$} & 0.563 &/99 & 0.776 &/77 & \multicolumn{2}{c}{$\cdots$} & 10.7 \\
J1741$+$1351 & 3.75 & 3.02 & 24.2 & 16.3 & \multicolumn{2}{c}{$\cdots$} & 0.154 &/90 & \multicolumn{2}{c}{$\cdots$} & 0.338 &/112 & 0.139 &/10 & \multicolumn{2}{c}{$\cdots$} & 11.0 \\
J1744$-$1134 & 4.07 & 0.89 & 3.1 & $\cdots$ & \multicolumn{2}{c}{$\cdots$} & \multicolumn{2}{c}{$\cdots$} & 0.161 &/158 & 0.258 &/159 & \multicolumn{2}{c}{$\cdots$} & \multicolumn{2}{c}{$\cdots$} & 15.7 \\
J1745$+$1017 & 2.65 & 0.26 & 24.0 & 0.7 & \multicolumn{2}{c}{$\cdots$} & \multicolumn{2}{c}{$\cdots$} & \multicolumn{2}{c}{$\cdots$} & 0.703 &/47 & 1.498 &/43 & \multicolumn{2}{c}{$\cdots$} & 4.5 \\
J1747$-$4036 & 1.65 & 1.31 & 153.0 & $\cdots$ & \multicolumn{2}{c}{$\cdots$} & \multicolumn{2}{c}{$\cdots$} & 1.092 &/94 & 1.285 &/93 & \multicolumn{2}{c}{$\cdots$} & \multicolumn{2}{c}{$\cdots$} & 8.1 \\
J1751$-$2857 & 3.91 & 1.12 & 42.8 & 110.7 & \multicolumn{2}{c}{$\cdots$} & \multicolumn{2}{c}{$\cdots$} & 1.825 &/25 & 1.824 &/38 & \multicolumn{2}{c}{$\cdots$} & \multicolumn{2}{c}{$\cdots$} & 3.5 \\
J1802$-$2124 & 12.65 & 7.29 & 149.6 & 0.7 & \multicolumn{2}{c}{$\cdots$} & \multicolumn{2}{c}{$\cdots$} & 0.955 &/35 & 0.978 &/33 & \multicolumn{2}{c}{$\cdots$} & \multicolumn{2}{c}{$\cdots$} & 3.5 \\
J1811$-$2405 & 2.66 & 1.34 & 60.6 & 6.3 & \multicolumn{2}{c}{$\cdots$} & \multicolumn{2}{c}{$\cdots$} & 0.375 &/38 & 0.657 &/37 & \multicolumn{2}{c}{$\cdots$} & \multicolumn{2}{c}{$\cdots$} & 3.5 \\
J1832$-$0836 & 2.72 & 0.83 & 28.2 & $\cdots$ & \multicolumn{2}{c}{$\cdots$} & \multicolumn{2}{c}{$\cdots$} & 0.605 &/85 & 0.585 &/88 & \multicolumn{2}{c}{$\cdots$} & \multicolumn{2}{c}{$\cdots$} & 7.1 \\
J1843$-$1113 & 1.85 & 0.96 & 60.0 & $\cdots$ & \multicolumn{2}{c}{$\cdots$} & \multicolumn{2}{c}{$\cdots$} & 0.578 &/38 & 0.580 &/39 & \multicolumn{2}{c}{$\cdots$} & \multicolumn{2}{c}{$\cdots$} & 3.5 \\
J1853$+$1303 & 4.09 & 0.87 & 30.6 & 115.7 & \multicolumn{2}{c}{$\cdots$} & 0.369 &/90 & \multicolumn{2}{c}{$\cdots$} & 0.691 &/97 & \multicolumn{2}{c}{$\cdots$} & \multicolumn{2}{c}{$\cdots$} & 9.1 \\
B1855$+$09\phantom{....} & 5.36 & 1.78 & 13.3 & 12.3 & \multicolumn{2}{c}{$\cdots$} & 0.227 &/136 & \multicolumn{2}{c}{$\cdots$} & 0.241 &/146 & \multicolumn{2}{c}{$\cdots$} & \multicolumn{2}{c}{$\cdots$} & 15.6 \\
J1903$+$0327 & 2.15 & 1.88 & 297.6 & 95.2 & \multicolumn{2}{c}{$\cdots$} & \multicolumn{2}{c}{$\cdots$} & \multicolumn{2}{c}{$\cdots$} & 0.561 &/98 & 0.591 &/96 & 2.004 &/12 & 10.7 \\
J1909$-$3744 & 2.95 & 1.40 & 10.4 & 1.5 & \multicolumn{2}{c}{$\cdots$} & \multicolumn{2}{c}{$\cdots$} & 0.072 &/154 & 0.131 &/390 & \multicolumn{2}{c}{$\cdots$} & 0.226 &/31 & 15.5 \\
J1910$+$1256 & 4.98 & 0.97 & 38.1 & 58.5 & \multicolumn{2}{c}{$\cdots$} & \multicolumn{2}{c}{$\cdots$} & \multicolumn{2}{c}{$\cdots$} & 0.406 &/105 & 0.791 &/96 & \multicolumn{2}{c}{$\cdots$} & 11.4 \\
J1911$+$1347 & 4.63 & 1.69 & 31.0 & $\cdots$ & \multicolumn{2}{c}{$\cdots$} & 0.635 &/60 & \multicolumn{2}{c}{$\cdots$} & 0.192 &/68 & \multicolumn{2}{c}{$\cdots$} & \multicolumn{2}{c}{$\cdots$} & 7.0 \\
J1918$-$0642 & 7.65 & 2.57 & 26.6 & 10.9 & \multicolumn{2}{c}{$\cdots$} & \multicolumn{2}{c}{$\cdots$} & 0.533 &/155 & 1.000 &/162 & \multicolumn{2}{c}{$\cdots$} & \multicolumn{2}{c}{$\cdots$} & 15.5 \\
J1923$+$2515 & 3.79 & 0.96 & 18.9 & $\cdots$ & \multicolumn{2}{c}{$\cdots$} & 0.282 &/73 & \multicolumn{2}{c}{$\cdots$} & 1.055 &/92 & \multicolumn{2}{c}{$\cdots$} & \multicolumn{2}{c}{$\cdots$} & 9.0 \\
B1937$+$21\phantom{....} & 1.56 & 10.51 & 71.1 & $\cdots$ & \multicolumn{2}{c}{$\cdots$} & \multicolumn{2}{c}{$\cdots$} & 0.008 &/161 & 0.015 &/279 & 0.022 &/98 & \multicolumn{2}{c}{$\cdots$} & 15.9 \\
J1944$+$0907 & 5.19 & 1.73 & 24.4 & $\cdots$ & \multicolumn{2}{c}{$\cdots$} & 0.307 &/80 & \multicolumn{2}{c}{$\cdots$} & 0.991 &/103 & 1.568 &/12 & \multicolumn{2}{c}{$\cdots$} & 12.5 \\
J1946$+$3417 & 3.17 & 0.31 & 110.2 & 27.0 & \multicolumn{2}{c}{$\cdots$} & \multicolumn{2}{c}{$\cdots$} & \multicolumn{2}{c}{$\cdots$} & 0.507 &/67 & 0.730 &/60 & \multicolumn{2}{c}{$\cdots$} & 5.7 \\
B1953$+$29\phantom{....} & 6.13 & 2.97 & 104.5 & 117.3 & \multicolumn{2}{c}{$\cdots$} & 0.321 &/78 & \multicolumn{2}{c}{$\cdots$} & 0.948 &/97 & 2.076 &/5 & \multicolumn{2}{c}{$\cdots$} & 11.1 \\
J2010$-$1323 & 5.22 & 0.48 & 22.2 & $\cdots$ & \multicolumn{2}{c}{$\cdots$} & \multicolumn{2}{c}{$\cdots$} & 0.430 &/125 & 1.069 &/123 & \multicolumn{2}{c}{$\cdots$} & \multicolumn{2}{c}{$\cdots$} & 10.5 \\
J2017$+$0603 & 2.90 & 0.80 & 23.9 & 2.2 & \multicolumn{2}{c}{$\cdots$} & 0.218 &/6 & \multicolumn{2}{c}{$\cdots$} & 0.464 &/85 & 0.595 &/60 & \multicolumn{2}{c}{$\cdots$} & 8.3 \\
J2033$+$1734 & 5.95 & 1.12 & 25.1 & 56.3 & \multicolumn{2}{c}{$\cdots$} & 0.233 &/59 & \multicolumn{2}{c}{$\cdots$} & 1.377 &/67 & \multicolumn{2}{c}{$\cdots$} & \multicolumn{2}{c}{$\cdots$} & 7.0 \\
J2043$+$1711 & 2.38 & 0.52 & 20.7 & 1.5 & \multicolumn{2}{c}{$\cdots$} & 0.096 &/205 & \multicolumn{2}{c}{$\cdots$} & 0.299 &/227 & \multicolumn{2}{c}{$\cdots$} & \multicolumn{2}{c}{$\cdots$} & 9.1 \\
J2124$-$3358 & 4.93 & 2.06 & 4.6 & $\cdots$ & \multicolumn{2}{c}{$\cdots$} & \multicolumn{2}{c}{$\cdots$} & 0.809 &/35 & 1.656 &/40 & \multicolumn{2}{c}{$\cdots$} & \multicolumn{2}{c}{$\cdots$} & 3.5 \\
J2145$-$0750 & 16.05 & 2.98 & 8.9 & 6.8 & \multicolumn{2}{c}{$\cdots$} & \multicolumn{2}{c}{$\cdots$} & 0.322 &/145 & 0.688 &/145 & \multicolumn{2}{c}{$\cdots$} & \multicolumn{2}{c}{$\cdots$} & 15.5 \\
J2214$+$3000 & 3.12 & 1.47 & 22.5 & 0.4 & \multicolumn{2}{c}{$\cdots$} & \multicolumn{2}{c}{$\cdots$} & \multicolumn{2}{c}{$\cdots$} & 0.844 &/91 & 1.203 &/61 & \multicolumn{2}{c}{$\cdots$} & 8.8 \\
J2229$+$2643 & 2.98 & 0.15 & 22.7 & 93.0 & \multicolumn{2}{c}{$\cdots$} & 0.280 &/69 & \multicolumn{2}{c}{$\cdots$} & 1.273 &/76 & \multicolumn{2}{c}{$\cdots$} & \multicolumn{2}{c}{$\cdots$} & 7.0 \\
J2234$+$0611 & 3.58 & 1.20 & 10.8 & 32.0 & \multicolumn{2}{c}{$\cdots$} & 0.449 &/55 & \multicolumn{2}{c}{$\cdots$} & 0.271 &/65 & \multicolumn{2}{c}{$\cdots$} & \multicolumn{2}{c}{$\cdots$} & 6.5 \\
J2234$+$0944 & 3.63 & 2.01 & 17.8 & 0.4 & \multicolumn{2}{c}{$\cdots$} & \multicolumn{2}{c}{$\cdots$} & \multicolumn{2}{c}{$\cdots$} & 0.444 &/62 & 0.840 &/60 & \multicolumn{2}{c}{$\cdots$} & 7.1 \\
J2302$+$4442 & 5.19 & 1.39 & 13.8 & 125.9 & \multicolumn{2}{c}{$\cdots$} & \multicolumn{2}{c}{$\cdots$} & 1.200 &/101 & 2.569 &/97 & \multicolumn{2}{c}{$\cdots$} & \multicolumn{2}{c}{$\cdots$} & 8.1 \\
J2317$+$1439 & 3.45 & 0.24 & 21.9 & 2.5 & 0.084 &/71 & 0.079 &/274 & \multicolumn{2}{c}{$\cdots$} & 0.714 &/241 & \multicolumn{2}{c}{$\cdots$} & \multicolumn{2}{c}{$\cdots$} & 15.6 \\
J2322$+$2057 & 4.81 & 0.97 & 13.4 & $\cdots$ & \multicolumn{2}{c}{$\cdots$} & 0.356 &/56 & \multicolumn{2}{c}{$\cdots$} & 1.148 &/58 & 1.481 &/8 & \multicolumn{2}{c}{$\cdots$} & 5.4 \\
\multicolumn{5}{r}{Nominal scaling factors\tablenotemark{b} for ASP/GASP:}
  & \multicolumn{2}{c}{0.58}
  & \multicolumn{2}{c}{0.45}
  & \multicolumn{2}{c}{0.80}
  & \multicolumn{2}{c}{0.80}
  & \multicolumn{2}{c}{0.80}
  & \multicolumn{2}{c}{$\cdots$}
  & \\
\multicolumn{5}{r}{GUPPI/PUPPI:}
  & \multicolumn{2}{c}{0.71}
  & \multicolumn{2}{c}{0.49}
  & \multicolumn{2}{c}{1.34}
  & \multicolumn{2}{c}{2.49}
  & \multicolumn{2}{c}{2.14}
  & \multicolumn{2}{c}{$\cdots$}
  & \\
\multicolumn{5}{r}{YUPPI:}
  & \multicolumn{2}{c}{$\cdots$}
  & \multicolumn{2}{c}{$\cdots$}
  & \multicolumn{2}{c}{$\cdots$}
  & \multicolumn{2}{c}{2.83}
  & \multicolumn{2}{c}{$\cdots$}
  & \multicolumn{2}{c}{4.12}
  & \\
\enddata
\tablenotetext{a}{Original narrowband TOA uncertainties were scaled by their
bandwidth-time product $\left( \frac{\Delta \nu}{100\,\mathrm{MHz}}
\frac{\tau}{1800\,\mathrm{s}} \right)^{1/2}$ to remove variation due to
different instrument bandwidths and integration times.}
\tablenotetext{b}{TOA uncertainties can be rescaled to the nominal full
instrumental bandwidth by dividing by these scaling factors.}
\end{deluxetable*}

\clearpage

\startlongtable
\begin{deluxetable*}{cr|rrrrrr|cc|ccc|c}
\tabletypesize{\footnotesize}
\tablecaption{Summary of Timing Model Fits\tablenotemark{a}\label{tab:models}}
\tablehead{
    \colhead{\phantom{XX}Source\phantom{XX}} &
    \colhead{\phantom{X}Number\phantom{X}} &
    \multicolumn{6}{c}{\phantom{XX}Number of Fit Parameters\tablenotemark{b}\phantom{XX}} &
    \multicolumn{2}{c}{\phantom{XX}rms\tablenotemark{c} ($\mu$s)\phantom{XX}} &
    \multicolumn{3}{c}{\phantom{XX}Red Noise\tablenotemark{d}\phantom{XX}} &
    \colhead{\phantom{XX}Figure\phantom{XX}} \\
    \cline{3-8} \cline{9-10} \cline{11-13}
    \colhead{\phantom{XXSourceXX}} &
    \colhead{of TOAs} &
    \colhead{S\phantom{X}} &
    \colhead{A\phantom{X}} &
    \colhead{B\phantom{X}} &
    \colhead{DM} &
    \colhead{FD} &
    \colhead{J\phantom{X}} &
    \colhead{Full} &
    \colhead{White} &
    \colhead{$A_{\mathrm{red}}$} &
    \colhead{$\gamma_{\mathrm{red}}$} &
    \colhead{log$_{10}B$} &
    \colhead{Number}}
\startdata
\multirow{2}{*}{J0023$+$0923} & 15896 & 3 & 5 & 10 & 92 & 4 & 1 & 0.326 & $\cdots$ & $\cdots$ & $\cdots$ & 0.24 & \multirow{2}{*}{\ref{fig:summary-J0023+0923}} \\
 & 824 & 3 & 5 & 10 & 89 & 0 & 1/2 & 0.320 & $\cdots$ & $\cdots$ & $\cdots$ & -0.11 &   \\
\hline 
\multirow{2}{*}{J0030$+$0451} & 19579 & 3 & 5 & 0 & 289 & 4 & 2 & 0.856 & 0.251 & 0.003 & -4.7 & $>$2 & \multirow{2}{*}{\ref{fig:summary-J0030+0451}} \\
 & 727 & 3 & 5 & 0 & 289 & 0 & 2/3 & 0.794 & 0.263 & 0.003 & -4.9 & $>$2 &   \\
\hline 
\multirow{2}{*}{J0340$+$4130} & 11093 & 3 & 5 & 0 & 108 & 2 & 1 & 0.597 & $\cdots$ & $\cdots$ & $\cdots$ & -0.22 & \multirow{2}{*}{\ref{fig:summary-J0340+4130}} \\
 & 228 & 3 & 5 & 0 & 108 & 0 & 1/2 & 0.591 & $\cdots$ & $\cdots$ & $\cdots$ & -0.12 &   \\
\hline 
\multirow{2}{*}{J0406$+$3039} & 2446 & 3 & 5 & 5 & 39 & 1 & 1 & 0.176 & $\cdots$ & $\cdots$ & $\cdots$ & -0.09 & \multirow{2}{*}{\ref{fig:summary-J0406+3039}} \\
 & 71 & 3 & 5 & 5 & 39 & 0 & 1/2 & 0.300 & $\cdots$ & $\cdots$ & $\cdots$ & -0.21 &   \\
\hline 
\multirow{2}{*}{J0437$-$4715} & 5830 & 3 & 5 & 7 & 30 & 3 & 1 & 0.186 & 0.110 & 0.027 & -0.1 & $>$2 & \multirow{2}{*}{\ref{fig:summary-J0437-4715}} \\
 & 117 & 3 & 5 & 6 & 31 & 1 & 1/2 & 0.123 & $\cdots$ & $\cdots$ & $\cdots$ & 0.27 &   \\
\hline 
\multirow{2}{*}{J0509$+$0856} & 2169 & 3 & 5 & 5 & 38 & 0 & 1 & 0.695 & $\cdots$ & $\cdots$ & $\cdots$ & 0.73 & \multirow{2}{*}{\ref{fig:summary-J0509+0856}} \\
 & 66 & 3 & 5 & 5 & 37 & 0 & 1/2 & 0.760 & $\cdots$ & $\cdots$ & $\cdots$ & 0.17 &   \\
\hline 
\multirow{2}{*}{J0557$+$1551} & 525 & 3 & 5 & 5 & 41 & 0 & 1 & 0.353 & $\cdots$ & $\cdots$ & $\cdots$ & 0.01 & \multirow{2}{*}{\ref{fig:summary-J0557+1551}} \\
 & 47 & 3 & 5 & 5 & 43 & 0 & 1/2 & 0.233 & $\cdots$ & $\cdots$ & $\cdots$ & -0.13 &   \\
\hline 
\multirow{2}{*}{J0605$+$3757} & 554 & 3 & 5 & 5 & 26 & 0 & 1 & 1.153 & $\cdots$ & $\cdots$ & $\cdots$ & -0.11 & \multirow{2}{*}{\ref{fig:summary-J0605+3757}} \\
 & 47 & 3 & 5 & 5 & 29 & 0 & 1/2 & 1.125 & $\cdots$ & $\cdots$ & $\cdots$ & -0.08 &   \\
\hline 
\multirow{2}{*}{J0610$-$2100} & 4885 & 3 & 5 & 5 & 38 & 4 & 1 & 1.703 & 1.059 & 0.229 & -2.6 & $>$2 & \multirow{2}{*}{\ref{fig:summary-J0610-2100}} \\
 & 214 & 3 & 5 & 5 & 38 & 2 & 1/2 & 2.162 & 1.025 & 0.105 & -4.5 & $>$2 &   \\
\hline 
\multirow{2}{*}{J0613$-$0200} & 17124 & 3 & 5 & 8 & 174 & 2 & 1 & 0.749 & 0.168 & 0.022 & -2.9 & $>$2 & \multirow{2}{*}{\ref{fig:summary-J0613-0200}} \\
 & 423 & 3 & 5 & 8 & 174 & 0 & 1/2 & 0.704 & 0.151 & 0.013 & -3.2 & $>$2 &   \\
\hline 
\multirow{2}{*}{J0614$-$3329} & 1714 & 3 & 5 & 5 & 30 & 1 & 1 & 0.276 & $\cdots$ & $\cdots$ & $\cdots$ & -0.08 & \multirow{2}{*}{\ref{fig:summary-J0614-3329}} \\
 & 56 & 3 & 5 & 5 & 30 & 0 & 1/2 & 0.350 & $\cdots$ & $\cdots$ & $\cdots$ & -0.06 &   \\
\hline 
\multirow{2}{*}{J0636$+$5128} & 32222 & 3 & 5 & 8 & 77 & 1 & 1 & 0.664 & $\cdots$ & $\cdots$ & $\cdots$ & -0.14 & \multirow{2}{*}{\ref{fig:summary-J0636+5128}} \\
 & 1221 & 3 & 5 & 8 & 79 & 0 & 1/2 & 0.645 & $\cdots$ & $\cdots$ & $\cdots$ & 0.24 &   \\
\hline 
\multirow{2}{*}{J0645$+$5158} & 17670 & 3 & 5 & 0 & 113 & 2 & 1 & 1.592 & $\cdots$ & $\cdots$ & $\cdots$ & 0.49 & \multirow{2}{*}{\ref{fig:summary-J0645+5158}} \\
 & 289 & 3 & 5 & 0 & 114 & 0 & 1/2 & 0.164 & $\cdots$ & $\cdots$ & $\cdots$ & 1.55 &   \\
\hline 
\multirow{2}{*}{J0709$+$0458} & 3030 & 3 & 5 & 5 & 51 & 1 & 1 & 1.165 & $\cdots$ & $\cdots$ & $\cdots$ & 0.01 & \multirow{2}{*}{\ref{fig:summary-J0709+0458}} \\
 & 91 & 3 & 5 & 5 & 52 & 0 & 1/2 & 1.151 & $\cdots$ & $\cdots$ & $\cdots$ & 0.00 &   \\
\hline 
\multirow{2}{*}{J0740$+$6620} & 13401 & 3 & 5 & 8 & 109 & 3 & 1 & 0.286 & $\cdots$ & $\cdots$ & $\cdots$ & -0.18 & \multirow{2}{*}{\ref{fig:summary-J0740+6620}} \\
 & 360 & 3 & 5 & 8 & 109 & 0 & 1/2 & 0.296 & $\cdots$ & $\cdots$ & $\cdots$ & -0.22 &   \\
\hline 
\multirow{2}{*}{J0931$-$1902} & 5473 & 3 & 5 & 0 & 91 & 0 & 1 & 0.415 & $\cdots$ & $\cdots$ & $\cdots$ & -0.29 & \multirow{2}{*}{\ref{fig:summary-J0931-1902}} \\
 & 190 & 3 & 5 & 0 & 91 & 0 & 1/2 & 0.382 & $\cdots$ & $\cdots$ & $\cdots$ & -0.11 &   \\
\hline 
\multirow{2}{*}{J1012$+$5307} & 25837 & 3 & 5 & 7 & 177 & 4 & 1 & 0.925 & 0.289 & 0.219 & -0.9 & $>$2 & \multirow{2}{*}{\ref{fig:summary-J1012+5307}} \\
 & 628 & 3 & 5 & 7 & 171 & 3 & 1/2 & 0.967 & 0.224 & 0.241 & -0.8 & $>$2 &   \\
\hline 
\multirow{2}{*}{J1012$-$4235} & 797 & 3 & 5 & 5 & 28 & 0 & 1 & 0.708 & $\cdots$ & $\cdots$ & $\cdots$ & -0.09 & \multirow{2}{*}{\ref{fig:summary-J1012-4235}} \\
 & 65 & 3 & 5 & 7 & 35 & 0 & 1/2 & 0.623 & $\cdots$ & $\cdots$ & $\cdots$ & -0.11 &   \\
\hline 
\multirow{2}{*}{J1022$+$1001} & 3978 & 3 & 5 & 6 & 55 & 5 & 2 & 2.835 & $\cdots$ & $\cdots$ & $\cdots$ & 0.15 & \multirow{2}{*}{\ref{fig:summary-J1022+1001}} \\
 & 116 & 3 & 5 & 6 & 56 & 0 & 2/3 & 3.050 & $\cdots$ & $\cdots$ & $\cdots$ & 0.09 &   \\
\hline 
\multirow{2}{*}{J1024$-$0719} & 12635 & 3 & 5 & 1 & 134 & 5 & 1 & 0.239 & $\cdots$ & $\cdots$ & $\cdots$ & -0.16 & \multirow{2}{*}{\ref{fig:summary-J1024-0719}} \\
 & 288 & 3 & 5 & 1 & 134 & 0 & 1/2 & 0.245 & $\cdots$ & $\cdots$ & $\cdots$ & -0.22 &   \\
\hline 
\multirow{2}{*}{J1125$+$7819} & 8723 & 3 & 5 & 6 & 78 & 2 & 1 & 0.688 & $\cdots$ & $\cdots$ & $\cdots$ & 0.10 & \multirow{2}{*}{\ref{fig:summary-J1125+7819}} \\
 & 206 & 3 & 5 & 6 & 79 & 0 & 1/2 & 0.619 & $\cdots$ & $\cdots$ & $\cdots$ & -0.11 &   \\
\hline 
\multirow{2}{*}{J1312$+$0051} & 1705 & 3 & 5 & 5 & 48 & 0 & 1 & 0.731 & $\cdots$ & $\cdots$ & $\cdots$ & -0.14 & \multirow{2}{*}{\ref{fig:summary-J1312+0051}} \\
 & 76 & 3 & 5 & 5 & 49 & 0 & 1/2 & 0.632 & $\cdots$ & $\cdots$ & $\cdots$ & -0.11 &   \\
\hline 
\multirow{2}{*}{J1453$+$1902} & 2551 & 3 & 5 & 0 & 69 & 0 & 1 & 0.783 & $\cdots$ & $\cdots$ & $\cdots$ & -0.18 & \multirow{2}{*}{\ref{fig:summary-J1453+1902}} \\
 & 122 & 3 & 5 & 0 & 71 & 0 & 1/2 & 0.895 & $\cdots$ & $\cdots$ & $\cdots$ & -0.09 &   \\
\hline 
\multirow{2}{*}{J1455$-$3330} & 10818 & 3 & 5 & 6 & 157 & 1 & 1 & 0.735 & $\cdots$ & $\cdots$ & $\cdots$ & -0.14 & \multirow{2}{*}{\ref{fig:summary-J1455-3330}} \\
 & 357 & 3 & 5 & 6 & 156 & 0 & 1/2 & 0.663 & $\cdots$ & $\cdots$ & $\cdots$ & -0.17 &   \\
\hline 
\multirow{2}{*}{J1600$-$3053} & 22955 & 3 & 5 & 9 & 181 & 3 & 2 & 0.271 & 0.202 & 0.037 & -0.9 & $>$2 & \multirow{2}{*}{\ref{fig:summary-J1600-3053}} \\
 & 481 & 3 & 5 & 8 & 182 & 1 & 2/3 & 0.338 & 0.145 & 0.052 & -1.3 & $>$2 &   \\
\hline 
\multirow{2}{*}{J1614$-$2230} & 18445 & 3 & 5 & 8 & 150 & 0 & 1 & 0.354 & 0.211 & 0.010 & -3.0 & $>$2 & \multirow{2}{*}{\ref{fig:summary-J1614-2230}} \\
 & 367 & 3 & 5 & 8 & 151 & 0 & 1/2 & 0.311 & 0.202 & 0.010 & -2.6 & $>$2 &   \\
\hline 
\multirow{2}{*}{J1630$+$3734} & 1815 & 3 & 5 & 8 & 34 & 1 & 1 & 0.271 & $\cdots$ & $\cdots$ & $\cdots$ & -0.09 & \multirow{2}{*}{\ref{fig:summary-J1630+3734}} \\
 & 71 & 3 & 5 & 7 & 38 & 0 & 1/2 & 0.473 & $\cdots$ & $\cdots$ & $\cdots$ & -0.04 &   \\
\hline 
\multirow{2}{*}{J1640$+$2224} & 14066 & 3 & 5 & 9 & 284 & 4 & 1 & 0.200 & $\cdots$ & $\cdots$ & $\cdots$ & -0.06 & \multirow{2}{*}{\ref{fig:summary-J1640+2224}} \\
 & 609 & 3 & 5 & 8 & 283 & 0 & 1/2 & 0.181 & $\cdots$ & $\cdots$ & $\cdots$ & -0.16 &   \\
\hline 
\multirow{2}{*}{J1643$-$1224} & 22144 & 3 & 5 & 7 & 193 & 5 & 2 & 2.335 & 0.898 & 0.543 & -0.7 & $>$2 & \multirow{2}{*}{\ref{fig:summary-J1643-1224}} \\
 & 478 & 3 & 5 & 6 & 186 & 1 & 2/3 & 2.598 & 0.675 & 0.538 & -1.2 & $>$2 &   \\
\hline 
\multirow{2}{*}{J1705$-$1903} & 9871 & 3 & 5 & 10 & 43 & 2 & 1 & 1.124 & 0.226 & 0.217 & -0.4 & $>$2 & \multirow{2}{*}{\ref{fig:summary-J1705-1903}} \\
 & 253 & 3 & 5 & 10 & 45 & 0 & 1/2 & 1.231 & 0.252 & 0.296 & -0.8 & $>$2 &   \\
\hline 
\multirow{2}{*}{J1713$+$0747} & 59389 & 3 & 5 & 8 & 398 & 4 & 5 & 0.201 & 0.095 & 0.011 & -2.2 & $>$2 & \multirow{2}{*}{\ref{fig:summary-J1713+0747}} \\
 & 1495 & 3 & 5 & 8 & 398 & 5 & 5/6 & 0.199 & 0.093 & 0.009 & -2.8 & $>$2 &   \\
\hline 
\multirow{2}{*}{J1719$-$1438} & 6356 & 3 & 5 & 5 & 42 & 1 & 1 & 2.471 & $\cdots$ & $\cdots$ & $\cdots$ & -0.10 & \multirow{2}{*}{\ref{fig:summary-J1719-1438}} \\
 & 463 & 3 & 5 & 6 & 41 & 0 & 1/2 & 2.703 & $\cdots$ & $\cdots$ & $\cdots$ & -0.08 &   \\
\hline 
\multirow{2}{*}{J1730$-$2304} & 4870 & 3 & 5 & 0 & 44 & 2 & 1 & 0.277 & $\cdots$ & $\cdots$ & $\cdots$ & 0.61 & \multirow{2}{*}{\ref{fig:summary-J1730-2304}} \\
 & 93 & 3 & 5 & 0 & 43 & 1 & 1/2 & 0.310 & $\cdots$ & $\cdots$ & $\cdots$ & -0.01 &   \\
\hline 
\multirow{2}{*}{J1738$+$0333} & 8790 & 3 & 5 & 5 & 104 & 1 & 1 & 0.262 & $\cdots$ & $\cdots$ & $\cdots$ & 1.57 & \multirow{2}{*}{\ref{fig:summary-J1738+0333}} \\
 & 336 & 3 & 5 & 5 & 102 & 0 & 1/2 & 1.022 & 0.259 & 0.000 & -6.9 & $>$2 &   \\
\hline 
\multirow{2}{*}{J1741$+$1351} & 5582 & 3 & 5 & 9 & 113 & 2 & 2 & 0.182 & $\cdots$ & $\cdots$ & $\cdots$ & 0.13 & \multirow{2}{*}{\ref{fig:summary-J1741+1351}} \\
 & 208 & 3 & 5 & 9 & 104 & 0 & 1/2 & 0.169 & $\cdots$ & $\cdots$ & $\cdots$ & 0.13 &   \\
\hline 
\multirow{2}{*}{J1744$-$1134} & 17745 & 3 & 5 & 0 & 169 & 4 & 1 & 0.505 & 0.287 & 0.022 & -2.5 & $>$2 & \multirow{2}{*}{\ref{fig:summary-J1744-1134}} \\
 & 433 & 3 & 5 & 0 & 169 & 0 & 1/2 & 1.042 & 0.271 & 0.006 & -4.7 & $>$2 &   \\
\hline 
\multirow{2}{*}{J1745$+$1017} & 3017 & 3 & 5 & 5 & 49 & 1 & 1 & 12.519 & 0.850 & 1.167 & -2.5 & $>$2 & \multirow{2}{*}{\ref{fig:summary-J1745+1017}} \\
 & 91 & 3 & 5 & 5 & 49 & 0 & 1/2 & 9.907 & 0.379 & 1.095 & -2.4 & $>$2 &   \\
\hline 
\multirow{2}{*}{J1747$-$4036} & 11055 & 3 & 5 & 0 & 106 & 1 & 1 & 3.214 & 1.480 & 0.268 & -2.4 & $>$2 & \multirow{2}{*}{\ref{fig:summary-J1747-4036}} \\
 & 222 & 3 & 5 & 0 & 105 & 0 & 1/2 & 2.802 & 0.756 & 0.364 & -2.0 & $>$2 &   \\
\hline 
\multirow{2}{*}{J1751$-$2857} & 2025 & 3 & 5 & 5 & 44 & 1 & 1 & 0.560 & $\cdots$ & $\cdots$ & $\cdots$ & -0.11 & \multirow{2}{*}{\ref{fig:summary-J1751-2857}} \\
 & 89 & 3 & 5 & 5 & 45 & 0 & 1/2 & 0.530 & $\cdots$ & $\cdots$ & $\cdots$ & -0.06 &   \\
\hline 
\multirow{2}{*}{J1802$-$2124} & 6796 & 3 & 5 & 5 & 45 & 1 & 1 & 2.247 & 0.878 & 0.544 & -1.5 & $>$2 & \multirow{2}{*}{\ref{fig:summary-J1802-2124}} \\
 & 126 & 3 & 5 & 7 & 45 & 5 & 1/2 & 0.974 & $\cdots$ & $\cdots$ & $\cdots$ & 1.09 &   \\
\hline 
\multirow{2}{*}{J1811$-$2405} & 5266 & 3 & 5 & 7 & 46 & 1 & 1 & 0.220 & $\cdots$ & $\cdots$ & $\cdots$ & -0.09 & \multirow{2}{*}{\ref{fig:summary-J1811-2405}} \\
 & 103 & 3 & 5 & 7 & 46 & 0 & 1/2 & 0.223 & $\cdots$ & $\cdots$ & $\cdots$ & -0.10 &   \\
\hline 
\multirow{2}{*}{J1832$-$0836} & 7739 & 3 & 5 & 0 & 93 & 1 & 1 & 0.214 & $\cdots$ & $\cdots$ & $\cdots$ & 0.07 & \multirow{2}{*}{\ref{fig:summary-J1832-0836}} \\
 & 207 & 3 & 5 & 0 & 93 & 1 & 1/2 & 0.208 & $\cdots$ & $\cdots$ & $\cdots$ & -0.07 &   \\
\hline 
\multirow{2}{*}{J1843$-$1113} & 4595 & 3 & 5 & 0 & 40 & 1 & 1 & 0.220 & $\cdots$ & $\cdots$ & $\cdots$ & -0.09 & \multirow{2}{*}{\ref{fig:summary-J1843-1113}} \\
 & 103 & 3 & 5 & 0 & 40 & 0 & 1/2 & 0.235 & $\cdots$ & $\cdots$ & $\cdots$ & 0.12 &   \\
\hline 
\multirow{2}{*}{J1853$+$1303} & 4570 & 3 & 5 & 8 & 98 & 0 & 1 & 0.337 & 0.163 & 0.034 & -2.0 & $>$2 & \multirow{2}{*}{\ref{fig:summary-J1853+1303}} \\
 & 184 & 3 & 5 & 8 & 96 & 0 & 1/2 & 0.354 & 0.171 & 0.088 & -5.4 & $>$2 &   \\
\hline 
\multirow{2}{*}{B1855$+$09} & 7758 & 3 & 5 & 7 & 147 & 3 & 1 & 0.829 & 0.330 & 0.011 & -3.7 & $>$2 & \multirow{2}{*}{\ref{fig:summary-B1855+09}} \\
 & 364 & 3 & 5 & 7 & 149 & 0 & 1/2 & 0.845 & 0.352 & 0.016 & -3.4 & $>$2 &   \\
\hline 
\multirow{2}{*}{J1903$+$0327} & 6856 & 3 & 5 & 8 & 114 & 3 & 2 & 3.737 & 0.711 & 0.627 & -1.4 & $>$2 & \multirow{2}{*}{\ref{fig:summary-J1903+0327}} \\
 & 226 & 3 & 5 & 8 & 114 & 0 & 2/3 & 1.670 & 0.207 & 0.318 & -1.1 & $>$2 &   \\
\hline 
\multirow{2}{*}{J1909$-$3744} & 35037 & 3 & 5 & 9 & 325 & 1 & 2 & 0.303 & 0.066 & 0.002 & -4.4 & $>$2 & \multirow{2}{*}{\ref{fig:summary-J1909-3744}} \\
 & 833 & 3 & 5 & 9 & 327 & 0 & 2/3 & 0.287 & 0.062 & 0.003 & -4.2 & $>$2 &   \\
\hline 
\multirow{2}{*}{J1910$+$1256} & 6486 & 3 & 5 & 6 & 114 & 2 & 1 & 0.413 & $\cdots$ & $\cdots$ & $\cdots$ & 0.06 & \multirow{2}{*}{\ref{fig:summary-J1910+1256}} \\
 & 216 & 3 & 5 & 6 & 114 & 0 & 1/2 & 0.438 & $\cdots$ & $\cdots$ & $\cdots$ & -0.16 &   \\
\hline 
\multirow{2}{*}{J1911$+$1347} & 3786 & 3 & 5 & 0 & 69 & 2 & 1 & 0.074 & $\cdots$ & $\cdots$ & $\cdots$ & -0.19 & \multirow{2}{*}{\ref{fig:summary-J1911+1347}} \\
 & 126 & 3 & 5 & 0 & 69 & 0 & 1/2 & 0.088 & $\cdots$ & $\cdots$ & $\cdots$ & -0.26 &   \\
\hline 
\multirow{2}{*}{J1918$-$0642} & 18875 & 3 & 5 & 8 & 166 & 1 & 1 & 0.338 & $\cdots$ & $\cdots$ & $\cdots$ & 1.67 & \multirow{2}{*}{\ref{fig:summary-J1918-0642}} \\
 & 487 & 3 & 5 & 7 & 168 & 0 & 1/2 & 0.398 & 0.296 & 0.032 & -2.0 & $>$2 &   \\
\hline 
\multirow{2}{*}{J1923$+$2515} & 4001 & 3 & 5 & 0 & 90 & 1 & 1 & 0.280 & $\cdots$ & $\cdots$ & $\cdots$ & 0.07 & \multirow{2}{*}{\ref{fig:summary-J1923+2515}} \\
 & 170 & 3 & 5 & 0 & 92 & 0 & 1/2 & 0.214 & $\cdots$ & $\cdots$ & $\cdots$ & -0.25 &   \\
\hline 
\multirow{2}{*}{B1937$+$21} & 23023 & 3 & 5 & 0 & 262 & 5 & 3 & 5.774 & 5.774 & 0.026 & -4.2 & $>$2 & \multirow{2}{*}{\ref{fig:summary-B1937+21}} \\
 & 660 & 3 & 5 & 0 & 267 & 2 & 3/4 & 5.958 & 0.057 & 0.037 & -3.6 & $>$2 &   \\
\hline 
\multirow{2}{*}{J1944$+$0907} & 5328 & 3 & 5 & 0 & 104 & 2 & 2 & 0.461 & $\cdots$ & $\cdots$ & $\cdots$ & 0.47 & \multirow{2}{*}{\ref{fig:summary-J1944+0907}} \\
 & 180 & 3 & 5 & 0 & 95 & 0 & 1/2 & 0.411 & $\cdots$ & $\cdots$ & $\cdots$ & -0.03 &   \\
\hline 
\multirow{2}{*}{J1946$+$3417} & 4743 & 3 & 5 & 8 & 72 & 1 & 1 & 1.387 & 0.305 & 0.286 & -1.4 & $>$2 & \multirow{2}{*}{\ref{fig:summary-J1946+3417}} \\
 & 129 & 3 & 5 & 8 & 71 & 0 & 1/2 & 0.740 & 0.162 & 0.154 & -1.8 & $>$2 &   \\
\hline 
\multirow{2}{*}{B1953$+$29} & 5126 & 3 & 5 & 6 & 98 & 2 & 2 & 1.158 & 0.378 & 0.216 & -1.4 & $>$2 & \multirow{2}{*}{\ref{fig:summary-B1953+29}} \\
 & 173 & 3 & 5 & 6 & 94 & 0 & 1/2 & 1.450 & 0.279 & 0.195 & -2.2 & $>$2 &   \\
\hline 
\multirow{2}{*}{J2010$-$1323} & 17077 & 3 & 5 & 0 & 142 & 1 & 1 & 0.274 & $\cdots$ & $\cdots$ & $\cdots$ & 0.01 & \multirow{2}{*}{\ref{fig:summary-J2010-1323}} \\
 & 350 & 3 & 5 & 0 & 143 & 0 & 1/2 & 0.271 & $\cdots$ & $\cdots$ & $\cdots$ & -0.07 &   \\
\hline 
\multirow{2}{*}{J2017$+$0603} & 3512 & 3 & 5 & 7 & 92 & 0 & 2 & 0.109 & $\cdots$ & $\cdots$ & $\cdots$ & -0.17 & \multirow{2}{*}{\ref{fig:summary-J2017+0603}} \\
 & 154 & 3 & 5 & 7 & 93 & 0 & 2/3 & 0.126 & $\cdots$ & $\cdots$ & $\cdots$ & -0.22 &   \\
\hline 
\multirow{2}{*}{J2033$+$1734} & 3847 & 3 & 5 & 5 & 68 & 2 & 1 & 0.468 & $\cdots$ & $\cdots$ & $\cdots$ & -0.09 & \multirow{2}{*}{\ref{fig:summary-J2033+1734}} \\
 & 133 & 3 & 5 & 5 & 70 & 0 & 1/2 & 0.399 & $\cdots$ & $\cdots$ & $\cdots$ & -0.22 &   \\
\hline 
\multirow{2}{*}{J2043$+$1711} & 7398 & 3 & 5 & 8 & 228 & 2 & 1 & 0.115 & $\cdots$ & $\cdots$ & $\cdots$ & 0.59 & \multirow{2}{*}{\ref{fig:summary-J2043+1711}} \\
 & 459 & 3 & 5 & 8 & 223 & 0 & 1/2 & 0.103 & $\cdots$ & $\cdots$ & $\cdots$ & 0.75 &   \\
\hline 
\multirow{2}{*}{J2124$-$3358} & 4982 & 3 & 5 & 0 & 40 & 1 & 1 & 0.338 & $\cdots$ & $\cdots$ & $\cdots$ & -0.12 & \multirow{2}{*}{\ref{fig:summary-J2124-3358}} \\
 & 104 & 3 & 5 & 0 & 39 & 0 & 1/2 & 0.531 & $\cdots$ & $\cdots$ & $\cdots$ & -0.17 &   \\
\hline 
\multirow{2}{*}{J2145$-$0750} & 18675 & 3 & 5 & 7 & 161 & 4 & 1 & 0.799 & 0.644 & 0.111 & -0.5 & $>$2 & \multirow{2}{*}{\ref{fig:summary-J2145-0750}} \\
 & 400 & 3 & 5 & 7 & 161 & 0 & 1/2 & 1.045 & 0.358 & 0.068 & -2.4 & $>$2 &   \\
\hline 
\multirow{2}{*}{J2214$+$3000} & 7425 & 3 & 5 & 5 & 96 & 2 & 1 & 0.407 & $\cdots$ & $\cdots$ & $\cdots$ & -0.13 & \multirow{2}{*}{\ref{fig:summary-J2214+3000}} \\
 & 293 & 3 & 5 & 8 & 102 & 4 & 1/2 & 0.456 & $\cdots$ & $\cdots$ & $\cdots$ & 1.48 &   \\
\hline 
\multirow{2}{*}{J2229$+$2643} & 3716 & 3 & 5 & 6 & 76 & 2 & 1 & 0.280 & $\cdots$ & $\cdots$ & $\cdots$ & 0.02 & \multirow{2}{*}{\ref{fig:summary-J2229+2643}} \\
 & 151 & 3 & 5 & 6 & 77 & 5 & 1/2 & 0.231 & $\cdots$ & $\cdots$ & $\cdots$ & -0.07 &   \\
\hline 
\multirow{2}{*}{J2234$+$0611} & 3566 & 3 & 5 & 8 & 66 & 2 & 1 & 0.200 & 0.071 & 0.038 & -1.2 & $>$2 & \multirow{2}{*}{\ref{fig:summary-J2234+0611}} \\
 & 133 & 3 & 5 & 8 & 66 & 0 & 1/2 & 0.061 & $\cdots$ & $\cdots$ & $\cdots$ & 1.90 &   \\
\hline 
\multirow{2}{*}{J2234$+$0944} & 7535 & 3 & 5 & 5 & 72 & 2 & 1 & 0.197 & $\cdots$ & $\cdots$ & $\cdots$ & -0.17 & \multirow{2}{*}{\ref{fig:summary-J2234+0944}} \\
 & 245 & 3 & 5 & 5 & 74 & 0 & 1/2 & 0.796 & 0.209 & 0.176 & -0.1 & $>$2 &   \\
\hline 
\multirow{2}{*}{J2302$+$4442} & 10211 & 3 & 5 & 7 & 108 & 3 & 1 & 0.764 & $\cdots$ & $\cdots$ & $\cdots$ & -0.05 & \multirow{2}{*}{\ref{fig:summary-J2302+4442}} \\
 & 236 & 3 & 5 & 7 & 108 & 0 & 1/2 & 0.710 & $\cdots$ & $\cdots$ & $\cdots$ & -0.03 &   \\
\hline 
\multirow{2}{*}{J2317$+$1439} & 13942 & 3 & 5 & 6 & 303 & 3 & 2 & 0.345 & $\cdots$ & $\cdots$ & $\cdots$ & -0.09 & \multirow{2}{*}{\ref{fig:summary-J2317+1439}} \\
 & 711 & 3 & 5 & 6 & 309 & 0 & 2/3 & 0.690 & $\cdots$ & $\cdots$ & $\cdots$ & 0.01 &   \\
\hline 
\multirow{2}{*}{J2322$+$2057} & 3088 & 3 & 5 & 0 & 59 & 1 & 2 & 0.255 & $\cdots$ & $\cdots$ & $\cdots$ & -0.25 & \multirow{2}{*}{\ref{fig:summary-J2322+2057}} \\
 & 130 & 3 & 5 & 0 & 59 & 0 & 2/3 & 0.262 & $\cdots$ & $\cdots$ & $\cdots$ & -0.13 &   \\
\enddata
\tablenotetext{a}{First line for each pulsar is from the narrowband analysis and the second line is from the wideband analysis.}
\tablenotetext{b}{Fit parameters: S=spin; A=astrometry; B=binary;
DM=dispersion measure; FD=frequency dependence; J=jump (two numbers indicate wideband data with JUMPs/DMJUMPs).}
\tablenotetext{c}{Weighted root-mean-square of epoch-averaged post-fit
timing residuals, calculated using the procedure described in Appendix D
of \nineyr. For sources with red noise, the ``Full'' rms value includes
the red noise contribution, while the ``White'' rms does not.}
\tablenotetext{d}{Red noise parameters: $A_{\mathrm{red}}$ = amplitude of
red noise spectrum at $f$=1~yr$^{-1}$ measured in $\mu$s yr$^{1/2}$;
$\gamma_{\mathrm{red}}$ = spectral index; $B$ = Bayes factor (``$>$2''
indicates a Bayes factor larger than our threshold log$_{10}$B~$>$~2, but
which could not be estimated using the Savage-Dickey ratio).  See
Eqn.~\ref{eqn:rn_spec} and Appendix~C of \nineyr\ for details.}
\end{deluxetable*}

\clearpage

\bibliographystyle{aasjournal}
\bibliography{nano15y.bbl}

\begin{thebibliography}{}
\expandafter\ifx\csname natexlab\endcsname\relax\def\natexlab#1{#1}\fi
\providecommand{\url}[1]{\href{#1}{#1}}
\providecommand{\dodoi}[1]{doi:~\href{http://doi.org/#1}{\nolinkurl{#1}}}
\providecommand{\doeprint}[1]{\href{http://ascl.net/#1}{\nolinkurl{http://ascl.net/#1}}}
\providecommand{\doarXiv}[1]{\href{https://arxiv.org/abs/#1}{\nolinkurl{https://arxiv.org/abs/#1}}}

\bibitem[{{Afzal} {et~al.}(2023){Afzal}, {Agazie}, {Alam}, Anumarlapudi,
  Archibald, Arzoumanian, \& Others}]{NG15cosmo}
{Afzal}, A., {Agazie}, G., {Alam}, M.~F., {et~al.} 2023

\bibitem[{{Agazie} {et~al.}(2023{\natexlab{a}}){Agazie}, {Alam}, Anumarlapudi,
  Archibald, Arzoumanian, \& Others}]{NG15interp}
{Agazie}, G., {Alam}, M.~F., Anumarlapudi, A., {et~al.} 2023{\natexlab{a}}

\bibitem[{{Agazie} {et~al.}(2023{\natexlab{b}}){Agazie}, Anumarlapudi,
  Archibald, Arzoumanian, \& Others}]{NG15gwb}
{Agazie}, G., Anumarlapudi, A., Archibald, A.~M., Arzoumanian, Z., \& Others.
  2023{\natexlab{b}}, \dodoi{10.3847/2041-8213/acdac6}

\bibitem[{{Agazie} {et~al.}(2023{\natexlab{c}}){Agazie}, Anumarlapudi,
  Archibald, \& Others}]{NG15detchar}
{Agazie}, G., Anumarlapudi, A., Archibald, A.~M., \& Others.
  2023{\natexlab{c}}, \dodoi{10.3847/2041-8213/acda88}

\bibitem[{{Alam} {et~al.}(2021{\natexlab{a}}){Alam}, {Arzoumanian}, {Baker},
  {Blumer}, {Bohler}, {Brazier}, {Brook}, {Burke-Spolaor}, {Caballero},
  {Camuccio}, {Chamberlain}, {Chatterjee}, {Cordes}, {Cornish}, {Crawford},
  {Cromartie}, {Decesar}, {Demorest}, {Dolch}, {Ellis}, {Ferdman}, {Ferrara},
  {Fiore}, {Fonseca}, {Garcia}, {Garver-Daniels}, {Gentile}, {Good},
  {Gusdorff}, {Halmrast}, {Hazboun}, {Islo}, {Jennings}, {Jessup}, {Jones},
  {Kaiser}, {Kaplan}, {Kelley}, {Key}, {Lam}, {Lazio}, {Lorimer}, {Luo},
  {Lynch}, {Madison}, {Maraccini}, {McLaughlin}, {Mingarelli}, {Ng}, {Nguyen},
  {Nice}, {Pennucci}, {Pol}, {Ramette}, {Ransom}, {Ray}, {Shapiro-Albert},
  {Siemens}, {Simon}, {Spiewak}, {Stairs}, {Stinebring}, {Stovall}, {Swiggum},
  {Taylor}, {Tripepi}, {Vallisneri}, {Vigeland}, {Witt}, {Zhu}, \& {Nanograv
  Collaboration}}]{Alam21}
{Alam}, M.~F., {Arzoumanian}, Z., {Baker}, P.~T., {et~al.} 2021{\natexlab{a}},
  \apjs, 252, 4, \dodoi{10.3847/1538-4365/abc6a0}

\bibitem[{{Alam} {et~al.}(2021{\natexlab{b}}){Alam}, {Arzoumanian}, {Baker},
  {Blumer}, {Bohler}, {Brazier}, {Brook}, {Burke-Spolaor}, {Caballero},
  {Camuccio}, {Chamberlain}, {Chatterjee}, {Cordes}, {Cornish}, {Crawford},
  {Cromartie}, {Decesar}, {Demorest}, {Dolch}, {Ellis}, {Ferdman}, {Ferrara},
  {Fiore}, {Fonseca}, {Garcia}, {Garver-Daniels}, {Gentile}, {Good},
  {Gusdorff}, {Halmrast}, {Hazboun}, {Islo}, {Jennings}, {Jessup}, {Jones},
  {Kaiser}, {Kaplan}, {Kelley}, {Key}, {Lam}, {Lazio}, {Lorimer}, {Luo},
  {Lynch}, {Madison}, {Maraccini}, {McLaughlin}, {Mingarelli}, {Ng}, {Nguyen},
  {Nice}, {Pennucci}, {Pol}, {Ramette}, {Ransom}, {Ray}, {Shapiro-Albert},
  {Siemens}, {Simon}, {Spiewak}, {Stairs}, {Stinebring}, {Stovall}, {Swiggum},
  {Taylor}, {Tripepi}, {Vallisneri}, {Vigeland}, {Witt}, {Zhu}, \& {Nanograv
  Collaboration}}]{alam21wb}
---. 2021{\natexlab{b}}, \apjs, 252, 5, \dodoi{10.3847/1538-4365/abc6a1}

\bibitem[{{Alpar} {et~al.}(1982){Alpar}, {Cheng}, {Ruderman}, \&
  {Shaham}}]{acr+82}
{Alpar}, M.~A., {Cheng}, A.~F., {Ruderman}, M.~A., \& {Shaham}, J. 1982, \nat,
  300, 728, \dodoi{10.1038/300728a0}

\bibitem[{{Arzoumanian} {et~al.}(1996){Arzoumanian}, {Joshi}, {Rasio}, \&
  {Thorsett}}]{Arzoumanian1996}
{Arzoumanian}, Z., {Joshi}, K., {Rasio}, F.~A., \& {Thorsett}, S.~E. 1996, in
  Astronomical Society of the Pacific Conference Series, Vol. 105, IAU Colloq.
  160: Pulsars: Problems and Progress, ed. S.~{Johnston}, M.~A. {Walker}, \&
  M.~{Bailes}, 525--530, \dodoi{10.48550/arXiv.astro-ph/9605141}

\bibitem[{{Arzoumanian} {et~al.}(2015){Arzoumanian}, {Brazier},
  {Burke-Spolaor}, {Chamberlin}, {Chatterjee}, {Christy}, {Cordes}, {Cornish},
  {Crowter}, {Demorest}, {Dolch}, {Ellis}, {Ferdman}, {Fonseca},
  {Garver-Daniels}, {Gonzalez}, {Jenet}, {Jones}, {Jones}, {Kaspi}, {Koop},
  {Lam}, {Lazio}, {Levin}, {Lommen}, {Lorimer}, {Luo}, {Lynch}, {Madison},
  {McLaughlin}, {McWilliams}, {Nice}, {Palliyaguru}, {Pennucci}, {Ransom},
  {Siemens}, {Stairs}, {Stinebring}, {Stovall}, {Swiggum}, {Vallisneri}, {van
  Haasteren}, {Wang}, \& {Zhu}}]{Arzoumanian2015b}
{Arzoumanian}, Z., {Brazier}, A., {Burke-Spolaor}, S., {et~al.} 2015, \apj,
  813, 65, \dodoi{10.1088/0004-637X/813/1/65}

\bibitem[{{Arzoumanian} {et~al.}(2018){Arzoumanian}, {Brazier},
  {Burke-Spolaor}, {Chamberlin}, {Chatterjee}, {Christy}, {Cordes}, {Cornish},
  {Crawford}, {Thankful Cromartie}, {Crowter}, {DeCesar}, {Demorest}, {Dolch},
  {Ellis}, {Ferdman}, {Ferrara}, {Fonseca}, {Garver-Daniels}, {Gentile},
  {Halmrast}, {Huerta}, {Jenet}, {Jessup}, {Jones}, {Jones}, {Kaplan}, {Lam},
  {Lazio}, {Levin}, {Lommen}, {Lorimer}, {Luo}, {Lynch}, {Madison}, {Matthews},
  {McLaughlin}, {McWilliams}, {Mingarelli}, {Ng}, {Nice}, {Pennucci}, {Ransom},
  {Ray}, {Siemens}, {Simon}, {Spiewak}, {Stairs}, {Stinebring}, {Stovall},
  {Swiggum}, {Taylor}, {Vallisneri}, {van Haasteren}, {Vigeland}, \&
  {Zhu}}]{Arzoumanian2018a}
---. 2018, ArXiv e-prints.
\newblock \doarXiv{1801.01837}

\bibitem[{{Arzoumanian} {et~al.}(2020){Arzoumanian}, {Baker}, {Blumer},
  {B{\'e}csy}, {Brazier}, {Brook}, {Burke-Spolaor}, {Chatterjee}, {Chen},
  {Cordes}, {Cornish}, {Crawford}, {Cromartie}, {Decesar}, {Demorest}, {Dolch},
  {Ellis}, {Ferrara}, {Fiore}, {Fonseca}, {Garver-Daniels}, {Gentile}, {Good},
  {Hazboun}, {Holgado}, {Islo}, {Jennings}, {Jones}, {Kaiser}, {Kaplan},
  {Kelley}, {Key}, {Laal}, {Lam}, {Lazio}, {Lorimer}, {Luo}, {Lynch},
  {Madison}, {McLaughlin}, {Mingarelli}, {Ng}, {Nice}, {Pennucci}, {Pol},
  {Ransom}, {Ray}, {Shapiro-Albert}, {Siemens}, {Simon}, {Spiewak}, {Stairs},
  {Stinebring}, {Stovall}, {Sun}, {Swiggum}, {Taylor}, {Turner}, {Vallisneri},
  {Vigeland}, {Witt}, \& {Nanograv Collaboration}}]{arzoumanian2020}
{Arzoumanian}, Z., {Baker}, P.~T., {Blumer}, H., {et~al.} 2020, \apjl, 905,
  L34, \dodoi{10.3847/2041-8213/abd401}

\bibitem[{{Arzoumanian} {et~al.}(2021{\natexlab{a}}){Arzoumanian}, {Baker},
  {Blumer}, {B{\'e}csy}, {Brazier}, {Brook}, {Burke-Spolaor}, {Charisi},
  {Chatterjee}, {Chen}, {Cordes}, {Cornish}, {Crawford}, {Cromartie},
  {Decesar}, {Degan}, {Demorest}, {Dolch}, {Drachler}, {Ellis}, {Ferrara},
  {Fiore}, {Fonseca}, {Garver-Daniels}, {Gentile}, {Good}, {Hazboun},
  {Holgado}, {Islo}, {Jennings}, {Jones}, {Kaiser}, {Kaplan}, {Kelley}, {Key},
  {Laal}, {Lam}, {W. Lazio}, {Lorimer}, {Liu}, {Luo}, {Lynch}, {Madison},
  {McEwen}, {McLaughlin}, {Mingarelli}, {Ng}, {Nice}, {Olum}, {Pennucci},
  {Pol}, {Ransom}, {Ray}, {Romano}, {Sardesai}, {Shapiro-Albert}, {Siemens},
  {Simon}, {Siwek}, {Spiewak}, {Stairs}, {Stinebring}, {Stovall}, {Sun},
  {Swiggum}, {Taylor}, {Turner}, {Vallisneri}, {Vigeland}, {Wahl}, {Witt}, \&
  {NANOGRAV Collaboration}}]{arzoumanian2021}
---. 2021{\natexlab{a}}, \apjl, 923, L22, \dodoi{10.3847/2041-8213/ac401c}

\bibitem[{{Arzoumanian} {et~al.}(2021{\natexlab{b}}){Arzoumanian}, {Baker},
  {Blumer}, {B{\'e}csy}, {Brazier}, {Brook}, {Burke-Spolaor}, {Charisi},
  {Chatterjee}, {Chen}, {Cordes}, {Cornish}, {Crawford}, {Cromartie},
  {Decesar}, {Demorest}, {Dolch}, {Ellis}, {Ferrara}, {Fiore}, {Fonseca},
  {Garver-Daniels}, {Gentile}, {Good}, {Hazboun}, {Holgado}, {Islo},
  {Jennings}, {Jones}, {Kaiser}, {Kaplan}, {Kelley}, {Key}, {Laal}, {Lam},
  {Lazio}, {Lee}, {Lorimer}, {Luo}, {Lynch}, {Madison}, {McLaughlin},
  {Mingarelli}, {Mitridate}, {Ng}, {Nice}, {Pennucci}, {Pol}, {Ransom}, {Ray},
  {Shapiro-Albert}, {Siemens}, {Simon}, {Spiewak}, {Stairs}, {Stinebring},
  {Stovall}, {Sun}, {Swiggum}, {Taylor}, {Turner}, {Vallisneri}, {Vigeland},
  {Witt}, {Zurek}, \& {Nanograv Collaboration}}]{arzoumanian2021b}
---. 2021{\natexlab{b}}, \prl, 127, 251302,
  \dodoi{10.1103/PhysRevLett.127.251302}

\bibitem[{{Arzoumanian} {et~al.}(2023){Arzoumanian}, {Baker}, {Blecha},
  {Blumer}, {Brazier}, {Brook}, {Burke-Spolaor}, {B{\'e}csy}, {Casey-Clyde},
  {Charisi}, {Chatterjee}, {Chen}, {Cordes}, {Cornish}, {Crawford},
  {Cromartie}, {DeCesar}, {Demorest}, {Dolch}, {Drachler}, {Ellis}, {Ferrara},
  {Fiore}, {Fonseca}, {Freedman}, {Garver-Daniels}, {Gentile}, {Glaser},
  {Good}, {G{\"u}ltekin}, {Hazboun}, {Jennings}, {Johnson}, {Jones}, {Kaiser},
  {Kaplan}, {Kelley}, {Shapiro Key}, {Laal}, {Lam}, {Lamb}, {Lazio},
  {Lewandowska}, {Liu}, {Lorimer}, {Luo}, {Lynch}, {Madison}, {McEwen},
  {McLaughlin}, {Mingarelli}, {Ng}, {Nice}, {Ocker}, {Olum}, {Pennucci}, {Pol},
  {Ransom}, {Ray}, {Romano}, {Shapiro-Albert}, {Siemens}, {Simon}, {Siwek},
  {Spiewak}, {Stairs}, {Stinebring}, {Stovall}, {Swiggum}, {Sydnor}, {Taylor},
  {Turner}, {Vallisneri}, {Vigeland}, {Wahl}, {Walsh}, {Witt}, \&
  {Young}}]{arzoumanian2023}
{Arzoumanian}, Z., {Baker}, P.~T., {Blecha}, L., {et~al.} 2023, arXiv e-prints,
  arXiv:2301.03608, \dodoi{10.48550/arXiv.2301.03608}

\bibitem[{{Astropy Collaboration} {et~al.}(2022){Astropy Collaboration},
  {Price-Whelan}, {Lim}, {Earl}, {Starkman}, {Bradley}, {Shupe}, {Patil},
  {Corrales}, {Brasseur}, {N{"o}the}, {Donath}, {Tollerud}, {Morris},
  {Ginsburg}, {Vaher}, {Weaver}, {Tocknell}, {Jamieson}, {van Kerkwijk},
  {Robitaille}, {Merry}, {Bachetti}, {G{"u}nther}, {Aldcroft},
  {Alvarado-Montes}, {Archibald}, {B{'o}di}, {Bapat}, {Barentsen}, {Baz{'a}n},
  {Biswas}, {Boquien}, {Burke}, {Cara}, {Cara}, {Conroy}, {Conseil}, {Craig},
  {Cross}, {Cruz}, {D'Eugenio}, {Dencheva}, {Devillepoix}, {Dietrich},
  {Eigenbrot}, {Erben}, {Ferreira}, {Foreman-Mackey}, {Fox}, {Freij}, {Garg},
  {Geda}, {Glattly}, {Gondhalekar}, {Gordon}, {Grant}, {Greenfield}, {Groener},
  {Guest}, {Gurovich}, {Handberg}, {Hart}, {Hatfield-Dodds}, {Homeier},
  {Hosseinzadeh}, {Jenness}, {Jones}, {Joseph}, {Kalmbach}, {Karamehmetoglu},
  {Ka{l}uszy{'n}ski}, {Kelley}, {Kern}, {Kerzendorf}, {Koch}, {Kulumani},
  {Lee}, {Ly}, {Ma}, {MacBride}, {Maljaars}, {Muna}, {Murphy}, {Norman},
  {O'Steen}, {Oman}, {Pacifici}, {Pascual}, {Pascual-Granado}, {Patil},
  {Perren}, {Pickering}, {Rastogi}, {Roulston}, {Ryan}, {Rykoff}, {Sabater},
  {Sakurikar}, {Salgado}, {Sanghi}, {Saunders}, {Savchenko}, {Schwardt},
  {Seifert-Eckert}, {Shih}, {Jain}, {Shukla}, {Sick}, {Simpson},
  {Singanamalla}, {Singer}, {Singhal}, {Sinha}, {Sip{H{o}}cz}, {Spitler},
  {Stansby}, {Streicher}, {{{S}}umak}, {Swinbank}, {Taranu}, {Tewary},
  {Tremblay}, {Val-Borro}, {Van Kooten}, {Vasovi{'c}}, {Verma}, {de Miranda
  Cardoso}, {Williams}, {Wilson}, {Winkel}, {Wood-Vasey}, {Xue}, {Yoachim},
  {Zhang}, {Zonca}, \& {Astropy Project Contributors}}]{astropy:2022}
{Astropy Collaboration}, {Price-Whelan}, A.~M., {Lim}, P.~L., {et~al.} 2022,
  apj, 935, 167, \dodoi{10.3847/1538-4357/ac7c74}

\bibitem[{{Backer} {et~al.}(1982){Backer}, {Kulkarni}, {Heiles}, {Davis}, \&
  {Goss}}]{bkh+82}
{Backer}, D.~C., {Kulkarni}, S.~R., {Heiles}, C., {Davis}, M.~M., \& {Goss},
  W.~M. 1982, \nat, 300, 615, \dodoi{10.1038/300615a0}

\bibitem[{{Barr} {et~al.}(2013){Barr}, {Guillemot}, {Champion}, {Kramer},
  {Eatough}, {Lee}, {Verbiest}, {Bassa}, {Camilo}, {{\c{C}}elik}, {Cognard},
  {Ferrara}, {Freire}, {Janssen}, {Johnston}, {Keith}, {Lyne}, {Michelson},
  {Parkinson}, {Ransom}, {Ray}, {Stappers}, \& {Wood}}]{bgc+13}
{Barr}, E.~D., {Guillemot}, L., {Champion}, D.~J., {et~al.} 2013, \mnras, 429,
  1633, \dodoi{10.1093/mnras/sts449}

\bibitem[{{Blandford} \& {Teukolsky}(1976)}]{BlandfordTeukolsky1976}
{Blandford}, R., \& {Teukolsky}, S.~A. 1976, \apj, 205, 580,
  \dodoi{10.1086/154315}

\bibitem[{{Camilo} {et~al.}(2015){Camilo}, {Kerr}, {Ray}, {Ransom},
  {Sarkissian}, {Cromartie}, {Johnston}, {Reynolds}, {Wolff}, {Freire},
  {Bhattacharyya}, {Ferrara}, {Keith}, {Michelson}, {Saz Parkinson}, \&
  {Wood}}]{ckr+15}
{Camilo}, F., {Kerr}, M., {Ray}, P.~S., {et~al.} 2015, \apj, 810, 85,
  \dodoi{10.1088/0004-637X/810/2/85}

\bibitem[{{Chen} {et~al.}(2021){Chen}, {Caballero}, {Guo}, {Chalumeau}, {Liu},
  {Shaifullah}, {Lee}, {Babak}, {Desvignes}, {Parthasarathy}, {Hu}, {van der
  Wateren}, {Antoniadis}, {Bak Nielsen}, {Bassa}, {Berthereau}, {Burgay},
  {Champion}, {Cognard}, {Falxa}, {Ferdman}, {Freire}, {Gair}, {Graikou},
  {Guillemot}, {Jang}, {Janssen}, {Karuppusamy}, {Keith}, {Kramer}, {Liu},
  {Lyne}, {Main}, {McKee}, {Mickaliger}, {Perera}, {Perrodin}, {Petiteau},
  {Porayko}, {Possenti}, {Samajdar}, {Sanidas}, {Sesana}, {Speri}, {Stappers},
  {Theureau}, {Tiburzi}, {Vecchio}, {Verbiest}, {Wang}, {Wang}, \&
  {Xu}}]{ccg+21}
{Chen}, S., {Caballero}, R.~N., {Guo}, Y.~J., {et~al.} 2021, \mnras, 508, 4970,
  \dodoi{10.1093/mnras/stab2833}

\bibitem[{{CHIME/Pulsar Collaboration} {et~al.}(2021){CHIME/Pulsar
  Collaboration}, {Amiri}, {Bandura}, {Boyle}, {Brar}, {Cliche}, {Crowter},
  {Cubranic}, {Demorest}, {Denman}, {Dobbs}, {Dong}, {Fandino}, {Fonseca},
  {Good}, {Halpern}, {Hill}, {H{\"o}fer}, {Kaspi}, {Landecker}, {Leung}, {Lin},
  {Luo}, {Masui}, {McKee}, {Mena-Parra}, {Meyers}, {Michilli}, {Naidu},
  {Newburgh}, {Ng}, {Patel}, {Pinsonneault-Marotte}, {Ransom}, {Renard},
  {Scholz}, {Shaw}, {Sikora}, {Stairs}, {Tan}, {Tendulkar}, {Tretyakov},
  {Vanderlinde}, {Wang}, \& {Wang}}]{chimepulsaroverview}
{CHIME/Pulsar Collaboration}, {Amiri}, M., {Bandura}, K.~M., {et~al.} 2021,
  \apjs, 255, 5, \dodoi{10.3847/1538-4365/abfdcb}

\bibitem[{conda-forge Community(2015)}]{conda_forge_community_2015_4774216}
conda-forge Community. 2015, {The conda-forge Project: Community-based Software
  Distribution Built on the conda Package Format and Ecosystem},  Zenodo,
  \dodoi{10.5281/zenodo.4774216}

\bibitem[{{Cromartie} {et~al.}(2020){Cromartie}, {Fonseca}, {Ransom},
  {Demorest}, {Arzoumanian}, {Blumer}, {Brook}, {DeCesar}, {Dolch}, {Ellis},
  {Ferdman}, {Ferrara}, {Garver-Daniels}, {Gentile}, {Jones}, {Lam}, {Lorimer},
  {Lynch}, {McLaughlin}, {Ng}, {Nice}, {Pennucci}, {Spiewak}, {Stairs},
  {Stovall}, {Swiggum}, \& {Zhu}}]{cfr+20}
{Cromartie}, H.~T., {Fonseca}, E., {Ransom}, S.~M., {et~al.} 2020, Nature
  Astronomy, 4, 72, \dodoi{10.1038/s41550-019-0880-2}

\bibitem[{{Damour} \& {Deruelle}(1985)}]{Damour85}
{Damour}, T., \& {Deruelle}, N. 1985, Ann.~Inst.~Henri Poincar{\'e}
  Phys.~Th{\'e}or., Vol.~43, No.~1, p.~107 - 132, 43, 107

\bibitem[{{Damour} \& {Deruelle}(1986)}]{Damour86}
---. 1986, Ann.~Inst.~Henri Poincar{\'e} Phys.~Th{\'e}or., Vol.~44, No.~3,
  p.~263 - 292, 44, 263

\bibitem[{{Damour} \& {Taylor}(1992)}]{Damour1992}
{Damour}, T., \& {Taylor}, J.~H. 1992, \prd, 45, 1840,
  \dodoi{10.1103/PhysRevD.45.1840}

\bibitem[{{Demorest}(2007)}]{Demorest2007}
{Demorest}, P.~B. 2007, PhD thesis, University of California, Berkeley

\bibitem[{{Demorest}(2018)}]{nanopipe}
---. 2018, {nanopipe: Calibration and data reduction pipeline for pulsar
  timing}, Astrophysics Source Code Library.
\newblock \doeprint{1803.004}

\bibitem[{{Demorest} {et~al.}(2010){Demorest}, {Pennucci}, {Ransom}, {Roberts},
  \& {Hessels}}]{dem10}
{Demorest}, P.~B., {Pennucci}, T., {Ransom}, S.~M., {Roberts}, M.~S.~E., \&
  {Hessels}, J.~W.~T. 2010, \nat, 467, 1081, \dodoi{10.1038/nature09466}

\bibitem[{{Demorest} {et~al.}(2013){Demorest}, {Ferdman}, {Gonzalez}, {Nice},
  {Ransom}, {Stairs}, {Arzoumanian}, {Brazier}, {Burke-Spolaor}, {Chamberlin},
  {Cordes}, {Ellis}, {Finn}, {Freire}, {Giampanis}, {Jenet}, {Kaspi}, {Lazio},
  {Lommen}, {McLaughlin}, {Palliyaguru}, {Perrodin}, {Shannon}, {Siemens},
  {Stinebring}, {Swiggum}, \& {Zhu}}]{Demorest2013}
{Demorest}, P.~B., {Ferdman}, R.~D., {Gonzalez}, M.~E., {et~al.} 2013, \apj,
  762, 94, \dodoi{10.1088/0004-637X/762/2/94}

\bibitem[{{Desvignes} {et~al.}(2016){Desvignes}, {Caballero}, {Lentati},
  {Verbiest}, {Champion}, {Stappers}, {Janssen}, {Lazarus}, {Os{\l}owski},
  {Babak}, {Bassa}, {Brem}, {Burgay}, {Cognard}, {Gair}, {Graikou},
  {Guillemot}, {Hessels}, {Jessner}, {Jordan}, {Karuppusamy}, {Kramer},
  {Lassus}, {Lazaridis}, {Lee}, {Liu}, {Lyne}, {McKee}, {Mingarelli},
  {Perrodin}, {Petiteau}, {Possenti}, {Purver}, {Rosado}, {Sanidas}, {Sesana},
  {Shaifullah}, {Smits}, {Taylor}, {Theureau}, {Tiburzi}, {van Haasteren}, \&
  {Vecchio}}]{Desvignes2016}
{Desvignes}, G., {Caballero}, R.~N., {Lentati}, L., {et~al.} 2016, \mnras, 458,
  3341, \dodoi{10.1093/mnras/stw483}

\bibitem[{{Detweiler}(1979)}]{detweiler79}
{Detweiler}, S. 1979, \apj, 234, 1100, \dodoi{10.1086/157593}

\bibitem[{{DuPlain} {et~al.}(2008){DuPlain}, {Ransom}, {Demorest}, {Brandt},
  {Ford}, \& {Shelton}}]{DuPlain2008}
{DuPlain}, R., {Ransom}, S., {Demorest}, P., {et~al.} 2008, in \procspie, Vol.
  7019, Advanced Software and Control for Astronomy II, 70191D,
  \dodoi{10.1117/12.790003}

\bibitem[{Ellis \& van Haasteren(2017)}]{ptmcmc}
Ellis, J., \& van Haasteren, R. 2017, jellis18/PTMCMCSampler: Official Release,
  \dodoi{10.5281/zenodo.1037579}

\bibitem[{{Ellis} {et~al.}(2013){Ellis}, {Siemens}, \& {van
  Haasteren}}]{esvh13}
{Ellis}, J.~A., {Siemens}, X., \& {van Haasteren}, R. 2013, \apj, 769, 63,
  \dodoi{10.1088/0004-637X/769/1/63}

\bibitem[{{Ellis} {et~al.}(2019){Ellis}, {Vallisneri}, {Taylor}, \&
  {Baker}}]{enterprise}
{Ellis}, J.~A., {Vallisneri}, M., {Taylor}, S.~R., \& {Baker}, P.~T. 2019,
  {ENTERPRISE: Enhanced Numerical Toolbox Enabling a Robust PulsaR Inference
  SuitE}.
\newblock \doeprint{1912.015}

\bibitem[{{Faulkner} {et~al.}(2004){Faulkner}, {Stairs}, {Kramer}, {Lyne},
  {Hobbs}, {Possenti}, {Lorimer}, {Manchester}, {McLaughlin}, {D'Amico},
  {Camilo}, \& {Burgay}}]{fsk+04}
{Faulkner}, A.~J., {Stairs}, I.~H., {Kramer}, M., {et~al.} 2004, \mnras, 355,
  147, \dodoi{10.1111/j.1365-2966.2004.08310.x}

\bibitem[{{Fonseca} {et~al.}(2016){Fonseca}, {Pennucci}, {Ellis}, {Stairs},
  {Nice}, {Ransom}, {Demorest}, {Arzoumanian}, {Crowter}, {Dolch}, {Ferdman},
  {Gonzalez}, {Jones}, {Jones}, {Lam}, {Levin}, {McLaughlin}, {Stovall},
  {Swiggum}, \& {Zhu}}]{Fonseca2016}
{Fonseca}, E., {Pennucci}, T.~T., {Ellis}, J.~A., {et~al.} 2016, \apj, 832,
  167, \dodoi{10.3847/0004-637X/832/2/167}

\bibitem[{{Fonseca} {et~al.}(2021){Fonseca}, {Cromartie}, {Pennucci}, {Ray},
  {Kirichenko}, {Ransom}, {Demorest}, {Stairs}, {Arzoumanian}, {Guillemot},
  {Parthasarathy}, {Kerr}, {Cognard}, {Baker}, {Blumer}, {Brook}, {DeCesar},
  {Dolch}, {Dong}, {Ferrara}, {Fiore}, {Garver-Daniels}, {Good}, {Jennings},
  {Jones}, {Kaspi}, {Lam}, {Lorimer}, {Luo}, {McEwen}, {McKee}, {McLaughlin},
  {McMann}, {Meyers}, {Naidu}, {Ng}, {Nice}, {Pol}, {Radovan},
  {Shapiro-Albert}, {Tan}, {Tendulkar}, {Swiggum}, {Wahl}, \& {Zhu}}]{fon21}
{Fonseca}, E., {Cromartie}, H.~T., {Pennucci}, T.~T., {et~al.} 2021, \apjl,
  915, L12, \dodoi{10.3847/2041-8213/ac03b8}

\bibitem[{{Foster} \& {Backer}(1990)}]{fb90}
{Foster}, R.~S., \& {Backer}, D.~C. 1990, \apj, 361, 300,
  \dodoi{10.1086/169195}

\bibitem[{Freedman {et~al.}(2021)Freedman, Johnson, \&
  Glaser}]{enterprise_outliers}
Freedman, G., Johnson, A.~D., \& Glaser, J.~P. 2021, enterprise\_outliers.
\newblock \url{https://github.com/nanograv/enterprise\_outliers}

\bibitem[{{Freire} \& {Wex}(2010)}]{Freire2010}
{Freire}, P.~C.~C., \& {Wex}, N. 2010, \mnras, 409, 199,
  \dodoi{10.1111/j.1365-2966.2010.17319.x}

\bibitem[{{Goncharov} {et~al.}(2021){Goncharov}, {Reardon}, {Shannon}, {Zhu},
  {Thrane}, {Bailes}, {Bhat}, {Dai}, {Hobbs}, {Kerr}, {Manchester},
  {Os{\l}owski}, {Parthasarathy}, {Russell}, {Spiewak}, {Thyagarajan}, \&
  {Wang}}]{grs+21}
{Goncharov}, B., {Reardon}, D.~J., {Shannon}, R.~M., {et~al.} 2021, \mnras,
  502, 478, \dodoi{10.1093/mnras/staa3411}

\bibitem[{Granger \& Pérez(2021)}]{Jupyter2021}
Granger, B.~E., \& Pérez, F. 2021, Computing in Science \& Engineering, 23, 7,
  \dodoi{10.1109/MCSE.2021.3059263}

\bibitem[{{Hallinan} {et~al.}(2019){Hallinan}, {Ravi}, {Weinreb}, {Kocz},
  {Huang}, {Woody}, {Lamb}, {D'Addario}, {Catha}, {Law}, {Kulkarni}, {Phinney},
  {Eastwood}, {Bouman}, {McLaughlin}, {Ransom}, {Siemens}, {Cordes}, {Lynch},
  {Kaplan}, {Brazier}, {Bhatnagar}, {Myers}, {Walter}, \&
  {Gaensler}}]{Hallinan2019}
{Hallinan}, G., {Ravi}, V., {Weinreb}, S., {et~al.} 2019, in Bulletin of the
  American Astronomical Society, Vol.~51, 255,
  \dodoi{10.48550/arXiv.1907.07648}

\bibitem[{Harris {et~al.}(2020)Harris, Millman, van~der Walt, Gommers,
  Virtanen, Cournapeau, Wieser, Taylor, Berg, Smith, Kern, Picus, Hoyer, van
  Kerkwijk, Brett, Haldane, del R{\'{i}}o, Wiebe, Peterson,
  G{\'{e}}rard-Marchant, Sheppard, Reddy, Weckesser, Abbasi, Gohlke, \&
  Oliphant}]{NumPy}
Harris, C.~R., Millman, K.~J., van~der Walt, S.~J., {et~al.} 2020, Nature, 585,
  357, \dodoi{10.1038/s41586-020-2649-2}

\bibitem[{Hobbs \& Dai(2017)}]{hd17}
Hobbs, G., \& Dai, S. 2017, National Science Review, 4, 707,
  \dodoi{10.1093/nsr/nwx126}

\bibitem[{{Hobbs} \& {Edwards}(2012)}]{tempo2}
{Hobbs}, G., \& {Edwards}, R. 2012, {Tempo2: Pulsar Timing Package}.
\newblock \doeprint{1210.015}

\bibitem[{{Hobbs} {et~al.}(2010){Hobbs}, {Archibald}, {Arzoumanian}, {Backer},
  {Bailes}, {Bhat}, {Burgay}, {Burke-Spolaor}, {Champion}, {Cognard}, {Coles},
  {Cordes}, {Demorest}, {Desvignes}, {Ferdman}, {Finn}, {Freire}, {Gonzalez},
  {Hessels}, {Hotan}, {Janssen}, {Jenet}, {Jessner}, {Jordan}, {Kaspi},
  {Kramer}, {Kondratiev}, {Lazio}, {Lazaridis}, {Lee}, {Levin}, {Lommen},
  {Lorimer}, {Lynch}, {Lyne}, {Manchester}, {McLaughlin}, {Nice}, {Oslowski},
  {Pilia}, {Possenti}, {Purver}, {Ransom}, {Reynolds}, {Sanidas}, {Sarkissian},
  {Sesana}, {Shannon}, {Siemens}, {Stairs}, {Stappers}, {Stinebring},
  {Theureau}, {van Haasteren}, {van Straten}, {Verbiest}, {Yardley}, \&
  {You}}]{IPTA2010}
{Hobbs}, G., {Archibald}, A., {Arzoumanian}, Z., {et~al.} 2010, Classical and
  Quantum Gravity, 27, 084013, \dodoi{10.1088/0264-9381/27/8/084013}

\bibitem[{{Hotan} {et~al.}(2004){Hotan}, {van Straten}, \&
  {Manchester}}]{Hotan04}
{Hotan}, A.~W., {van Straten}, W., \& {Manchester}, R.~N. 2004, Proc. Astron.
  Soc. Aust., 21, 302, \dodoi{10.1071/AS04022}

\bibitem[{Hunter(2007)}]{matplotlib}
Hunter, J.~D. 2007, Computing In Science \& Engineering, 9, 90,
  \dodoi{10.1109/MCSE.2007.55}

\bibitem[{{Jennings} {et~al.}(2022){Jennings}, {Cordes}, {Chatterjee},
  {McLaughlin}, {Demorest}, {Arzoumanian}, {Baker}, {Blumer}, {Brook}, {Cohen},
  {Crawford}, {Cromartie}, {DeCesar}, {Dolch}, {Ferrara}, {Fonseca}, {Good},
  {Hazboun}, {Jones}, {Kaplan}, {Lam}, {Lazio}, {Lorimer}, {Luo}, {Lynch},
  {McKee}, {Madison}, {Meyers}, {Mingarelli}, {Nice}, {Pennucci}, {Perera},
  {Pol}, {Ransom}, {Ray}, {Shapiro-Albert}, {Siemens}, {Stairs}, {Stinebring},
  {Swiggum}, {Tan}, {Taylor}, {Vigeland}, \& {Witt}}]{Jennings2022}
{Jennings}, R.~J., {Cordes}, J.~M., {Chatterjee}, S., {et~al.} 2022, arXiv
  e-prints, arXiv:2210.12266, \dodoi{10.48550/arXiv.2210.12266}

\bibitem[{{Jones} {et~al.}(2017){Jones}, {McLaughlin}, {Lam}, {Cordes},
  {Levin}, {Chatterjee}, {Arzoumanian}, {Crowter}, {Demorest}, {Dolch},
  {Ellis}, {Ferdman}, {Fonseca}, {Gonzalez}, {Jones}, {Lazio}, {Nice},
  {Pennucci}, {Ransom}, {Stinebring}, {Stairs}, {Stovall}, {Swiggum}, \&
  {Zhu}}]{Jones2017}
{Jones}, M.~L., {McLaughlin}, M.~A., {Lam}, M.~T., {et~al.} 2017, \apj, 841,
  125, \dodoi{10.3847/1538-4357/aa73df}

\bibitem[{Kluyver {et~al.}(2016)Kluyver, Ragan-Kelley, P{\'e}rez, Granger,
  Bussonnier, Frederic, Kelley, Hamrick, Grout, Corlay, Ivanov, Avila, Abdalla,
  \& Willing}]{Kluyver2016jupyter}
Kluyver, T., Ragan-Kelley, B., P{\'e}rez, F., {et~al.} 2016, in Positioning and
  Power in Academic Publishing: Players, Agents and Agendas, ed. F.~Loizides \&
  B.~Schmidt, IOS Press, 87 -- 90

\bibitem[{{Kopeikin}(1995)}]{Kopeikin1995}
{Kopeikin}, S.~M. 1995, \apjl, 439, L5, \dodoi{10.1086/187731}

\bibitem[{{Kopeikin}(1996)}]{Kopeikin1996}
---. 1996, \apjl, 467, L93, \dodoi{10.1086/310201}

\bibitem[{Kurtzer {et~al.}(2021)Kurtzer, cclerget, Bauer, Kaneshiro, Trudgian,
  \& Godlove}]{singularity:2021}
Kurtzer, G.~M., cclerget, Bauer, M., {et~al.} 2021, hpcng/singularity:
  Singularity 3.7.3, v3.7.3,  Zenodo, \dodoi{10.5281/zenodo.4667718}

\bibitem[{Kurtzer {et~al.}(2017)Kurtzer, Sochat, \& Bauer}]{Singularity}
Kurtzer, G.~M., Sochat, V., \& Bauer, M.~W. 2017, PLOS ONE, 12, 1,
  \dodoi{10.1371/journal.pone.0177459}

\bibitem[{{Lam}(2017)}]{pypulse}
{Lam}, M.~T. 2017, {PyPulse: PSRFITS handler}.
\newblock \doeprint{1706.011}

\bibitem[{{Lam} {et~al.}(2018{\natexlab{a}}){Lam}, {McLaughlin}, {Cordes},
  {Chatterjee}, \& {Lazio}}]{Lam18b}
{Lam}, M.~T., {McLaughlin}, M.~A., {Cordes}, J.~M., {Chatterjee}, S., \&
  {Lazio}, T.~J.~W. 2018{\natexlab{a}}, \apj, 861, 12,
  \dodoi{10.3847/1538-4357/aac48d}

\bibitem[{{Lam} {et~al.}(2017){Lam}, {Cordes}, {Chatterjee}, {Arzoumanian},
  {Crowter}, {Demorest}, {Dolch}, {Ellis}, {Ferdman}, {Fonseca}, {Gonzalez},
  {Jones}, {Jones}, {Levin}, {Madison}, {McLaughlin}, {Nice}, {Pennucci},
  {Ransom}, {Shannon}, {Siemens}, {Stairs}, {Stovall}, {Swiggum}, \&
  {Zhu}}]{lcc+17}
{Lam}, M.~T., {Cordes}, J.~M., {Chatterjee}, S., {et~al.} 2017, \apj, 834, 35,
  \dodoi{10.3847/1538-4357/834/1/35}

\bibitem[{{Lam} {et~al.}(2018{\natexlab{b}}){Lam}, {Ellis}, {Grillo}, {Jones},
  {Hazboun}, {Brook}, {Turner}, {Chatterjee}, {Cordes}, {Lazio}, {DeCesar},
  {Arzoumanian}, {Blumer}, {Cromartie}, {Demorest}, {Dolch}, {Ferdman},
  {Ferrara}, {Fonseca}, {Garver-Daniels}, {Gentile}, {Gupta}, {Lorimer},
  {Lynch}, {Madison}, {McLaughlin}, {Ng}, {Nice}, {Pennucci}, {Ransom},
  {Spiewak}, {Stairs}, {Stinebring}, {Stovall}, {Swiggum}, {Vigeland}, \&
  {Zhu}}]{Lam2018}
{Lam}, M.~T., {Ellis}, J.~A., {Grillo}, G., {et~al.} 2018{\natexlab{b}}, \apj,
  861, 132, \dodoi{10.3847/1538-4357/aac770}

\bibitem[{{Lange} {et~al.}(2001){Lange}, {Camilo}, {Wex}, {Kramer}, {Backer},
  {Lyne}, \& {Doroshenko}}]{Lange2001}
{Lange}, C., {Camilo}, F., {Wex}, N., {et~al.} 2001, \mnras, 326, 274,
  \dodoi{10.1046/j.1365-8711.2001.04606.x}

\bibitem[{{Lommen} \& {Demorest}(2013)}]{Lommen2013}
{Lommen}, A.~N., \& {Demorest}, P. 2013, Classical and Quantum Gravity, 30,
  224001, \dodoi{10.1088/0264-9381/30/22/224001}

\bibitem[{{Lorimer} \& {Kramer}(2012)}]{LorimerKramer2012}
{Lorimer}, D.~R., \& {Kramer}, M. 2012, {Handbook of Pulsar Astronomy}
  (Cambridge University Press)

\bibitem[{{Luo} {et~al.}(2019){Luo}, {Ransom}, {Demorest}, {van Haasteren},
  {Ray}, {Stovall}, {Bachetti}, {Archibald}, {Kerr}, {Colen}, \&
  {Jenet}}]{pint}
{Luo}, J., {Ransom}, S., {Demorest}, P., {et~al.} 2019, {PINT: High-precision
  pulsar timing analysis package}.
\newblock \doeprint{1902.007}

\bibitem[{{Luo} {et~al.}(2021){Luo}, {Ransom}, {Demorest}, {Ray}, {Archibald},
  {Kerr}, {Jennings}, {Bachetti}, {van Haasteren}, {Champagne}, {Colen},
  {Phillips}, {Zimmerman}, {Stovall}, {Lam}, \& {Jenet}}]{2021PINT}
---. 2021, \apj, 911, 45, \dodoi{10.3847/1538-4357/abe62f}

\bibitem[{{Martinez} {et~al.}(2019){Martinez}, {Gentile}, {Freire}, {Stovall},
  {Deneva}, {Desvignes}, {Jenet}, {McLaughlin}, {Bagchi}, \&
  {Devine}}]{Martinez2019}
{Martinez}, J.~G., {Gentile}, P., {Freire}, P.~C.~C., {et~al.} 2019, \apj, 881,
  166, \dodoi{10.3847/1538-4357/ab2877}

\bibitem[{Merkel(2014)}]{merkel2014docker}
Merkel, D. 2014, Linux journal, 2014, 2

\bibitem[{{Meyers} \& {Chime/Pulsar Collaboration}(2021)}]{Meyers2021}
{Meyers}, B., \& {Chime/Pulsar Collaboration}. 2021, The Astronomer's Telegram,
  14652, 1

\bibitem[{{Miles} {et~al.}(2023){Miles}, {Shannon}, {Bailes}, {Reardon},
  {Keith}, {Cameron}, {Parthasarathy}, {Shamohammadi}, {Spiewak}, {van
  Straten}, {Buchner}, {Camilo}, {Geyer}, {Karastergiou}, {Kramer}, {Serylak},
  {Theureau}, \& {Venkatraman Krishnan}}]{meerkat2023}
{Miles}, M.~T., {Shannon}, R.~M., {Bailes}, M., {et~al.} 2023, \mnras, 519,
  3976, \dodoi{10.1093/mnras/stac3644}

\bibitem[{{Miller} {et~al.}(2021){Miller}, {Lamb}, {Dittmann}, {Bogdanov},
  {Arzoumanian}, {Gendreau}, {Guillot}, {Ho}, {Lattimer}, {Loewenstein},
  {Morsink}, {Ray}, {Wolff}, {Baker}, {Cazeau}, {Manthripragada}, {Markwardt},
  {Okajima}, {Pollard}, {Cognard}, {Cromartie}, {Fonseca}, {Guillemot}, {Kerr},
  {Parthasarathy}, {Pennucci}, {Ransom}, \& {Stairs}}]{mil21}
{Miller}, M.~C., {Lamb}, F.~K., {Dittmann}, A.~J., {et~al.} 2021, \apjl, 918,
  L28, \dodoi{10.3847/2041-8213/ac089b}

\bibitem[{{Morello} {et~al.}(2019){Morello}, {Barr}, {Cooper}, {Bailes},
  {Bates}, {Bhat}, {Burgay}, {Burke-Spolaor}, {Cameron}, {Champion}, {Eatough},
  {Flynn}, {Jameson}, {Johnston}, {Keith}, {Keane}, {Kramer}, {Levin}, {Ng},
  {Petroff}, {Possenti}, {Stappers}, {van Straten}, \& {Tiburzi}}]{mbc+19}
{Morello}, V., {Barr}, E.~D., {Cooper}, S., {et~al.} 2019, \mnras, 483, 3673,
  \dodoi{10.1093/mnras/sty3328}

\bibitem[{{Ng} {et~al.}(2014){Ng}, {Bailes}, {Bates}, {Bhat}, {Burgay},
  {Burke-Spolaor}, {Champion}, {Coster}, {Johnston}, {Keith}, {Kramer},
  {Levin}, {Petroff}, {Possenti}, {Stappers}, {van Straten}, {Thornton},
  {Tiburzi}, {Bassa}, {Freire}, {Guillemot}, {Lyne}, {Tauris}, {Shannon}, \&
  {Wex}}]{nbb+14}
{Ng}, C., {Bailes}, M., {Bates}, S.~D., {et~al.} 2014, \mnras, 439, 1865,
  \dodoi{10.1093/mnras/stu067}

\bibitem[{{Nice} {et~al.}(2015){Nice}, {Demorest}, {Stairs}, {Manchester},
  {Taylor}, {Peters}, {Weisberg}, {Irwin}, {Wex}, \& {Huang}}]{tempo}
{Nice}, D., {Demorest}, P., {Stairs}, I., {et~al.} 2015, {Tempo: Pulsar timing
  data analysis}.
\newblock \doeprint{1509.002}

\bibitem[{Park {et~al.}(2021)Park, Folkner, Williams, \& Boggs}]{Park_2021}
Park, R.~S., Folkner, W.~M., Williams, J.~G., \& Boggs, D.~H. 2021, The
  Astronomical Journal, 161, 105, \dodoi{10.3847/1538-3881/abd414}

\bibitem[{{Pennucci}(2019)}]{Pennucci19}
{Pennucci}, T.~T. 2019, \apj, 871, 34, \dodoi{10.3847/1538-4357/aaf6ef}

\bibitem[{{Pennucci} {et~al.}(2014){Pennucci}, {Demorest}, \& {Ransom}}]{PDR14}
{Pennucci}, T.~T., {Demorest}, P.~B., \& {Ransom}, S.~M. 2014, \apj, 790, 93,
  \dodoi{10.1088/0004-637X/790/2/93}

\bibitem[{{Pennucci} {et~al.}(2016){Pennucci}, {Demorest}, \&
  {Ransom}}]{pulseportraiture}
---. 2016, {Pulse Portraiture: Pulsar Timing}, Astrophysics Source Code
  Library.
\newblock \doeprint{1606.013}

\bibitem[{{Perera} {et~al.}(2019){Perera}, {DeCesar}, {Demorest}, {Kerr},
  {Lentati}, {Nice}, {Os{\l}owski}, {Ransom}, {Keith}, {Arzoumanian}, {Bailes},
  {Baker}, {Bassa}, {Bhat}, {Brazier}, {Burgay}, {Burke-Spolaor}, {Caballero},
  {Champion}, {Chatterjee}, {Chen}, {Cognard}, {Cordes}, {Crowter}, {Dai},
  {Desvignes}, {Dolch}, {Ferdman}, {Ferrara}, {Fonseca}, {Goldstein},
  {Graikou}, {Guillemot}, {Hazboun}, {Hobbs}, {Hu}, {Islo}, {Janssen},
  {Karuppusamy}, {Kramer}, {Lam}, {Lee}, {Liu}, {Luo}, {Lyne}, {Manchester},
  {McKee}, {McLaughlin}, {Mingarelli}, {Parthasarathy}, {Pennucci}, {Perrodin},
  {Possenti}, {Reardon}, {Russell}, {Sanidas}, {Sesana}, {Shaifullah},
  {Shannon}, {Siemens}, {Simon}, {Spiewak}, {Stairs}, {Stappers}, {Swiggum},
  {Taylor}, {Theureau}, {Tiburzi}, {Vallisneri}, {Vecchio}, {Wang}, {Zhang},
  {Zhang}, {Zhu}, \& {Zhu}}]{IPTADR2}
{Perera}, B.~B.~P., {DeCesar}, M.~E., {Demorest}, P.~B., {et~al.} 2019, \mnras,
  490, 4666, \dodoi{10.1093/mnras/stz2857}

\bibitem[{{Ransom} {et~al.}(2019){Ransom}, {Brazier}, {Chatterjee}, {Cohen},
  {Cordes}, {DeCesar}, {Demorest}, {Hazboun}, {Lam}, {Lynch}, {McLaughlin},
  {Ransom}, {Siemens}, {Taylor}, \& {Vigeland}}]{Ransom2019}
{Ransom}, S., {Brazier}, A., {Chatterjee}, S., {et~al.} 2019, in \baas,
  Vol.~51, 195.
\newblock \doarXiv{1908.05356}

\bibitem[{{Ransom} {et~al.}(2011){Ransom}, {Ray}, {Camilo}, {Roberts},
  {{\c{C}}elik}, {Wolff}, {Cheung}, {Kerr}, {Pennucci}, {DeCesar}, {Cognard},
  {Lyne}, {Stappers}, {Freire}, {Grove}, {Abdo}, {Desvignes}, {Donato},
  {Ferrara}, {Gehrels}, {Guillemot}, {Gwon}, {Harding}, {Johnston}, {Keith},
  {Kramer}, {Michelson}, {Parent}, {Saz Parkinson}, {Romani}, {Smith},
  {Theureau}, {Thompson}, {Weltevrede}, {Wood}, \& {Ziegler}}]{rrc+11}
{Ransom}, S.~M., {Ray}, P.~S., {Camilo}, F., {et~al.} 2011, \apjl, 727, L16,
  \dodoi{10.1088/2041-8205/727/1/L16}

\bibitem[{{Reardon} {et~al.}(2021){Reardon}, {Shannon}, {Cameron}, {Goncharov},
  {Hobbs}, {Middleton}, {Shamohammadi}, {Thyagarajan}, {Bailes}, {Bhat}, {Dai},
  {Kerr}, {Manchester}, {Russell}, {Spiewak}, {Wang}, \& {Zhu}}]{rsc+21}
{Reardon}, D.~J., {Shannon}, R.~M., {Cameron}, A.~D., {et~al.} 2021, \mnras,
  507, 2137, \dodoi{10.1093/mnras/stab1990}

\bibitem[{{Riley} {et~al.}(2021){Riley}, {Watts}, {Ray}, {Bogdanov}, {Guillot},
  {Morsink}, {Bilous}, {Arzoumanian}, {Choudhury}, {Deneva}, {Gendreau},
  {Harding}, {Ho}, {Lattimer}, {Loewenstein}, {Ludlam}, {Markwardt}, {Okajima},
  {Prescod-Weinstein}, {Remillard}, {Wolff}, {Fonseca}, {Cromartie}, {Kerr},
  {Pennucci}, {Parthasarathy}, {Ransom}, {Stairs}, {Guillemot}, \&
  {Cognard}}]{ril21}
{Riley}, T.~E., {Watts}, A.~L., {Ray}, P.~S., {et~al.} 2021, \apjl, 918, L27,
  \dodoi{10.3847/2041-8213/ac0a81}

\bibitem[{{Sanpa-Arsa}(2016)}]{SanpaArsa2016}
{Sanpa-Arsa}, S. 2016, PhD thesis, University of Virginia

\bibitem[{{Sazhin}(1978)}]{sazhin78}
{Sazhin}, M.~V. 1978, \sovast, 22, 36

\bibitem[{{Scholz} {et~al.}(2015){Scholz}, {Kaspi}, {Lyne}, {Stappers},
  {Bogdanov}, {Cordes}, {Crawford}, {Ferdman}, {Freire}, {Hessels}, {Lorimer},
  {Stairs}, {Allen}, {Brazier}, {Camilo}, {Cardoso}, {Chatterjee}, {Deneva},
  {Jenet}, {Karako-Argaman}, {Knispel}, {Lazarus}, {Lee}, {van Leeuwen},
  {Lynch}, {Madsen}, {McLaughlin}, {Ransom}, {Siemens}, {Spitler}, {Stovall},
  {Swiggum}, {Venkataraman}, \& {Zhu}}]{skl+15}
{Scholz}, P., {Kaspi}, V.~M., {Lyne}, A.~G., {et~al.} 2015, \apj, 800, 123,
  \dodoi{10.1088/0004-637X/800/2/123}

\bibitem[{{Siemens} {et~al.}(2013){Siemens}, {Ellis}, {Jenet}, \&
  {Romano}}]{Siemens2013}
{Siemens}, X., {Ellis}, J., {Jenet}, F., \& {Romano}, J.~D. 2013, Classical and
  Quantum Gravity, 30, 224015, \dodoi{10.1088/0264-9381/30/22/224015}

\bibitem[{{Singha} {et~al.}(2021){Singha}, {Surnis}, {Joshi}, {Tarafdar},
  {Rana}, {Susobhanan}, {Girgaonkar}, {Kolhe}, {Agarwal}, {Desai}, {Prabu},
  {Bathula}, {Dandapat}, {Dey}, {Hisano}, {Kato}, {Kharbanda}, {Kikunaga},
  {Marmat}, {Susarla}, {Bagchi}, {Dhanda Batra}, {Choudhury}, {Gopakumar},
  {Gupta}, {Krishnakumar}, {Maan}, {Manoharan}, {Nobleson}, {Pandian},
  {Pathak}, \& {Takahashi}}]{Singha2021}
{Singha}, J., {Surnis}, M.~P., {Joshi}, B.~C., {et~al.} 2021, \mnras, 507, L57,
  \dodoi{10.1093/mnrasl/slab098}

\bibitem[{{Splaver} {et~al.}(2005){Splaver}, {Nice}, {Stairs}, {Lommen}, \&
  {Backer}}]{Splaver2005}
{Splaver}, E.~M., {Nice}, D.~J., {Stairs}, I.~H., {Lommen}, A.~N., \& {Backer},
  D.~C. 2005, \apj, 620, 405, \dodoi{10.1086/426804}

\bibitem[{{Stairs} {et~al.}(2004){Stairs}, {Thorsett}, \&
  {Arzoumanian}}]{Stairs04}
{Stairs}, I.~H., {Thorsett}, S.~E., \& {Arzoumanian}, Z. 2004, \prl, 93,
  141101, \dodoi{10.1103/PhysRevLett.93.141101}

\bibitem[{{Swihart} {et~al.}(2022){Swihart}, {Strader}, {Chomiuk}, {Aydi},
  {Sokolovsky}, {Ray}, \& {Kerr}}]{Swihart22}
{Swihart}, S.~J., {Strader}, J., {Chomiuk}, L., {et~al.} 2022, \apj, 941, 199,
  \dodoi{10.3847/1538-4357/aca2ac}

\bibitem[{{Tarafdar} {et~al.}(2022){Tarafdar}, {Nobleson}, {Rana}, {Singha},
  {Krishnakumar}, {Joshi}, {Paladi}, {Kolhe}, {Batra}, {Agarwal}, {Bathula},
  {Dandapat}, {Desai}, {Dey}, {Hisano}, {Ingale}, {Kato}, {Kharbanda},
  {Kikunaga}, {Marmat}, {Pandian}, {Prabu}, {Srivastava}, {Surnis}, {Susarla},
  {Susobhanan}, {Takahashi}, {Arumugam}, {Bagchi}, {Banik}, {De}, {Girgaonkar},
  {Gopakumar}, {Gupta}, {Maan}, {Manoharan}, {Naidu}, \& {Pathak}}]{inpta22}
{Tarafdar}, P., {Nobleson}, K., {Rana}, P., {et~al.} 2022, \pasa, 39, e053,
  \dodoi{10.1017/pasa.2022.46}

\bibitem[{{Taylor}(1992)}]{Taylor92}
{Taylor}, J.~H. 1992, Royal Society of London Philosophical Transactions Series
  A, 341, 117, \dodoi{10.1098/rsta.1992.0088}

\bibitem[{Taylor(2021)}]{taylor2021}
Taylor, S.~R. 2021, Nanohertz Gravitational Wave Astronomy (CRC Press),
  \dodoi{10.48550/arXiv.2105.13270}

\bibitem[{Taylor {et~al.}(2021)Taylor, Baker, Hazboun, Simon, \&
  Vigeland}]{enterprise_ext}
Taylor, S.~R., Baker, P.~T., Hazboun, J.~S., Simon, J., \& Vigeland, S.~J.
  2021, enterprise\_extensions.
\newblock \url{https://github.com/nanograv/enterprise\_extensions}

\bibitem[{{Tiburzi}(2018)}]{Tiburzi2018}
{Tiburzi}, C. 2018, \pasa, 35, e013, \dodoi{10.1017/pasa.2018.7}

\bibitem[{{Vallisneri}(2020)}]{libstempo}
{Vallisneri}, M. 2020, {libstempo: Python wrapper for Tempo2}.
\newblock \doeprint{2002.017}

\bibitem[{{Vallisneri} \& {van Haasteren}(2017)}]{Vallisneri2017}
{Vallisneri}, M., \& {van Haasteren}, R. 2017, \mnras, 466, 4954,
  \dodoi{10.1093/mnras/stx069}

\bibitem[{{van Straten}(2003)}]{vanstraten03}
{van Straten}, W. 2003, PhD thesis, Swinburne University of Technology,
  Australia

\bibitem[{{van Straten} {et~al.}(2001){van Straten}, {Bailes}, {Britton},
  {Kulkarni}, {Anderson}, {Manchester}, \& {Sarkissian}}]{vanStraten01}
{van Straten}, W., {Bailes}, M., {Britton}, M., {et~al.} 2001, \nat, 412, 158,
  \dodoi{10.1038/35084015}

\bibitem[{{Verbiest} {et~al.}(2021){Verbiest}, {Os{\l}owski}, \&
  {Burke-Spolaor}}]{vob21}
{Verbiest}, J.~P.~W., {Os{\l}owski}, S., \& {Burke-Spolaor}, S. 2021, in
  Handbook of Gravitational Wave Astronomy (Springer), 4,
  \dodoi{10.1007/978-981-15-4702-7_4-1}

\bibitem[{Virtanen {et~al.}(2020)Virtanen, Gommers, Oliphant, Haberland, Reddy,
  Cournapeau, Burovski, Peterson, Weckesser, Bright, {van der Walt}, Brett,
  Wilson, Millman, Mayorov, Nelson, Jones, Kern, Larson, Carey, Polat, Feng,
  Moore, {VanderPlas}, Laxalde, Perktold, Cimrman, Henriksen, Quintero, Harris,
  Archibald, Ribeiro, Pedregosa, {van Mulbregt}, \& {SciPy 1.0
  Contributors}}]{SciPy}
Virtanen, P., Gommers, R., Oliphant, T.~E., {et~al.} 2020, Nature Methods, 17,
  261, \dodoi{10.1038/s41592-019-0686-2}

\bibitem[{{Wang} \& {Taylor}(2022)}]{wan22}
{Wang}, Q., \& {Taylor}, S.~R. 2022, \mnras, 516, 5874,
  \dodoi{10.1093/mnras/stac2679}

\bibitem[{{Xu} {et~al.}(2021){Xu}, {Huang}, {Burgay}, {Champion}, {Cognard},
  {Guillemot}, {Jang}, {Karuppusamy}, {Kramer}, {Lackeos}, {Lee}, {Liu},
  {Perrodin}, {Possenti}, {Stappers}, \& {Theureau}}]{Xu2021}
{Xu}, H., {Huang}, Y.~X., {Burgay}, M., {et~al.} 2021, The Astronomer's
  Telegram, 14642, 1

\end{thebibliography}

\clearpage
\appendix
\section{Further Details on TOA Removal}
\label{sec:cutdetail}
\subsection{Poor Frontend/Backend Cut (\texttt{poorfebe})}
In most cases, the \texttt{poorfebe} cut was applied in cases where the vast majority of, or all, TOAs from a given frontend/backend combination were being cut by other means. It was only used for six MSPs and mostly to indicate poor performance at higher observing frequencies (e.g., 3/6\,GHz with the VLA and 2\,GHz with Arecibo). In one case, a pulsar had been observed more than three times with a non-standard frontend due to RFI issues at Arecibo Observatory and ad hoc scheduling changes. These TOAs were not caught by our orphan data cut, and for consistency, we removed them with \texttt{poorfebe}.

\subsection{Bad Range Cut (\texttt{badrange})\label{sec:badrange}}
In preliminary stages of our analysis, we discovered that a significant portion of data collected at Arecibo exclusively between 2017 August and 2018 November were corrupted and the issue was traced back to a malfunctioning local oscillator (LO).\footnote{Diagnosing this issue was aided by private communication with Shriharsh Tendulkar and others.} During this time period, the LO's reference frequency exhibited sudden shifts by 5--10\,MHz, and sometimes erratic wandering across this range on sub-millisecond timescales. This sort of behavior manifests itself as what looks like improperly dedispersed archives in the NANOGrav data set, and causes often unpredictable amounts of smearing and phase variations within individual observations. To determine the scope of the problem, we inspected phase versus frequency residuals, after subtracting the pulse portrait from the corresponding band. Corrupted scans were easily detectable by eye for our brightest pulsars (i.e., PSR~J1713+0747), but due to some concern about low-level effects on high-precision timing for all of our pulsars monitored at Arecibo, we took a conservative approach and excised all scans within the time range MJD 57984--58447, where corruption was apparent. TOAs from this MJD range were assigned the cut flag \texttt{badrange}. After identifying this issue in \fifteenyr, we have implemented additional quality assurance measures to ensure that we catch similar issues more quickly in the future.

\subsection{Orbital Phase Range Cut (\texttt{eclipsing})\label{sec:phasecut}}
Several black-widow pulsars included in NANOGrav's regular observing schedule appear in \fifteenyr\ and two of these, PSRs J1705$-$1903 and J1802$-$2124, exhibit eclipses. By examining timing residuals as a function of orbital phase and looking at individual scans, we imposed an \texttt{eclipsing} cut so that TOAs generated within 10--15\% of an orbit from superior conjunction (when the pulsar's signal is eclipsed by its companion) are automatically removed.

\subsection{Gibbs Outlier Cut (\texttt{outlier10}, \texttt{maxout}, \texttt{epochdrop}) \label{sec:outlier}} 
While \twelveyr\ employed the \cite{Vallisneri2017} outlier identification algorithm, the growing TOA volume of our narrowband data set, as well as the desire to iteratively perform outlier analyses in response to timing model modifications, necessitated a more computationally efficient approach. In this work, we use a Gibbs sampler to determine TOA outliers in a fully Bayesian manner that is more computationally efficient than acceptance/rejection-based samplers. \cite{wan22} provides a thorough overview of this technique, including a demonstration of its efficacy when applied to the NANOGrav data set. Both outlier methods \citep{Vallisneri2017,wan22} are available in \texttt{enterprise\_outliers}\footnote{\url{https://github.com/nanograv/enterprise_outliers}}, which is a dependency for our timing analysis pipeline (see Section \ref{sec:timing} for more details). As in \twelveyr, any narrowband TOAs with outlier probability $p>0.1$ were removed and assigned the cut flag \texttt{outlier10}. Through experimentation, we found that in situations where more than five narrowband TOAs were flagged as outliers, all TOAs from that file were often corrupted. Therefore, if more than 8\% of TOAs (usually 5/64, but there are fewer total TOAs generated for some bands) from a single observing epoch were flagged for removal by the Gibbs algorithm, the remaining TOAs were also cut using the \texttt{maxout} flag. Neither of these steps were used for our wideband data set, but since there are 1-2 wideband TOAs generated per file, the epoch $F$-test cut (see \twelveyr, Section 2.5.8) provides a similar per-TOA outlier assessment in that case (\texttt{epochdrop}). This final stage of our automated outlier analysis is also applied to narrowband TOAs.

\subsection{Bad TOA/File Cut (\texttt{badtoa}, \texttt{badfile})\label{sec:badfiletoa}}
After the aforementioned outlier removal techniques were applied and the timing model was fully fit, a comparatively small number of TOAs/files were removed manually with \texttt{badtoa} and \texttt{badfile} cut flags, respectively. For every TOA/file removed this way, corresponding data cubes were inspected and the exact TOA/file and reason for removal were recorded in timing configuration files for transparency and posterity. Reasons for manual TOA/file cuts include: \texttt{snr}/\texttt{non-detection} when TOAs barely exceeded the S/N threshold and/or where no signal was visible by eye; \texttt{rfi} when RFI was still clearly present; \texttt{few\_chans} when only one or several channels remained post-zapping, suggesting almost the entire band had been affected by RFI; \texttt{smeared} when archives had been folded improperly or significant pulse broadening was apparent for other reasons; and \texttt{isolated} when one or several observations were separated from the rest by a long timespan (e.g.\ test observations for PSR~J1730$-$2304 were separated from the regular monitoring by 12 years). A total of 14 narrowband and 22 wideband TOAs were removed despite there being no obvious reason for removal (\texttt{unknown}), but these have also been explicitly recorded in the configuration files.

\section{Kopeikin Parameter Implementations}
\label{sec:kopeikin}
For PSR~J1713+0747, which is one of the most precisely timed pulsars, we use parameters which ultimately relate to the orientation of the binary orbit on the sky.  These are often called the ``Kopeikin parameters" after two papers \citep{Kopeikin1995,Kopeikin1996} which laid out the relevant expressions.  

We begin with a recap of parameters used in pulsar timing codes:
\begin{itemize}
\item $x$ -- projected semi-major axis of the pulsar's orbit
\item $\omega$ -- longitude of periastron
\item $d$ -- distance to the pulsar (derived from the parallax parameter)
\item $\mu_\alpha$ -- proper motion in Right Ascension, $\dot \alpha \cos\delta$
\item $\mu_\delta$ -- proper motion in Declination, $\dot \delta$
\item $i$ -- orbital inclination angle (generally has a quadrant ambiguity as often only $\sin i$ is accessible through timing; has different definitions)
\item $\Omega$ -- longitude of the ascending node (angle between the orbital plane and a reference line on the sky; this has different definitions as discussed below)
\end{itemize}

\citet{Kopeikin1995} defines the annual-orbital parallax (AOP) term as follows (eq.\ 17):
\begin{eqnarray}
\Delta_{\pi M} & = & \frac{x}{d}\left[(\Delta_{I_0} \sin \Omega - \Delta_{J_0}\cos \Omega) R(u) \cot i  -(\Delta_{I_0} \cos \Omega + \Delta_{J_0}\sin \Omega) Q(u) \csc i \right] \label{eq:kop95eq17}
\end{eqnarray}
\noindent where $Q(u)$ is an orbital ``cosine" term often designated as $C$ in timing codes, $R(u)$ is an orbital ``sine" term often designated as $S$ in timing codes, $u$ is the eccentric anomaly, and $\Delta_{I_0}$ and $\Delta_{J_0}$ are dot products of the Earth's position with unit vectors pointing East and North, respectively, on the sky at the position of the pulsar system.  \citet{Kopeikin1995} points out that this term can be broken into effects on $x$ and  $\omega$:
\begin{eqnarray} 
x^{\rm obs}  & = & x^{\rm intrinsic} \left[ 1+ \frac{\cot i}{d }(\Delta_{I_0} \sin \Omega - \Delta_{J_0}\cos \Omega) \right]  \label{eq:kop95eq18} \\
\omega^{\rm obs}  & = & \omega^{\rm intrinsic} - \frac{\csc i}{d }(\Delta_{I_0} \cos \Omega + \Delta_{J_0}\sin \Omega)  \label{eq:kop95eq19}
\end{eqnarray}

\citet{Kopeikin1996} (see also \citet{Arzoumanian1996}) presents the excess R{\/o}mer delay due to the effects of changing projected orbit due to proper motion:
\begin{eqnarray}
\frac{(t-t_0)}{c}(\mu \cdot \bf{r}_p) & = & x(t-t_0) \csc i (\mu_\alpha \cos \Omega + \mu_\delta \sin\Omega)(1-e \cos u) \cos(\omega+A_e(u)) +  \nonumber \\
 & & x(t-t_0)\cot i(-\mu_\alpha \sin \Omega + \mu_\delta \cos\Omega) (1-e \cos u)\sin(\omega+A_e(u)), \label{eq:kop96eq6}
\end{eqnarray} 
\noindent where $\bf{r}_p$ is the changing radius vector of the pulsar's orbit about the binary center of mass and $A_e(u)$ is the corresponding true anomaly.  The resulting equations for the changes in observables are:
\begin{eqnarray} 
\delta i &=& (-\mu_\alpha \sin \Omega + \mu_\delta \cos \Omega) (t-t_0)  \label{eq:kop96eq10}\\
\frac{\delta x}{x} & = & \cot i (-\mu_\alpha \sin \Omega + \mu_\delta \cos \Omega) (t-t_0) \label{eq:kop96eq8} \\
\delta \omega & = &  \csc i (\mu_\alpha \cos \Omega + \mu_\delta \sin \Omega) (t-t_0) \label{eq:kop96eq9}
\end{eqnarray}
\noindent where $t-t_0$ is the time elapsed since the reference epoch.

Between one and four of these effects can be observed in many systems, and all can be parameterized using $i$ and $\Omega$.  Typically $\chi^2$ values are computed for a grid of $i$ and $\Omega$ values; only strong ``Kopeikin effects" will break the natural four-fold degeneracy in the parameters and narrow down the possible values of $i$ and $\Omega$.

Unfortunately there are two different sign conventions in use to describe $i$ and $\Omega$, with literature and software packages (\tempo, \tempotwo, and \pint) using both.  Use of the incorrect sign convention when updating pulsar parameters will result in poor fits, and possible settling in a local rather than global minimum in the $i$--$\Omega$ space.  

To forestall such errors,  we summarize the two conventions in use:
\begin{enumerate}
 \item IAU convention 
  \begin{itemize}
  \item  $i=0^\circ$ means the orbital angular momentum vector points toward the Earth, and $i = 180^\circ$ means the orbital angular momentum vector points away from the Earth.
  \item $\Omega$ is $0^\circ$ toward the North and increases counter-clockwise on the sky; it is measured ``North through East."
  \end{itemize}
  \item \citet{Damour1992} (DT92) convention
  \begin{itemize}
  \item $i=180^\circ$ means the orbital angular momentum vector points toward the Earth, and $i = 0^\circ$ means the orbital angular momentum vector points away from the Earth.
  \item $\Omega$ is $0^\circ$ toward the East and increases clockwise on the sky; it is measured ``East through North."
  \end{itemize}
\end{enumerate}

These conventions are related by:
\begin{enumerate} 
\item $i_{\rm DT92} = 180^\circ - i_{\rm IAU}$; $\sin i_{\rm DT92}  = \sin  i_{\rm IAU}$; $\cos  i_{\rm DT92} = -\cos  i_{\rm IAU}$
\item $\Omega_{\rm DT92} = 90^\circ - \Omega_{\rm IAU}$; $ \sin \Omega_{\rm DT92} = \cos \Omega_{\rm IAU}$ ; $\cos \Omega_{\rm DT92} = \sin \Omega_{\rm IAU}$
\end{enumerate}

\begin{figure*}
    \centering
    \includegraphics[scale=0.84]{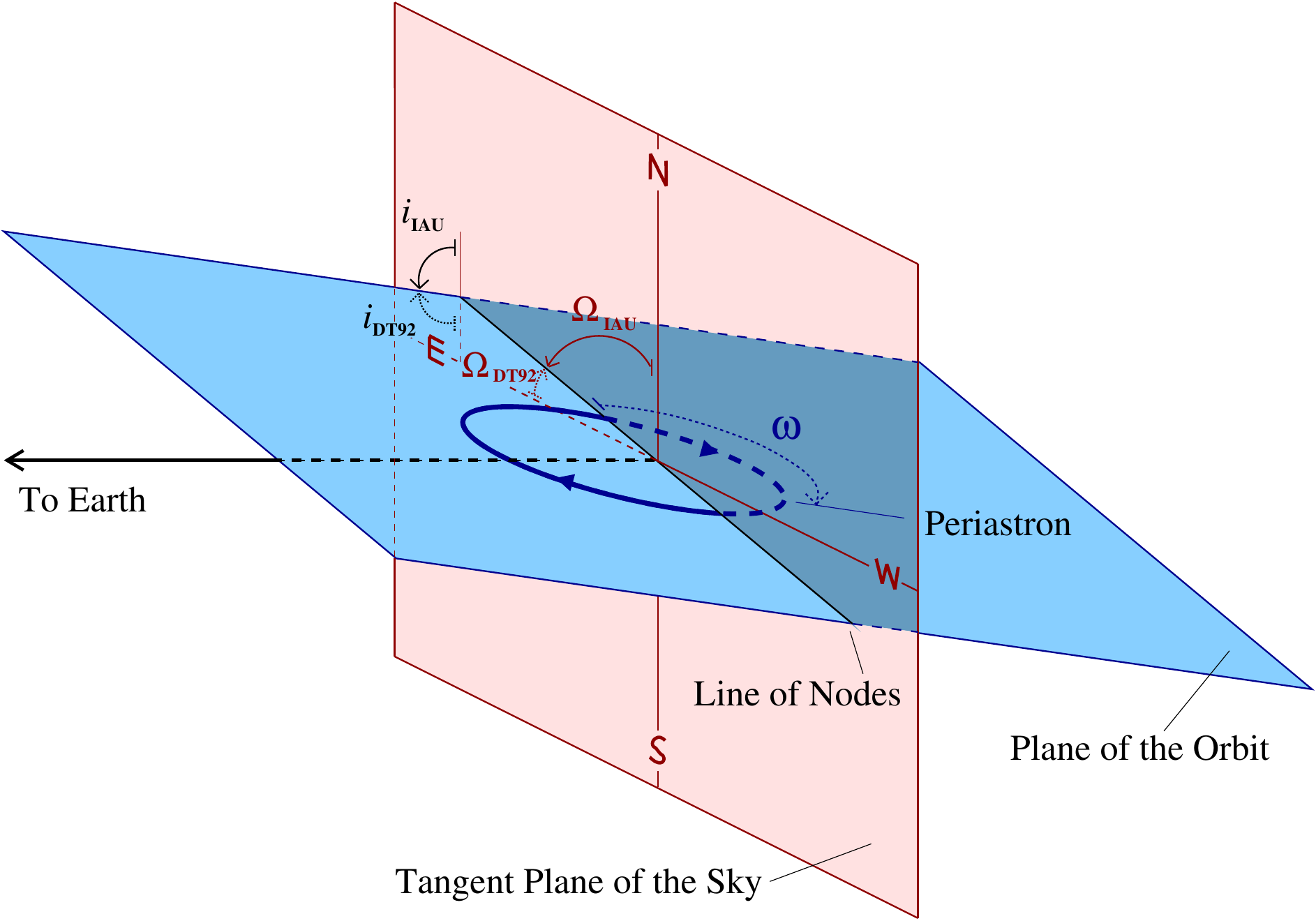}
    \caption{Geometry of the binary system, showing the position angle of ascending node in the IAU and DT92 conventions, $\Omega_{\rm IAU}$ and $\Omega_{\rm DT92}$; the inclination angle in the IAU and DT92 conventions, $i_{\rm IAU}$ and $i_{\rm DT92}$; and the angle of periastron, $\omega$, which is the same in both conventions.  The cardinal directions---N, S, E and W for north, south, east, and west--- are relative to whichever astrometric coordinate system is used for the pulsar position fit (equatorial or ecliptic).  Red portions of the figure are in the plane of the sky; blue portions are in the orbital plane. This figure is adapted from \citet{Splaver2005}.}
    \label{fig:binary_angles}
\end{figure*}

The IAU convention has been used in \citet{Splaver2005} and NANOGrav \tempo-derived papers such as \citet{Fonseca2016} and the 12.5\,yr data release \citep{Alam21}.
The DT92 convention was used to derive the equations in \citet{Kopeikin1995} and \citet{Kopeikin1996} and has been used for measurements in \citet{vanStraten01} and \citet{Stairs04}. We include a diagram illustrating both conventions in Figure \ref{fig:binary_angles}, which is adapted from \citet{Splaver2005}.

The conventions used in the various pulsar timing codes are as follows:
\begin{itemize}
    \item  \tempo\ DDK Model \citep{vanstraten03}: The code used to read and write parfiles uses the {\bf IAU convention}.  Internally to the code, the input $i$ (``KIN") and $\Omega$ (``KOM") values are immediately transformed to the DT92 convention, and the Kopeikin equations are used directly in that convention.  For each TOA, this code first computes the locally adjusted $x$ and $\omega$ based on equations~\ref{eq:kop96eq8} and \ref{eq:kop96eq9}, followed by the annual-orbital parallax corrections to $x$ and $\omega$ based on  equations~\ref{eq:kop95eq18} and \ref{eq:kop95eq19}.  It then proceeds through the standard computation of orbital delays, including Shapiro delay based on $i$, calculation of all relevant derivatives, and parameter adjustment. The fit results are transformed back to the IAU convention before ouput. Note that \tempo\ requires a parfile line {\tt "K96  1"} to ensure that the proper-motion corrections are done. 
    \item \tempo2\ DDK model: The code used to read and write parfiles uses the {\bf DT92 convention} for KIN and KOM.  The delay/derivative code closely follows the logic of the \tempo\ DDK code, including the use of the DT92 convention in the Kopeikin equations. We have recently debugged this routine and confirmed that it returns equivalent results to the tempo DDK model.
    \item \tempo2\ T2 model: The code used to read and write parfiles uses the {\bf IAU convention} for KIN and KOM.  This model provides a superset of multiple binary models, and uses equation~\ref{eq:kop95eq17} directly rather than breaking it down. The fitting code uses an IAU-convention implementation of the Kopeikin equations.  We note that only the base values of $i$, $x$ and $\omega$, not values adjusted via equations~\ref{eq:kop96eq10} --\ref{eq:kop96eq9}, are used in determining the annual-orbital parallax terms and also the R\/omer delay. This introduces small discrepancies between its results and those of DDK, with the T2 model parameters being incorrect by up to roughly 0.2 sigma, according to our simulations.
    \item \pint\ DDK model: The code used to read and write parfiles uses the {\bf DT92 convention} for KIN and KOM. Internally the code uses the equations from \citet{Kopeikin1995} and \citet{Kopeikin1996} directly.  The \pint\ code does apply corrections based on equations~\ref{eq:kop96eq10} --\ref{eq:kop96eq9} before computing the annual-orbital parallax terms.  
\end{itemize}

In this paper, we present astrometric results in ecliptic coordinates rather than equatorial.  \tempo\ and \pint\ are written to fit for $\Omega$ relative to ecliptic North (\tempo) or East (\pint) rather than equatorial North/East in this case, and use the proper motions in ecliptic coordinates ($\mu_\lambda,\mu_\beta$) to calculate the relevant parameters.  

\begin{figure}
\centering
\includegraphics[width=0.85\linewidth]{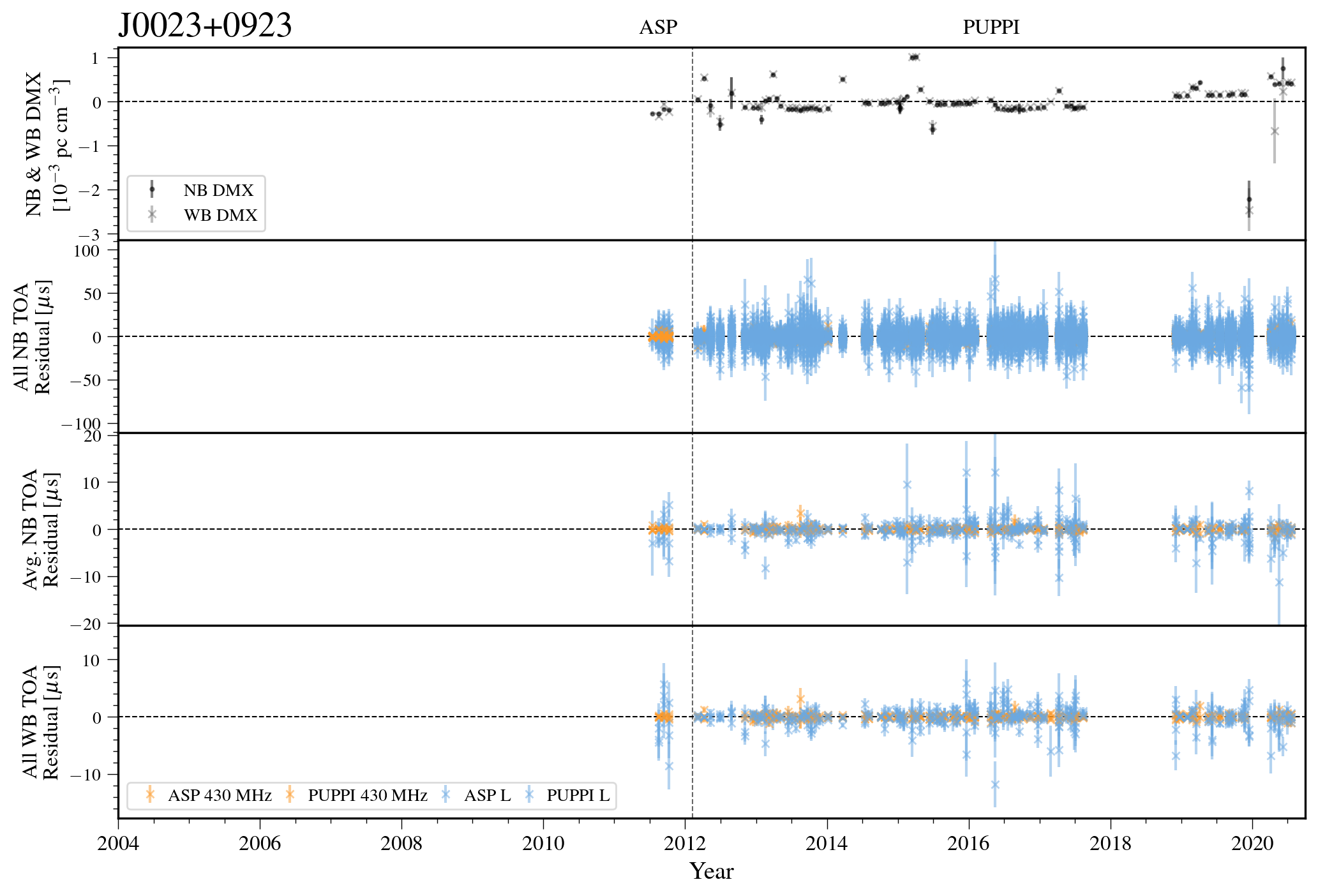}
\caption{Narrowband and wideband timing residuals and DMX timeseries for J0023+0923. The wideband data set (and therefore, the wideband residuals and DMX timeseries presented here) are only used for pipeline development and the mass determinations presented in Section~\ref{subsec:binaries}. Panel 1: Narrowband and wideband DMX variations. Panel 2: Residual arrival times for all TOAs. Points are semi-transparent; dark regions arise from the overlap of many points. Panel 3: Average residual arrival times. Panel 4 (in 4-panel plots): All wideband TOA residuals. Receivers and backends corresponding to each TOA are shown in the bottom panel's legend. Dashed vertical line(s) denote the divide between backend systems.}
\label{fig:summary-J0023+0923}
\end{figure}
\clearpage

\begin{figure}
\centering
\includegraphics[width=0.85\linewidth]{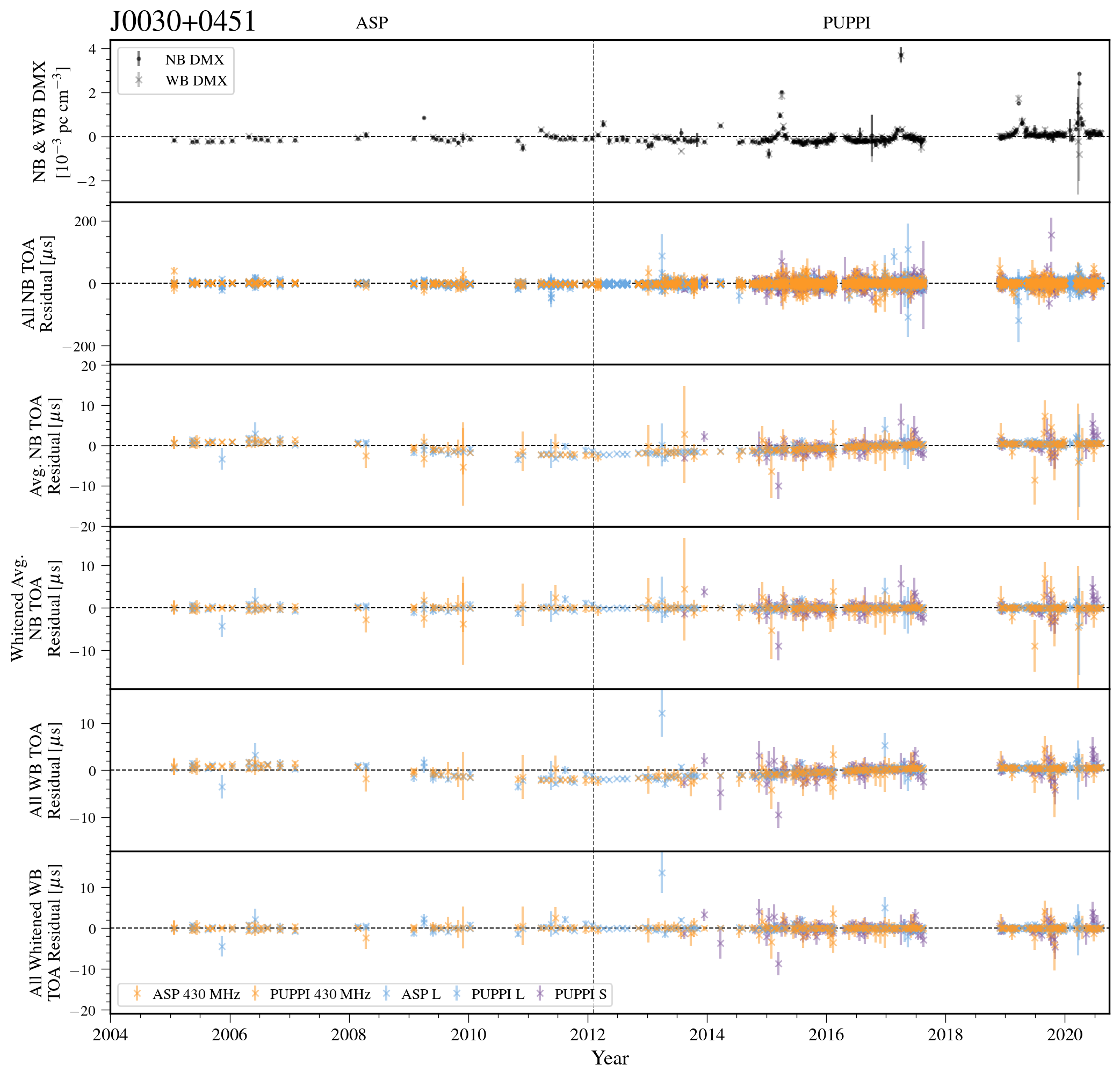}
\caption{Narrowband and wideband timing residuals and DMX timeseries for J0030+0451. The wideband data set (and therefore, the wideband residuals and DMX timeseries presented here) are only used for pipeline development and the mass determinations presented in Section~\ref{subsec:binaries}. Panel 1: Narrowband and wideband DMX variations. Panel 2: Residual arrival times for all TOAs. Points are semi-transparent; dark regions arise from the overlap of many points. Panel 3: Average residual arrival times. Panel 4 (in 6-panel plots): Whitened average narrowband TOA residuals. Panel 5: All wideband TOA residuals. Panel 6: All whitened wideband TOA residuals. Receivers and backends corresponding to each TOA are shown in the bottom panel's legend. Dashed vertical line(s) denote the divide between backend systems.}
\label{fig:summary-J0030+0451}
\end{figure}
\clearpage

\begin{figure}
\centering
\includegraphics[width=0.85\linewidth]{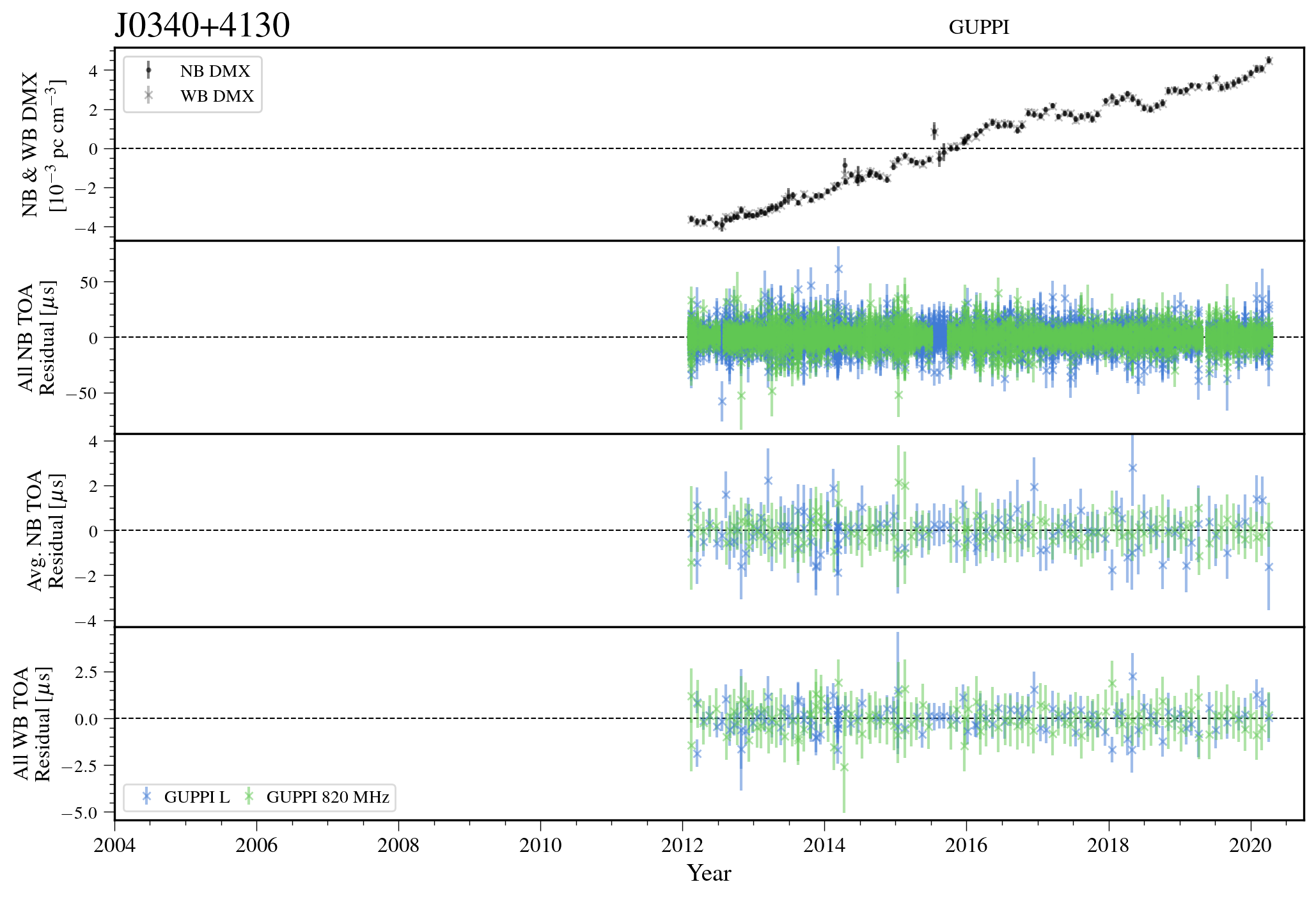}
\caption{Narrowband and wideband timing residuals and DMX timeseries for J0340+4130. See Figure~\ref{fig:summary-J0023+0923} for details.}
\label{fig:summary-J0340+4130}
\end{figure}

\begin{figure}
\centering
\includegraphics[width=0.85\linewidth]{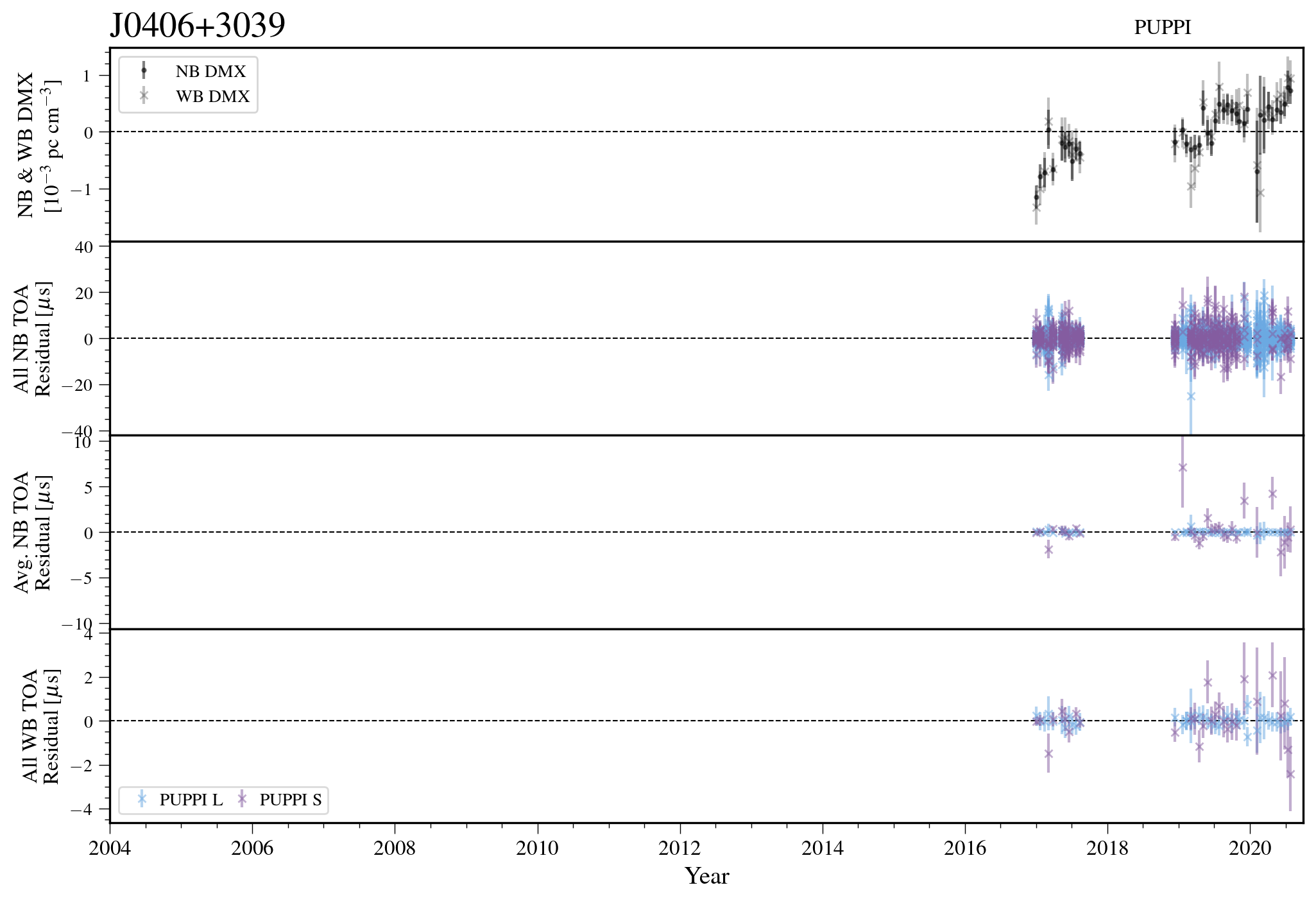}
\caption{Narrowband and wideband timing residuals and DMX timeseries for J0406+3039. See Figure~\ref{fig:summary-J0023+0923} for details.}
\label{fig:summary-J0406+3039}
\end{figure}

\begin{figure}
\centering
\includegraphics[width=0.85\linewidth]{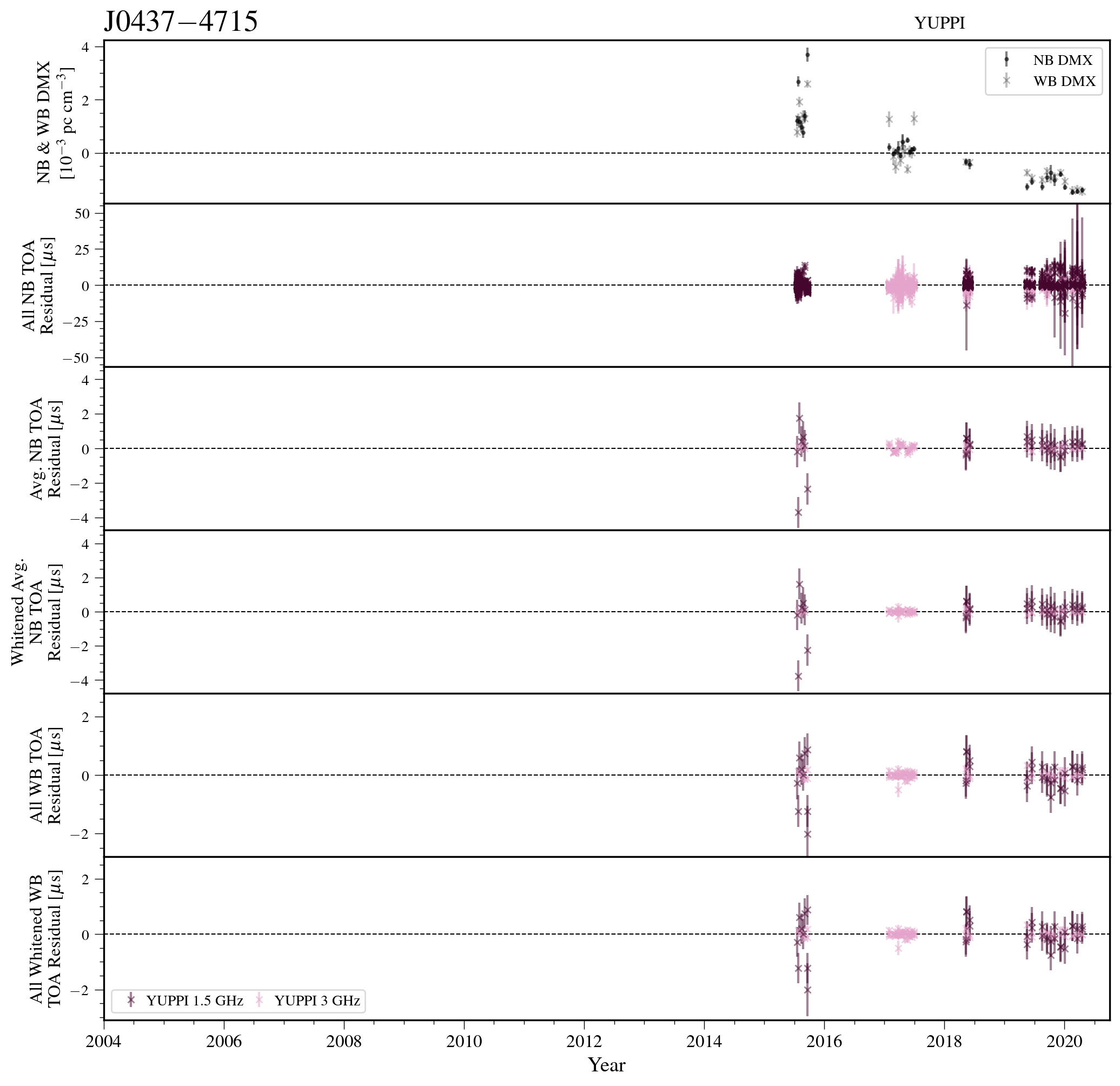}
\caption{Narrowband and wideband timing residuals and DMX timeseries for J0437-4715. See Figure~\ref{fig:summary-J0030+0451} for details.}
\label{fig:summary-J0437-4715}
\end{figure}
\clearpage

\begin{figure}
\centering
\includegraphics[width=0.85\linewidth]{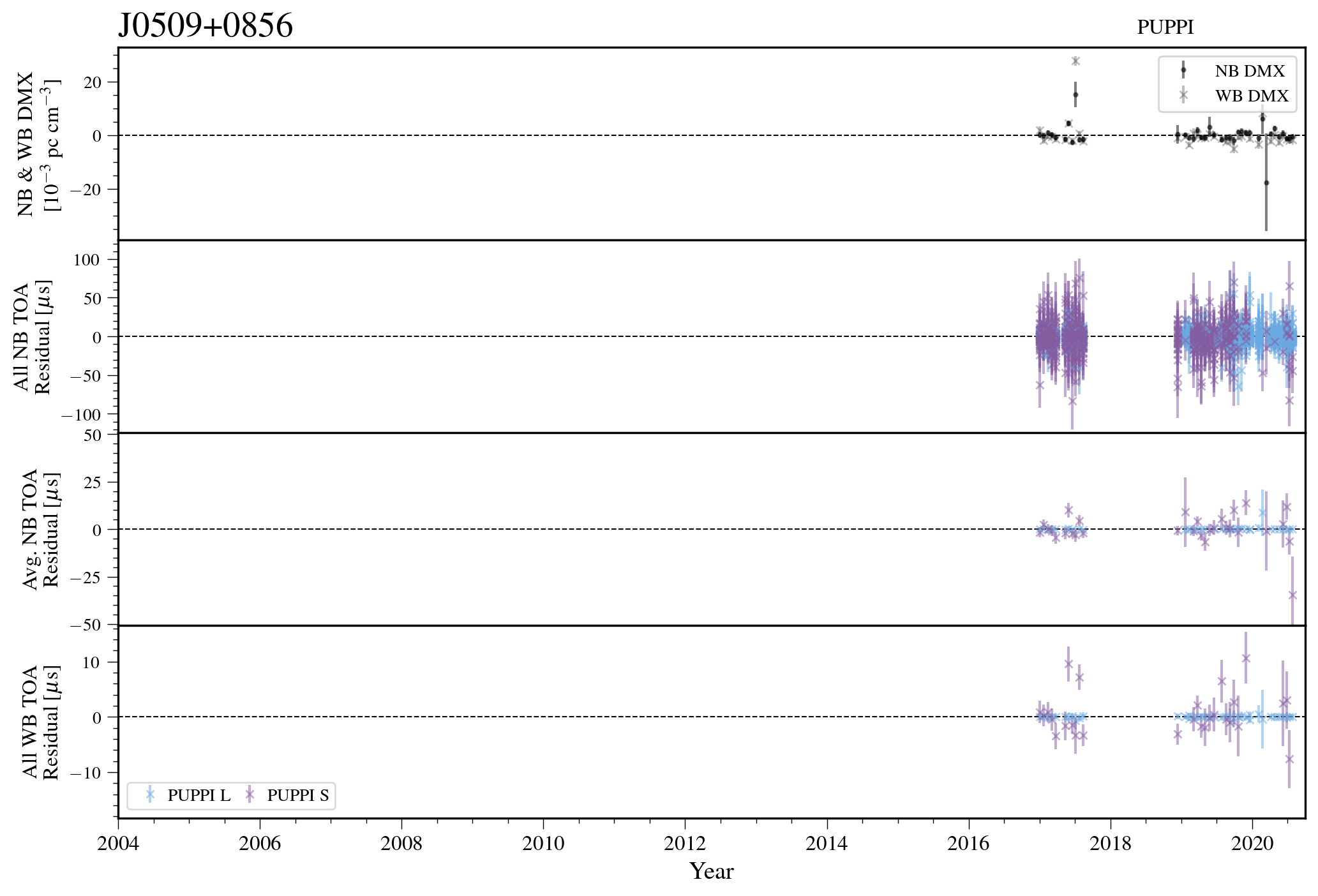}
\caption{Narrowband and wideband timing residuals and DMX timeseries for J0509+0856. See Figure~\ref{fig:summary-J0023+0923} for details.}
\label{fig:summary-J0509+0856}
\end{figure}

\begin{figure}
\centering
\includegraphics[width=0.85\linewidth]{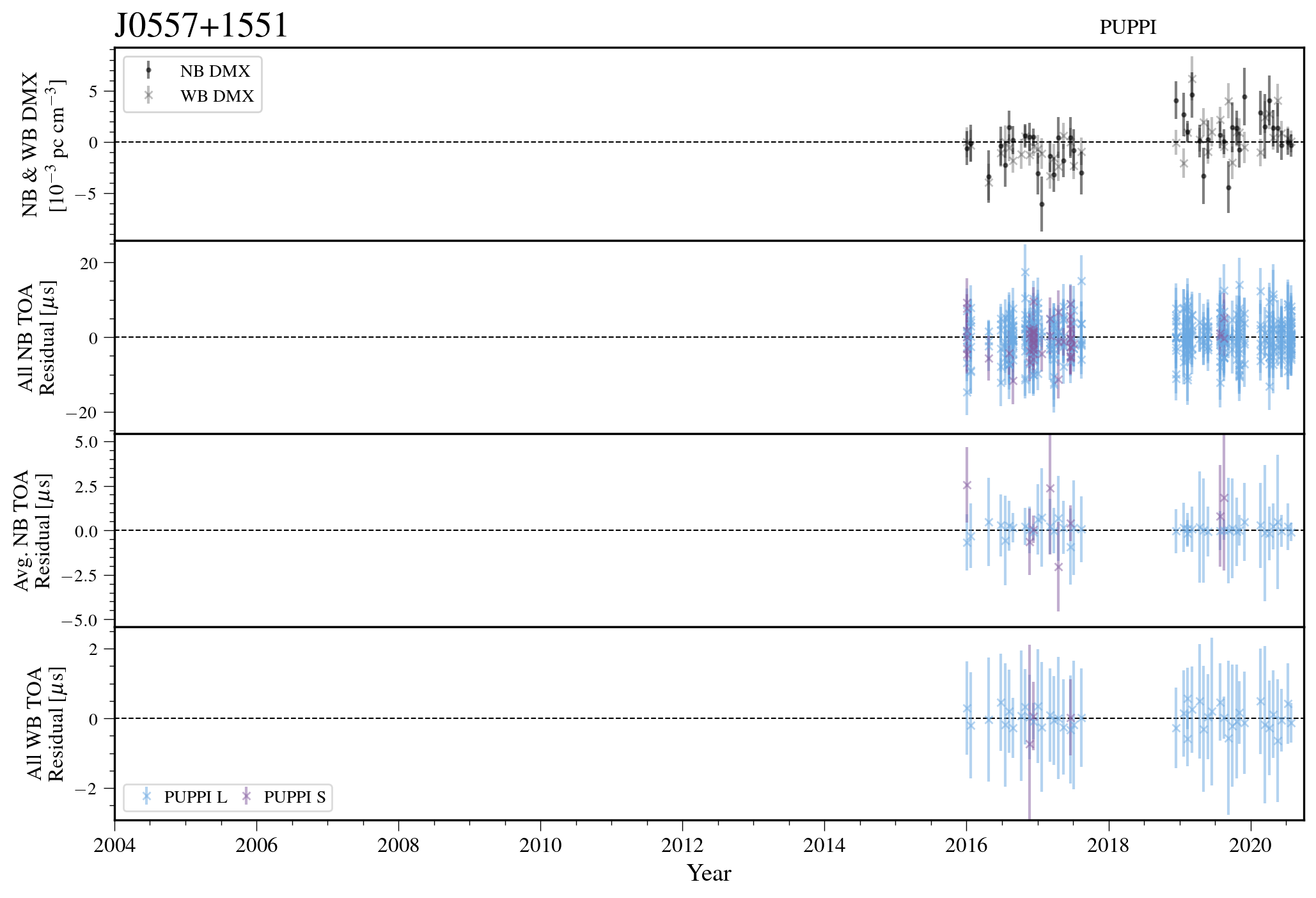}
\caption{Narrowband and wideband timing residuals and DMX timeseries for J0557+1551. See Figure~\ref{fig:summary-J0023+0923} for details.}
\label{fig:summary-J0557+1551}
\end{figure}
\clearpage

\begin{figure}
\centering
\includegraphics[width=0.85\linewidth]{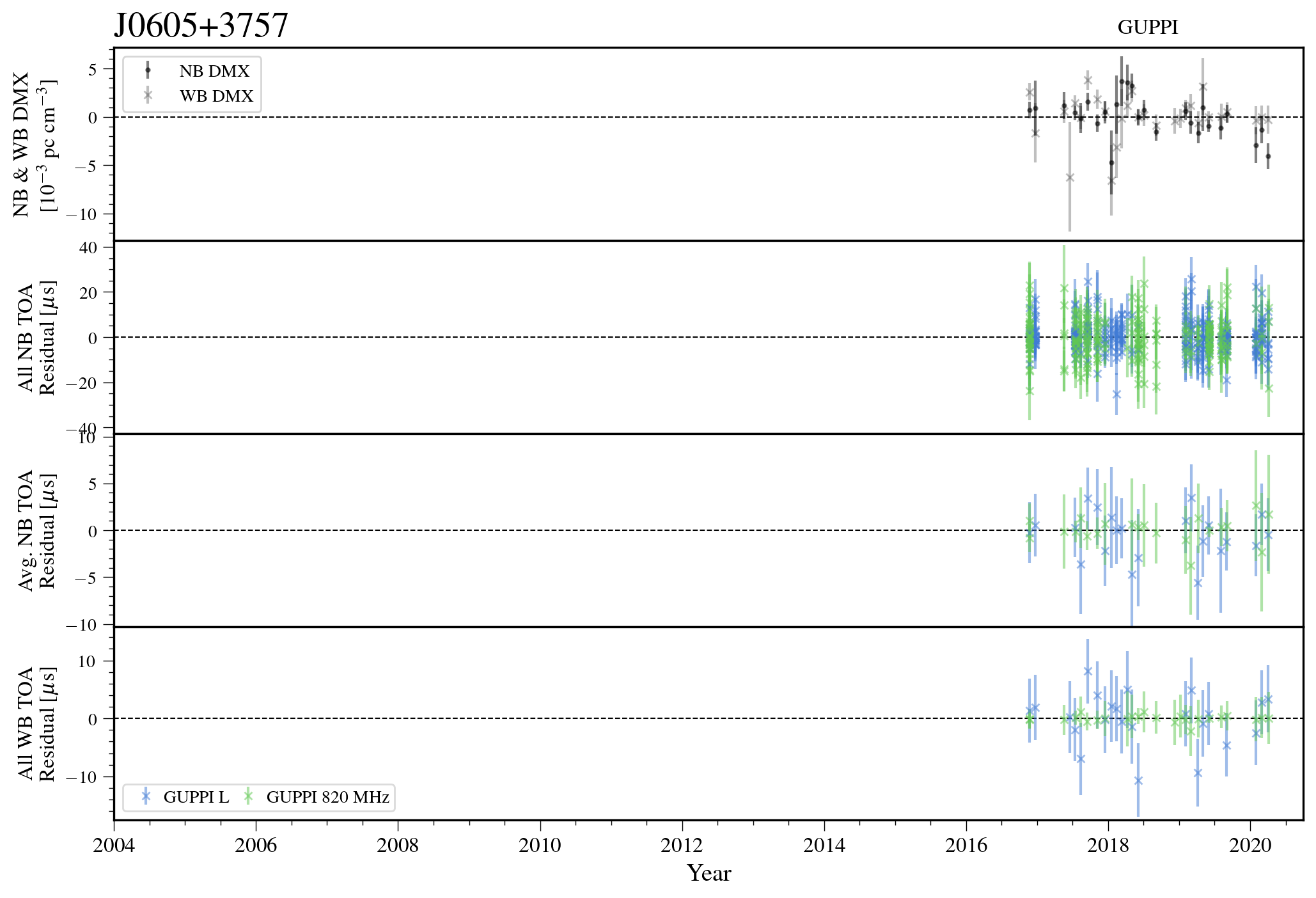}
\caption{Narrowband and wideband timing residuals and DMX timeseries for J0605+3757. See Figure~\ref{fig:summary-J0023+0923} for details.}
\label{fig:summary-J0605+3757}
\end{figure}
\clearpage

\begin{figure}
\centering
\includegraphics[width=0.85\linewidth]{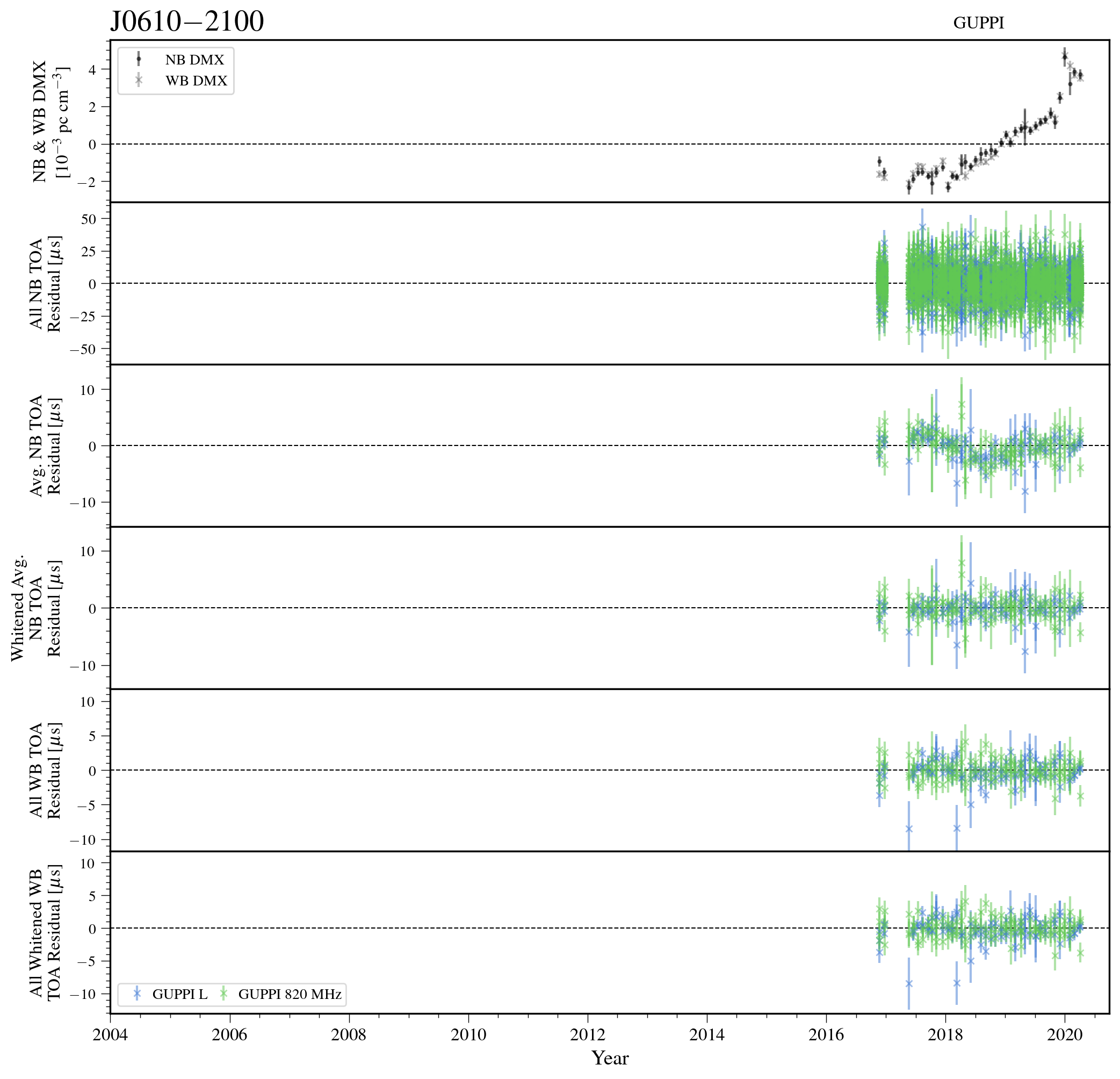}
\caption{Narrowband and wideband timing residuals and DMX timeseries for J0610-2100. See Figure~\ref{fig:summary-J0030+0451} for details.}
\label{fig:summary-J0610-2100}
\end{figure}
\clearpage

\begin{figure}
\centering
\includegraphics[width=0.85\linewidth]{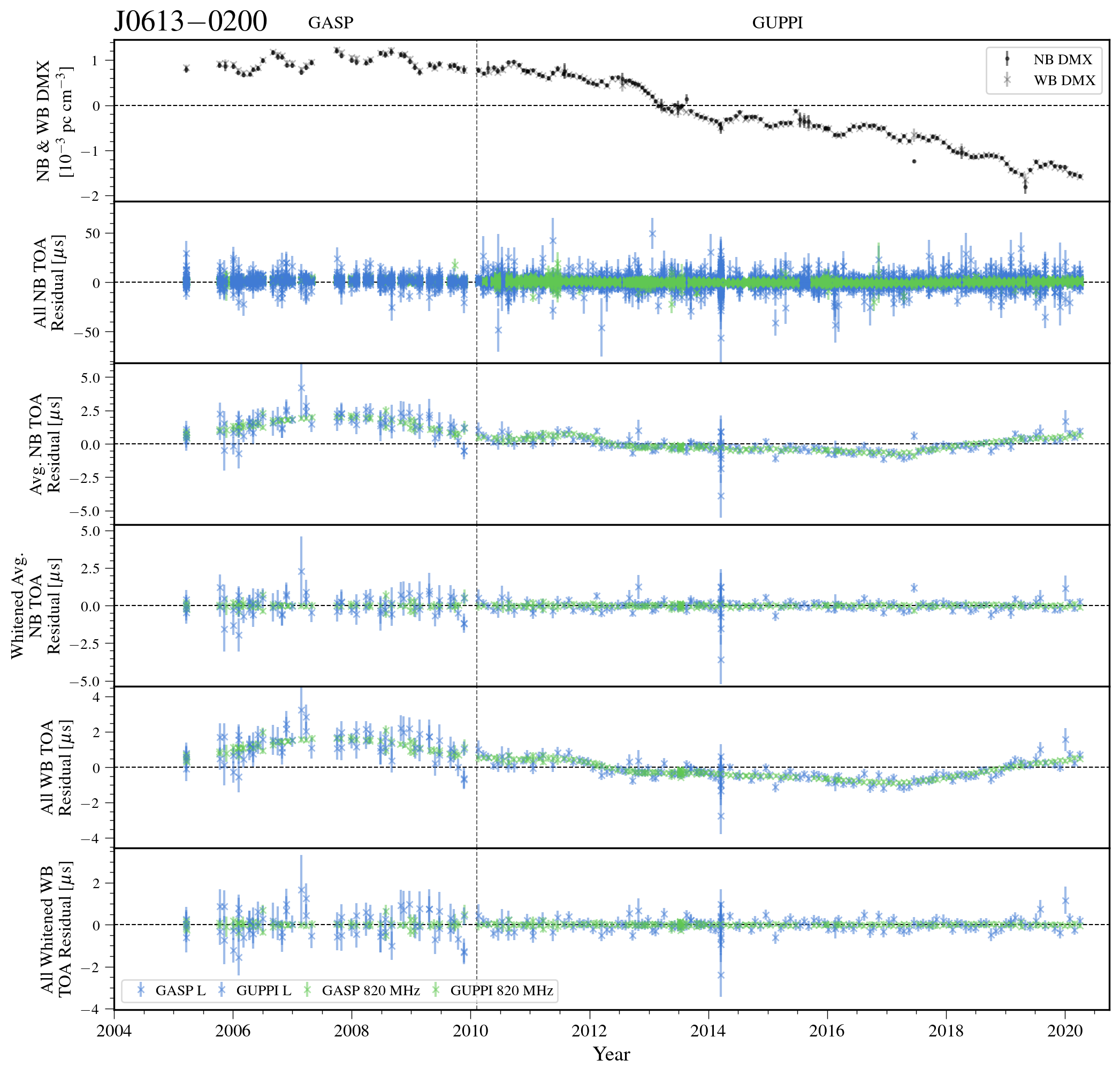}
\caption{Narrowband and wideband timing residuals and DMX timeseries for J0613-0200. See Figure~\ref{fig:summary-J0030+0451} for details.}
\label{fig:summary-J0613-0200}
\end{figure}
\clearpage

\begin{figure}
\centering
\includegraphics[width=0.85\linewidth]{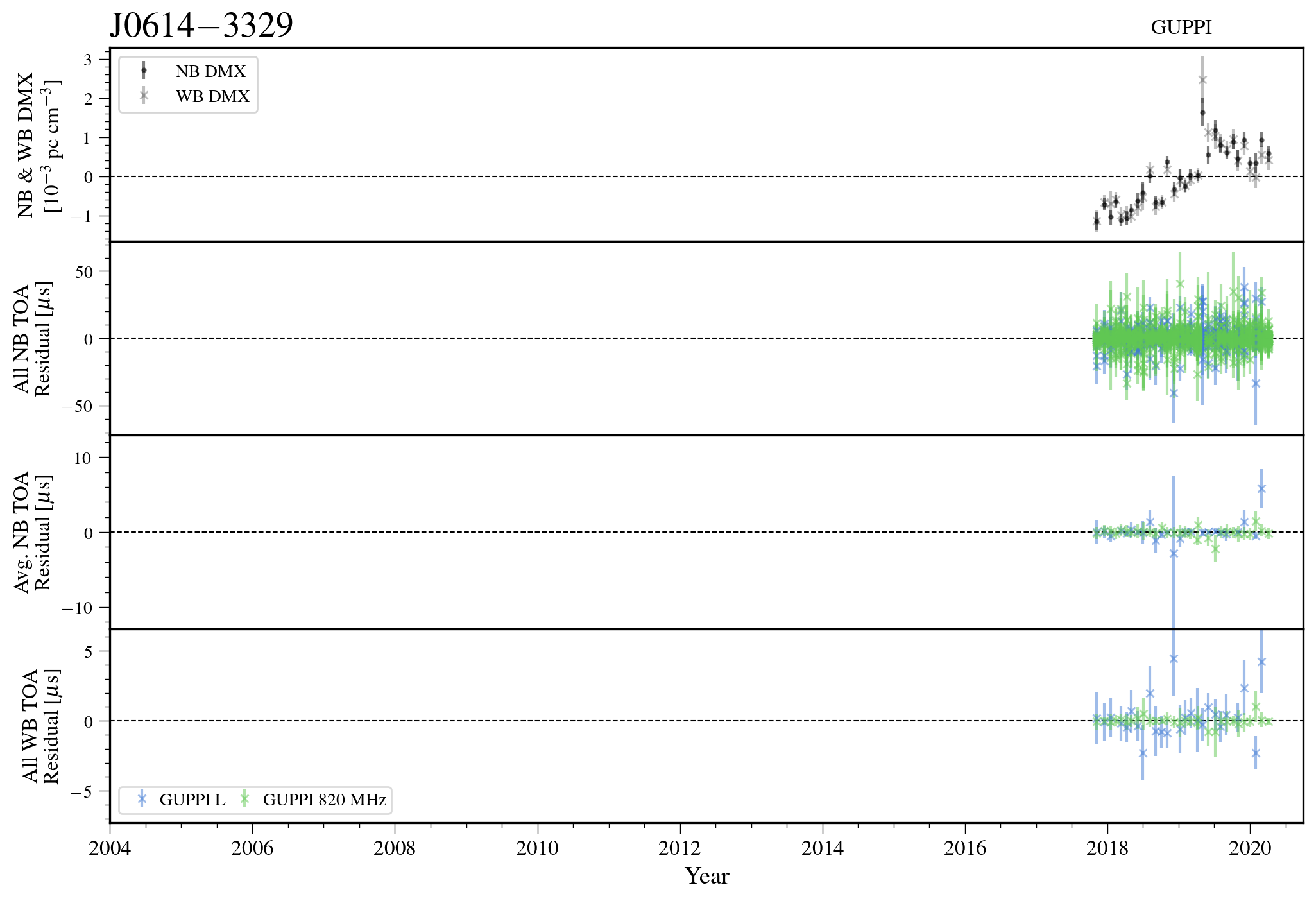}
\caption{Narrowband and wideband timing residuals and DMX timeseries for J0614-3329. See Figure~\ref{fig:summary-J0023+0923} for details.}
\label{fig:summary-J0614-3329}
\end{figure}

\begin{figure}
\centering
\includegraphics[width=0.85\linewidth]{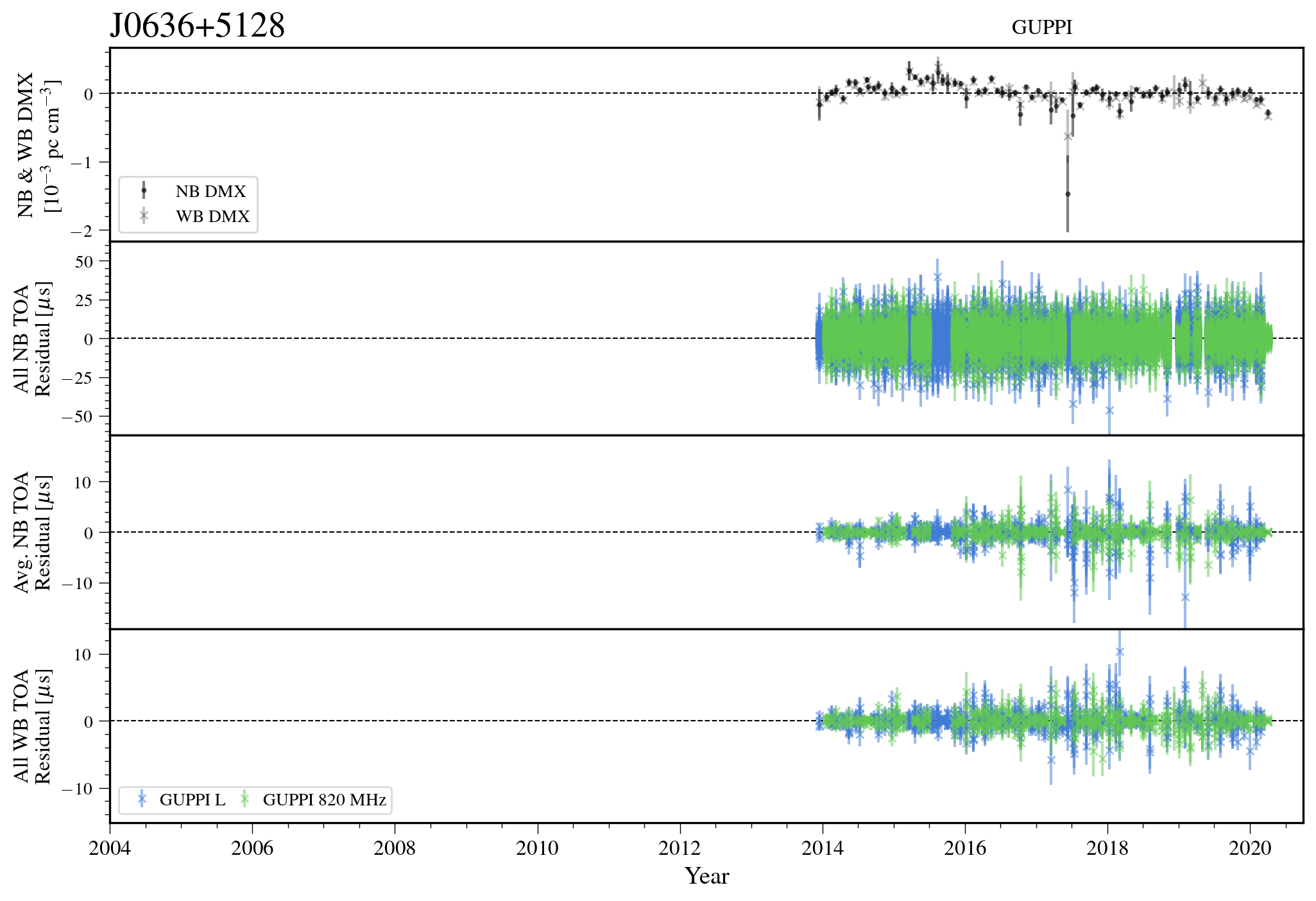}
\caption{Narrowband and wideband timing residuals and DMX timeseries for J0636+5128. See Figure~\ref{fig:summary-J0023+0923} for details.}
\label{fig:summary-J0636+5128}
\end{figure}
\clearpage

\begin{figure}
\centering
\includegraphics[width=0.85\linewidth]{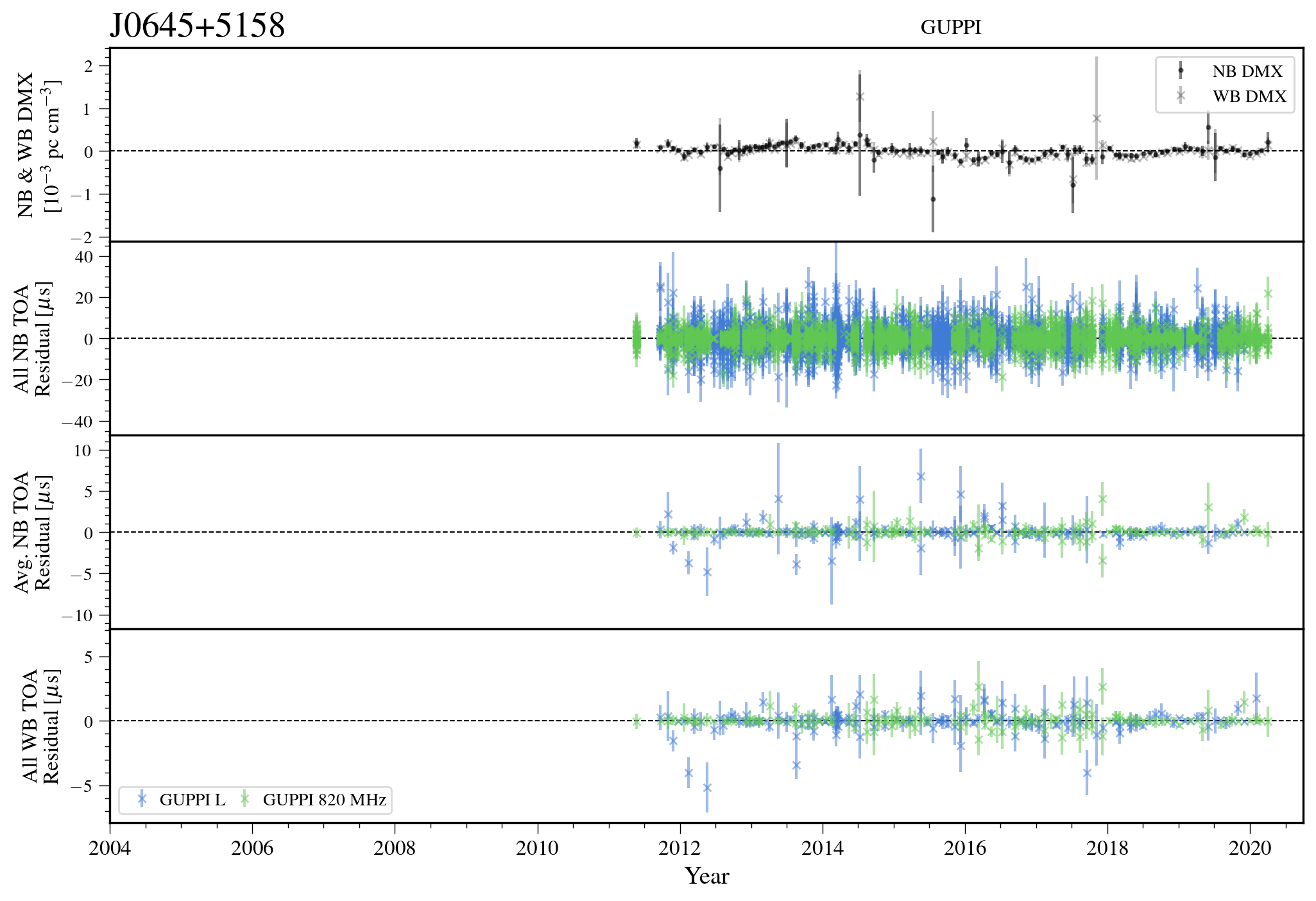}
\caption{Narrowband and wideband timing residuals and DMX timeseries for J0645+5158. See Figure~\ref{fig:summary-J0023+0923} for details.}
\label{fig:summary-J0645+5158}
\end{figure}
\clearpage

\begin{figure}
\centering
\includegraphics[width=0.85\linewidth]{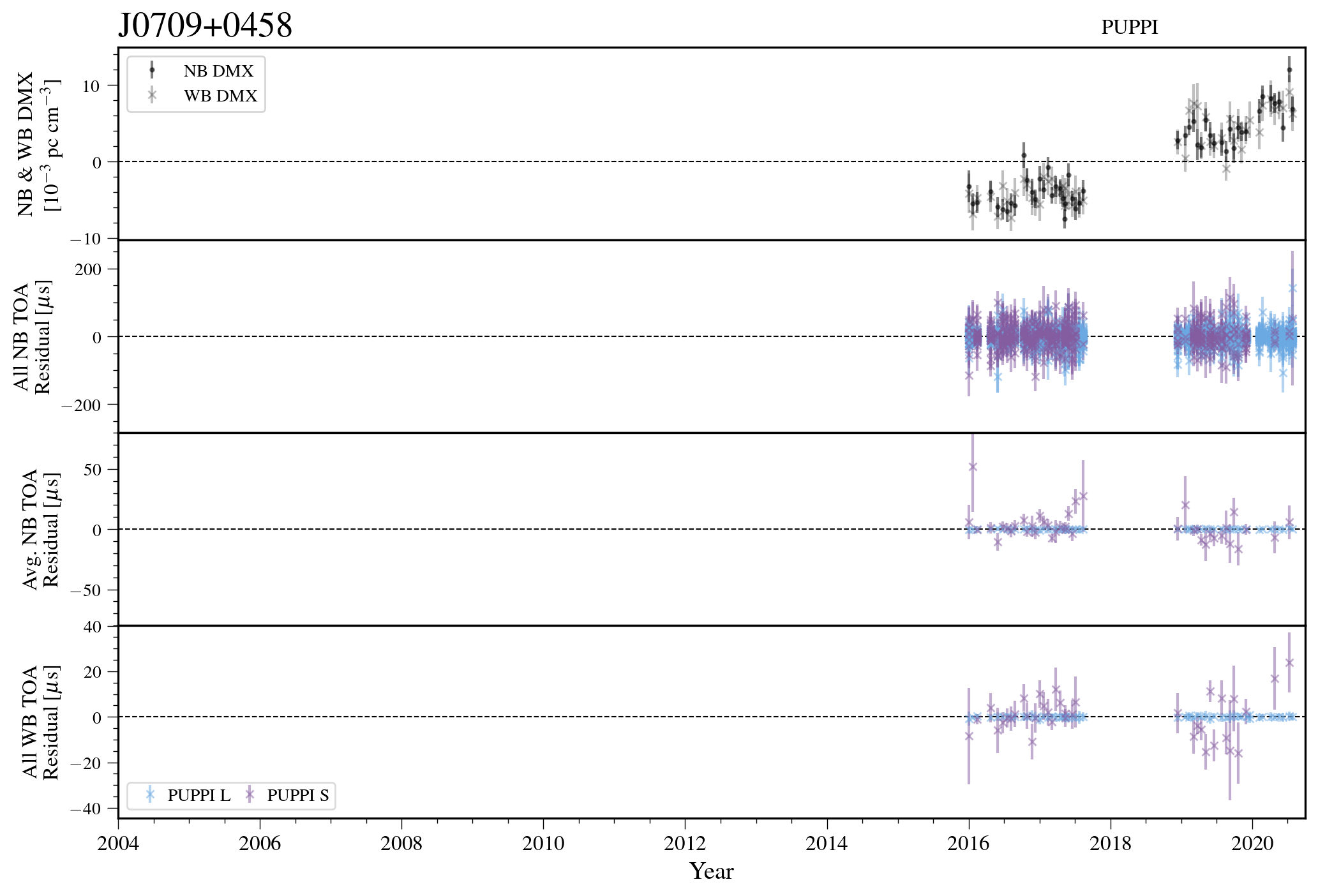}
\caption{Narrowband and wideband timing residuals and DMX timeseries for J0709+0458. See Figure~\ref{fig:summary-J0023+0923} for details.}
\label{fig:summary-J0709+0458}
\end{figure}

\begin{figure}
\centering
\includegraphics[width=0.85\linewidth]{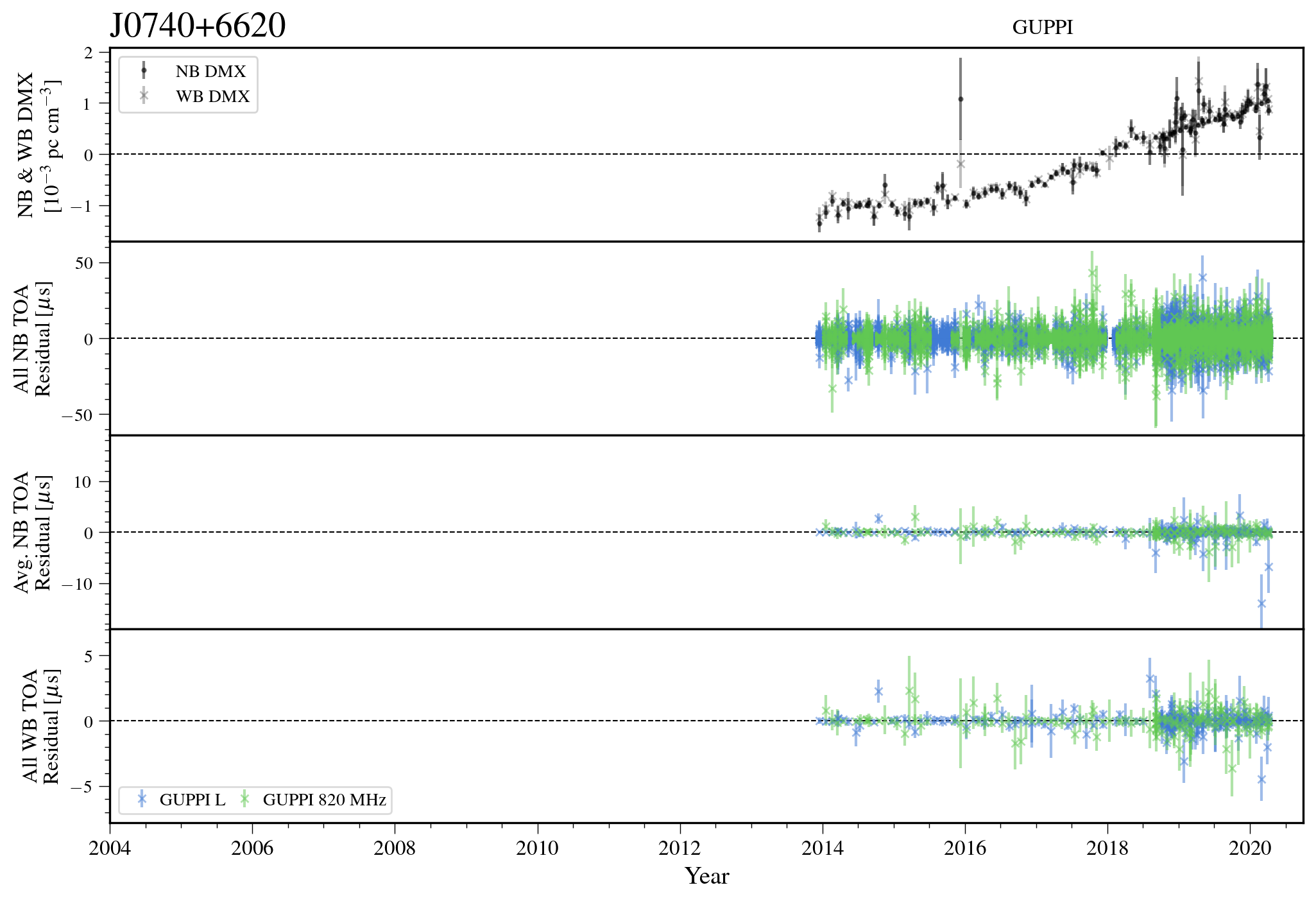}
\caption{Narrowband and wideband timing residuals and DMX timeseries for J0740+6620. See Figure~\ref{fig:summary-J0023+0923} for details.}
\label{fig:summary-J0740+6620}
\end{figure}
\clearpage

\begin{figure}
\centering
\includegraphics[width=0.85\linewidth]{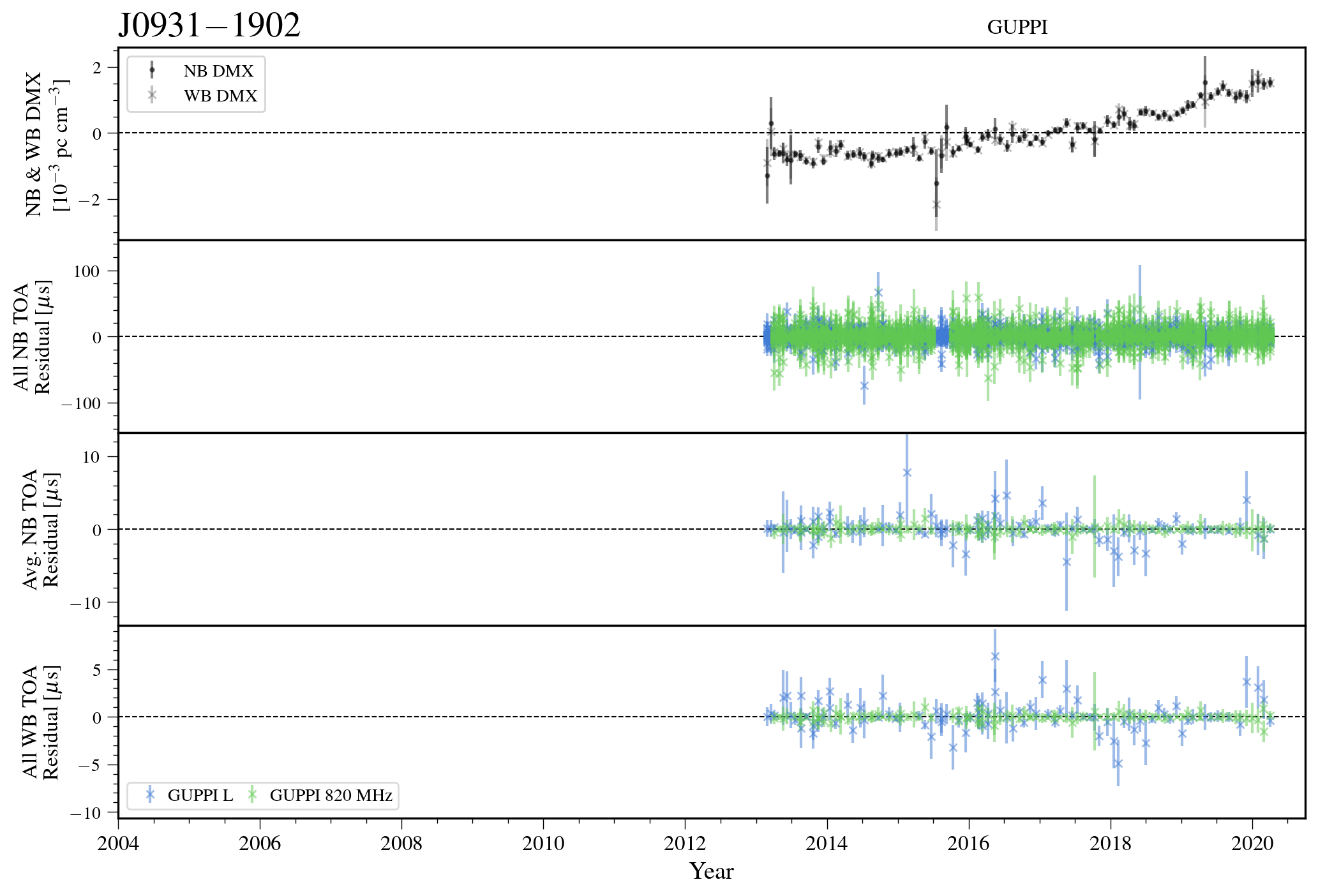}
\caption{Narrowband and wideband timing residuals and DMX timeseries for J0931-1902. See Figure~\ref{fig:summary-J0023+0923} for details.}
\label{fig:summary-J0931-1902}
\end{figure}
\clearpage

\begin{figure}
\centering
\includegraphics[width=0.85\linewidth]{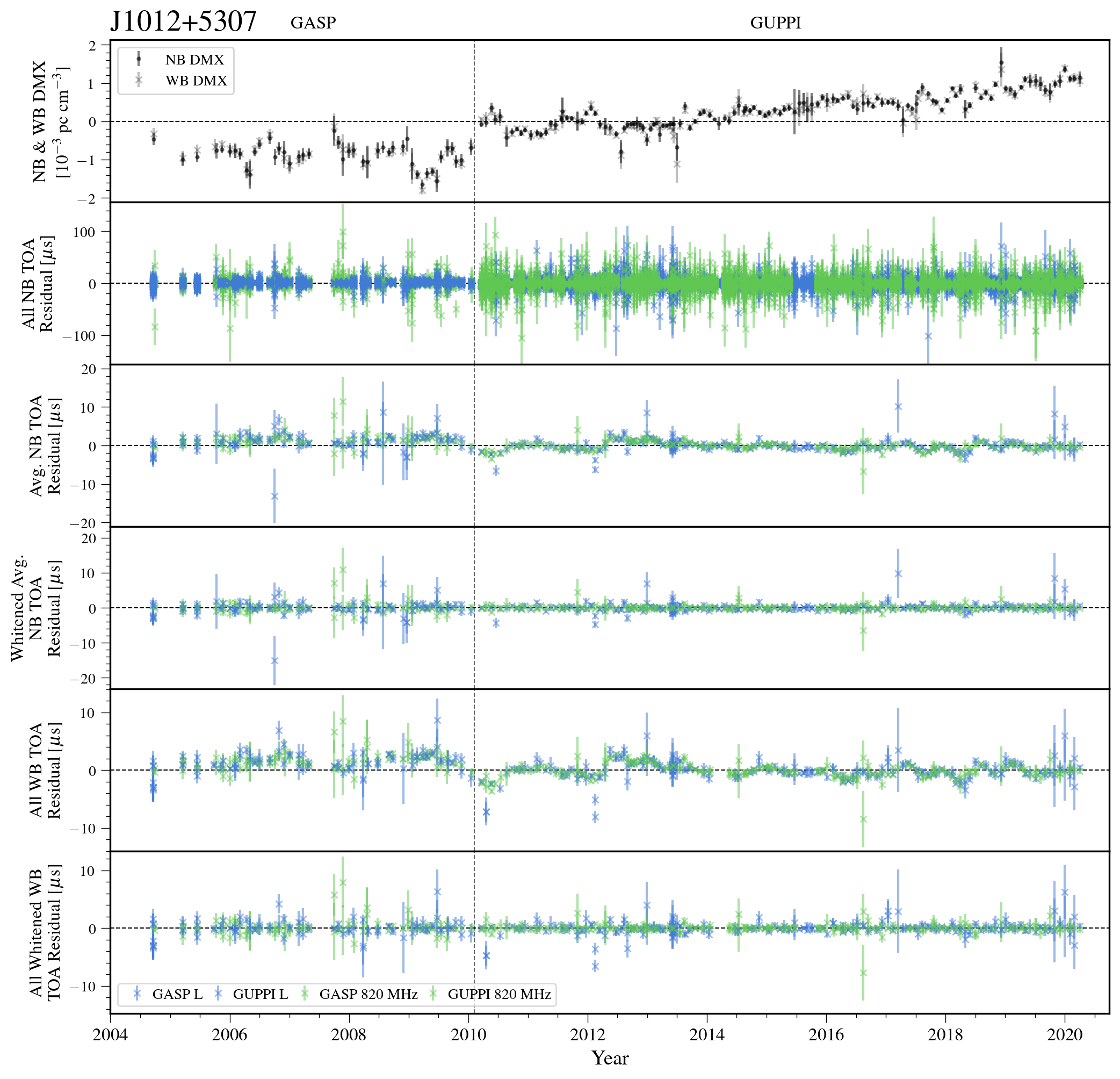}
\caption{Narrowband and wideband timing residuals and DMX timeseries for J1012+5307. See Figure~\ref{fig:summary-J0030+0451} for details.}
\label{fig:summary-J1012+5307}
\end{figure}
\clearpage

\begin{figure}
\centering
\includegraphics[width=0.85\linewidth]{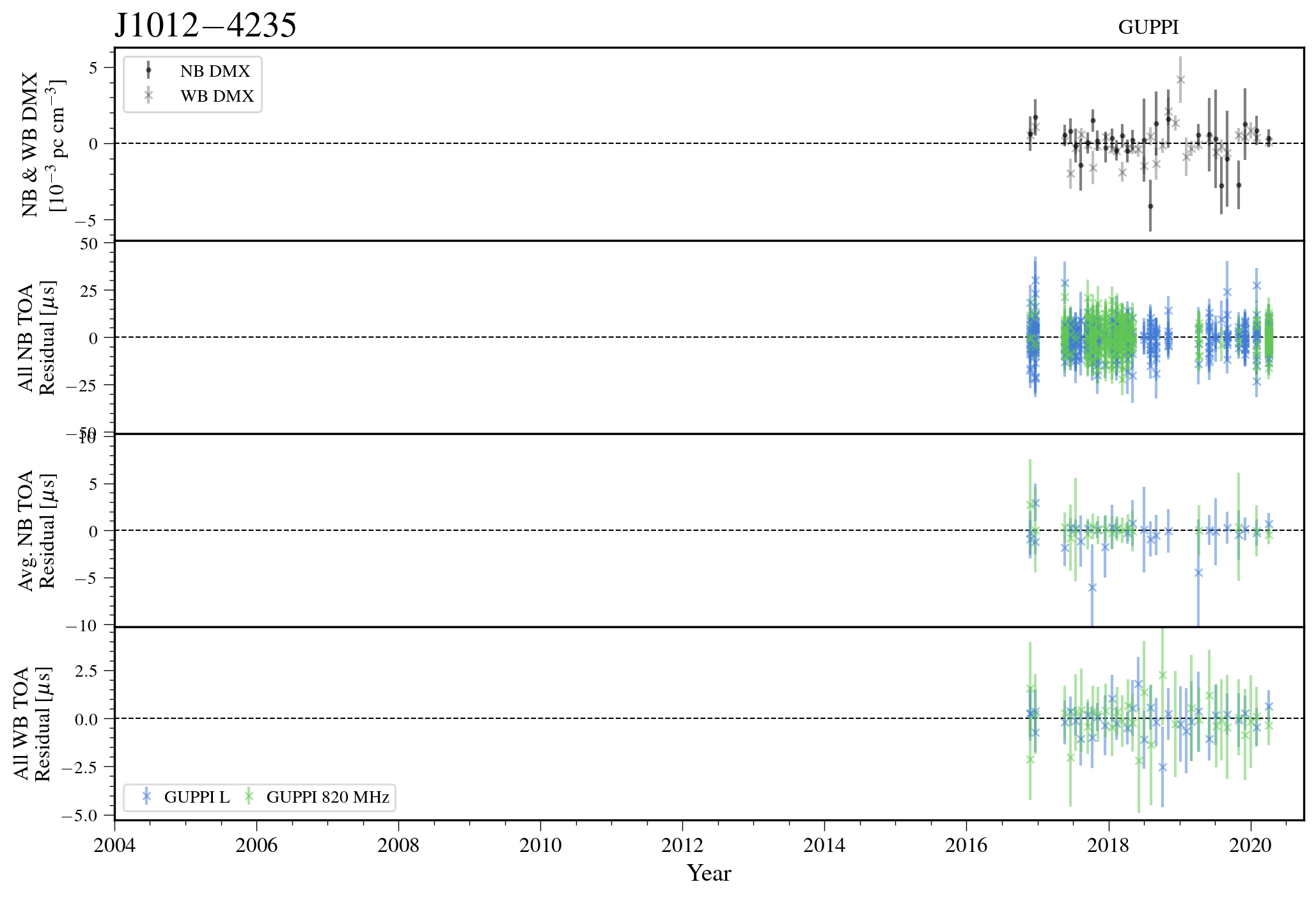}
\caption{Narrowband and wideband timing residuals and DMX timeseries for J1012-4235. See Figure~\ref{fig:summary-J0023+0923} for details.}
\label{fig:summary-J1012-4235}
\end{figure}

\begin{figure}
\centering
\includegraphics[width=0.85\linewidth]{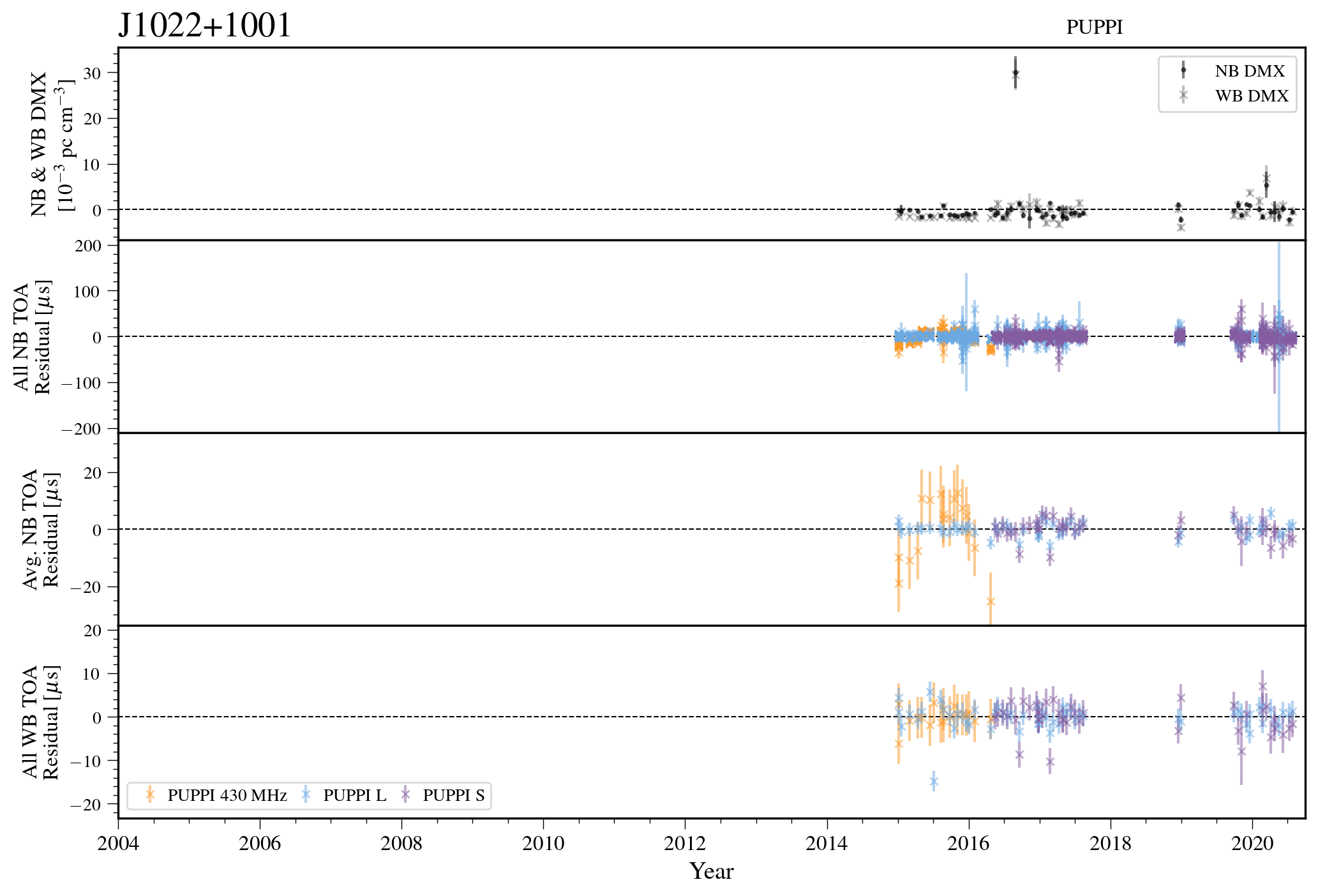}
\caption{Narrowband and wideband timing residuals and DMX timeseries for J1022+1001. See Figure~\ref{fig:summary-J0023+0923} for details.}
\label{fig:summary-J1022+1001}
\end{figure}
\clearpage

\begin{figure}
\centering
\includegraphics[width=0.85\linewidth]{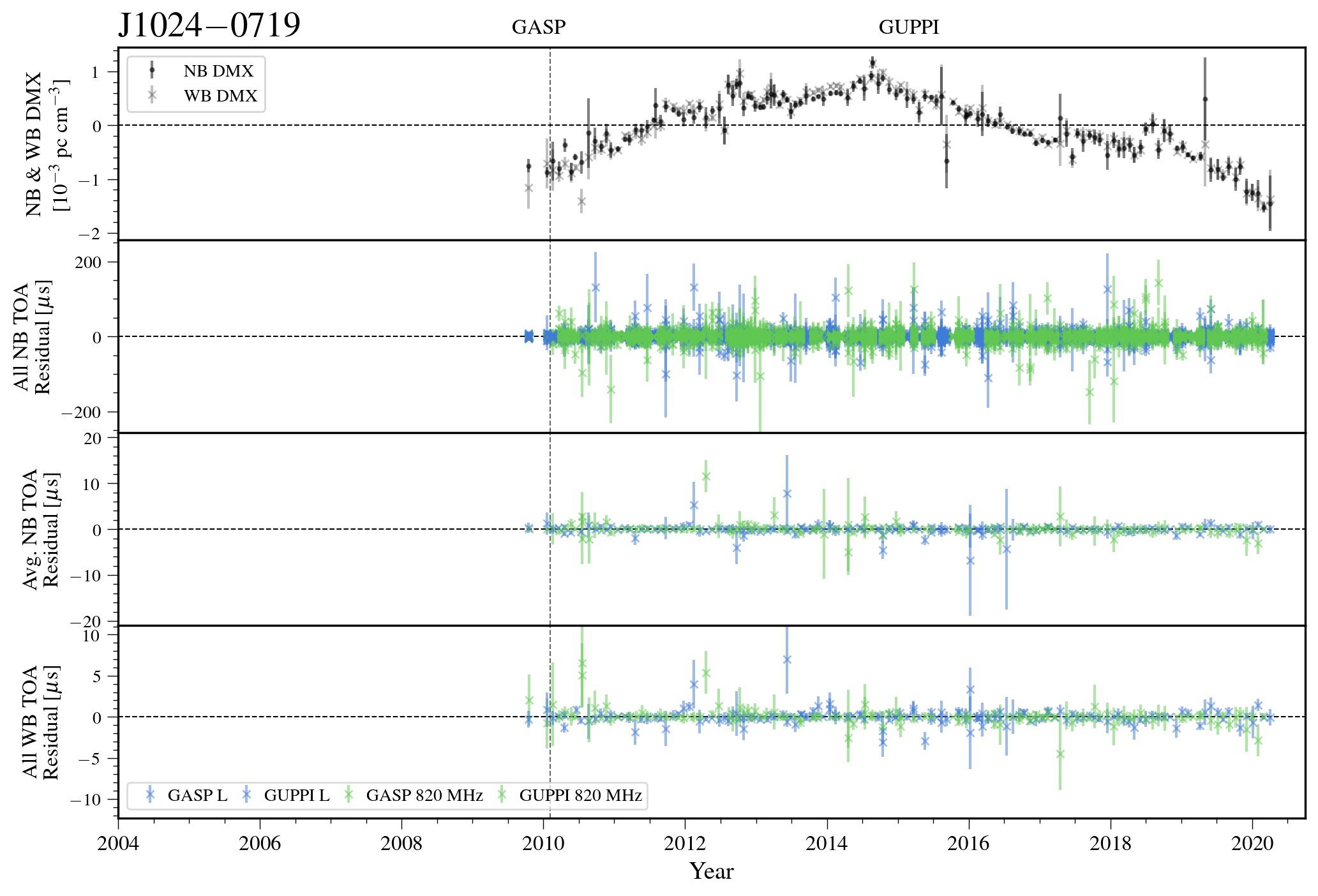}
\caption{Narrowband and wideband timing residuals and DMX timeseries for J1024-0719. See Figure~\ref{fig:summary-J0023+0923} for details.}
\label{fig:summary-J1024-0719}
\end{figure}

\begin{figure}
\centering
\includegraphics[width=0.85\linewidth]{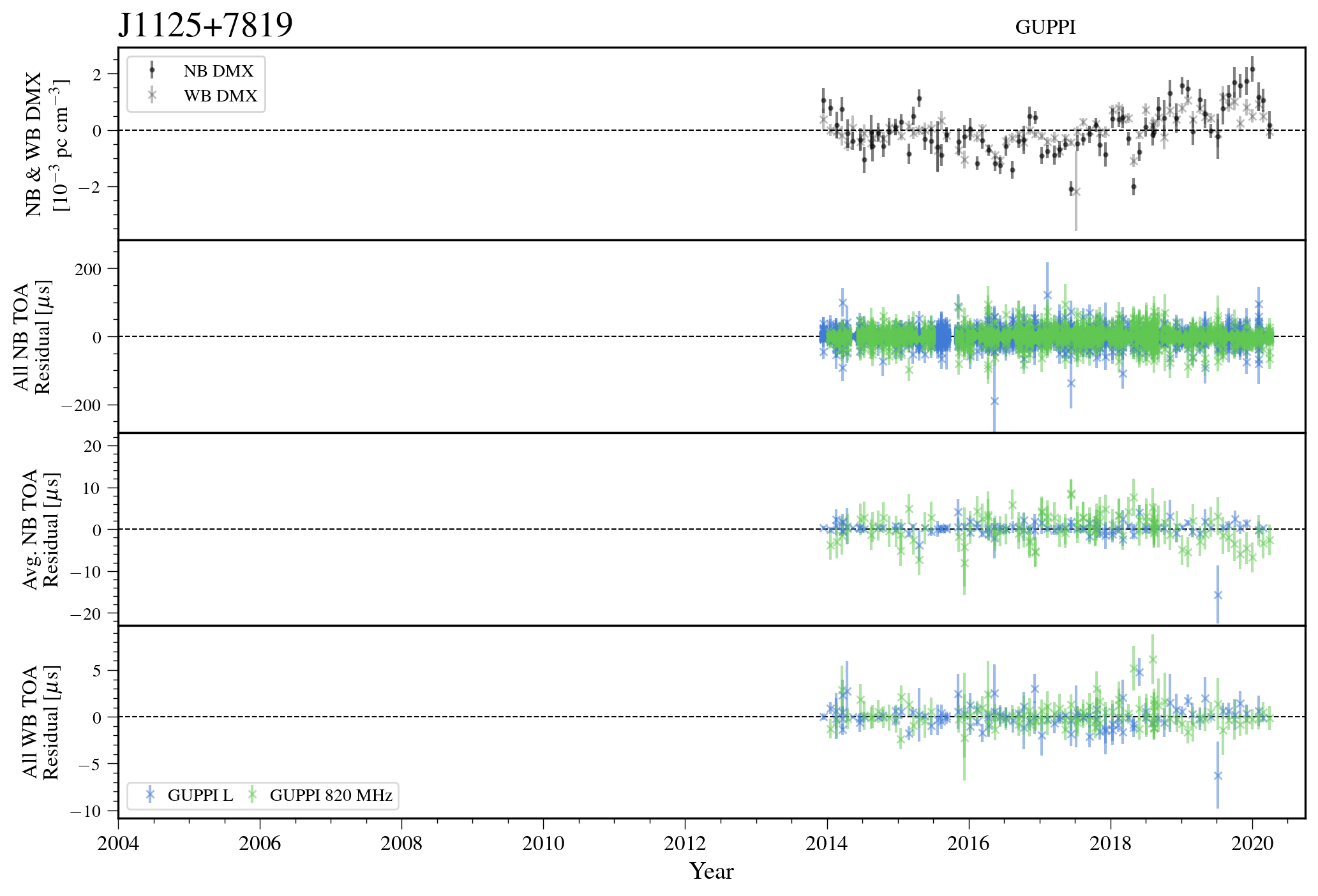}
\caption{Narrowband and wideband timing residuals and DMX timeseries for J1125+7819. See Figure~\ref{fig:summary-J0023+0923} for details.}
\label{fig:summary-J1125+7819}
\end{figure}
\clearpage

\begin{figure}
\centering
\includegraphics[width=0.85\linewidth]{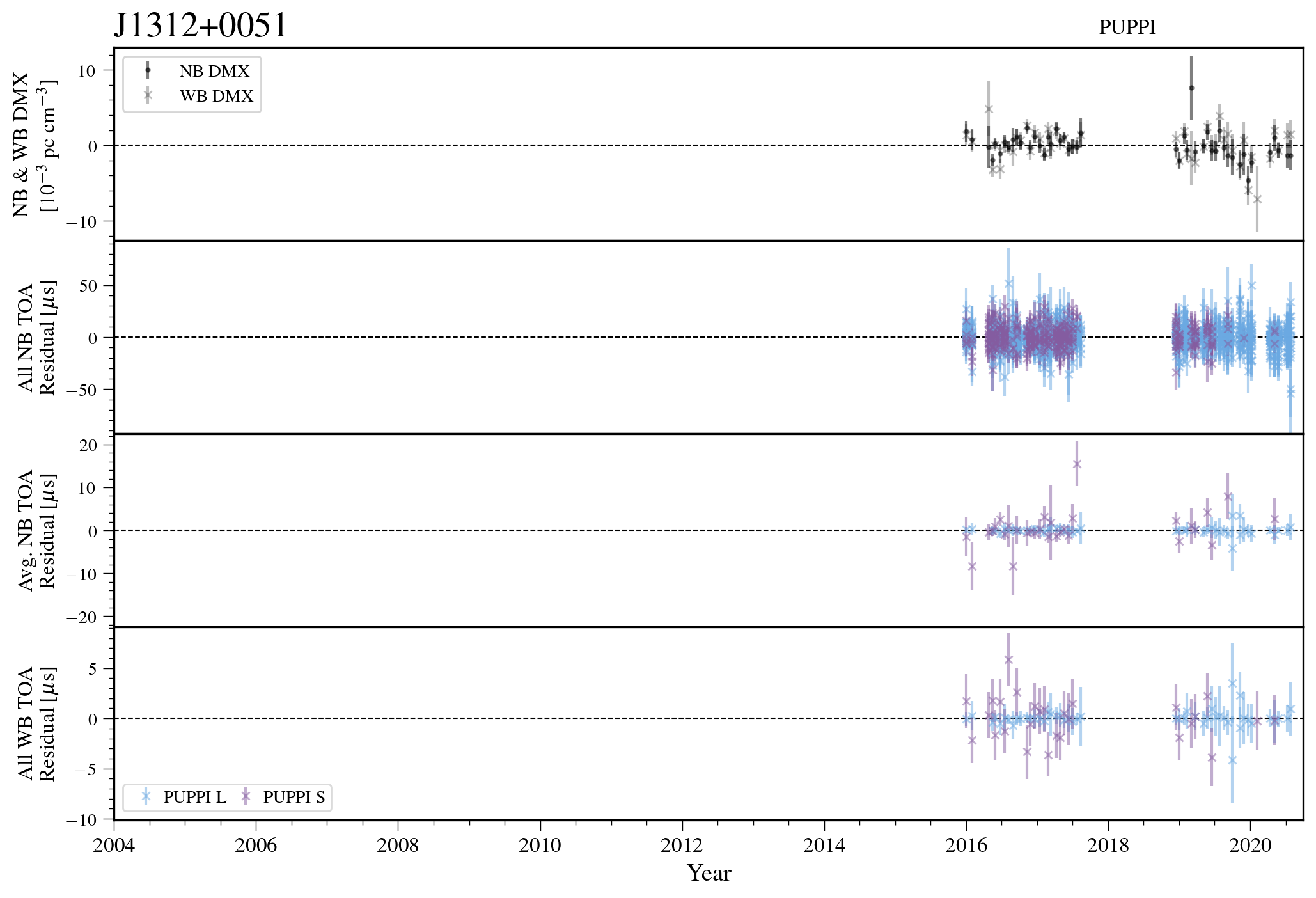}
\caption{Narrowband and wideband timing residuals and DMX timeseries for J1312+0051. See Figure~\ref{fig:summary-J0023+0923} for details.}
\label{fig:summary-J1312+0051}
\end{figure}

\begin{figure}
\centering
\includegraphics[width=0.85\linewidth]{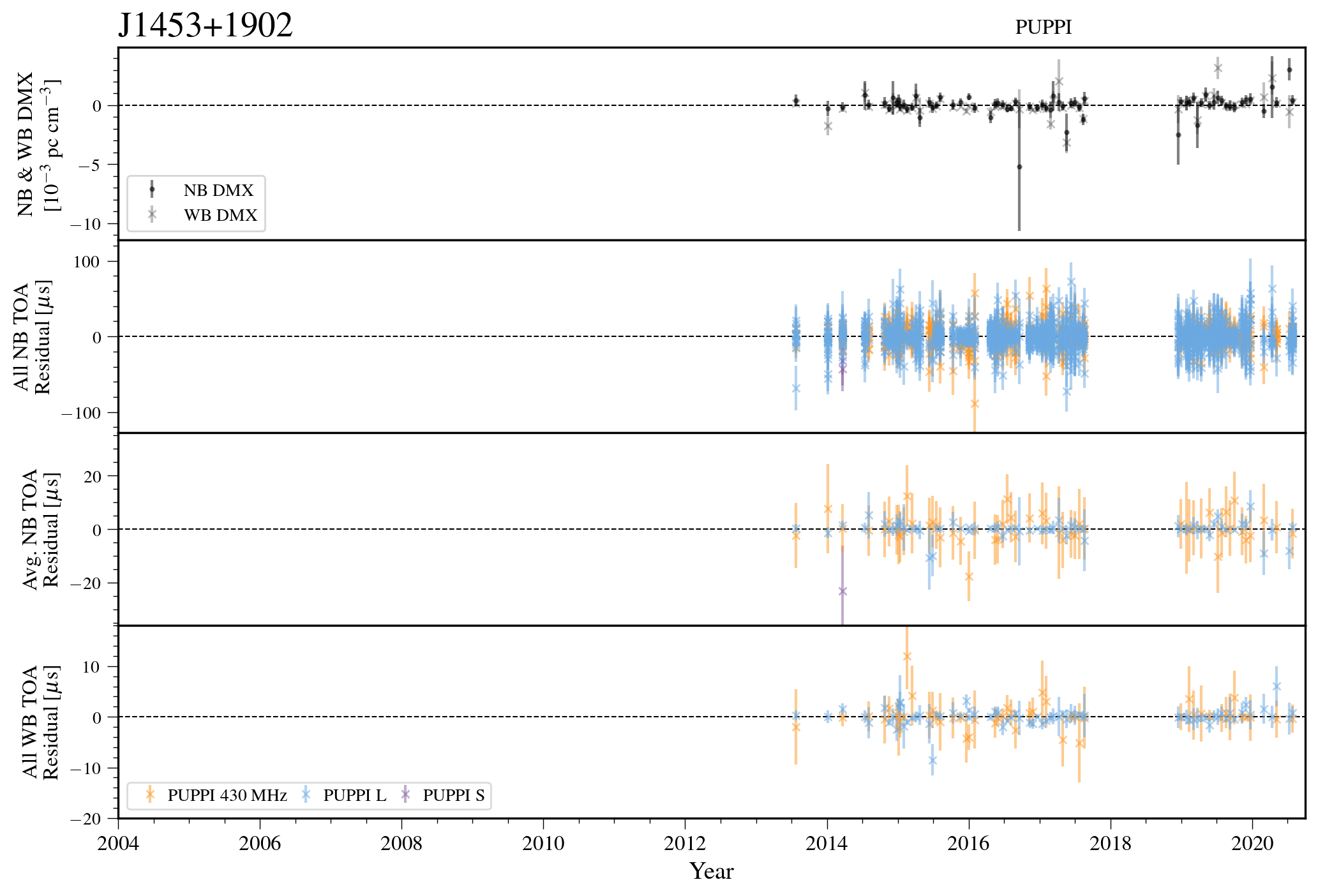}
\caption{Narrowband and wideband timing residuals and DMX timeseries for J1453+1902. See Figure~\ref{fig:summary-J0023+0923} for details.}
\label{fig:summary-J1453+1902}
\end{figure}
\clearpage

\begin{figure}
\centering
\includegraphics[width=0.85\linewidth]{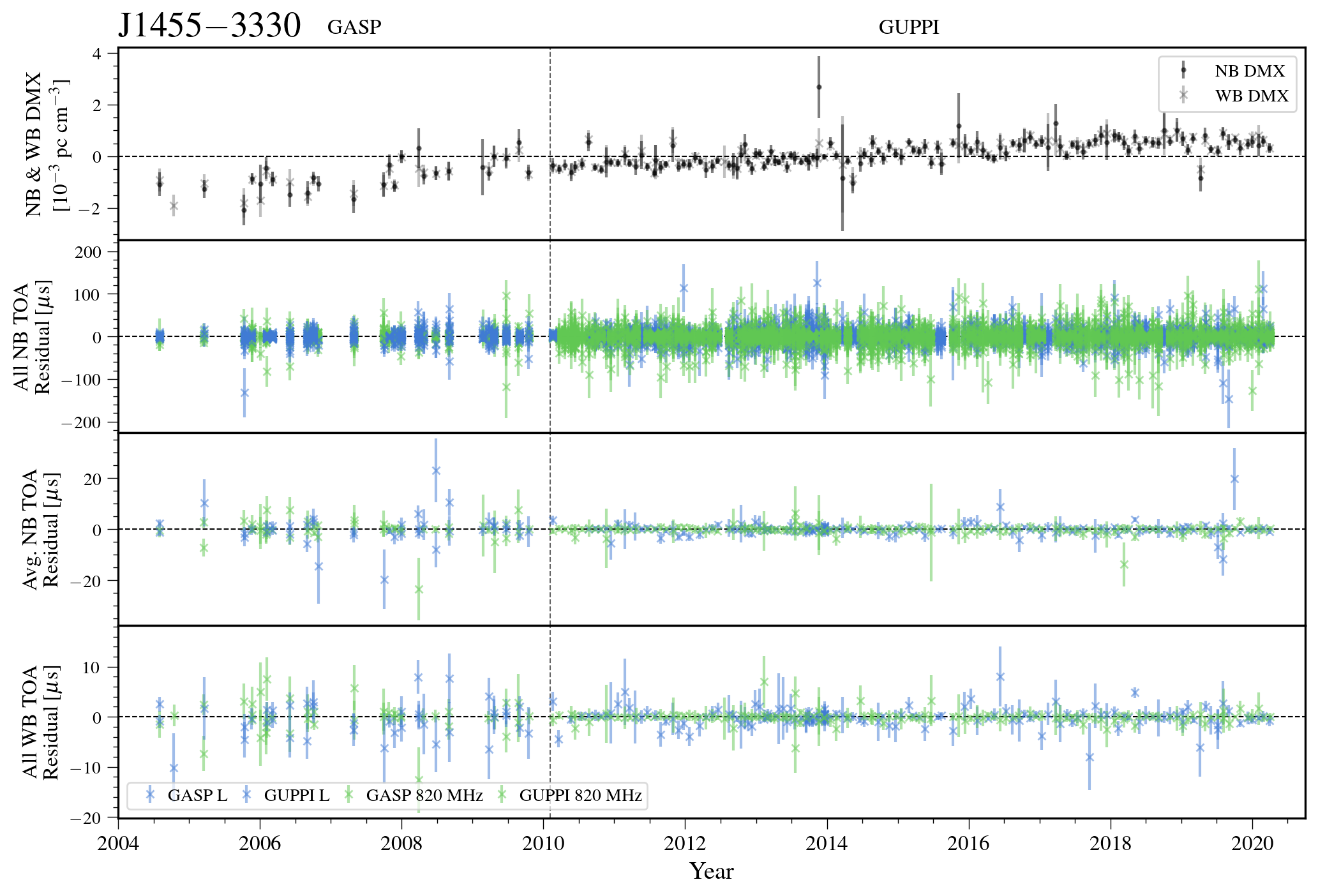}
\caption{Narrowband and wideband timing residuals and DMX timeseries for J1455-3330. See Figure~\ref{fig:summary-J0023+0923} for details.}
\label{fig:summary-J1455-3330}
\end{figure}
\clearpage

\begin{figure}
\centering
\includegraphics[width=0.85\linewidth]{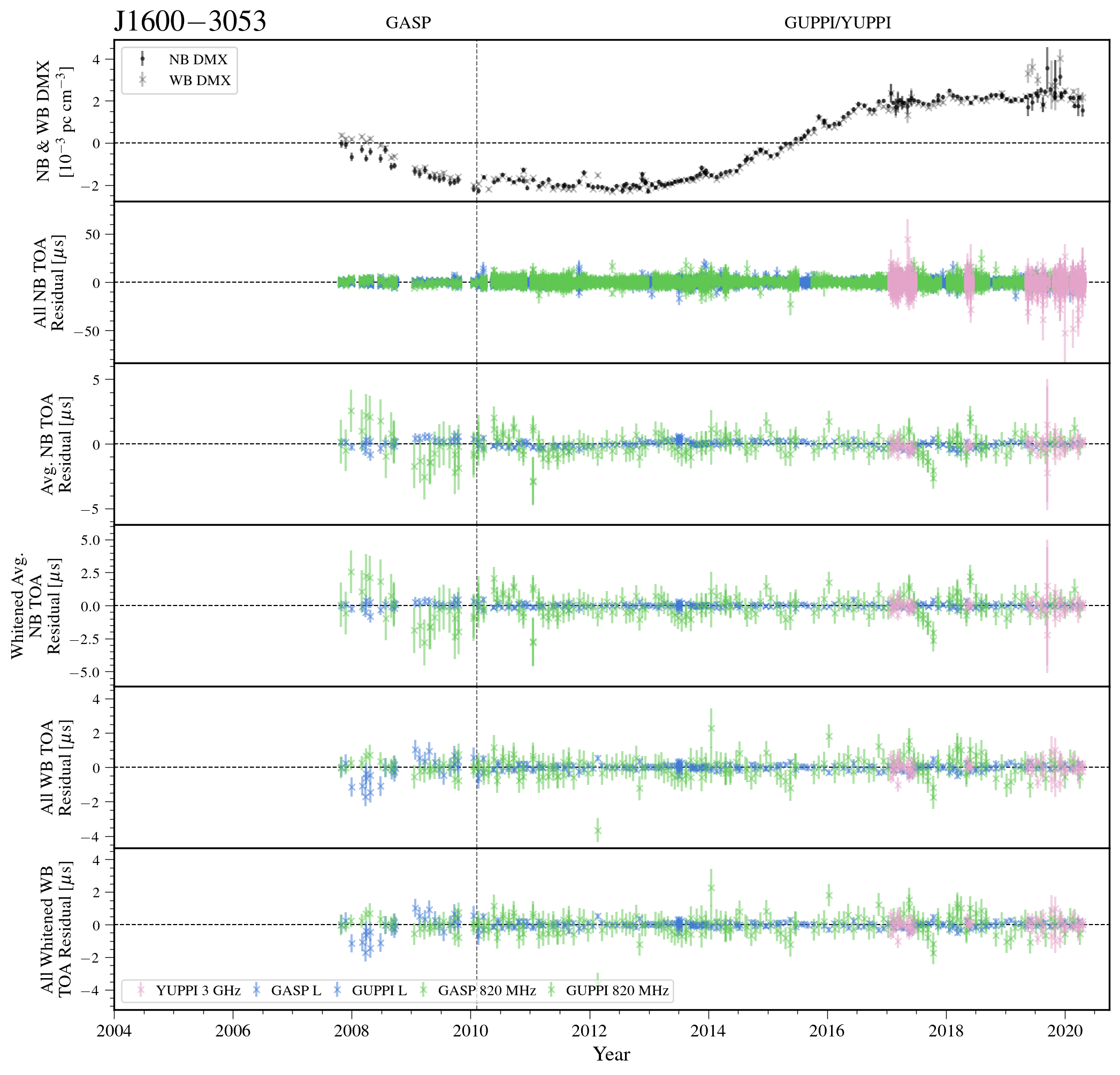}
\caption{Narrowband and wideband timing residuals and DMX timeseries for J1600-3053. See Figure~\ref{fig:summary-J0030+0451} for details.}
\label{fig:summary-J1600-3053}
\end{figure}
\clearpage

\begin{figure}
\centering
\includegraphics[width=0.85\linewidth]{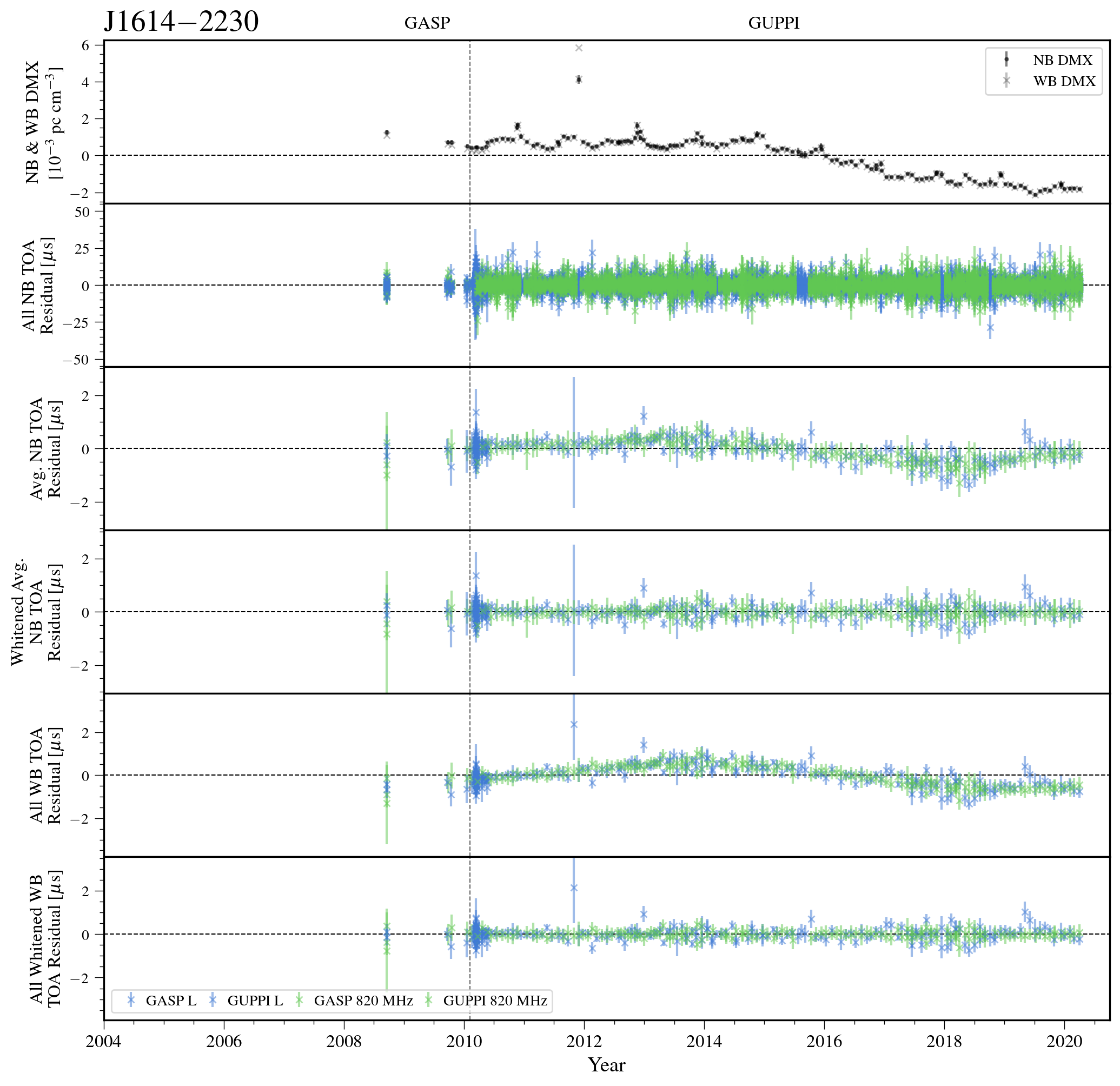}
\caption{Narrowband and wideband timing residuals and DMX timeseries for J1614-2230. See Figure~\ref{fig:summary-J0030+0451} for details.}
\label{fig:summary-J1614-2230}
\end{figure}
\clearpage

\begin{figure}
\centering
\includegraphics[width=0.85\linewidth]{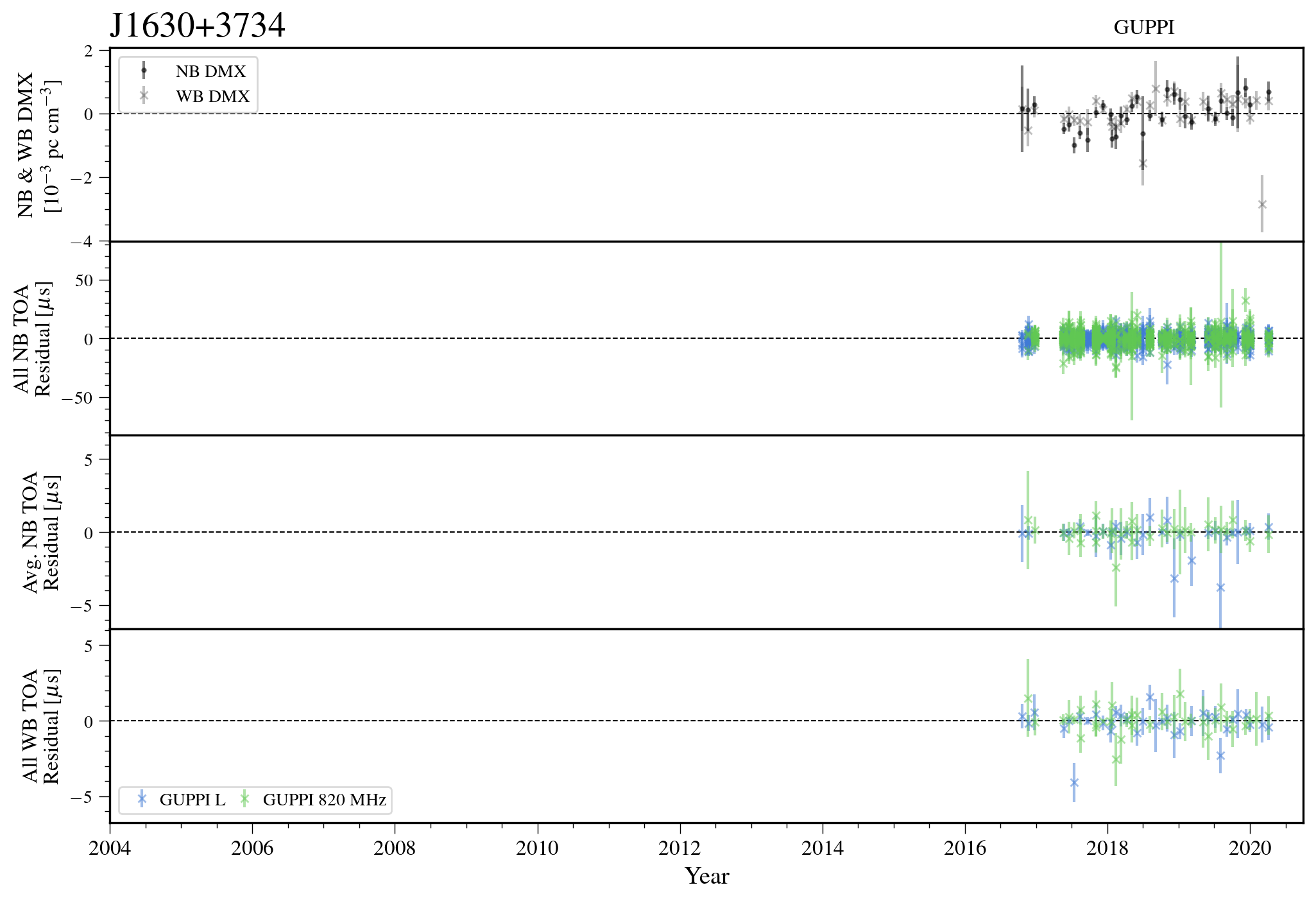}
\caption{Narrowband and wideband timing residuals and DMX timeseries for J1630+3734. See Figure~\ref{fig:summary-J0023+0923} for details.}
\label{fig:summary-J1630+3734}
\end{figure}

\begin{figure}
\centering
\includegraphics[width=0.85\linewidth]{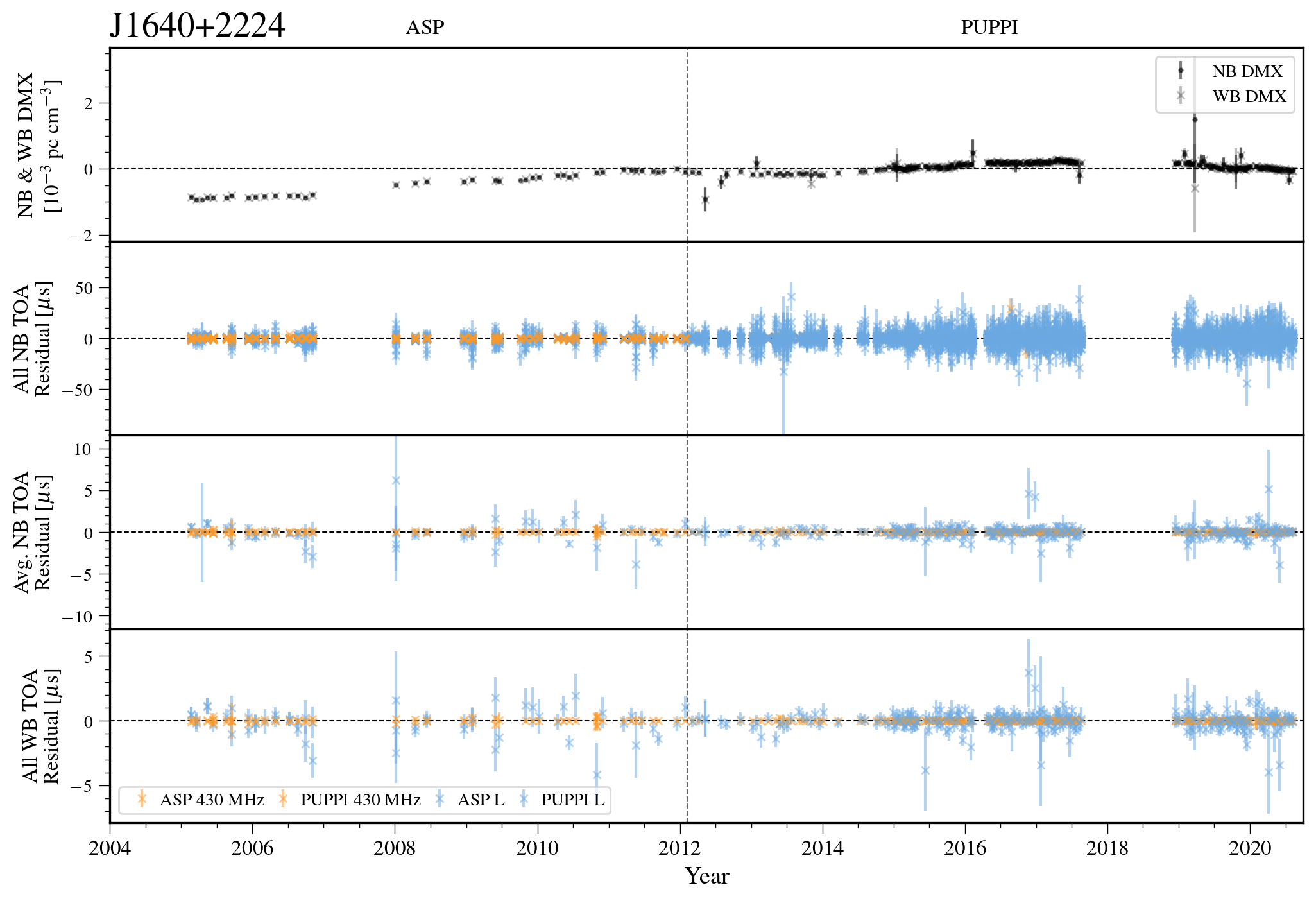}
\caption{Narrowband and wideband timing residuals and DMX timeseries for J1640+2224. See Figure~\ref{fig:summary-J0023+0923} for details.}
\label{fig:summary-J1640+2224}
\end{figure}
\clearpage

\begin{figure}
\centering
\includegraphics[width=0.85\linewidth]{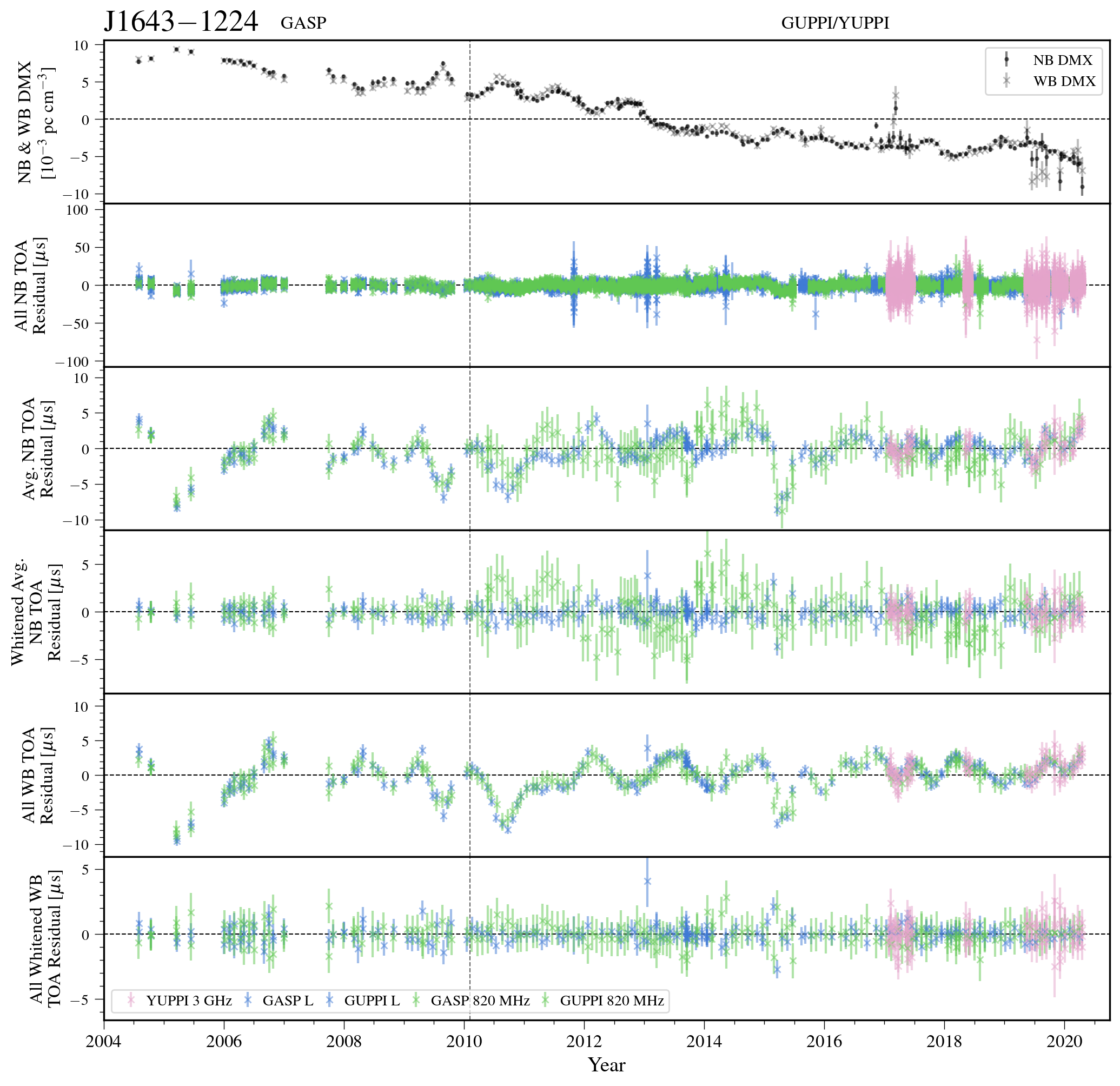}
\caption{Narrowband and wideband timing residuals and DMX timeseries for J1643-1224. See Figure~\ref{fig:summary-J0030+0451} for details.}
\label{fig:summary-J1643-1224}
\end{figure}
\clearpage

\begin{figure}
\centering
\includegraphics[width=0.85\linewidth]{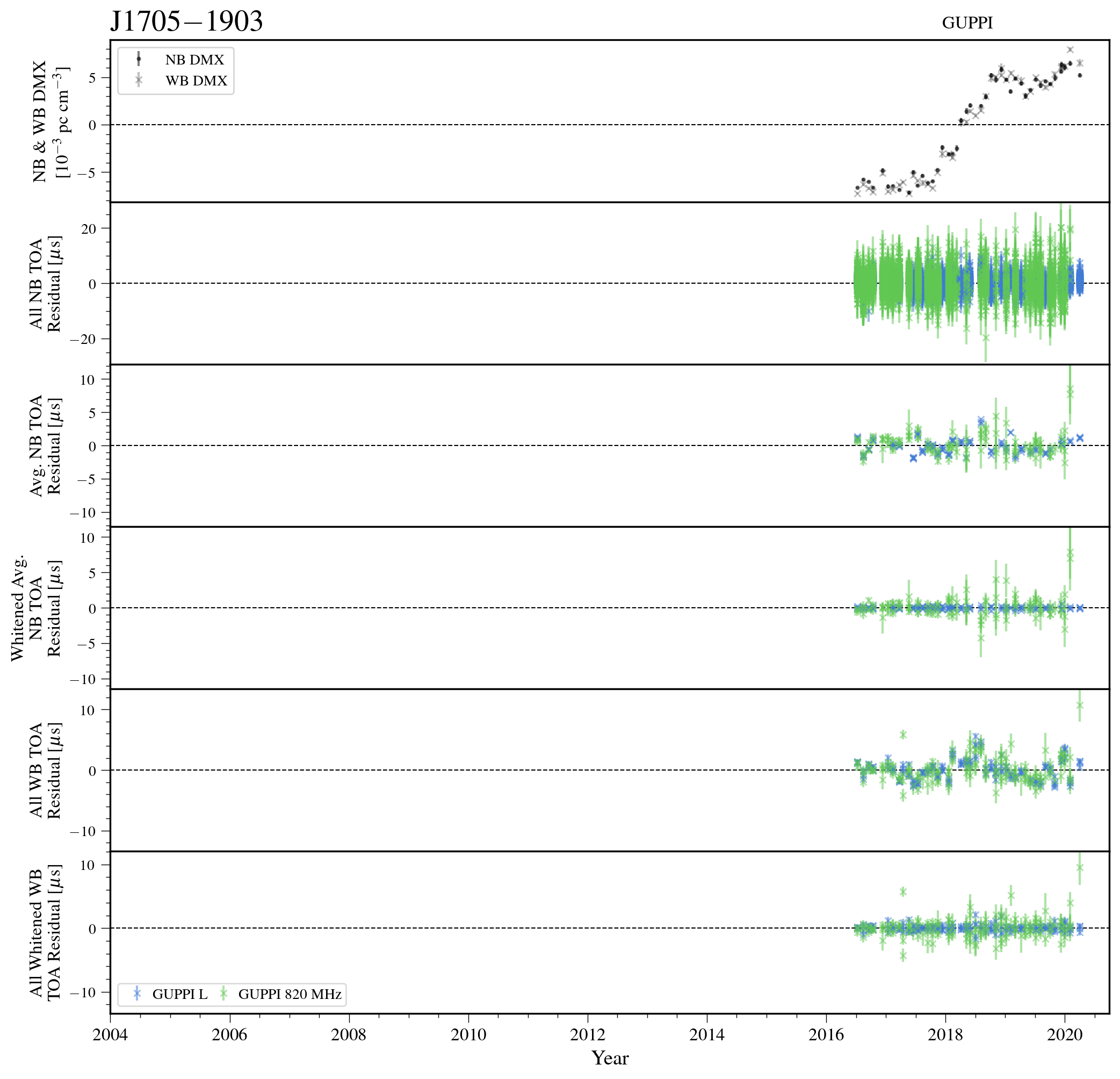}
\caption{Narrowband and wideband timing residuals and DMX timeseries for J1705-1903. See Figure~\ref{fig:summary-J0030+0451} for details.}
\label{fig:summary-J1705-1903}
\end{figure}
\clearpage

\begin{figure}
\centering
\includegraphics[width=0.85\linewidth]{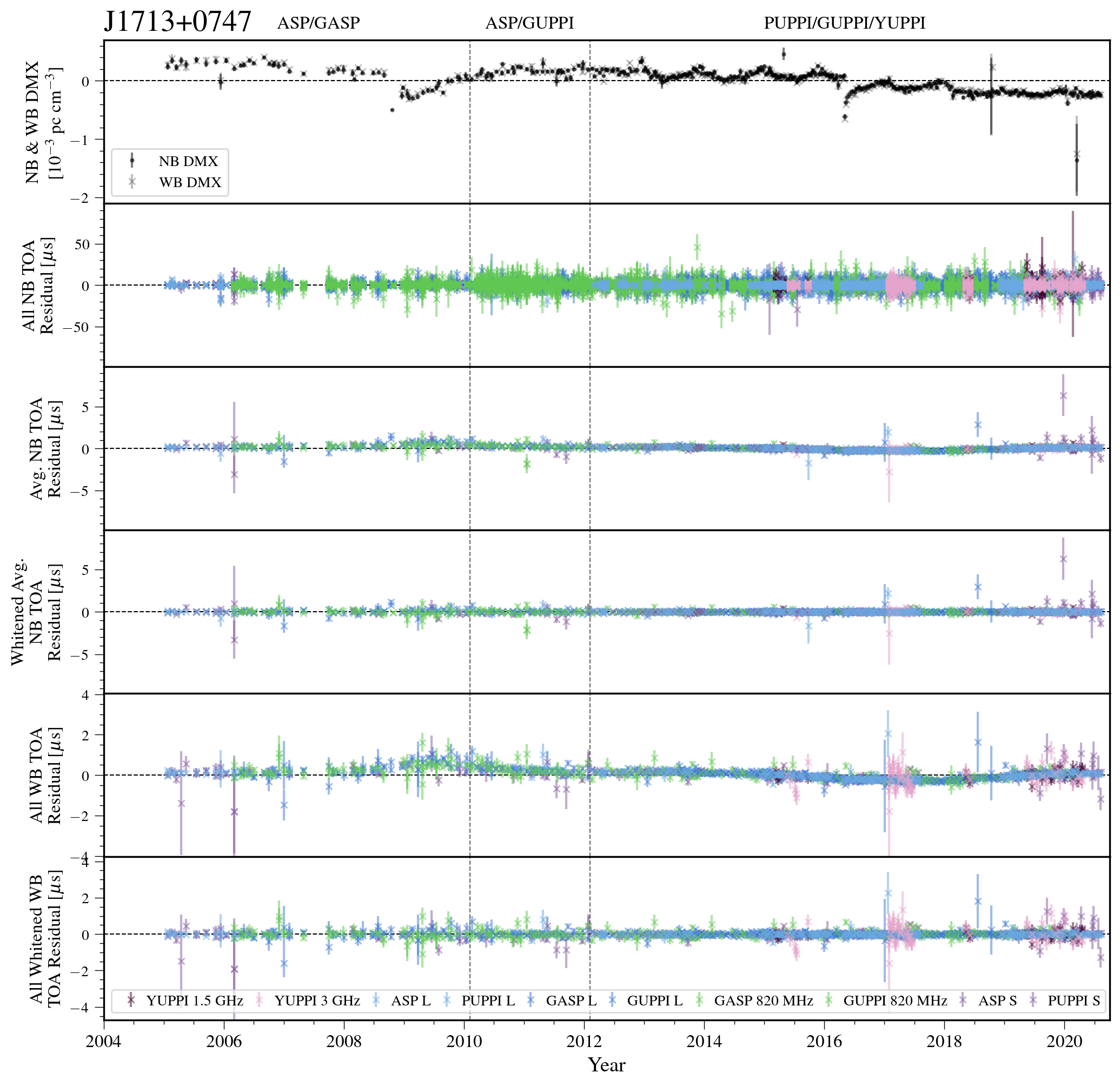}
\caption{Narrowband and wideband timing residuals and DMX timeseries for J1713+0747. See Figure~\ref{fig:summary-J0030+0451} for details.}
\label{fig:summary-J1713+0747}
\end{figure}
\clearpage

\begin{figure}
\centering
\includegraphics[width=0.85\linewidth]{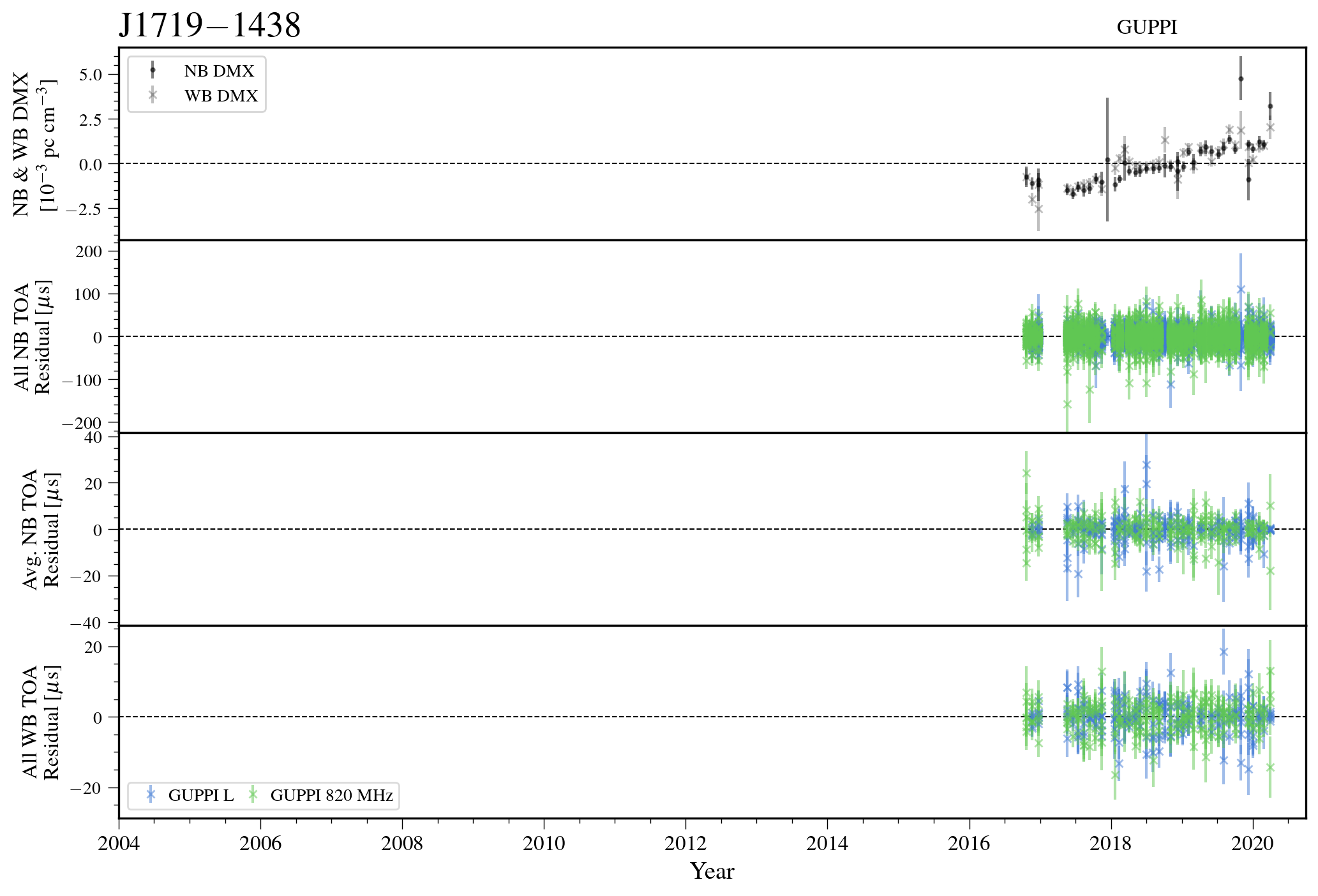}
\caption{Narrowband and wideband timing residuals and DMX timeseries for J1719-1438. See Figure~\ref{fig:summary-J0023+0923} for details.}
\label{fig:summary-J1719-1438}
\end{figure}

\begin{figure}
\centering
\includegraphics[width=0.85\linewidth]{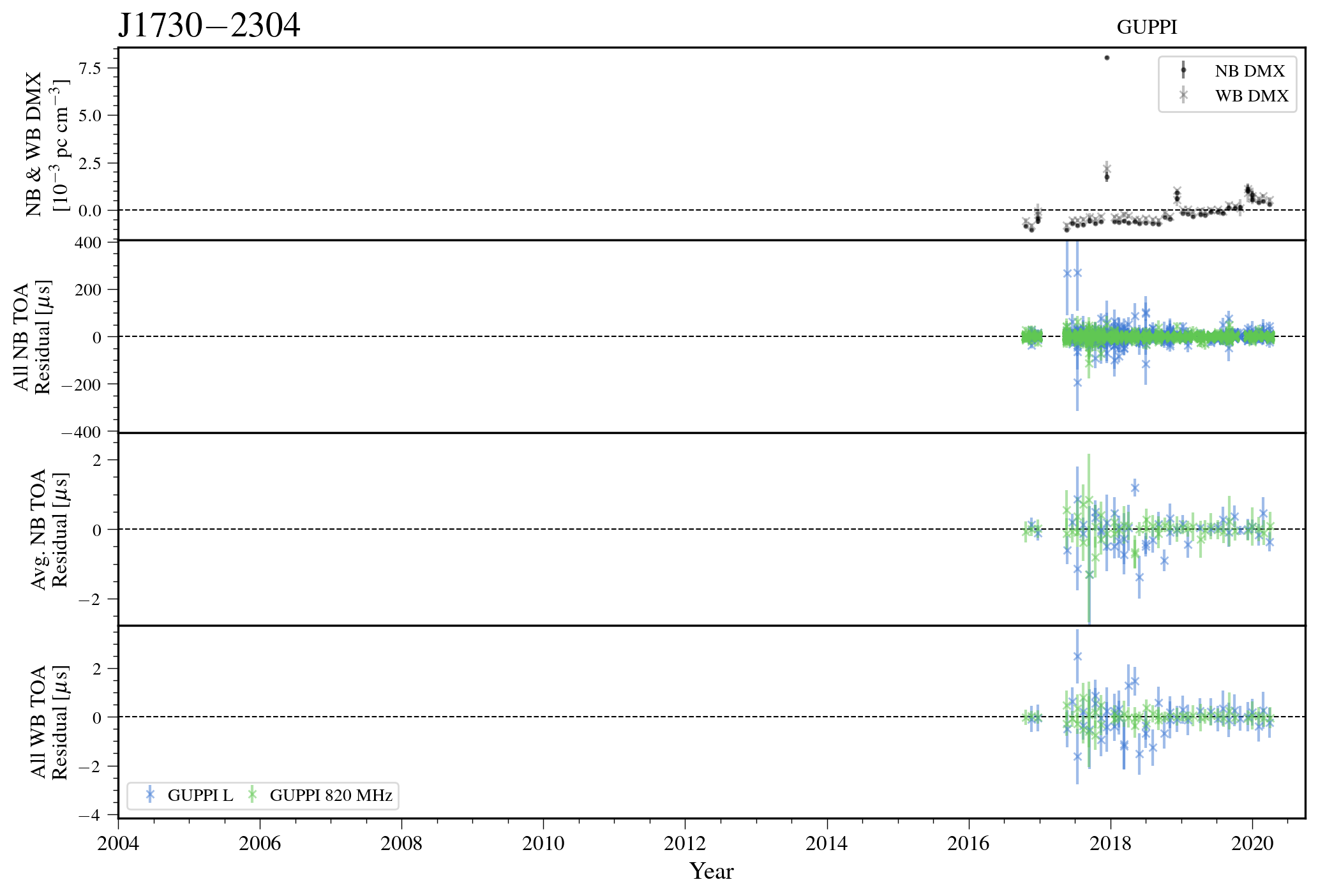}
\caption{Narrowband and wideband timing residuals and DMX timeseries for J1730-2304. See Figure~\ref{fig:summary-J0023+0923} for details.}
\label{fig:summary-J1730-2304}
\end{figure}
\clearpage

\begin{figure}
\centering
\includegraphics[width=0.85\linewidth]{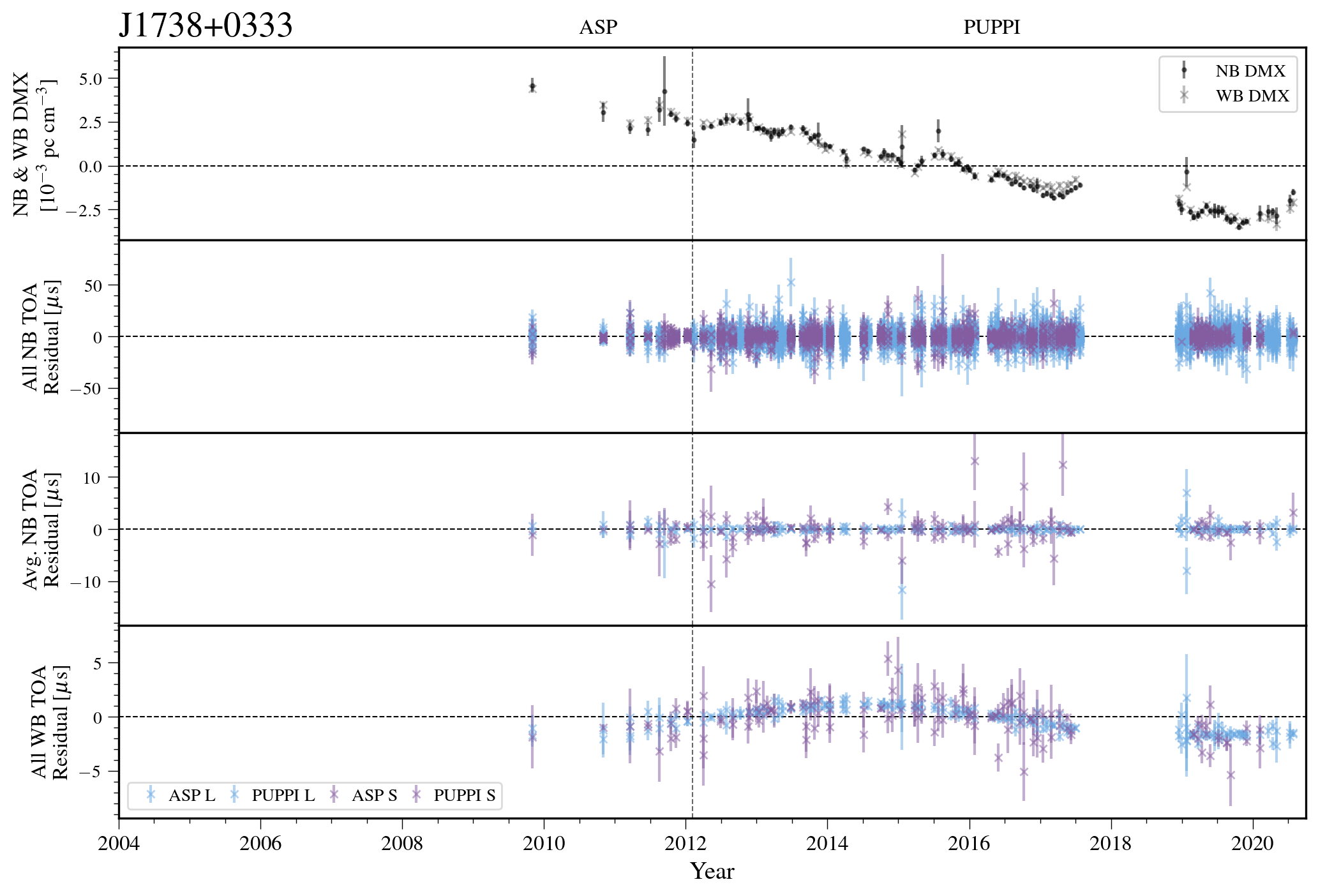}
\caption{Narrowband and wideband timing residuals and DMX timeseries for J1738+0333. See Figure~\ref{fig:summary-J0023+0923} for details.}
\label{fig:summary-J1738+0333}
\end{figure}

\begin{figure}
\centering
\includegraphics[width=0.85\linewidth]{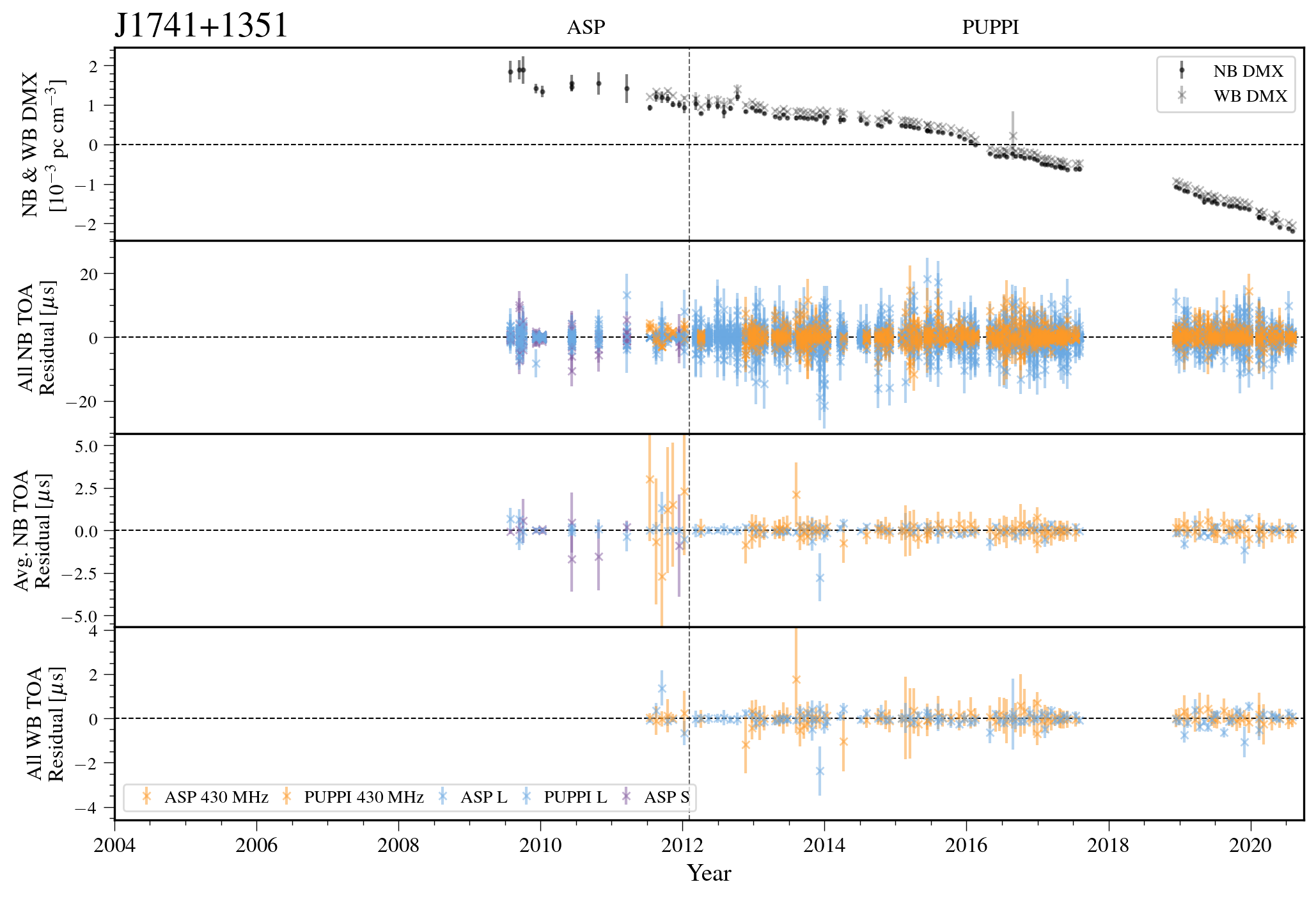}
\caption{Narrowband and wideband timing residuals and DMX timeseries for J1741+1351. See Figure~\ref{fig:summary-J0023+0923} for details.}
\label{fig:summary-J1741+1351}
\end{figure}
\clearpage

\begin{figure}
\centering
\includegraphics[width=0.85\linewidth]{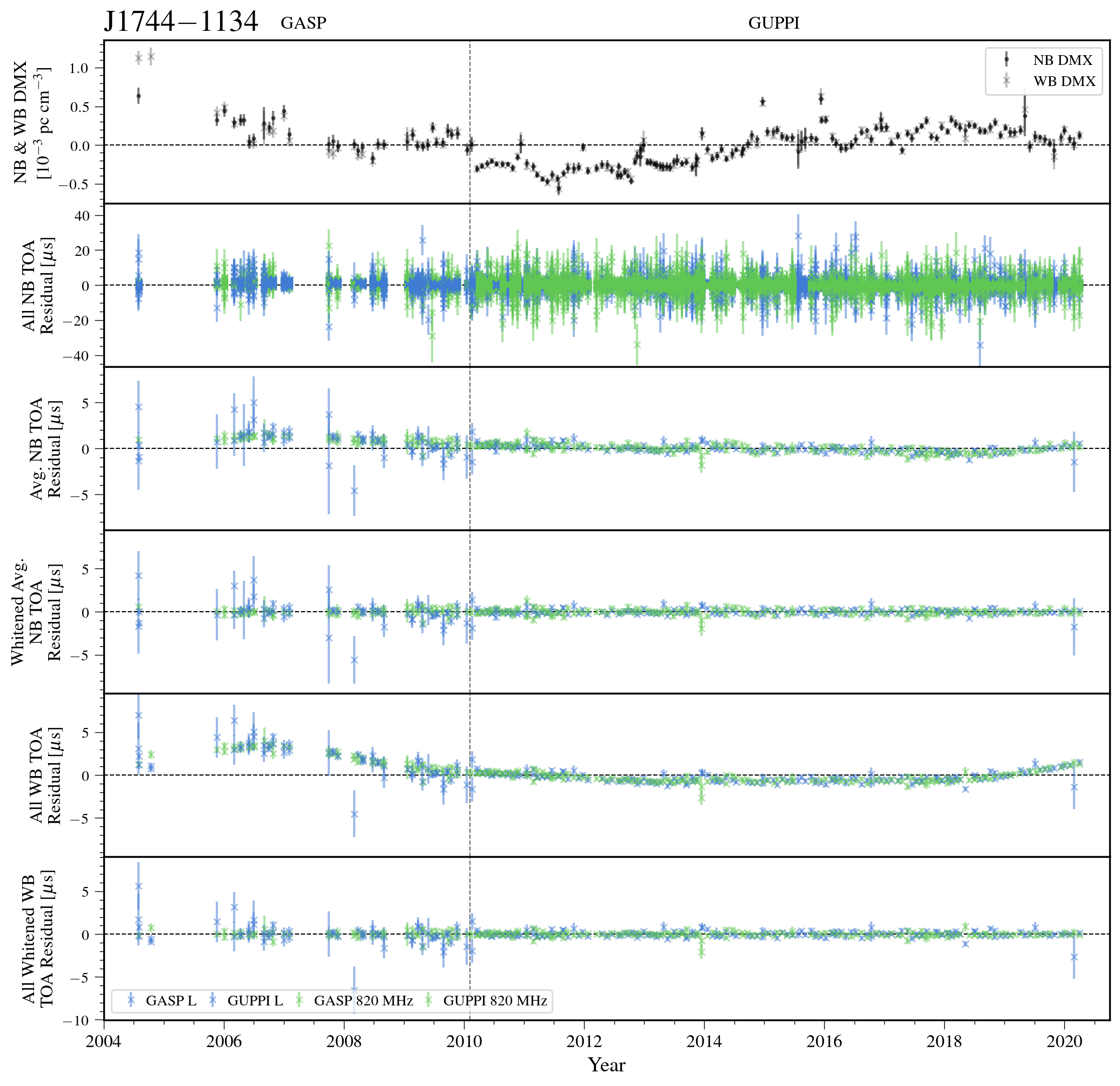}
\caption{Narrowband and wideband timing residuals and DMX timeseries for J1744-1134. See Figure~\ref{fig:summary-J0030+0451} for details.}
\label{fig:summary-J1744-1134}
\end{figure}
\clearpage

\begin{figure}
\centering
\includegraphics[width=0.85\linewidth]{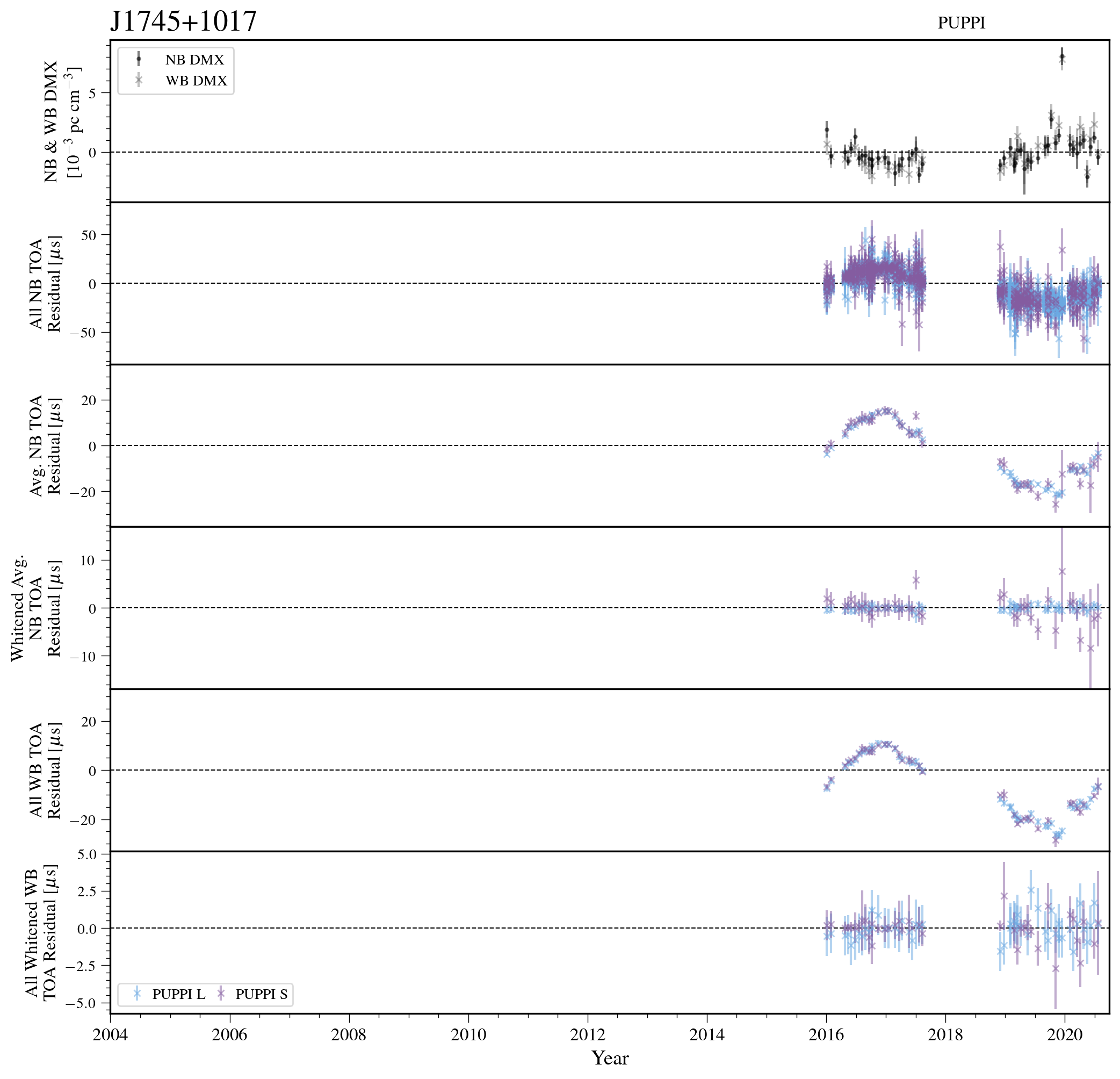}
\caption{Narrowband and wideband timing residuals and DMX timeseries for J1745+1017. See Figure~\ref{fig:summary-J0030+0451} for details.}
\label{fig:summary-J1745+1017}
\end{figure}
\clearpage

\begin{figure}
\centering
\includegraphics[width=0.85\linewidth]{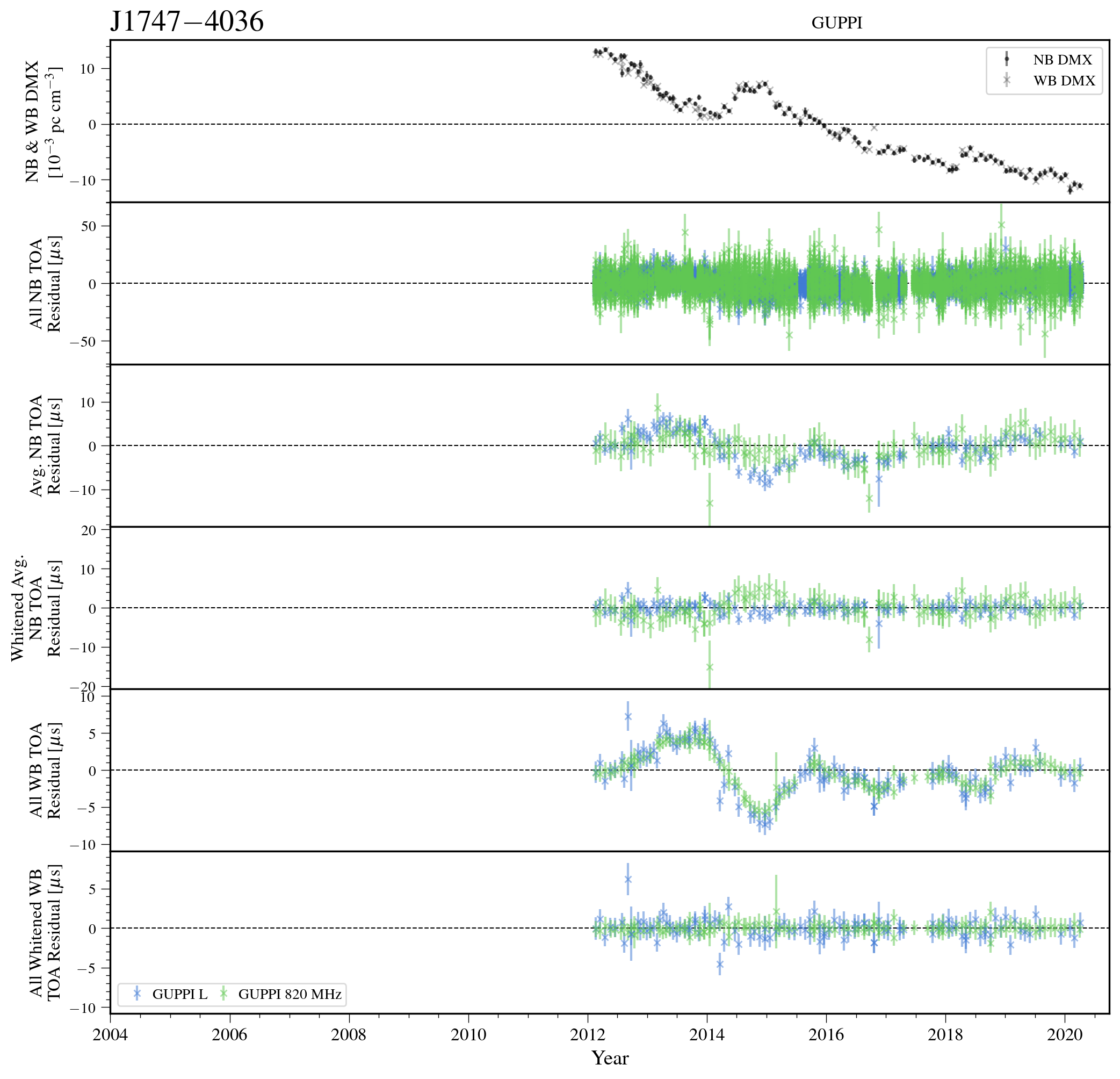}
\caption{Narrowband and wideband timing residuals and DMX timeseries for J1747-4036. See Figure~\ref{fig:summary-J0030+0451} for details.}
\label{fig:summary-J1747-4036}
\end{figure}
\clearpage

\begin{figure}
\centering
\includegraphics[width=0.85\linewidth]{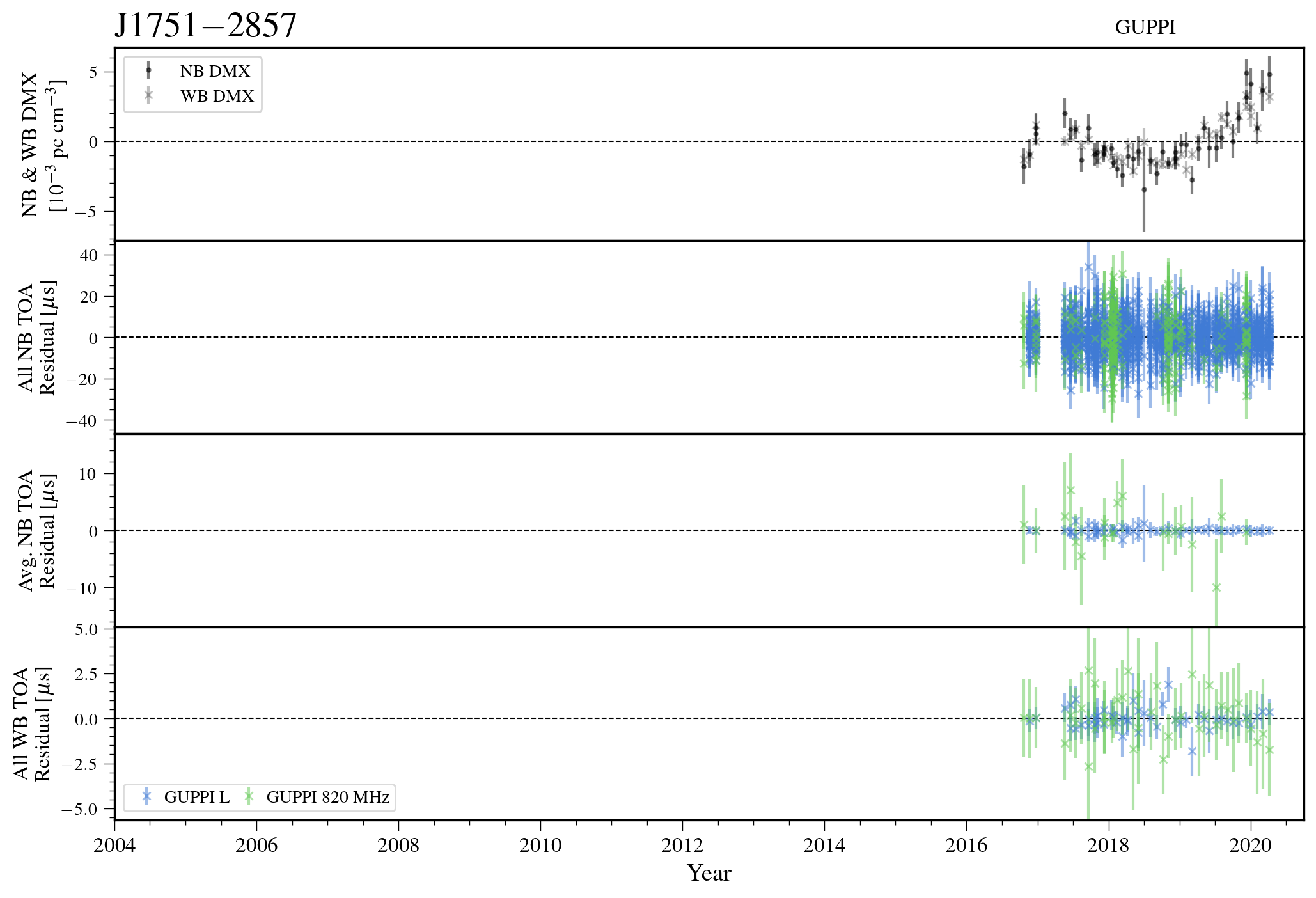}
\caption{Narrowband and wideband timing residuals and DMX timeseries for J1751-2857. See Figure~\ref{fig:summary-J0023+0923} for details.}
\label{fig:summary-J1751-2857}
\end{figure}
\clearpage

\begin{figure}
\centering
\includegraphics[width=0.85\linewidth]{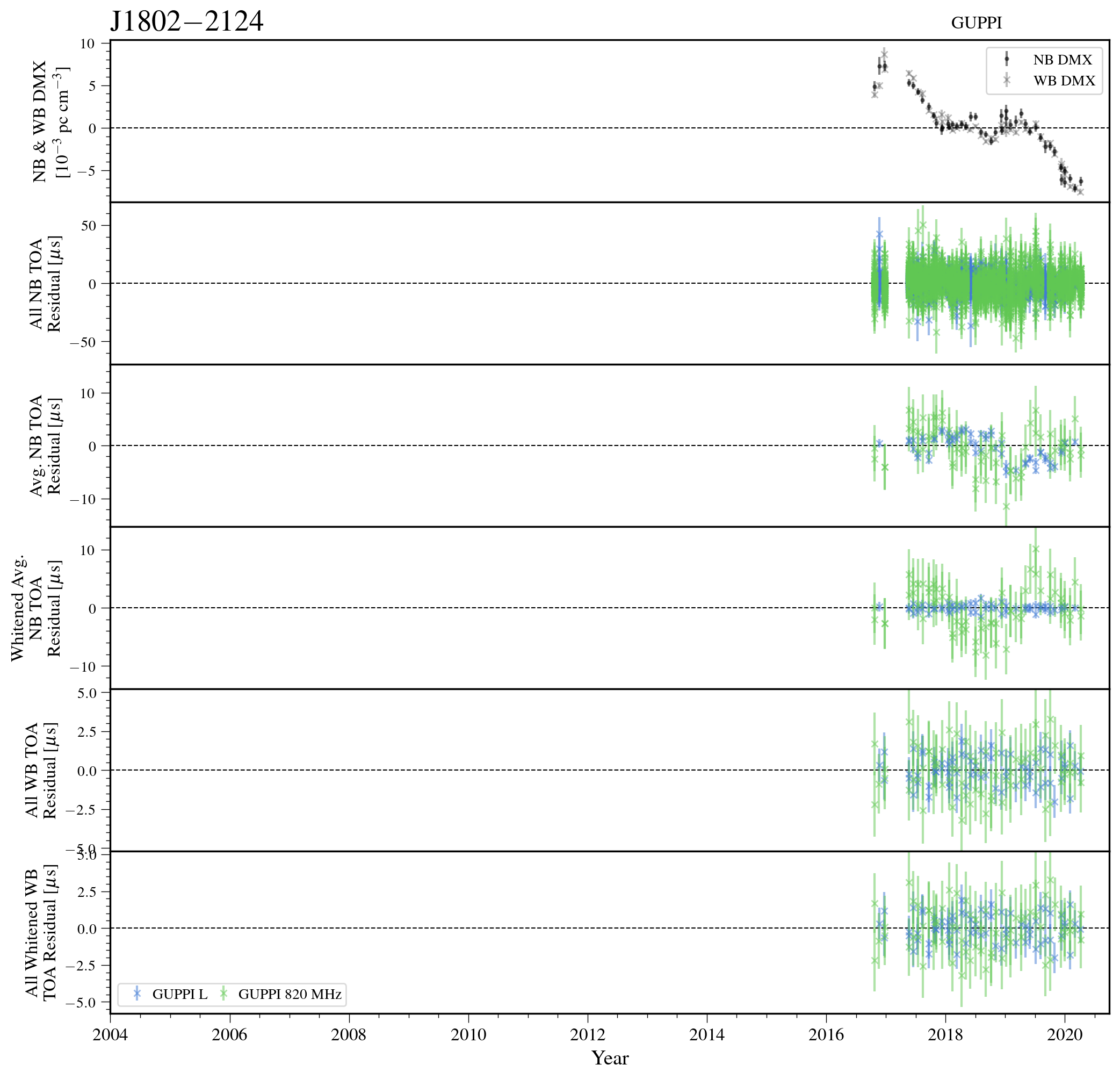}
\caption{Narrowband and wideband timing residuals and DMX timeseries for J1802-2124. See Figure~\ref{fig:summary-J0030+0451} for details.}
\label{fig:summary-J1802-2124}
\end{figure}
\clearpage

\begin{figure}
\centering
\includegraphics[width=0.85\linewidth]{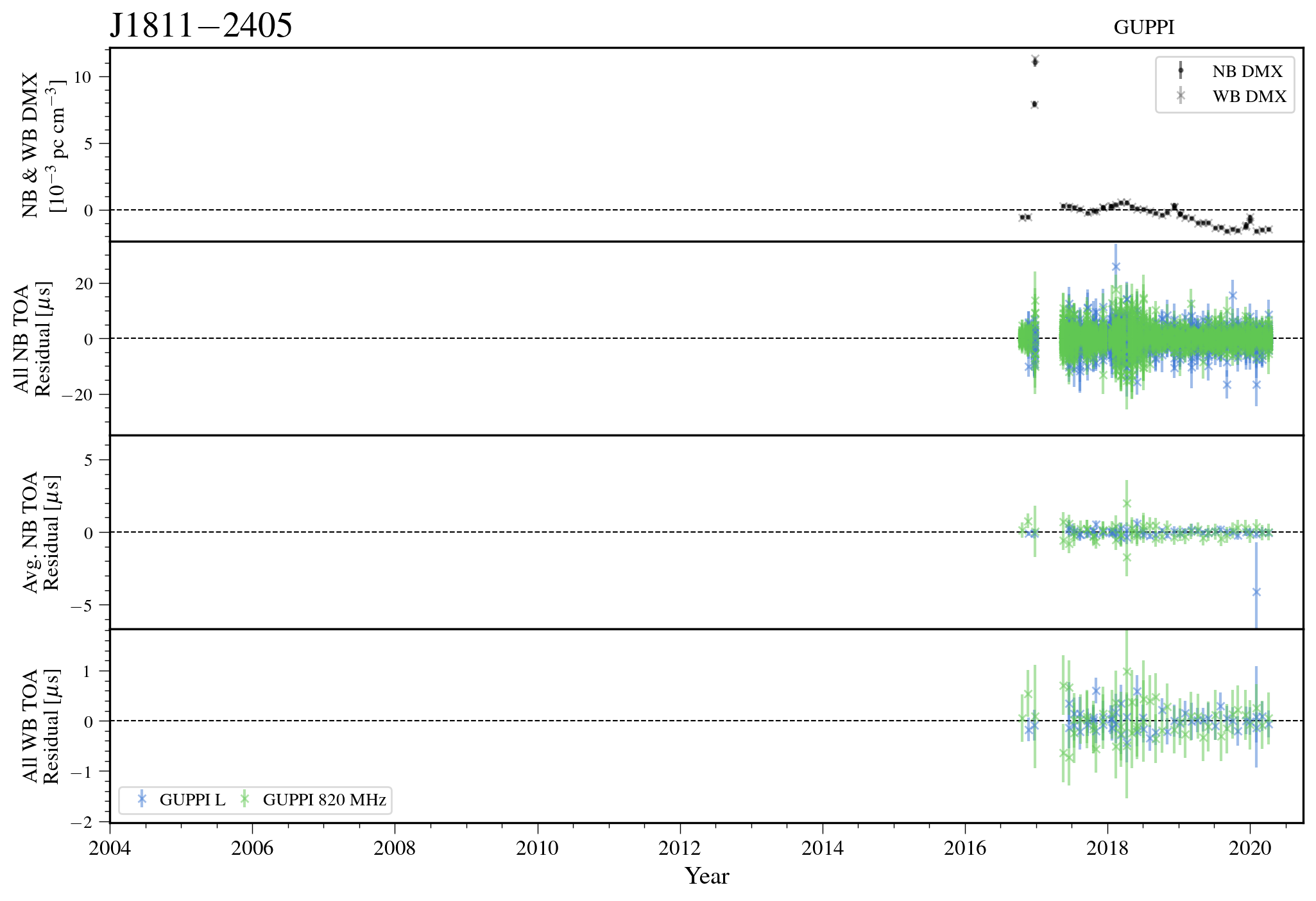}
\caption{Narrowband and wideband timing residuals and DMX timeseries for J1811-2405. See Figure~\ref{fig:summary-J0023+0923} for details.}
\label{fig:summary-J1811-2405}
\end{figure}

\begin{figure}
\centering
\includegraphics[width=0.85\linewidth]{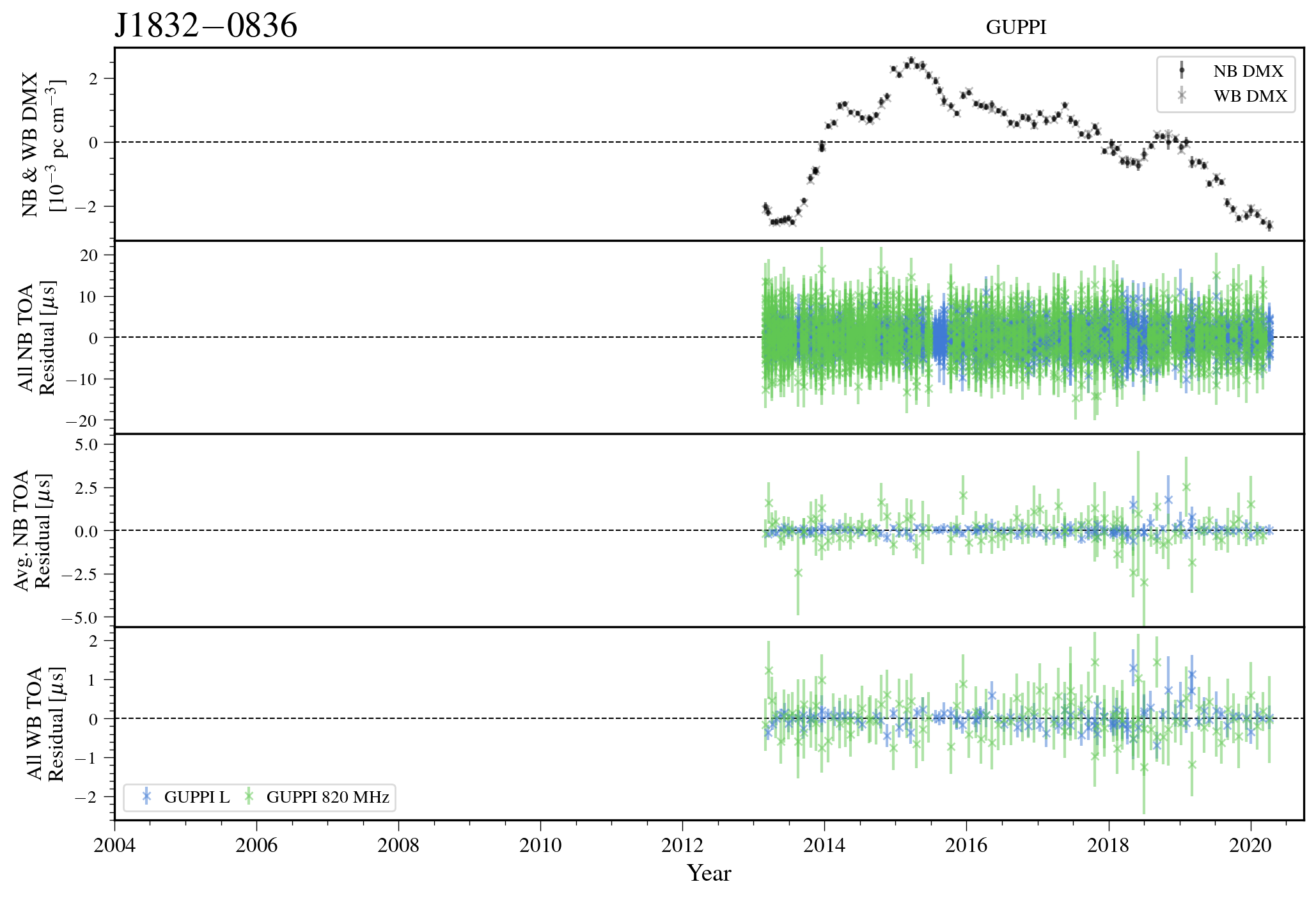}
\caption{Narrowband and wideband timing residuals and DMX timeseries for J1832-0836. See Figure~\ref{fig:summary-J0023+0923} for details.}
\label{fig:summary-J1832-0836}
\end{figure}
\clearpage

\begin{figure}
\centering
\includegraphics[width=0.85\linewidth]{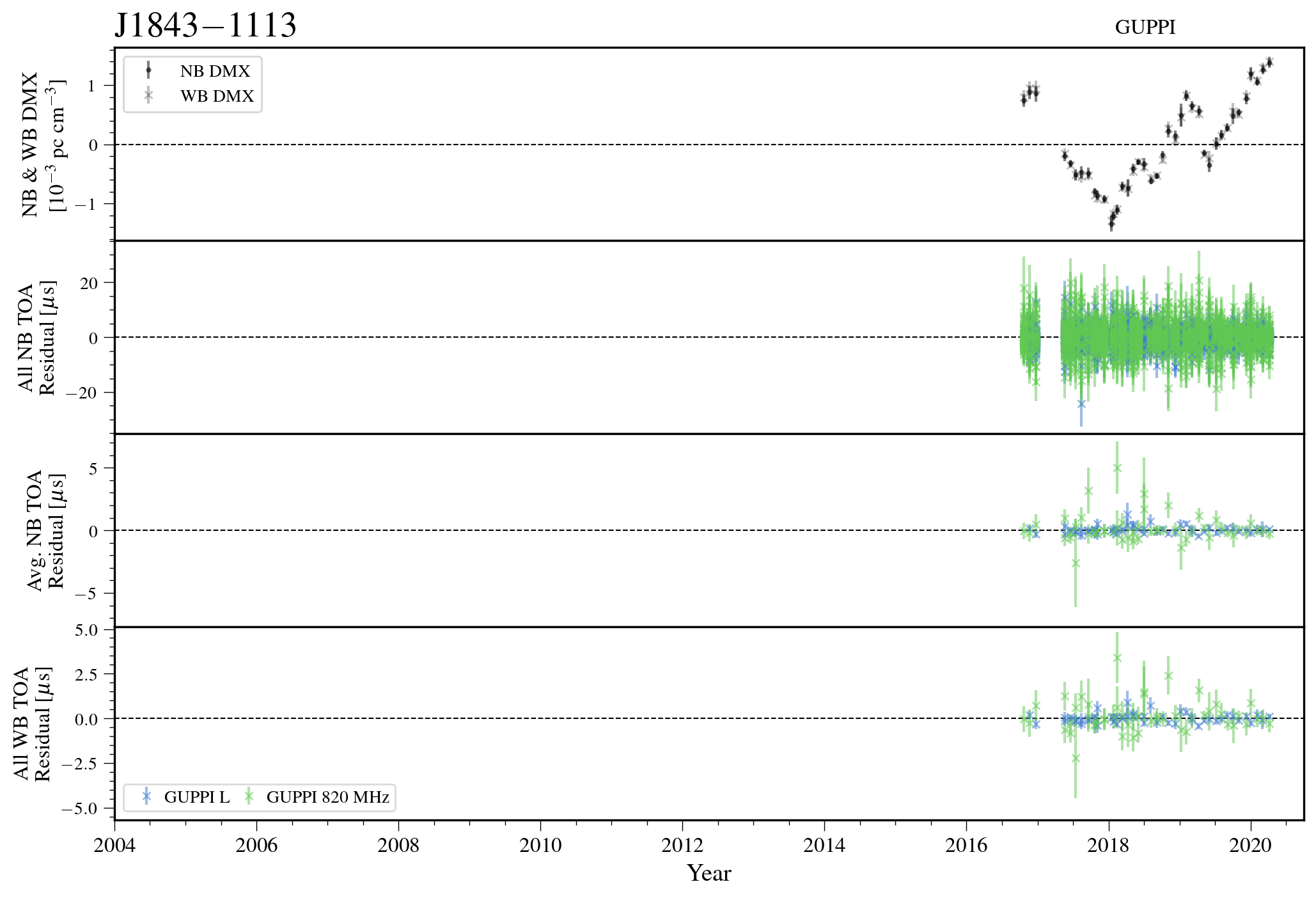}
\caption{Narrowband and wideband timing residuals and DMX timeseries for J1843-1113. See Figure~\ref{fig:summary-J0023+0923} for details.}
\label{fig:summary-J1843-1113}
\end{figure}
\clearpage

\begin{figure}
\centering
\includegraphics[width=0.85\linewidth]{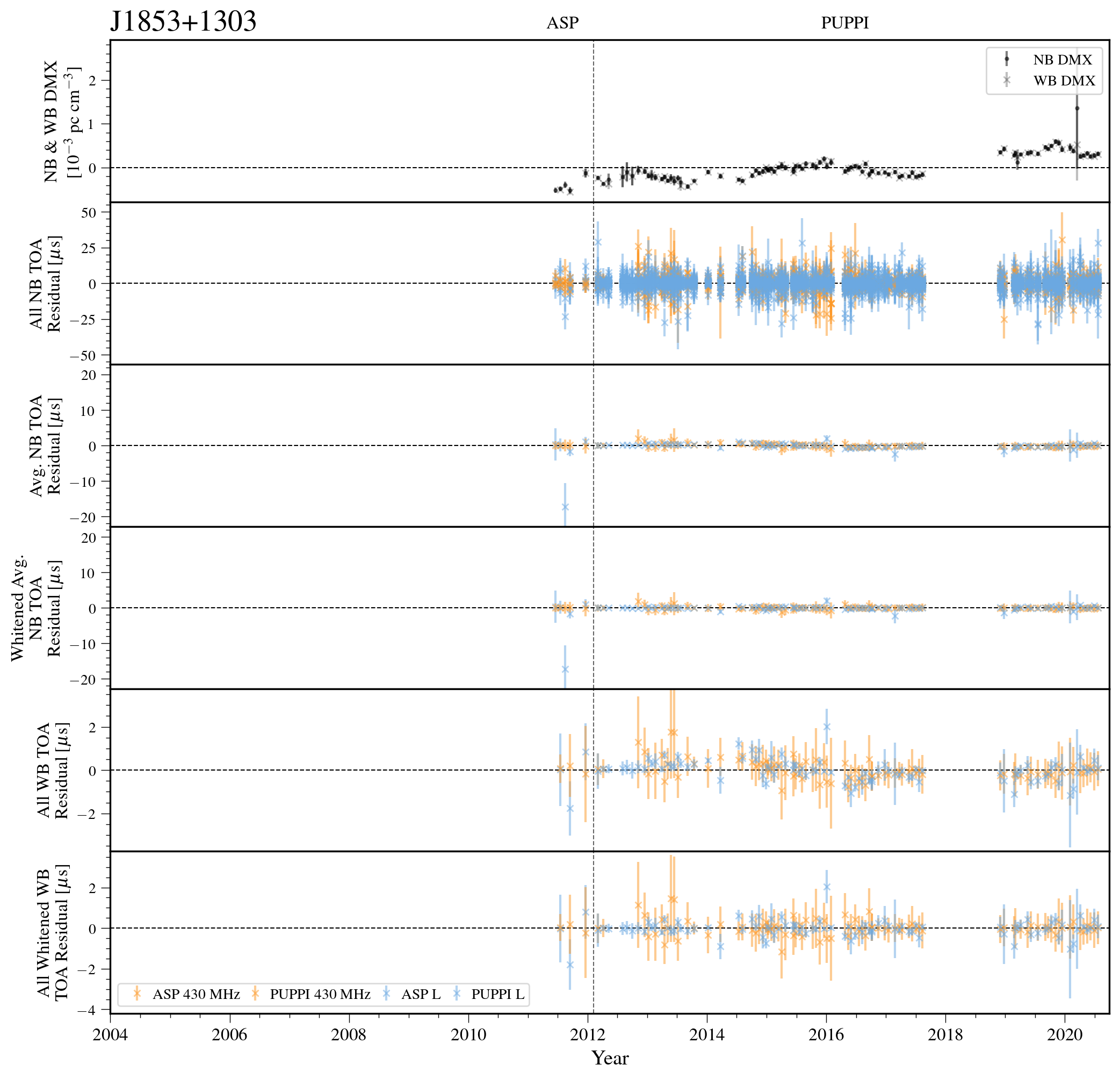}
\caption{Narrowband and wideband timing residuals and DMX timeseries for J1853+1303. See Figure~\ref{fig:summary-J0030+0451} for details.}
\label{fig:summary-J1853+1303}
\end{figure}
\clearpage

\begin{figure}
\centering
\includegraphics[width=0.85\linewidth]{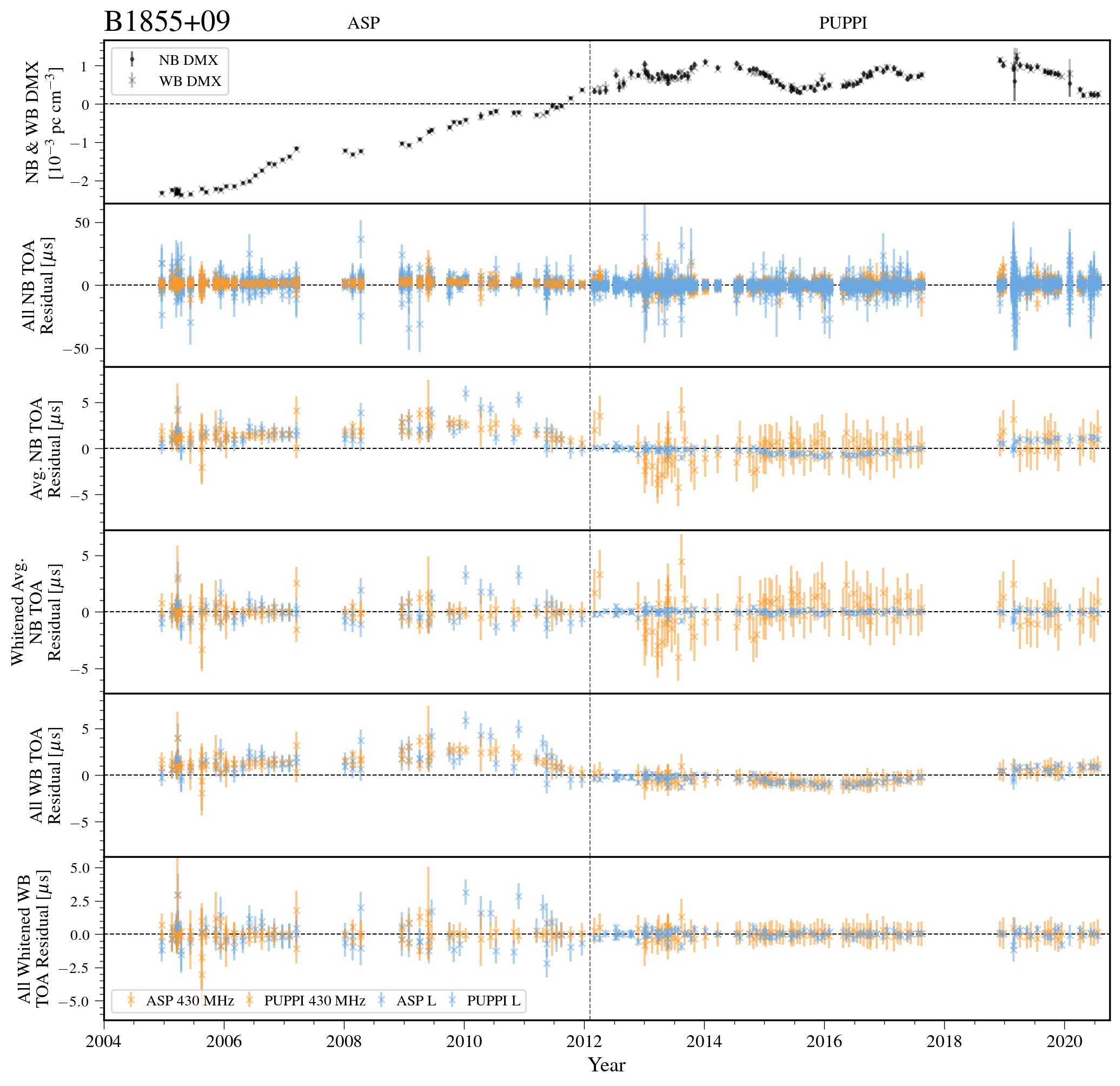}
\caption{Narrowband and wideband timing residuals and DMX timeseries for B1855+09. See Figure~\ref{fig:summary-J0030+0451} for details.}
\label{fig:summary-B1855+09}
\end{figure}
\clearpage

\begin{figure}
\centering
\includegraphics[width=0.85\linewidth]{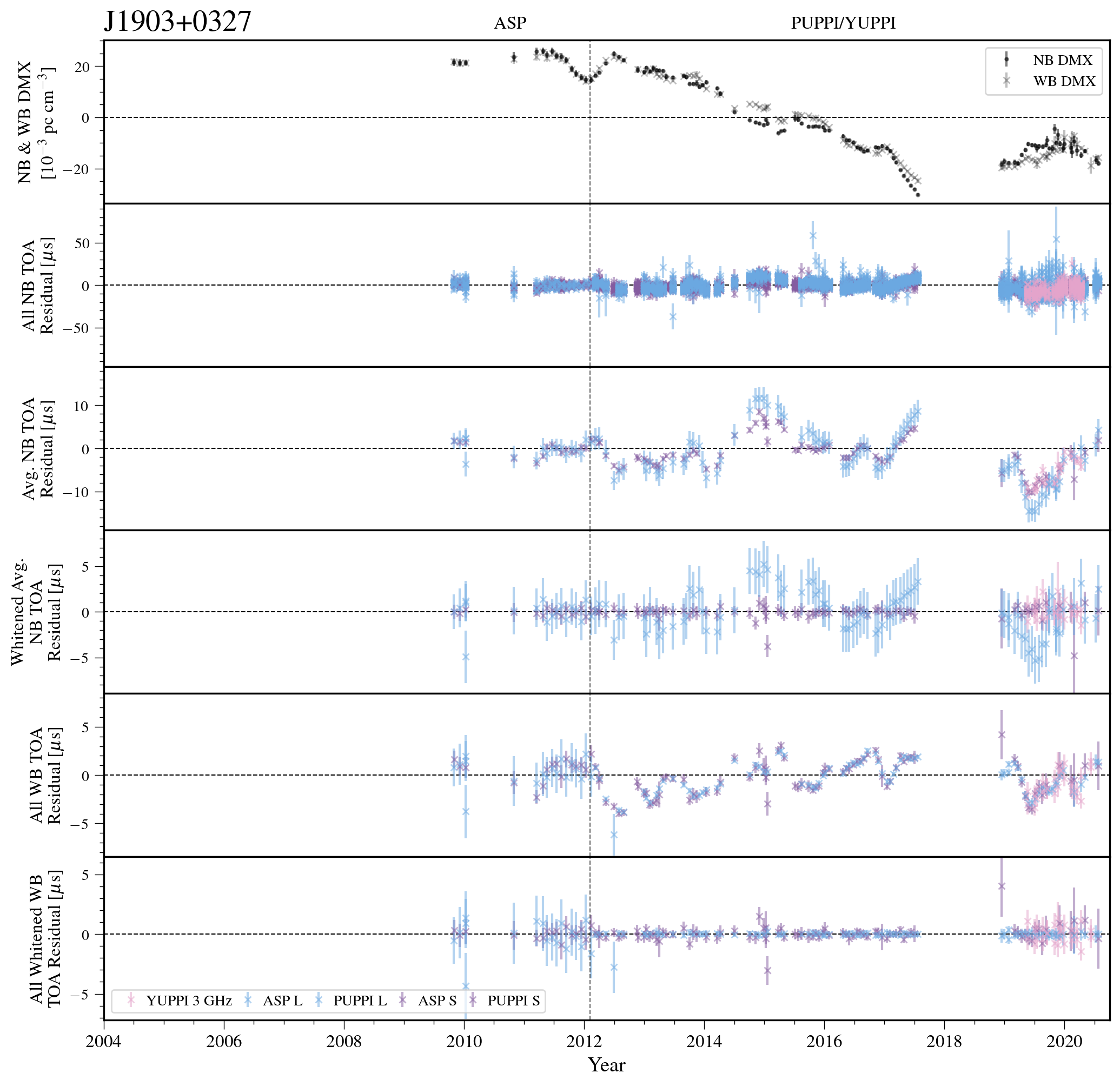}
\caption{Narrowband and wideband timing residuals and DMX timeseries for J1903+0327. See Figure~\ref{fig:summary-J0030+0451} for details.}
\label{fig:summary-J1903+0327}
\end{figure}
\clearpage

\begin{figure}
\centering
\includegraphics[width=0.85\linewidth]{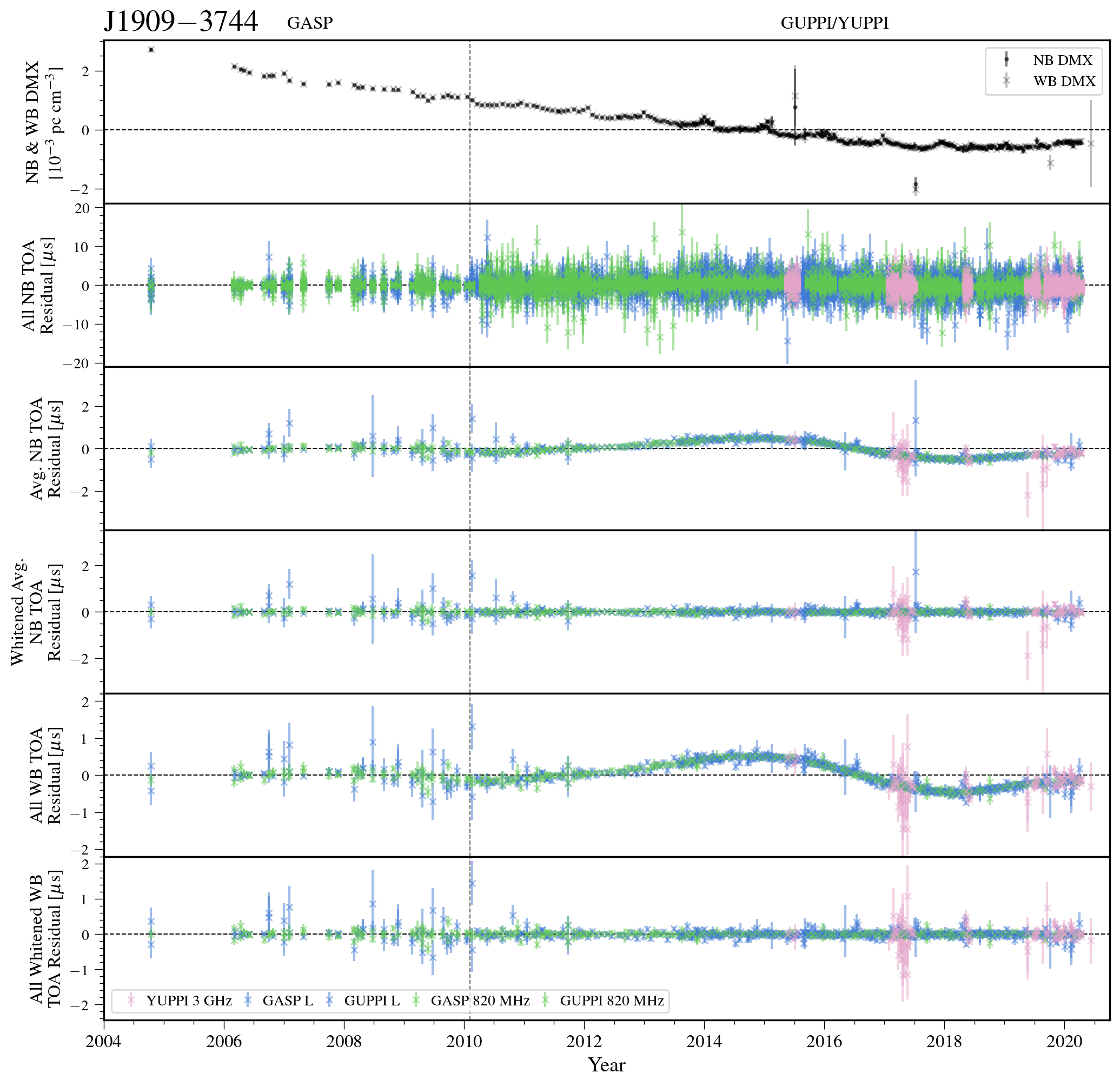}
\caption{Narrowband and wideband timing residuals and DMX timeseries for J1909-3744. See Figure~\ref{fig:summary-J0030+0451} for details.}
\label{fig:summary-J1909-3744}
\end{figure}
\clearpage

\begin{figure}
\centering
\includegraphics[width=0.85\linewidth]{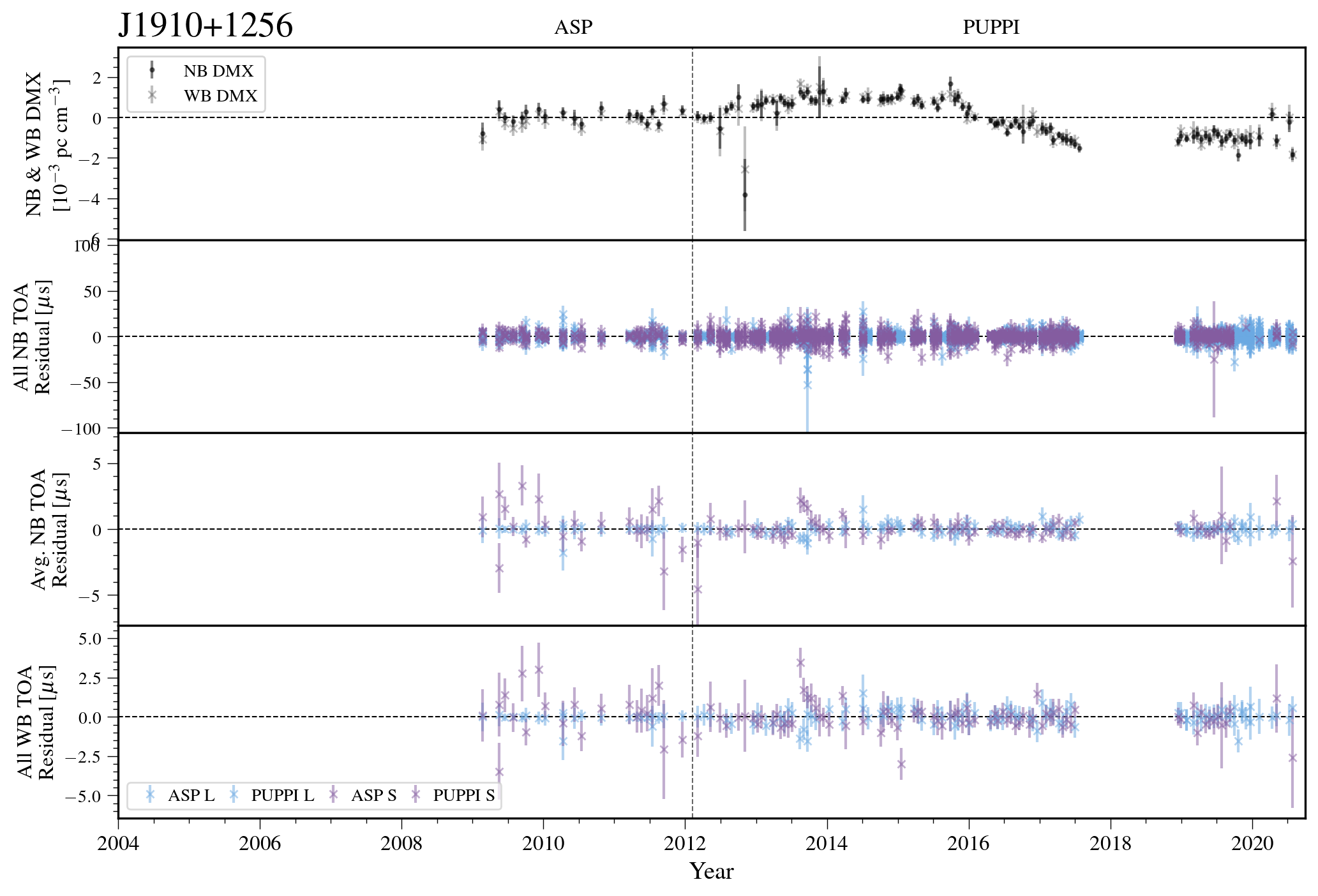}
\caption{Narrowband and wideband timing residuals and DMX timeseries for J1910+1256. See Figure~\ref{fig:summary-J0023+0923} for details.}
\label{fig:summary-J1910+1256}
\end{figure}
\clearpage

\begin{figure}
\centering
\includegraphics[width=0.85\linewidth]{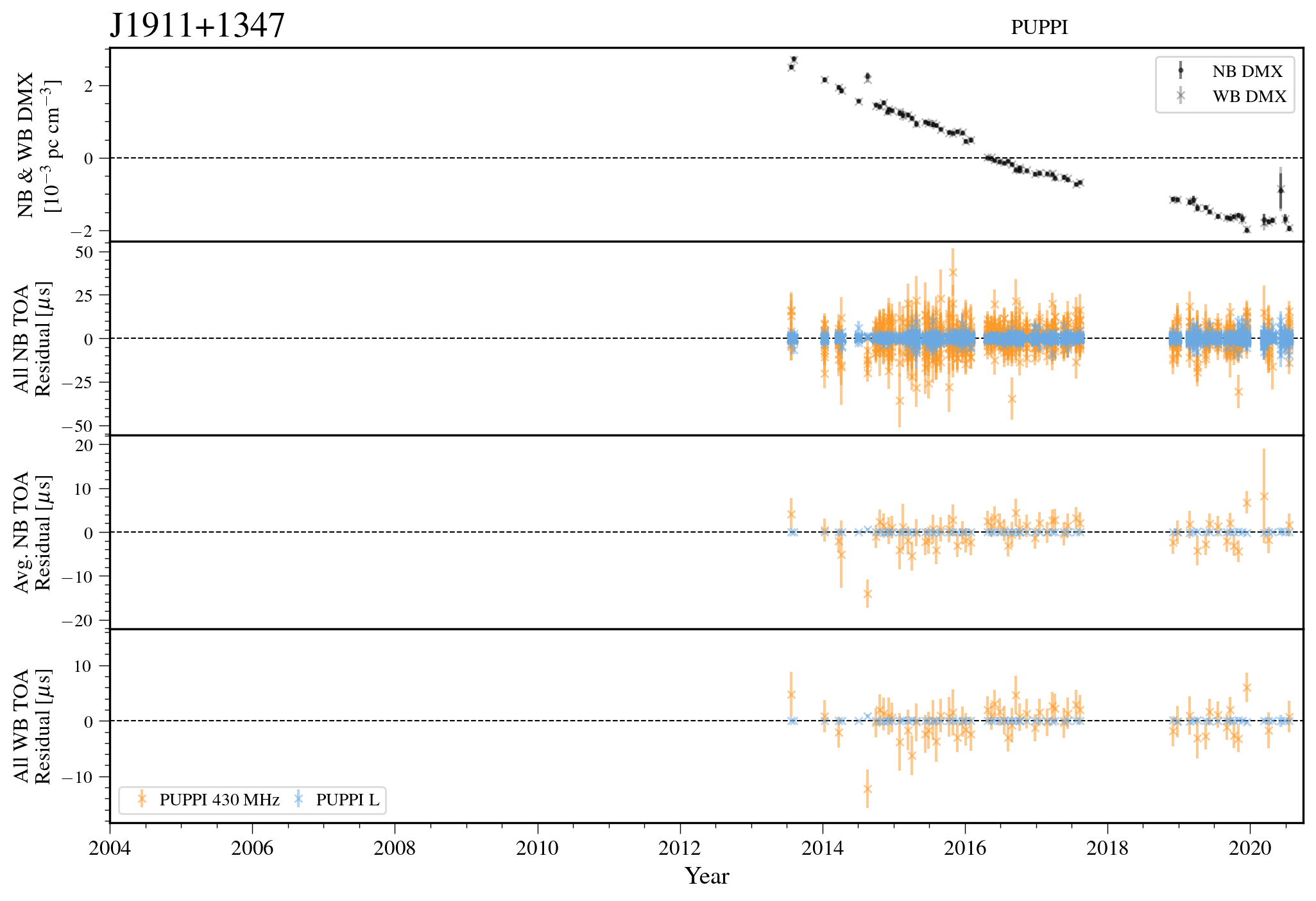}
\caption{Narrowband and wideband timing residuals and DMX timeseries for J1911+1347. See Figure~\ref{fig:summary-J0023+0923} for details.}
\label{fig:summary-J1911+1347}
\end{figure}

\begin{figure}
\centering
\includegraphics[width=0.85\linewidth]{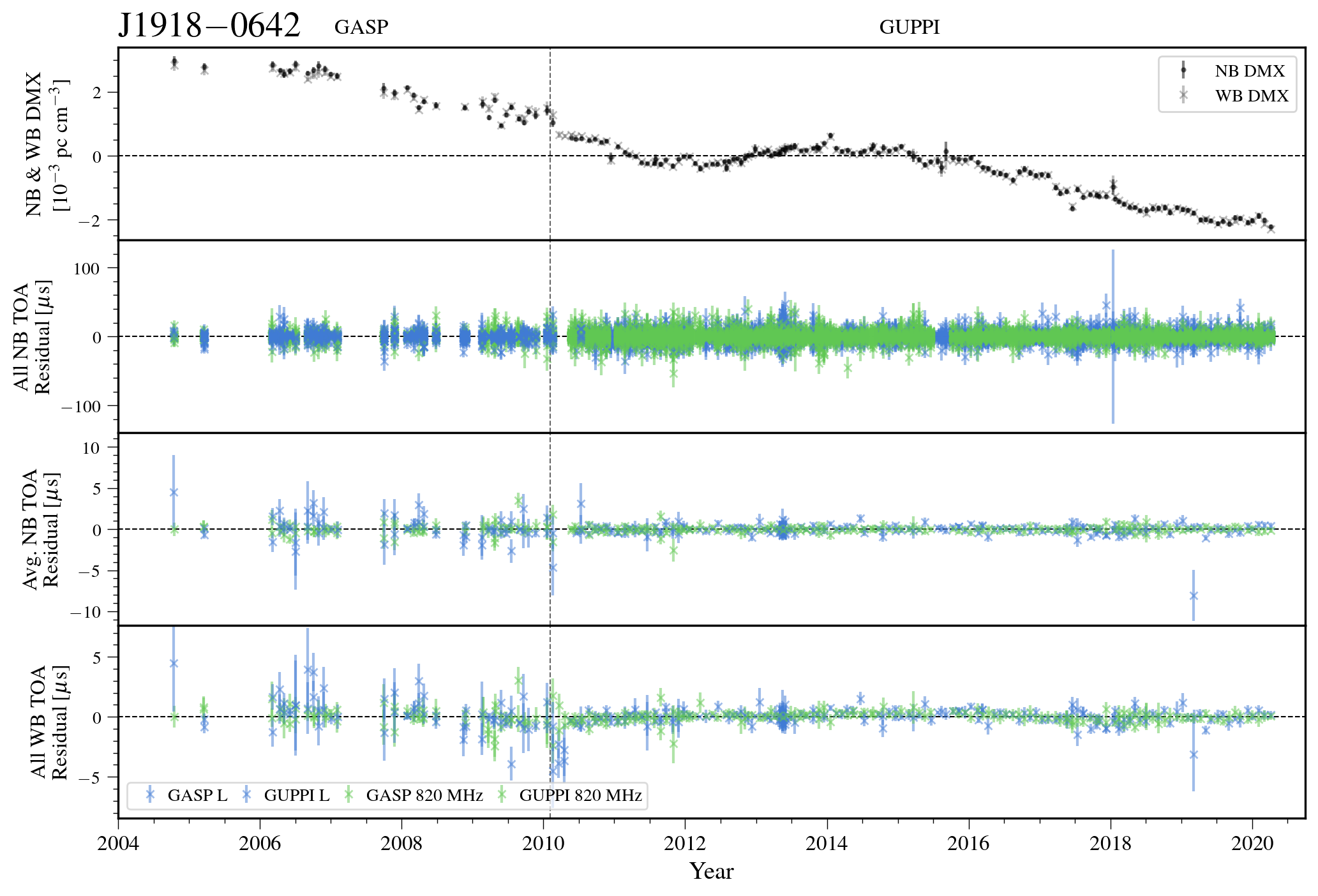}
\caption{Narrowband and wideband timing residuals and DMX timeseries for J1918-0642. See Figure~\ref{fig:summary-J0023+0923} for details.}
\label{fig:summary-J1918-0642}
\end{figure}
\clearpage

\begin{figure}
\centering
\includegraphics[width=0.85\linewidth]{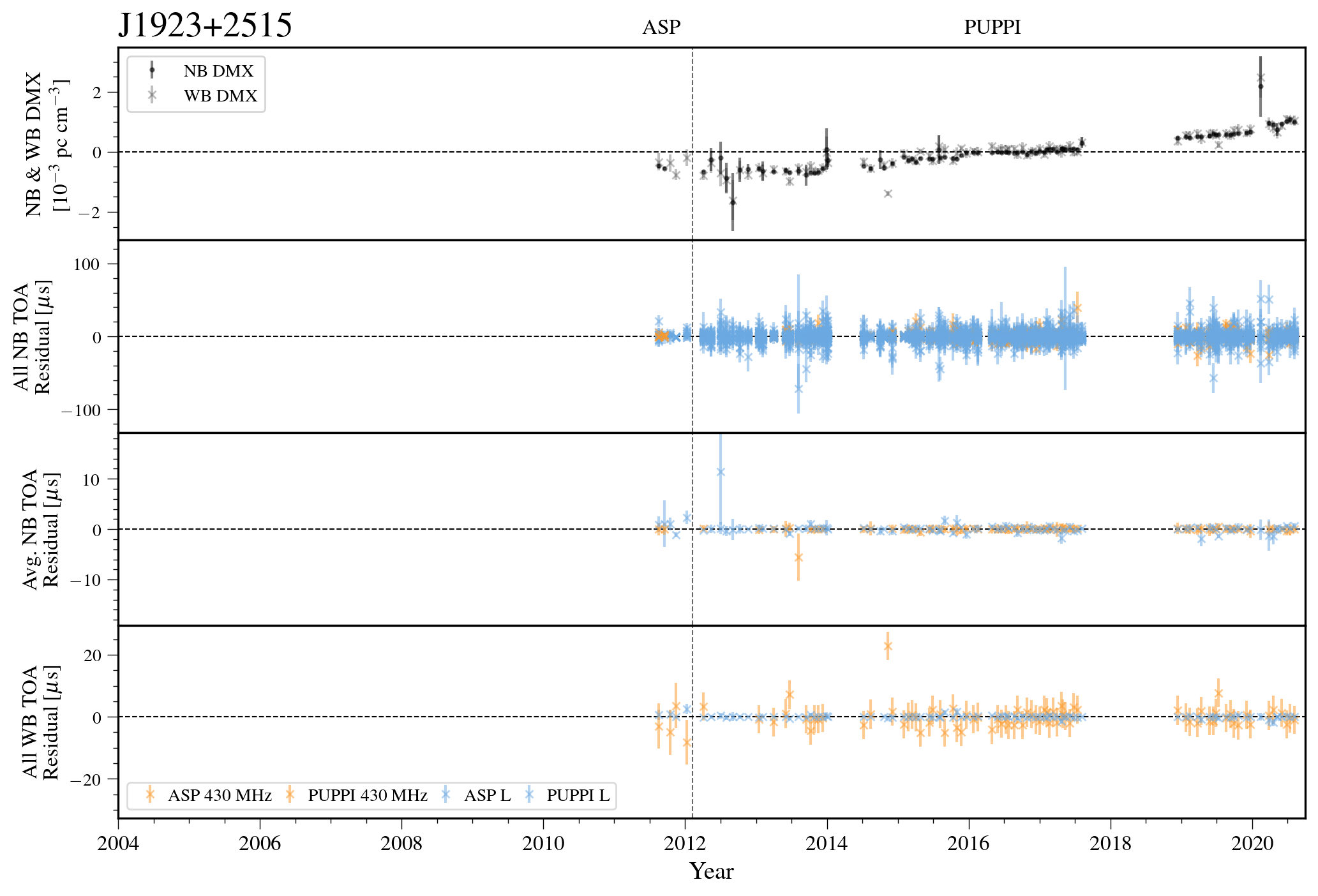}
\caption{Narrowband and wideband timing residuals and DMX timeseries for J1923+2515. See Figure~\ref{fig:summary-J0023+0923} for details.}
\label{fig:summary-J1923+2515}
\end{figure}
\clearpage

\begin{figure}
\centering
\includegraphics[width=0.85\linewidth]{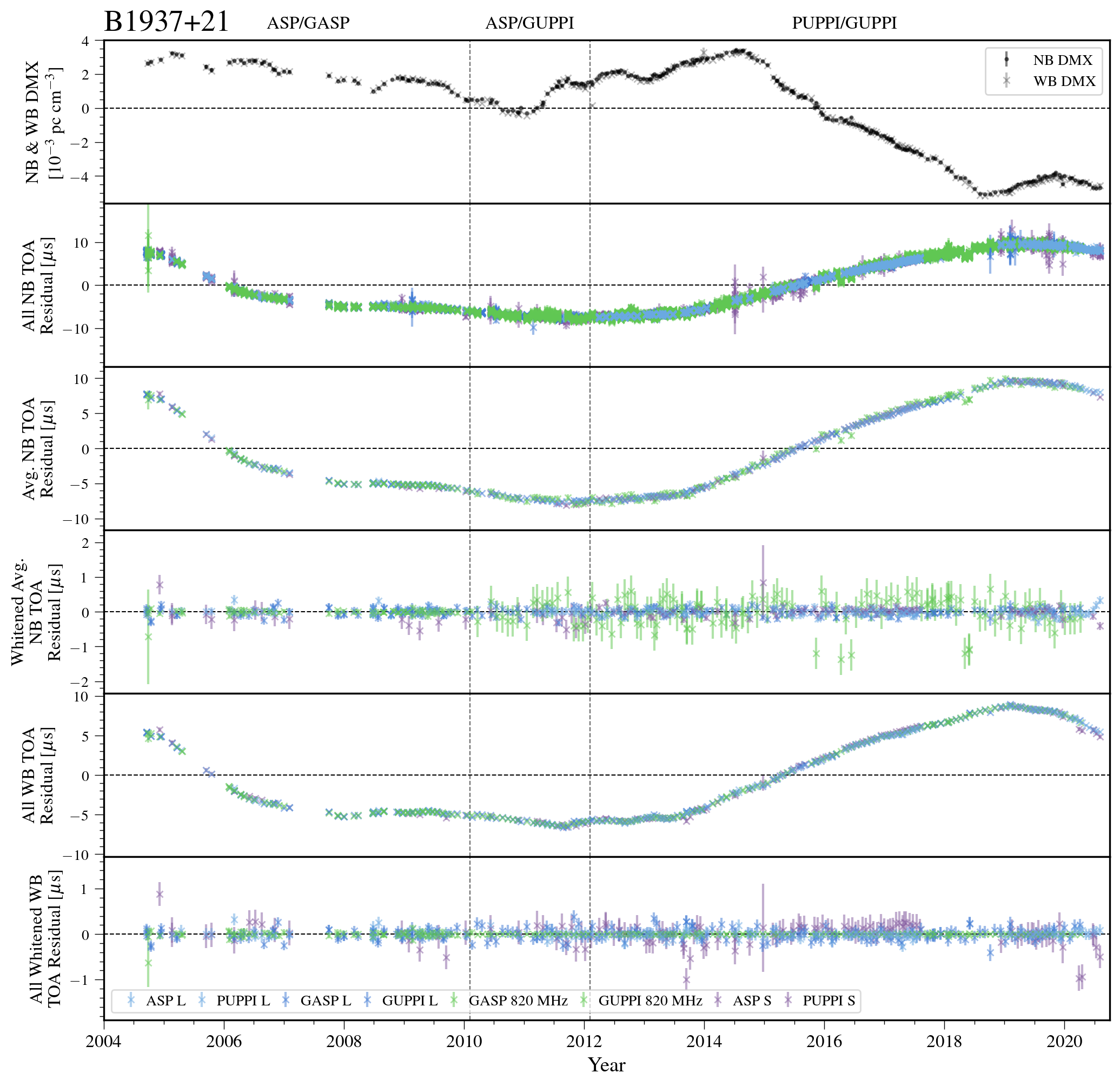}
\caption{Narrowband and wideband timing residuals and DMX timeseries for B1937+21. See Figure~\ref{fig:summary-J0030+0451} for details.}
\label{fig:summary-B1937+21}
\end{figure}
\clearpage

\begin{figure}
\centering
\includegraphics[width=0.85\linewidth]{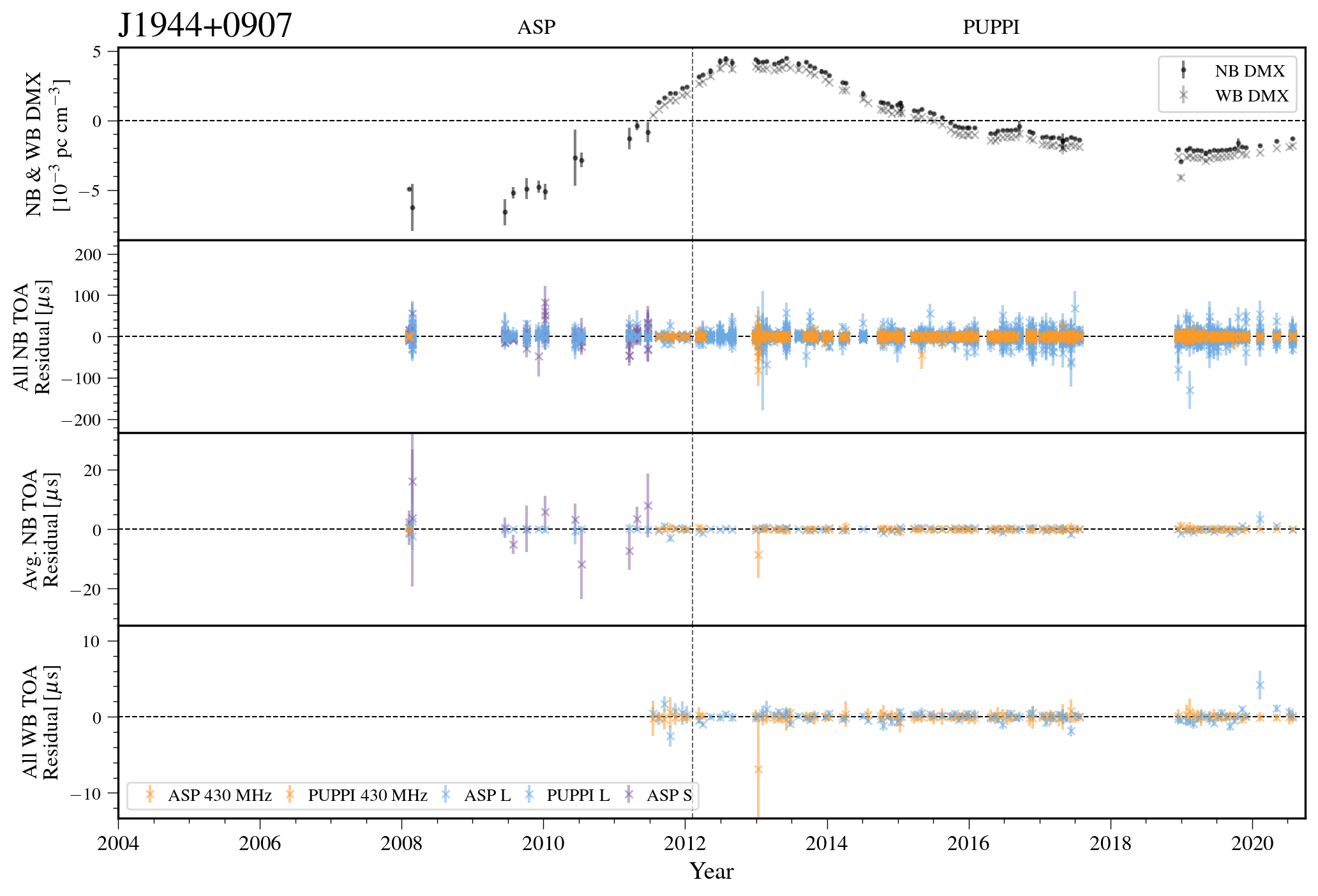}
\caption{Narrowband and wideband timing residuals and DMX timeseries for J1944+0907. See Figure~\ref{fig:summary-J0023+0923} for details.}
\label{fig:summary-J1944+0907}
\end{figure}
\clearpage

\begin{figure}
\centering
\includegraphics[width=0.85\linewidth]{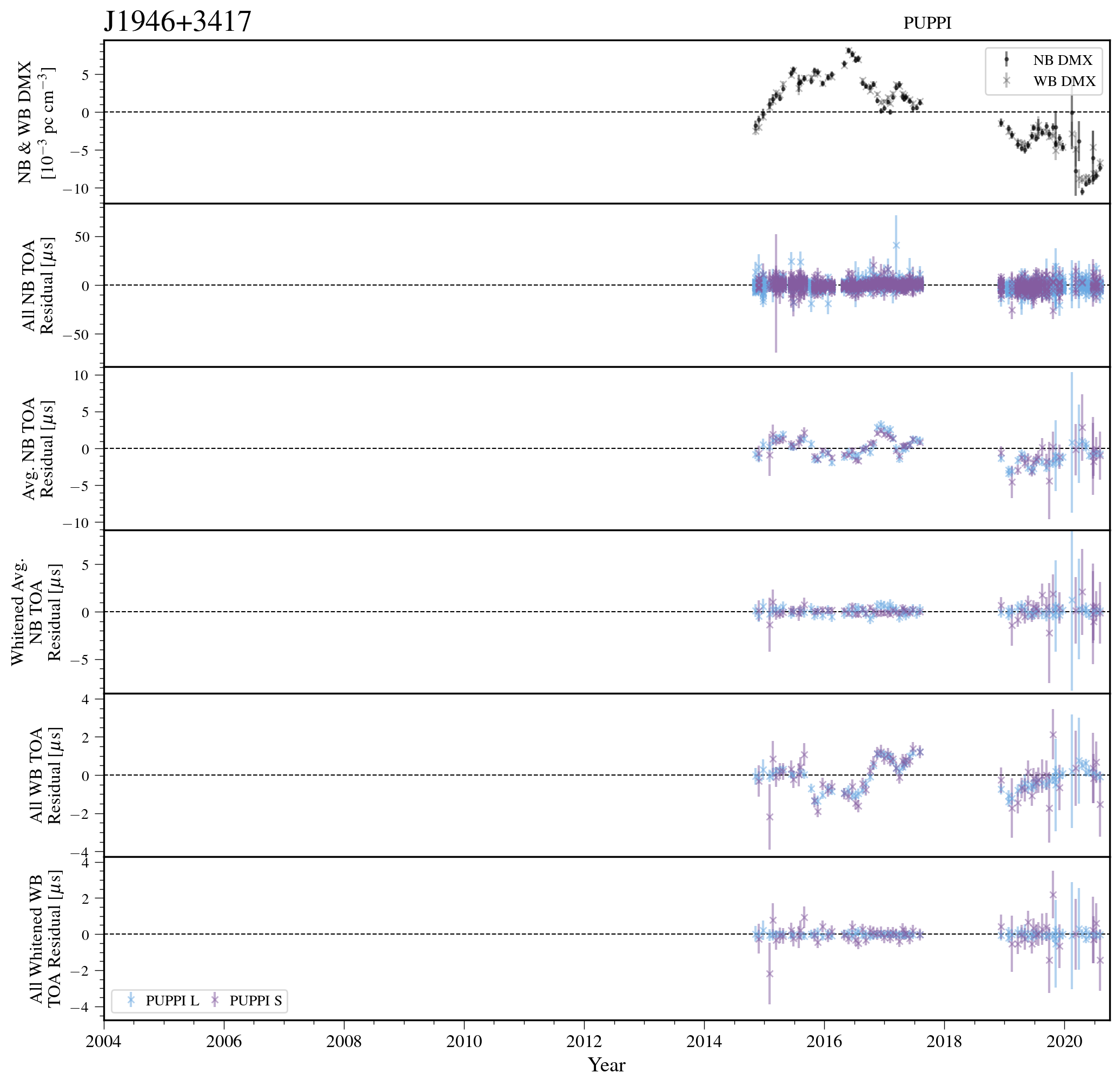}
\caption{Narrowband and wideband timing residuals and DMX timeseries for J1946+3417. See Figure~\ref{fig:summary-J0030+0451} for details.}
\label{fig:summary-J1946+3417}
\end{figure}
\clearpage

\begin{figure}
\centering
\includegraphics[width=0.85\linewidth]{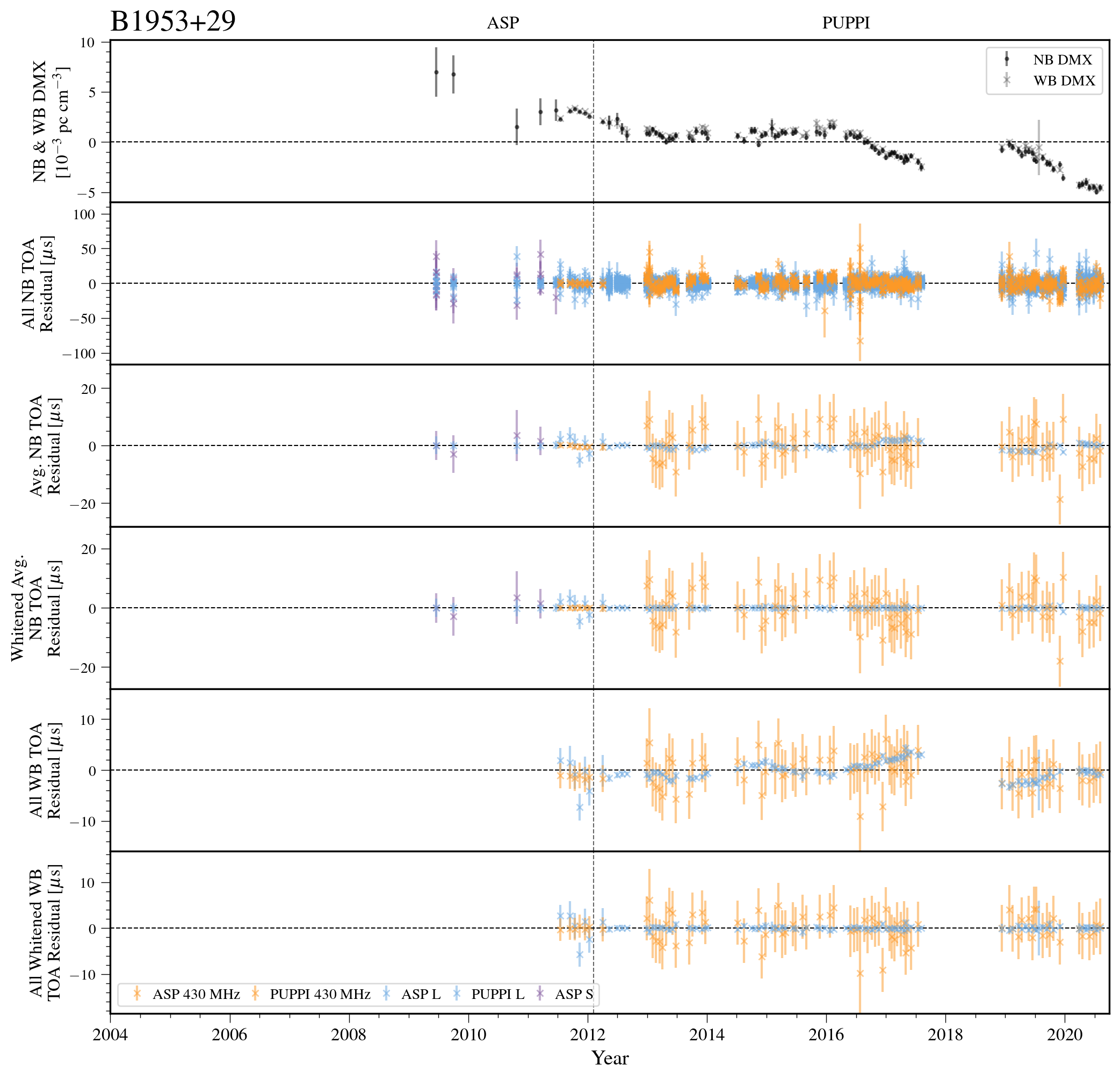}
\caption{Narrowband and wideband timing residuals and DMX timeseries for B1953+29. See Figure~\ref{fig:summary-J0030+0451} for details.}
\label{fig:summary-B1953+29}
\end{figure}
\clearpage

\begin{figure}
\centering
\includegraphics[width=0.85\linewidth]{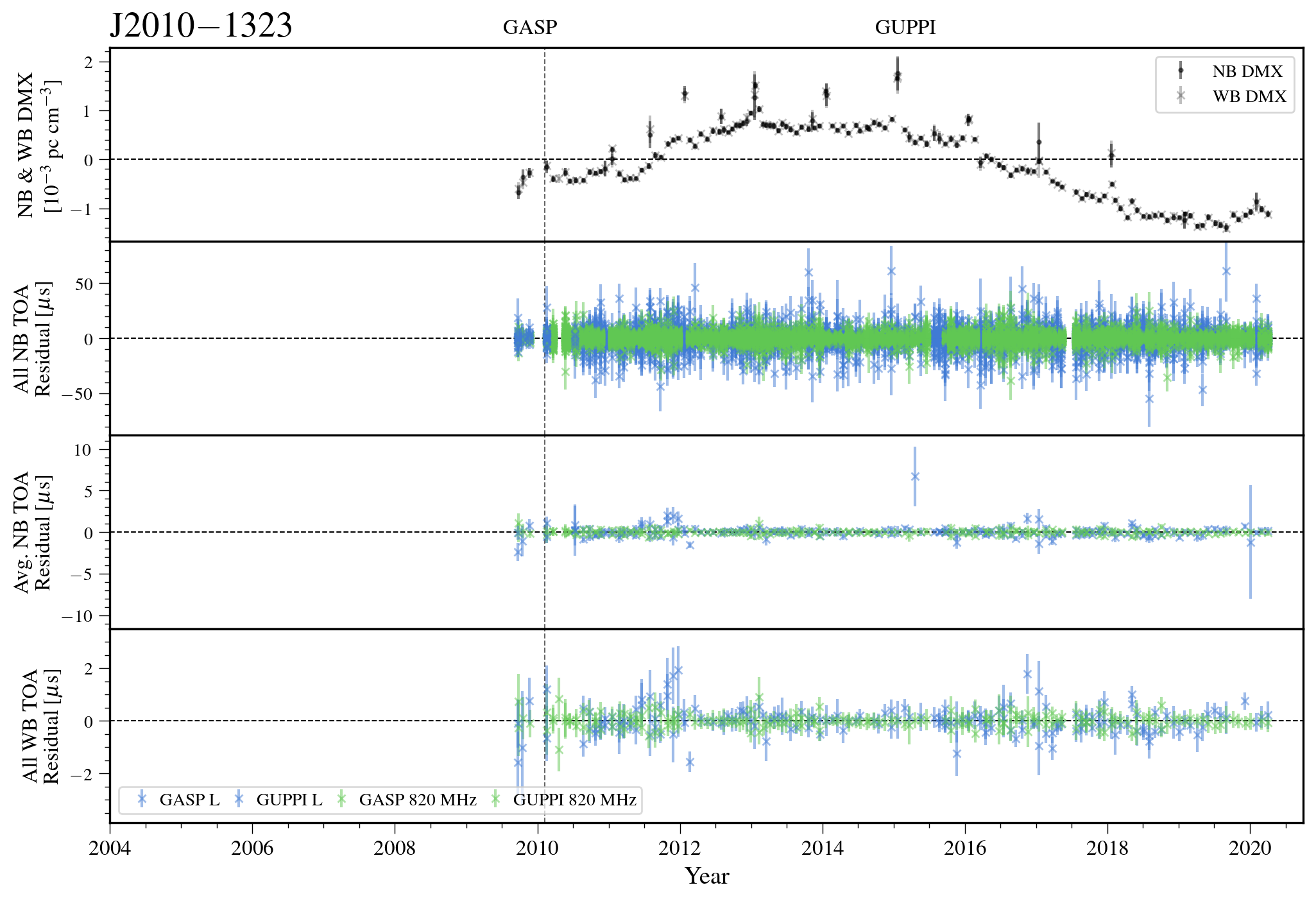}
\caption{Narrowband and wideband timing residuals and DMX timeseries for J2010-1323. See Figure~\ref{fig:summary-J0023+0923} for details.}
\label{fig:summary-J2010-1323}
\end{figure}

\begin{figure}
\centering
\includegraphics[width=0.85\linewidth]{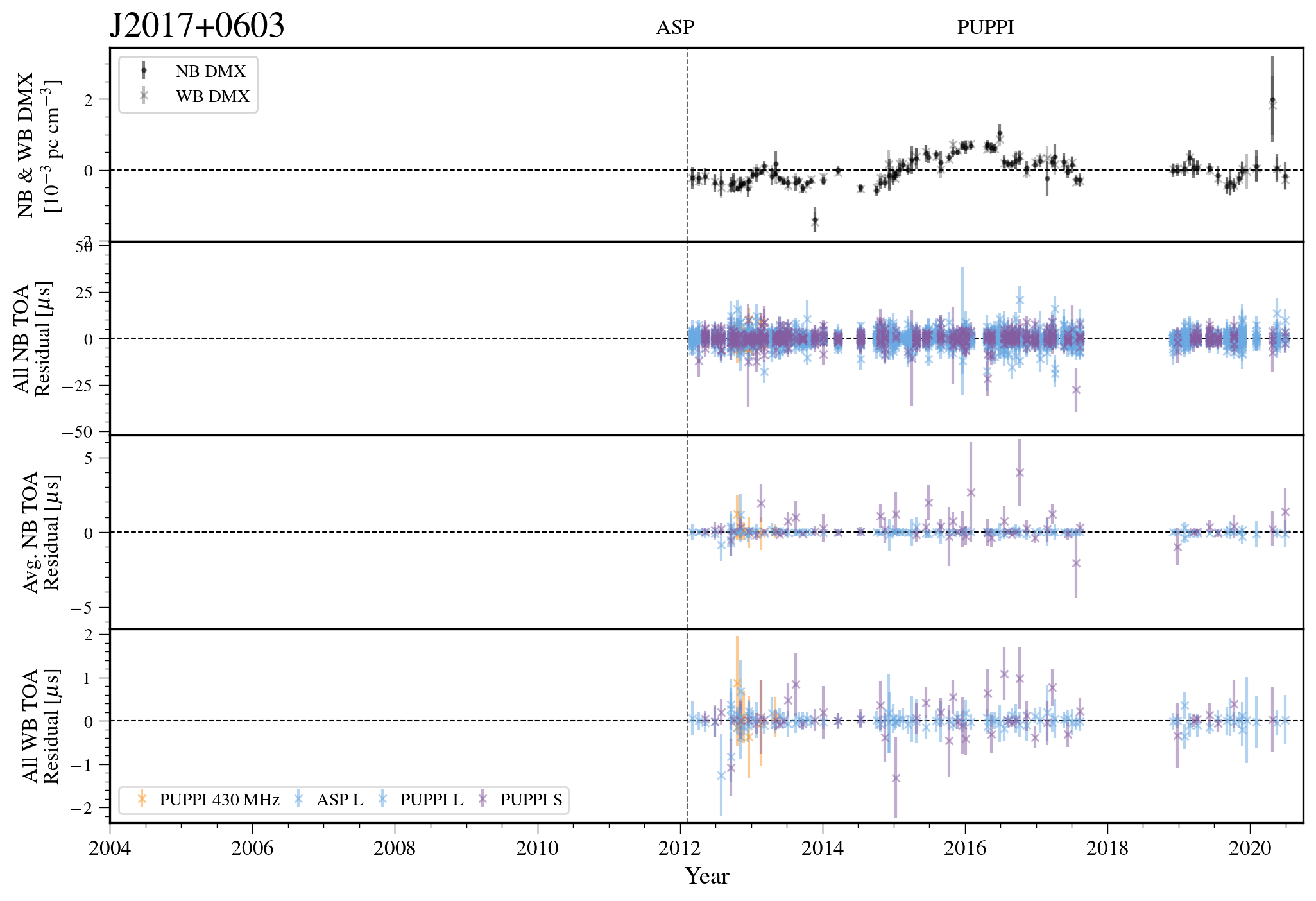}
\caption{Narrowband and wideband timing residuals and DMX timeseries for J2017+0603. See Figure~\ref{fig:summary-J0023+0923} for details.}
\label{fig:summary-J2017+0603}
\end{figure}
\clearpage

\begin{figure}
\centering
\includegraphics[width=0.85\linewidth]{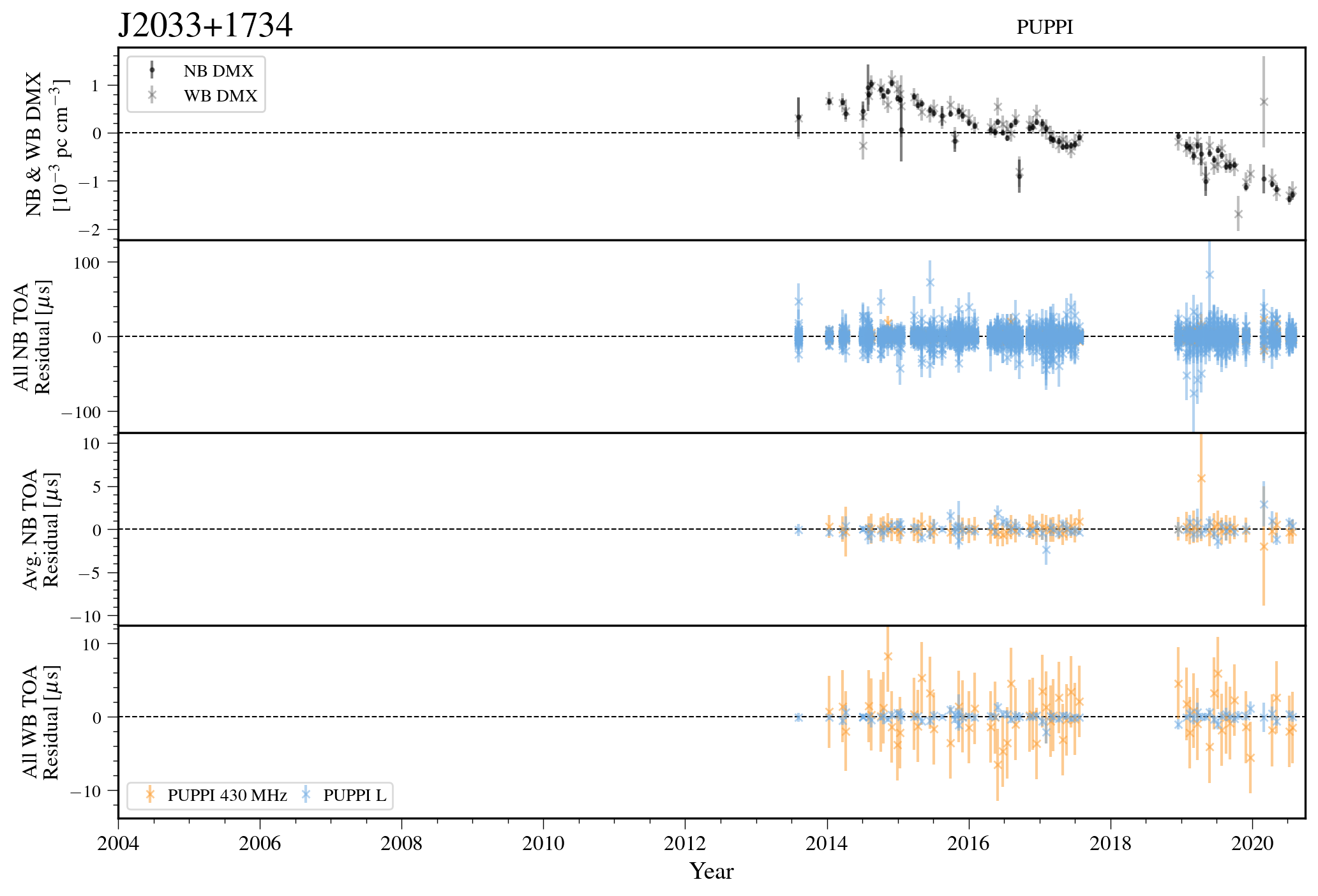}
\caption{Narrowband and wideband timing residuals and DMX timeseries for J2033+1734. See Figure~\ref{fig:summary-J0023+0923} for details.}
\label{fig:summary-J2033+1734}
\end{figure}

\begin{figure}
\centering
\includegraphics[width=0.85\linewidth]{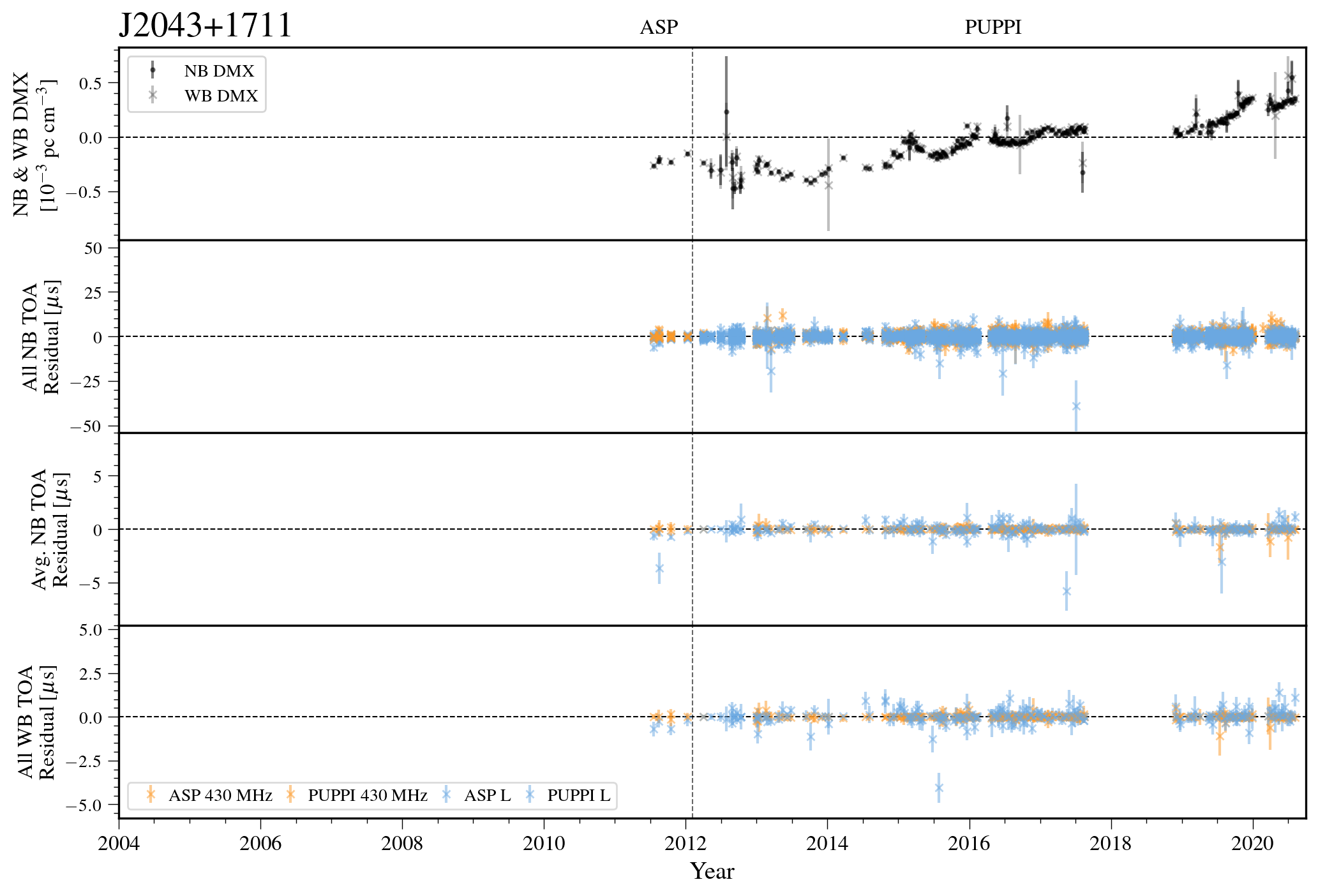}
\caption{Narrowband and wideband timing residuals and DMX timeseries for J2043+1711. See Figure~\ref{fig:summary-J0023+0923} for details.}
\label{fig:summary-J2043+1711}
\end{figure}
\clearpage

\begin{figure}
\centering
\includegraphics[width=0.85\linewidth]{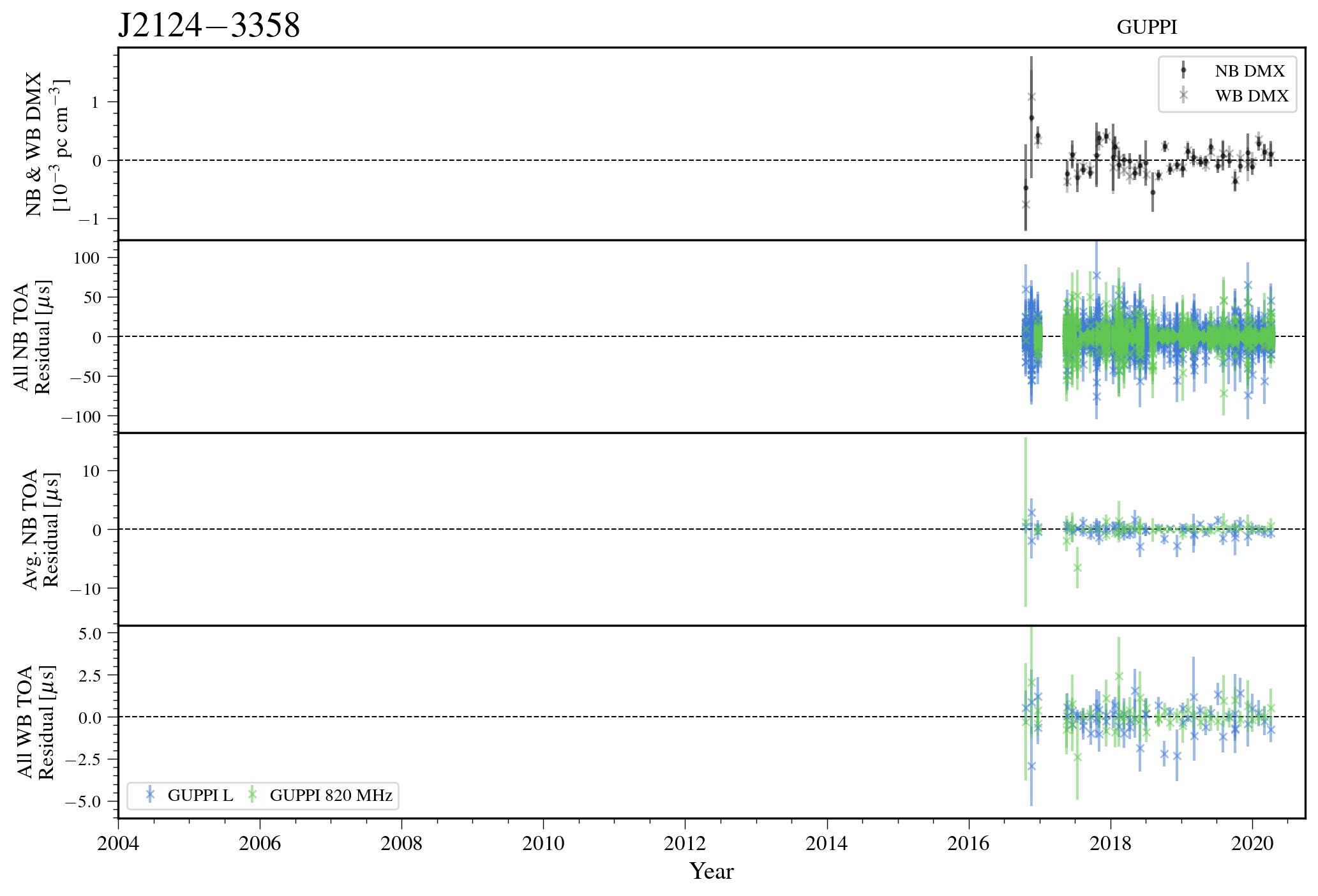}
\caption{Narrowband and wideband timing residuals and DMX timeseries for J2124-3358. See Figure~\ref{fig:summary-J0023+0923} for details.}
\label{fig:summary-J2124-3358}
\end{figure}
\clearpage

\begin{figure}
\centering
\includegraphics[width=0.85\linewidth]{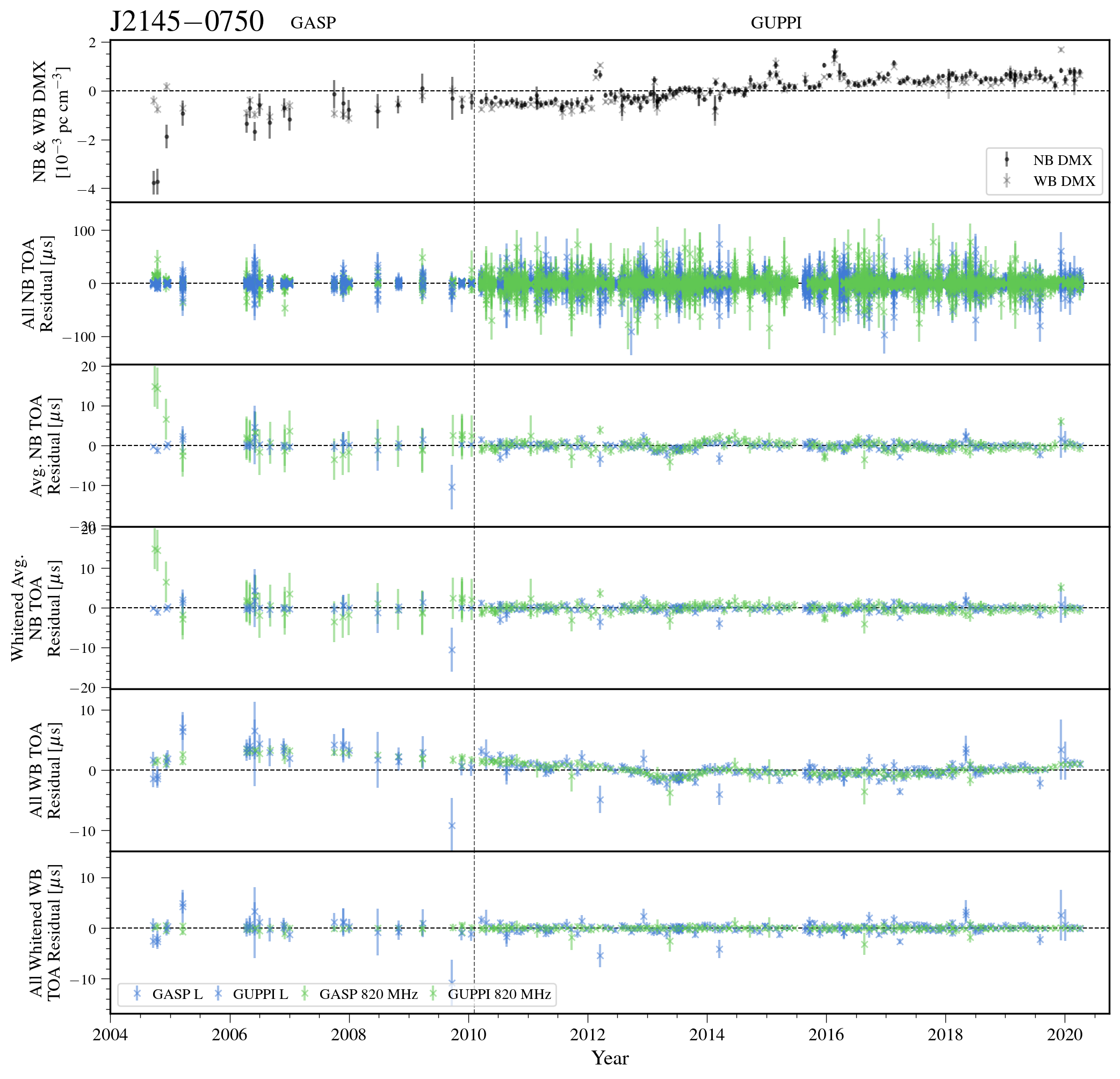}
\caption{Narrowband and wideband timing residuals and DMX timeseries for J2145-0750. See Figure~\ref{fig:summary-J0030+0451} for details.}
\label{fig:summary-J2145-0750}
\end{figure}
\clearpage

\begin{figure}
\centering
\includegraphics[width=0.85\linewidth]{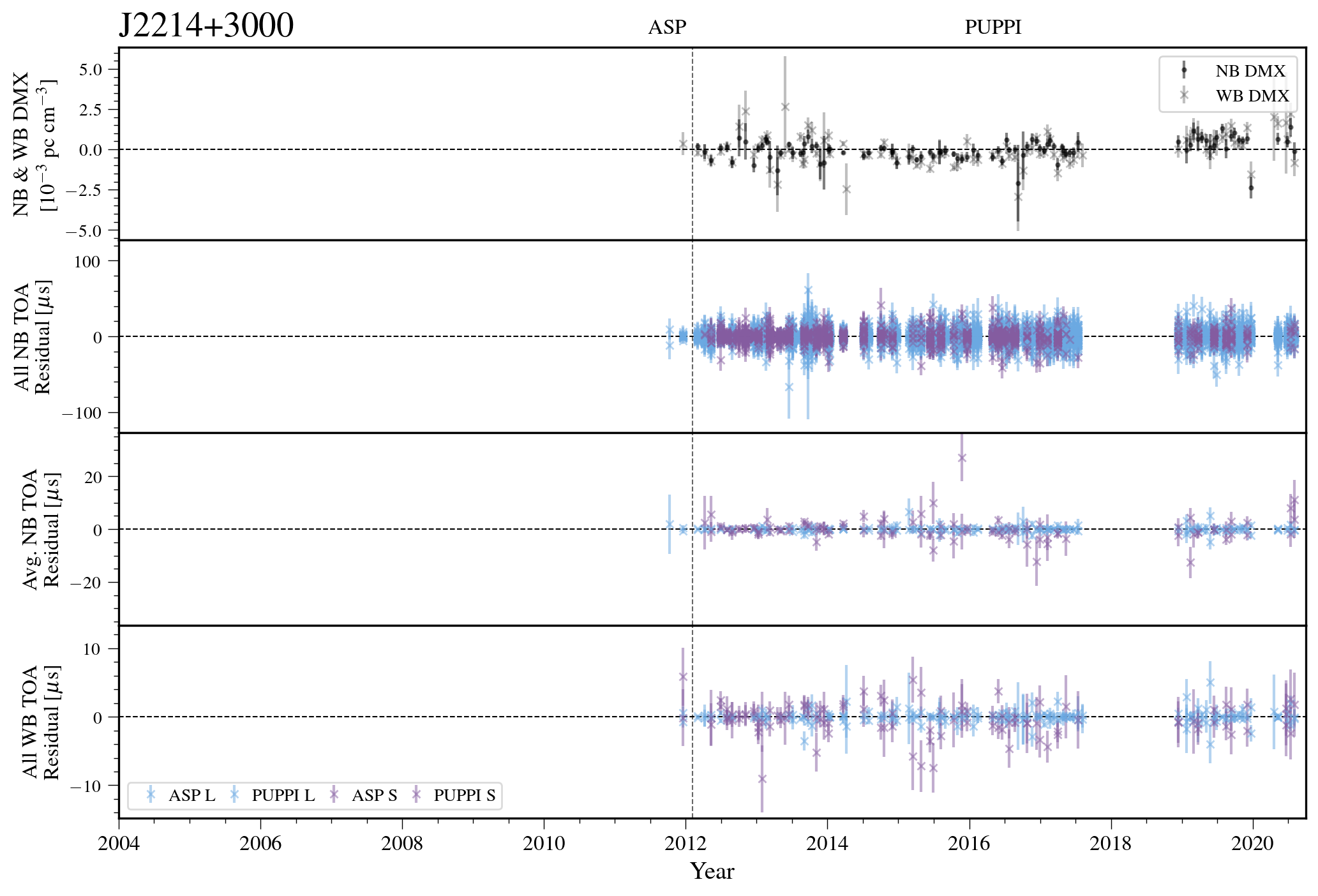}
\caption{Narrowband and wideband timing residuals and DMX timeseries for J2214+3000. See Figure~\ref{fig:summary-J0023+0923} for details.}
\label{fig:summary-J2214+3000}
\end{figure}

\begin{figure}
\centering
\includegraphics[width=0.85\linewidth]{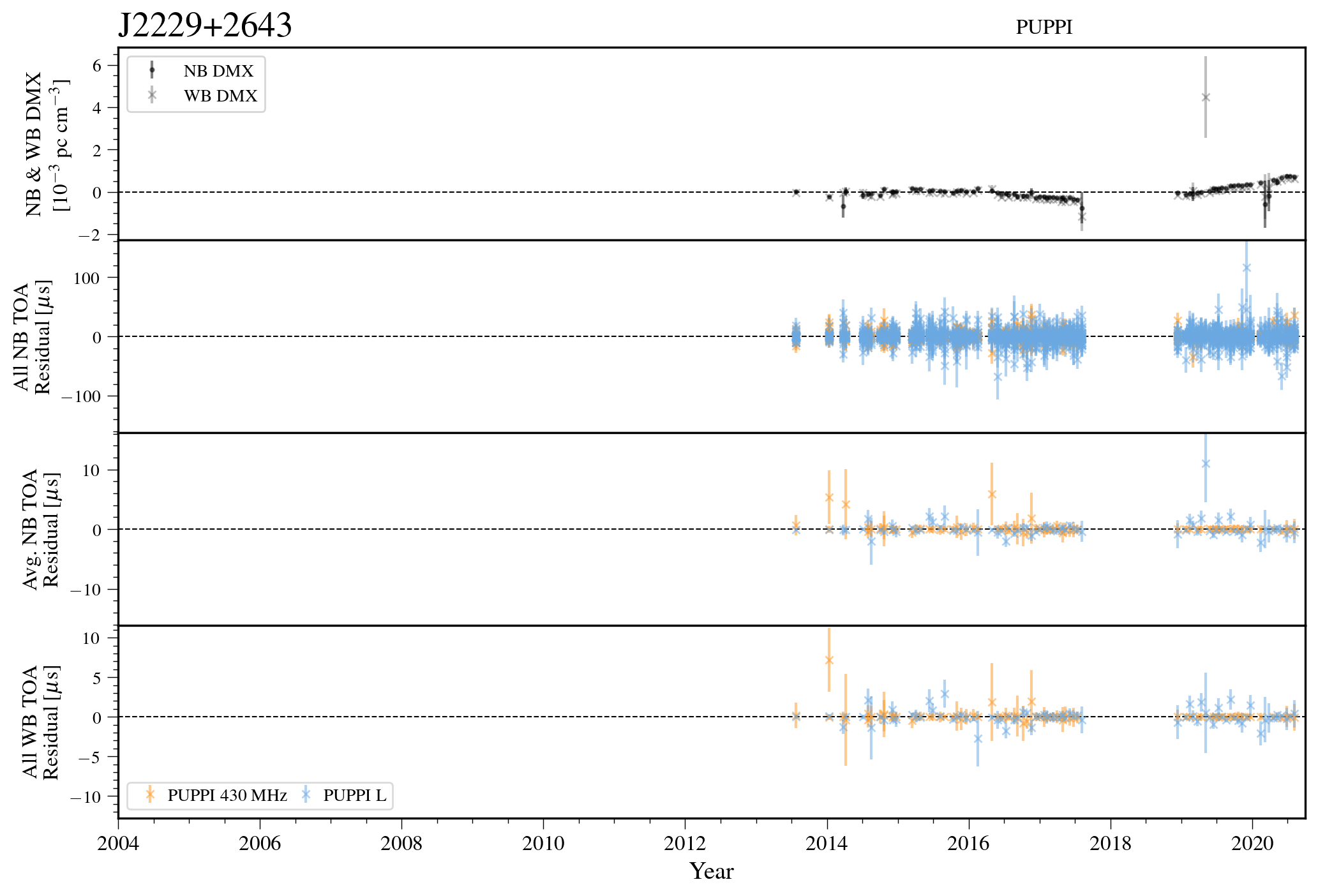}
\caption{Narrowband and wideband timing residuals and DMX timeseries for J2229+2643. See Figure~\ref{fig:summary-J0023+0923} for details.}
\label{fig:summary-J2229+2643}
\end{figure}
\clearpage

\begin{figure}
\centering
\includegraphics[width=0.85\linewidth]{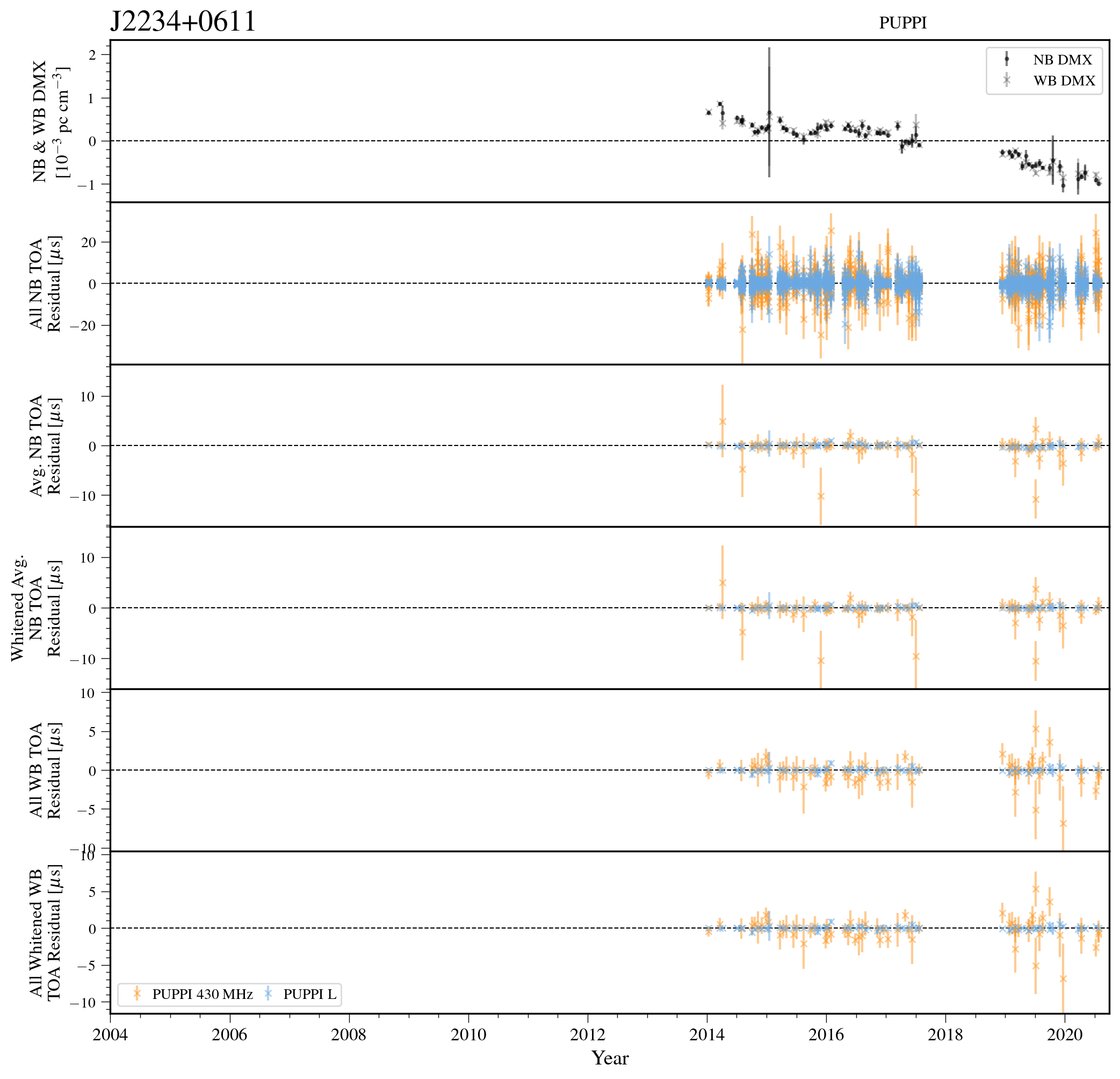}
\caption{Narrowband and wideband timing residuals and DMX timeseries for J2234+0611. See Figure~\ref{fig:summary-J0030+0451} for details.}
\label{fig:summary-J2234+0611}
\end{figure}
\clearpage

\begin{figure}
\centering
\includegraphics[width=0.85\linewidth]{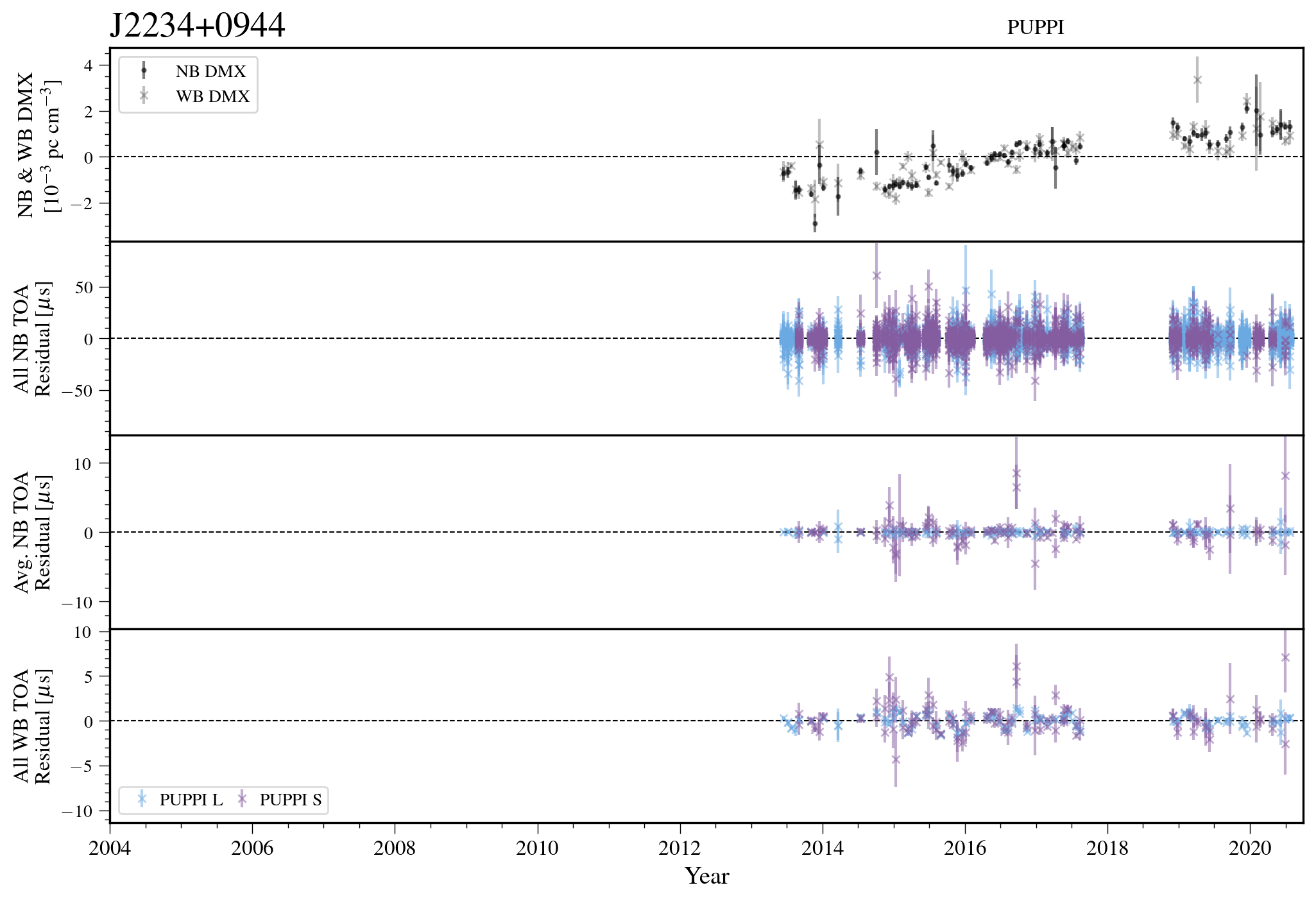}
\caption{Narrowband and wideband timing residuals and DMX timeseries for J2234+0944. See Figure~\ref{fig:summary-J0023+0923} for details.}
\label{fig:summary-J2234+0944}
\end{figure}

\begin{figure}
\centering
\includegraphics[width=0.85\linewidth]{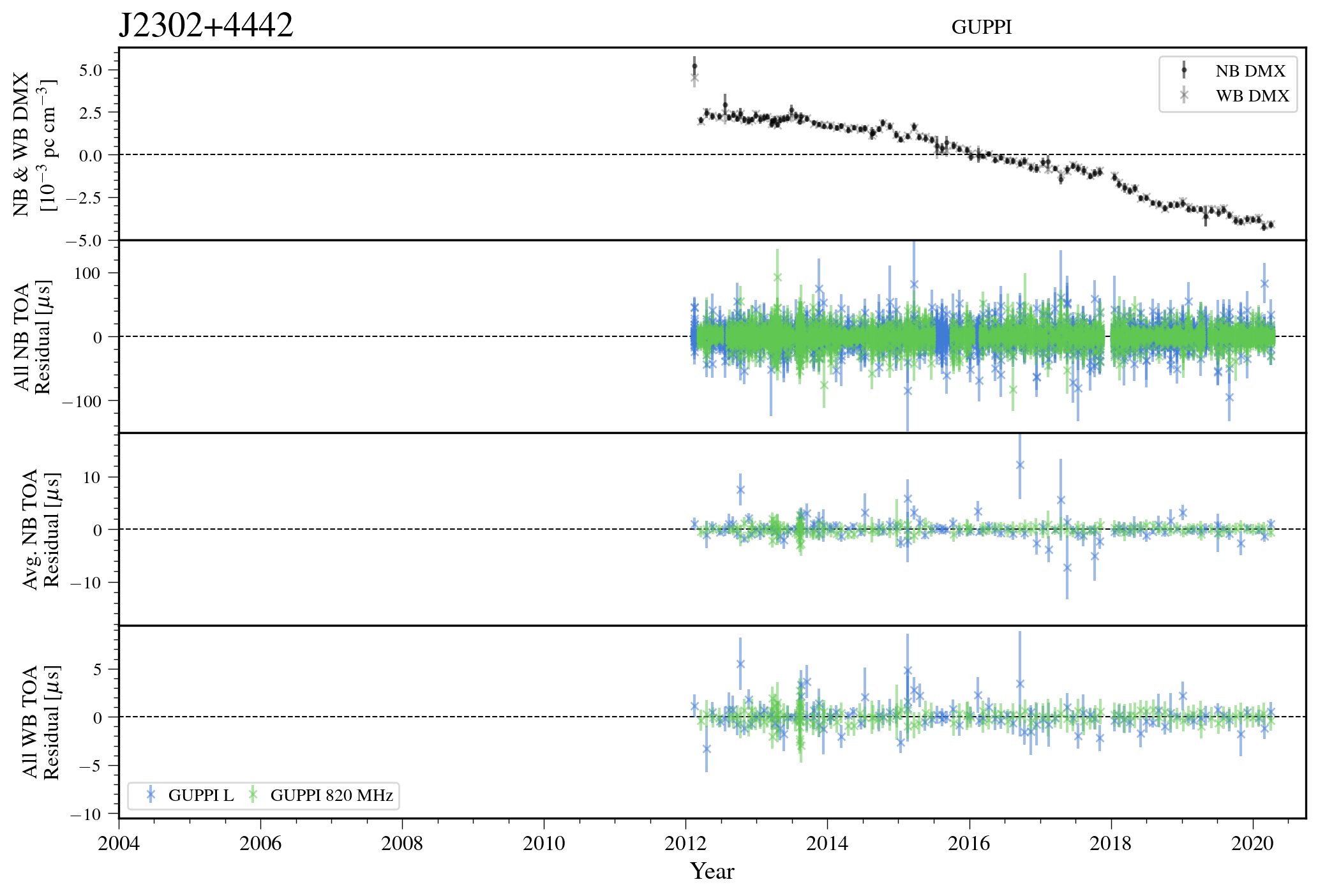}
\caption{Narrowband and wideband timing residuals and DMX timeseries for J2302+4442. See Figure~\ref{fig:summary-J0023+0923} for details.}
\label{fig:summary-J2302+4442}
\end{figure}
\clearpage

\begin{figure}
\centering
\includegraphics[width=0.85\linewidth]{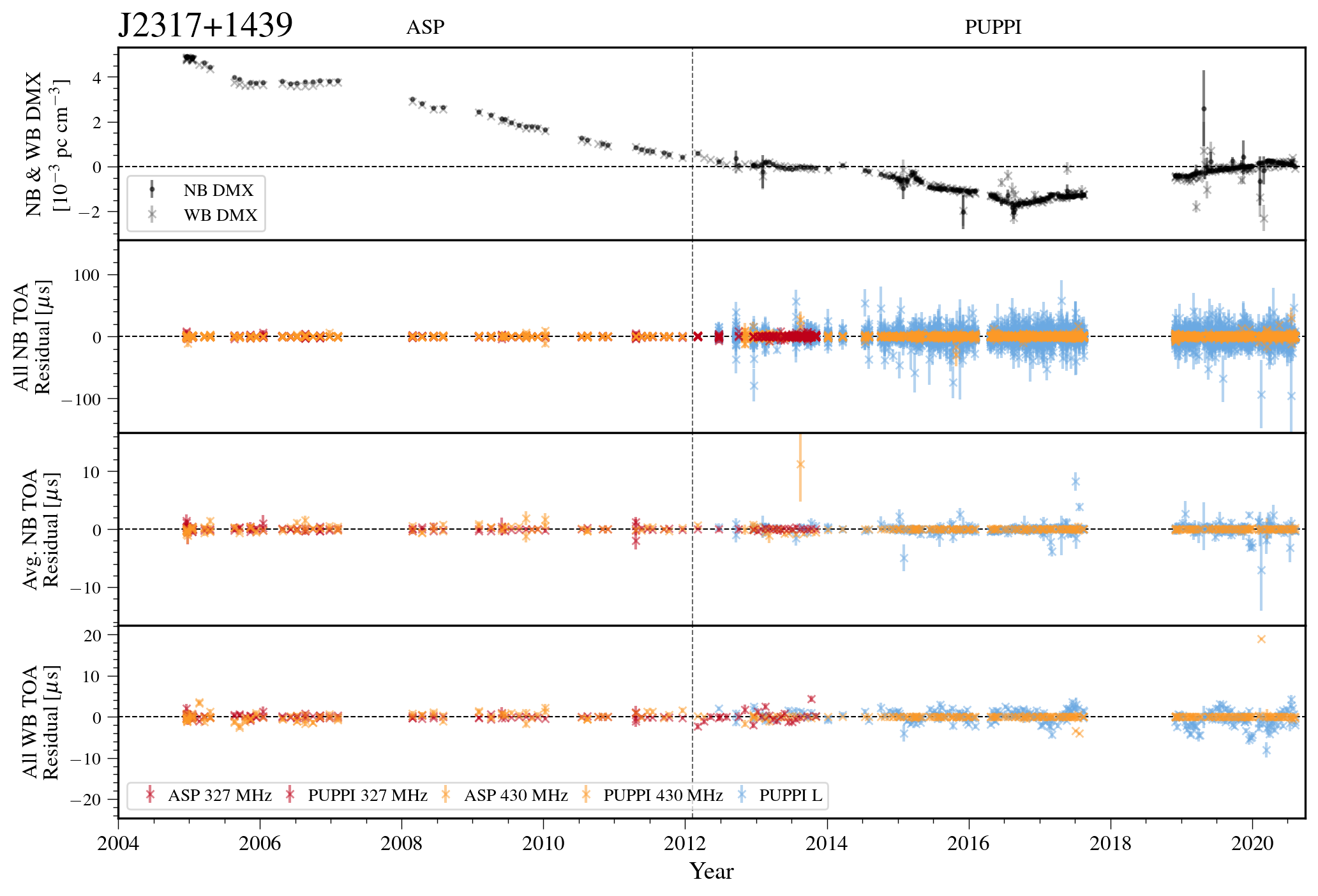}
\caption{Narrowband and wideband timing residuals and DMX timeseries for J2317+1439. See Figure~\ref{fig:summary-J0023+0923} for details.}
\label{fig:summary-J2317+1439}
\end{figure}

\begin{figure}
\centering
\includegraphics[width=0.85\linewidth]{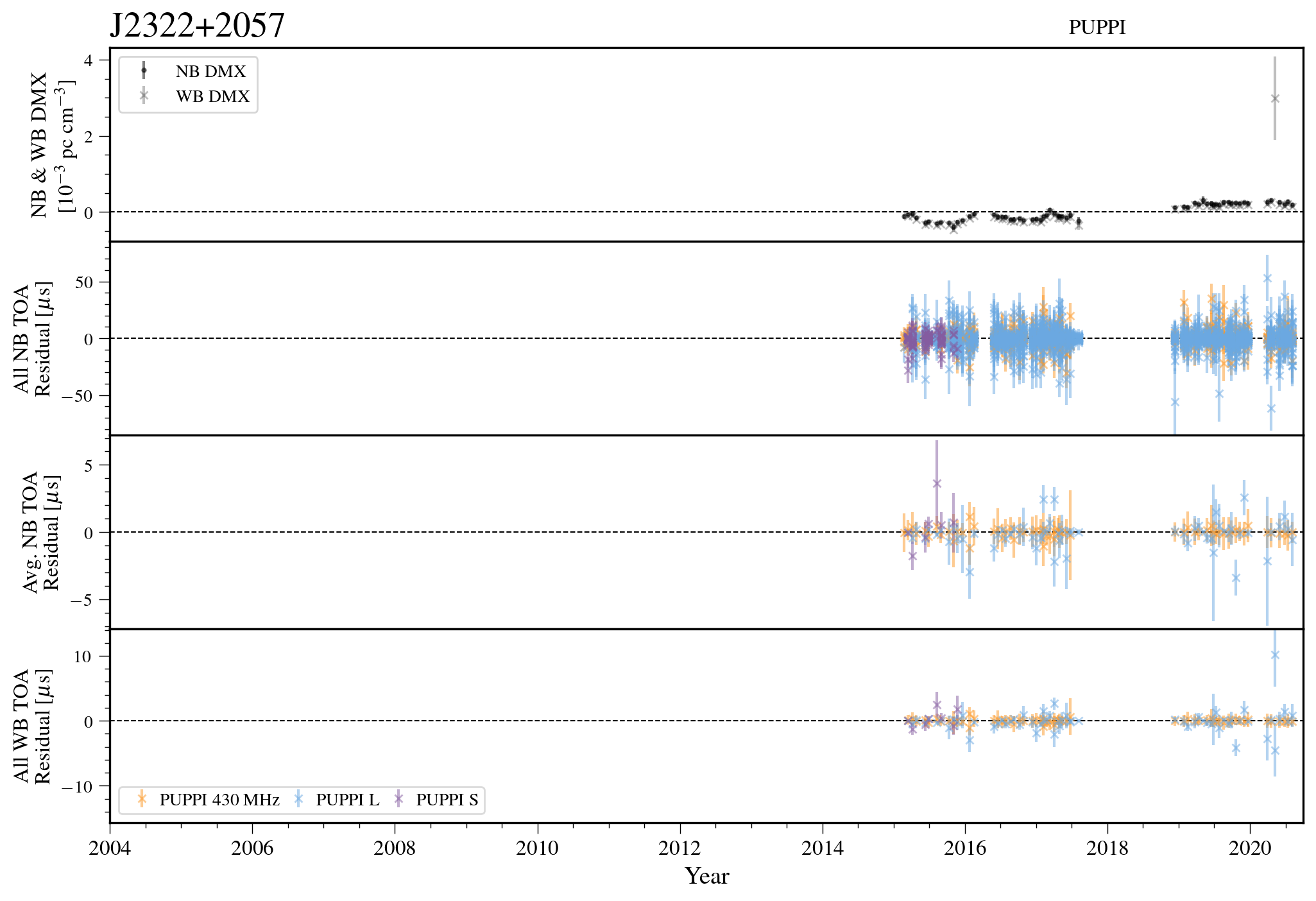}
\caption{Narrowband and wideband timing residuals and DMX timeseries for J2322+2057. See Figure~\ref{fig:summary-J0023+0923} for details.}
\label{fig:summary-J2322+2057}
\end{figure}
\clearpage

\end{document}